\documentclass[%
 reprint,
superscriptaddress,
nofootinbib,
 amsmath,amssymb,
 aps,
 prd,
floatfix,
]{revtex4-2}

\usepackage{color}
\usepackage{amsmath}
\allowdisplaybreaks

\usepackage{amssymb}
\usepackage{aas_macros}
\usepackage{graphicx}%
\usepackage{dcolumn}%
\usepackage{bm}%
\usepackage{hyperref}%
\hypersetup{pdfborder={0 0 0}}
\usepackage{cleveref}
\usepackage{pbox}
\usepackage[T1]{fontenc}
\usepackage[utf8]{inputenc}
\usepackage[]{times}
\usepackage{soul} %
\usepackage[normalem]{ulem}

\definecolor{orange}{rgb}{1.0, 0.5, 0.0}
\definecolor{pink}{rgb}{1.0, 0.0, 0.5}

\setstcolor{blue}

\newcommand{\cc}[1]{\textcolor{red}{#1}}

\definecolor{twofluidaudit}{rgb}{0.0,0.50,0.0}

\definecolor{inversionaudit}{rgb}{0.55, 0.0, 0.55}

\definecolor{darkblue}{rgb}{0.0, 0.0, 0.55}
\newcommand{\imp}[1]{\textcolor{darkblue}{#1}}
\definecolor{auditedit}{rgb}{0.0,0.42,0.42}
\definecolor{auditcomment}{rgb}{0.58,0.0,0.42}

\DeclareRobustCommand{\auditDelete}[1]{{\color{auditedit}\sout{#1}}}
\DeclareRobustCommand{\auditDeleteMath}[1]{\text{\color{auditedit}\sout{\ensuremath{\displaystyle #1}}}}

\newcommand{\an}[2]{\textcolor{auditcomment}{\textbf{[Audit #1:} #2\textbf{]}}}

\newcommand{\ju}{j}
\newcommand{\jn}{j_{\mathrm{n}}}

\newcommand{\dPiPi}{\delta_{\Pi\Pi}}
\newcommand{\dpipi}{\delta_{\pi\pi}}
\newcommand{\dJB}{\delta_{\textsc{jb}}}
\newcommand{\dqB}{\delta_{\textrm{qB}}}
\newcommand{\dpiB}{\delta_{\pi \textrm{B}}}
\newcommand{\dqq}{\delta_{\textrm{qq}}}
\newcommand{\dtI}{\Delta t_{\mathrm{I}}}

\newcommand{\rhob}{\rho_{\mathrm{b}}}
\newcommand{\Db}{D_{\mathrm{b}}}
\newcommand{\rhoq}{\rho_{\mathrm{q}}}
\newcommand{\Dq}{D_{\mathrm{q}}}
\newcommand{\tauq}{\tau_{\mathrm{q}}}
\newcommand{\kappaq}{\kappa_{\mathrm{q}}}
\newcommand{\tauJ}{\tau_{\textsc{j}}}
\newcommand{\omegaJ}{\omega_{\textsc{j}}}
\newcommand{\omegaq}{\omega_{\textrm{q}}}
\newcommand{\cfl}{C_{\textsc{cfl}}}

\newcommand{\T}{\top} %

\newcommand{\myeqref}[1]{Eq.~\eqref{#1}}
\newcommand{\figref}[1]{Fig.~\ref{#1}}

\newcommand{\starF}{^{\ast} \! F}
\newcommand{\levciv}{\varepsilon}
\newcommand{\bs}{\boldsymbol}
\newcommand{\mc}{\mathcal}
\newcommand{\p}{\partial}
\newcommand{\akk}{\tilde{a}_{kk}}

\graphicspath{{figs/}}

\begin{document}

\interfootnotelinepenalty=10000
\preprint{APS/123-QED}

\title{Nonlinearly Causal General-Relativistic Two-Fluid Dissipative Magnetohydrodynamics}

\author{Elias R. Most}
\email{emost@caltech.edu}
\affiliation{TAPIR, Mailcode 350-17, California Institute of Technology,
Pasadena, CA 91125, USA}
\affiliation{Walter Burke Institute for Theoretical Physics,
California Institute of Technology, Pasadena, CA 91125, USA}

\author{Samuel J. Dunham}
\affiliation{TAPIR, Mailcode 350-17, California Institute of Technology,
Pasadena, CA 91125, USA}

\begin{abstract}
  {We present a formulation} of general-relativistic (GR) {19-moment} dissipative magnetohydrodynamics (MHD), capable of handling all first-order dissipative terms (viscosity, heat conductivity, resistivity and Hall terms), as well as all ideal electron degrees of freedom (number density, momentum and energy) in a GR astrophysical two-fluid plasma. {We derive necessary and sufficient nonlinear causality conditions for the constrained first-order 19-moment system, as well as both necessary and sufficient conditions for strong hyperbolicity.}
 To assess the formulation numerically, we develop a high-resolution shock-capturing scheme that solves these equations in a performance-portable fashion. {The scheme expresses all evolution equations, including the dissipative and electron sectors, in flux-divergence form, allowing us to model systems with scales separated by orders of magnitude without resolving kinetic scales everywhere on the grid.}
 In addition, we use implicit integration methods to systematically overstep kinetic scales in MHD regions, such as cyclotron and plasma frequencies. To make the scheme as robust as GRMHD codes, {we construct necessary and sufficient conditions for a conserved state to be physically admissible, and based on these construct a new physicality-enforcement scheme.}
 {As an exact validation comparison, we formulate and derive a full solution to dissipative two-fluid Bondi accretion in general relativity.}
  We then validate the equations against a series of results from kinetic particle-in-cell models of black hole accretion and magnetospheric dynamics, demonstrating that our formulation and scheme can correctly capture major features relevant for feedback on global scales of these solutions, {including dimensionless reconnection rates of order $0.1$} and Braginskii-like anisotropic pressures and heat fluxes.
\end{abstract}

\maketitle

\section{Introduction}

Relativistic fluids and plasmas are ubiquitous in astrophysics. For example, in accreting black holes and jets \cite{Davis:2020wea}, neutron star magnetospheres \cite{2022ARA&A..60..495P}, pulsar wind termination shocks \cite{arons2012pulsar}, and neutron star interiors \citep{Andersson:2020phh},
they describe a large variety of astrophysical phenomena.
They can also be used to study nuclear collisions \cite{Shen:2025unr,Romatschke:2017ejr}.

Depending on the system modeled, the requirements for modeling a relativistic
plasma can vary substantially.
The global dynamics of these systems is usually well described using
relativistic hydrodynamics \cite{rezzolla2013relativistic} or, when magnetic fields play a role,
magnetohydrodynamics \cite{Anile1989}.
One challenge in modeling these systems is accounting for nonideal effects arising from kinetic-scale \cite{Sironi:2014jfa} or multispecies physics \cite{Andersson:2020phh}.
For example, heavy-ion collision experiments have demonstrated the need to model non-Maxwellian distribution functions \cite{Heinz:2013th}. Ideal MHD also fails to fully describe reconnection in astrophysical plasmas, especially in the collisionless regime, both in the solar system \cite{gershman2024magnetic} and in high-energy astrophysical environments \cite{Sironi:2014jfa}. Many of the latter are relativistic in the sense that the Alfv\'en speed approaches the speed of light.
 For example, in low-luminosity accretion flows, such as those present in the Event Horizon Telescope sources M87* \cite{EventHorizonTelescope:2019dse} and Sgr A* \cite{EventHorizonTelescope:2022wkp}, it has been argued that wave-particle instabilities \cite{Kunz:2014qha} drive anisotropic pressures and heat fluxes \cite{Kempski:2019tvo} (though it is unclear to what level these leave observable imprints \cite{Foucart:2017axc,Dhruv:2025isy}).
Depending on the context and process modeled, the study of {nonideal} effects in relativistic plasmas is usually split across scales.
Their large-scale evolution is commonly modeled using ideal
general-relativistic magnetohydrodynamics (GRMHD) \cite{Gammie:2003rj,DeVilliers:2002ab,EventHorizonTelescope:2019pcy}. This approximation has been remarkably successful, but it eliminates the microscopic plasma scales and assumes that all charge carriers are locked to a single perfectly conducting
fluid. This approximation breaks down whenever kinetic scale effects become relevant. Important examples include resistive dissipation in current sheets \cite{Zenitani:2007,barkov2014twofluid}, Hall and electron-inertia effects \cite{Zenitani:2009bj,Koide:2009yx,BarkovKomissarov:2016}, field-aligned heat conduction, and anisotropic pressures in weakly collisional accretion flows
\cite{Chandra:2015iza,Foucart:2015cws,Foucart:2017axc}.

This scale-separation problem is particularly acute in relativistic magnetic
reconnection \cite{2025ARA&A..63..127S}. The dilute magnetospheric plasma surrounding a black hole or neutron star is often collisionless, which in this example means that the Coulomb mean free path is much larger than the electron skin depth of the current sheet \cite{Uzdensky:2010ts}. Kinetic approaches, such as particle-in-cell (PIC) simulations, can then
determine the charge carrier dynamics and {nonideal} electric field from first
principles \cite{Sironi:2014jfa,Guo:2015cua}. However, the scale separation between the skin depth or Larmor radius
and the global compact-object scale makes a fully kinetic calculation at
realistic scale separation computationally prohibitive. Fluid descriptions can access the global scales,
but a single-fluid Ohm law must supply the physics that breaks flux freezing \cite{Palenzuela:2008sf}.
With a small uniform scalar resistivity, plasmoid-mediated relativistic MHD
{reconnection proceeds at a characteristic rate of order $0.01c$}  (e.g., \cite{Ripperda:2016sxe}), whereas
{collisionless kinetic calculations robustly find rates of order $0.1c$}
\cite{Sironi:2014jfa,Guo:2015cua}. Capturing such fast reconnection processes requires physics beyond that included in single-fluid MHD \cite{Goodbred:2022,Moran:2025aqb,Ripperda:2026EffectiveResistivity}.

In the context of reconnection, multiple options have been proposed. Using PIC simulations of relativistic pair reconnection, several studies have proposed effective nonuniform-resistivity closures \cite{Selvi:2022uup,Bugli:2024fby,Moran:2025aqb,Ripperda:2026EffectiveResistivity}, which aim to mimic the behavior of a full two-species Ohm's law \cite{2007PhPl...14e6503B}. Depending on the regime modeled, these closures can be physically motivated, e.g., by invoking a charge-starved regime at X-points \cite{Goodbred:2022,Moran:2025aqb,Ripperda:2026EffectiveResistivity}, although such approaches must break down at low Alfv\'en speeds \cite{Robbins:2025ofe}. Most resistive GRMHD simulations have used only simple uniform resistivity prescriptions \cite{Palenzuela:2012my,Dionysopoulou:2012zv,Dionysopoulou:2015tda,Ripperda:2019lsi,Wright:2019blb,Ripperda:2020bpz,Shibata:2021xmo,Mattia:2023klx,Bugli:2024fby,Azizi:2025nkk,Beaudoin:2026kka,Miravet-Tenes:2026gno}.

At the same time, weakly collisional effects have long been studied in non-relativistic plasmas \cite{braginskii1965transport,1956RSPSA.236..112C} (see also Refs. \cite{Ley:2026DoubleAdiabatic,Wierzchucka:2026DoubleAdiabatic} for recent extensions to the relativistic regime).
In this highly anisotropic limit, these can be modeled as anisotropic viscosities \cite{Chandra:2015iza,Chandra:2017auj} (see also Refs. \cite{Chatterjee:2014qsa,Pimentel:2018uwh,Most:2025kqf} for related applications to neutron star matter).
A limited number of numerical simulations with these effects have been performed \cite{Foucart:2015cws,Foucart:2017axc,Dhruv:2025isy}.
However, while idealized, full kinetic models (using idealized scale separations and electron-ion mass ratios) have also become available \cite{Galishnikova:2023}, currently no model exists that is fully calibrated in this regime.

Similarly, while a limited number of relativistic two-fluid approaches have been attempted for improved modeling of relativistic reconnection \cite{Zenitani:2009bj,Zenitani:2009di,barkov2014twofluid,BarkovKomissarov:2016,Gorard:2025grmhd},
there is currently no code or even algorithm available that can model weakly collisional and resistive plasmas together, which would be required to capture both the weakly collisional nature \cite{Galishnikova:2023,Foucart:2017axc,Koide:2020xts} and flaring aspects \cite{Ripperda:2020bpz,Ripperda:2021zpn} for low-luminosity accretion flows, occurring in different parts of the same system.

Multi-fluid dynamics in relativity is broadly modeled using two approaches: Carter's multi-fluid approach \cite{Carter:1987qr}, and multi-moment approaches (see, e.g., \cite{Rocha:2023ilf} for a recent review). Carter's formulation starts from the thermodynamics of a multi-component system and then constructs a coupled multi-component stress-energy tensor \cite{Andersson:2020phh}. This has been very successful in describing relativistic superfluids \cite{Carter:1993aq,Carter:1998rn,Langlois:1997bz}, and it can be shown that this approach can (after proper redefinition of variables) be recast into a full 14-moment system \cite{Gavassino:2019wzd}. Similarly, Refs. \cite{Andersson:2016wnc,Andersson:2016fva} have proposed deriving multi-component plasma systems from this formulation, although they have never been demonstrated to work numerically. It is also possible to derive effective, Newtonian-like limits from this formulation \cite{Andersson:2021kfk}.
On the other hand, most Newtonian plasma-physics modeling has focused on using effective moment methods. These can either be constructed in the gyrotropic limit \cite{1956RSPSA.236..112C,Wierzchucka:2026DoubleAdiabatic,Ley:2026DoubleAdiabatic,HegadeKR:2026azf}, or, more generally, as multi-moment schemes \cite{CercignaniKremer2002}. The latter approach is appealing as it can be systematically constructed as a reduction from the Boltzmann-Vlasov equation \cite{Denicol:2012cn,Denicol:2018rbw,Denicol:2019iyh,Kushwah:2024zgd,Kushwah:2025jsb}, and has been extensively compared in the Newtonian literature \cite{1992PhFlB...4.2052H,2015PhPl...22a2108W}.
The relativistic case comes with multiple pitfalls, though over the past decade most of these have been addressed and resolved through progress driven by the nuclear physics community \cite{Romatschke:2009im,Rocha:2023ilf}.
First, systematic derivations of moment equations from kinetic theory have been developed for hydrodynamics \cite{Denicol:2012cn}, non-resistive \cite{Denicol:2018rbw}, resistive \cite{Denicol:2019iyh}, and two-fluid plasmas \cite{Kushwah:2025jsb}. The minimum set required in relativity contains 14 moments, and the corresponding theories are therefore known as 14-moment formulations. Kinetic-theory reductions can provide effective transport coefficients in different limits \cite{Kushwah:2025jsb}.
Some effects, including the relativistic drift-kink instability \cite{Zenitani:2009bj}, require true two-fluid behavior to enable counter-streaming momenta. Formulations of relativistic two-fluid MHD have also been proposed for ideal \cite{Zenitani:2009bj,Koide:2009yx,barkov2014twofluid} and dissipative plasmas \cite{Most:2021uck}. We follow the latter approach here.
However, these theories come with several caveats that are important to highlight. The most important one is the meaning of causality in these theories. In short, while kinetic theory is naturally causal, truncated moment reductions need not be \cite{Gavassino:2026fil}. Indeed, naive Navier-Stokes-type approaches have long been shown to be acausal \cite{Hiscock:1983zz,Hiscock:1985zz}, and either second-order theories, like M\"uller-Israel-Stewart (MIS) \cite{muller1967paradoxon,Israel:1979wp}, or generalized-frame theories like Bemfica-Disconzi-Noronha-Kovtun (BDNK) \cite{Bemfica:2017wps,Kovtun:2019hdm,Bemfica:2020zjp} (including for magnetic fields \cite{Armas:2022wvb,Lier:2025wfw}) have been proposed. The main insight is that all these theories have their natural regime of validity in the sense of an effective field theory \cite{Figueras:2024bba}, and it is therefore necessary to impose conditions such as causality as explicit limits on the theory. Indeed, for MIS theories these have been computed in some cases in a series of papers \cite{Bemfica:2019cop,Bemfica:2019knx,Bemfica:2020zjp,Cordeiro:2023ljz,Cordeiro:2026} over the past years. However, there is currently no complete set of causality conditions for a general MIS theory with bulk viscosity, shear viscosity, and heat flux combined, let alone for magnetic fields, except in special cases \cite{Cordeiro:2023ljz}. While massively unphysical ion states are likely rare in astrophysical applications (much unlike in heavy-ion collisions \cite{Plumberg:2021bme,Gavassino:2025mcq}), electrons will be relativistic and causality (as we will show) does become an important issue.
Similarly, only for a subset of these theories is formal well-posedness understood, which in addition to causality also requires strong hyperbolicity, which is not necessarily guaranteed (e.g., standard resistive GRMHD has been proven to be only weakly hyperbolic \cite{Schoepe:2018}). Other terms like the relativistic Hall effect remain almost entirely unmodeled \cite{Dumbser:2009hw}.
Electrons have been modeled in an ad hoc way by including their heat conduction to model electron thermodynamics more accurately \cite{Ressler:2015ipa,Sadowski:2016fdh,Chael:2018gzl}, but without including them in a resistive Ohm's law or heating them self-consistently through modeled dissipation.

Moreover, solving these equations stably in astrophysical applications, where scales spanning many orders of magnitude coexist in one numerical domain, is challenging and may require specialized implicit algorithms \cite{Most:2021rhr} or other advances not previously investigated in the literature.

Here, we address these issues by proposing {a two-fluid {19-moment} model for dissipative GRMHD and analyzing the conditions for its nonlinear causality and strong hyperbolicity.} This model includes all dissipative channels (bulk, heat, shear, and resistivity), as well as Hall terms and all ideal electron degrees of freedom (number density, momentum, and energy), which are self-consistently coupled to Ohm's law and the dissipation of the system. {We identify explicit causal, strongly hyperbolic domains of the constrained first-order system and develop a performance-portable numerical algorithm for general-relativistic spacetimes.}

We illustrate that the {19-moment} system is fully able to reproduce all macroscopic features of previously studied {general-relativistic particle-in-cell (GRPIC)} simulations, apart from non-thermal particle acceleration.

{Section~\ref{sec:methods} presents the two-fluid  {19-moment} equations and
their causality and strong-hyperbolicity conditions.  Section~\ref{sec:numerics}
describes the conservative formulation, implicit integration, and primitive
recovery.  Section~\ref{sec:grplasmas} tests the scheme with dissipative Bondi
accretion and black-hole magnetospheres, and Sec.~\ref{sec:conclusions}
summarizes the results.  The appendices give the characteristic analysis,
{numerical details and a dissipative two-fluid Bondi reference solution.}}
We use Heaviside--Lorentz units with $G=c=1$.

\section{Mathematical formulation}\label{sec:methods}

In this work, we adopt the 3+1 formulation of general-relativistic
spacetime \cite{Arnowitt:1962hi}.
We introduce a four-metric, $g_{\mu\nu}$, which we decompose relative to a
normal observer, $n_{\mu} = (-\alpha, 0,0,0)$, such that
\begin{align}
  {\rm d}s^{2} &= g_{\mu\nu} \, {\rm d}x^{\mu} \, {\rm d} x^{\nu} \\
  &= \left( -\alpha^{2} + \beta^{i} \, \beta_{i} \right) {\rm d}t^{2}
  + 2 \, \beta_{j} \, {\rm d}t \, {\rm d}x^{j}
  + \gamma_{ij} \, {\rm d}x^{i} \, {\rm d}x^{j} \,,
\end{align}
where $x^{\mu} := (t, x^i)$ is the coordinate vector of a spacetime point,
$\alpha$ is the lapse function,
$\beta^{i}$ is the shift vector,
and $\gamma_{\mu\nu} := g_{\mu\nu} + n_{\mu} \, n_{\nu}$
is the spatial metric.
In addition, we will find it convenient later on to also define the time
derivative of the metric, relative to a normal observer.
This is naturally facilitated by defining the extrinsic curvature
$K_{\mu\nu}$, such that
\begin{align}
  K_{\mu\nu} = -\frac{1}{2} \, \mathcal{L}_{n} \, \gamma_{\mu\nu} \, .
\end{align}
We will also find it convenient to define a vector of spacetime variables,
$G := \left( \alpha, \beta^{i}, \gamma_{ij}, K_{ij} \right)^{\T}$,
where $\T$ denotes the transpose. {Densitized quantities are formed
throughout with $\sqrt{\gamma}$, where
$\gamma := \mathrm{det}\left(\gamma_{ij}\right)$ is the determinant of the
spatial metric; no conformal factor is introduced, so $\psi$ denotes the
electric divergence-cleaning scalar of Eq.~(\ref{eq.maxwellE}) alone.}

\subsection{Relativistic Maxwell equations}
\label{sec:maxwell}
Here, we briefly review the covariant description of electrodynamics used
in this work following Refs. \cite{Baumgarte:2002vv,Palenzuela:2012my}.\\

This plasma couples to an electromagnetic field,
the evolution of which is governed by Maxwell's equations,
\begin{align}
  \nabla_{\mu} \left( F^{\mu\nu} + g^{\mu\nu} \, \psi \right)
  &= {\kappa_{\rm dc}} \, n^{\nu} \, \psi -\mathcal{J}^{\nu} \, ,
  \label{eq.maxwellE} \\
  \nabla_{\mu} \left( \starF^{\mu\nu} + g^{\mu\nu} \, \phi \right)
  &= {\kappa_{\rm dc}} \, n^{\nu} \, \phi \, ,
\end{align}
where $F^{\mu\nu}$ is the Faraday tensor,
\begin{equation}
  \starF^{\mu\nu}
  := \frac{1}{2} \, \levciv^{\mu\nu\alpha\beta} \, F_{\alpha\beta}
  \label{eqn:dual}
\end{equation}
is the dual to $F^{\mu\nu}$,
with $\levciv_{\mu\nu\alpha\beta} := -\sqrt{-g} \, \eta_{\mu\nu\alpha\beta}$,
which implies
$\levciv^{\mu\nu\alpha\beta} = 1 / \sqrt{-g} \, \eta^{\mu\nu\alpha\beta}$,
with $\eta^{\mu\nu\alpha\beta} = \eta_{\mu\nu\alpha\beta}$,
where $\eta_{\mu\nu\alpha\beta}$ is the completely antisymmetric symbol,
with $\eta_{0123} = 1$,
$\mathcal{J}^\mu$ the electric current density four-vector,
and $g := \mathrm{det}\left(g_{\mu\nu}\right)$.
In order to more easily handle the monopole and charge constraints present in
the system, the set of equations has been augmented to include hyperbolic
divergence cleaning scalars \cite{dedner2002hyperbolic} in covariant form
\cite{Komissarov:2007wk,Palenzuela:2008sf}.
Here $\kappa_{\rm dc}$ denotes the constant damping rate of the
divergence-cleaning scalars $\psi$ and $\phi$, which are transported at the speed of
light in the normal frame.
The Faraday tensor and its dual can be expressed as
\begin{align}
       F^{\mu\nu} &= 2 \, u^{\left[\mu\right.} e^{\left.\nu\right]}
  + \sqrt{b^{2}} \, b^{\mu\nu} \, , \\
  \starF^{\mu\nu} &= 2 \, u^{\left[\mu\right.} b^{\left.\nu\right]}
  - \sqrt{e^{2}} \, e^{\mu\nu} \, ,
\end{align}
where $A^{\left[\mu\nu\right]} := \left( A^{\mu\nu} - A^{\nu\mu} \right) / 2$,
{$A^{\left(\mu\nu\right)} := \left( A^{\mu\nu} + A^{\nu\mu} \right) / 2$,
and $\Delta^{\mu\nu} := g^{\mu\nu} + u^{\mu} \, u^{\nu}$ projects
orthogonally to the fluid four-velocity,}
and where we have introduced the notation
\begin{align}\label{eqn:b_projector}
  b^{\mu\nu}
  &:= \frac{1}{\sqrt{b^{2}}} \,
  \levciv^{\mu\nu\alpha\beta} \, b_{\alpha} \, u_{\beta} \, , \\
  e^{\mu\nu}
  &:= \frac{1}{\sqrt{e^{2}}} \,
  \levciv^{\mu\nu\alpha\beta} \, e_{\alpha} \, u_{\beta} \, ,
\end{align}
with $e^{2} := e_{\mu} \, e^{\mu}$ and $b^{2} := b_{\mu} \, b^{\mu}$,
and where we define the electric ($e^{\mu}$) and magnetic ($b^{\mu}$) fields
as seen by an observer comoving with the fluid as
\begin{align}
  e^{\mu} &:=         F^{\mu\nu} \, u_{\nu} \, , \\
  b^{\mu} &:= \, \starF^{\mu\nu} \, u_{\nu} \, .
\end{align}
Similarly, we can {express the Faraday tensor
and its dual in terms of the normal observer,}
\begin{align}
       F^{\mu\nu} &= 2 \, n^{\left[\mu\right.} E^{\left.\nu\right]}
  + \sqrt{B^{2}} \, B^{\mu\nu} \, , \label{eq.Fmunun}\\
  \starF^{\mu\nu} &= 2 \, n^{\left[\mu\right.} B^{\left.\nu\right]}
  - \sqrt{E^{2}} \, E^{\mu\nu} \, , \label{eq.sFmunun}
\end{align}
where we defined the electric ($E^{\mu}$) and magnetic ($B^{\mu}$) fields
as seen by a normal observer as
\begin{align}
  E^{\mu} &:=         F^{\mu\nu} \, n_{\nu} \, , \\
  B^{\mu} &:= \, \starF^{\mu\nu} \, n_{\nu} \, ,
\end{align}
and
\begin{align}
  B^{\mu\nu}
  &:= \frac{1}{\sqrt{B^{2}}} \,
  \levciv^{\mu\nu\alpha\beta} \, B_{\alpha} \, n_{\beta}
  = \frac{1}{\sqrt{B^{2}}} \,
  \levciv^{\mu\nu\alpha} \, B_{\alpha} \, , \\
  E^{\mu\nu}
  &:= \frac{1}{\sqrt{E^{2}}} \,
  \levciv^{\mu\nu\alpha\beta} \, E_{\alpha} \, n_{\beta}
  = \frac{1}{\sqrt{E^{2}}} \,
  \levciv^{\mu\nu\alpha} \, E_{\alpha} \, ,
\end{align}
with $E^{2} := E_{\mu} \, E^{\mu}$ and $B^{2} := B_{\mu} \, B^{\mu}$, and
\begin{align}
  \levciv^{\mu\nu\alpha} := \levciv^{\mu\nu\alpha\beta} \, n_{\beta}
  = \frac{1}{\sqrt{\gamma}} \, \eta^{0\mu\nu\alpha} \, .
\end{align}
The energy-momentum tensor of the electromagnetic field can then be written as
\begin{align}
  T^{\mu\nu}_{\rm EM}
  &:= F^{\mu\alpha} \, F^{\nu}_{\phantom{\nu}\alpha}
  - \frac{1}{4} \, g^{\mu\nu} \, F^{\alpha\beta} \, F_{\alpha\beta}
  \label{eq.TEM} \\
  &\phantom{:}=
  \frac{1}{2} \left( u^{\mu} \, u^{\nu} + \Delta^{\mu\nu} \right)
  \left( e^2 + b^2 \right) \nonumber \\
  &\phantom{:==}
  + 2 \, \sqrt{b^{2}} \,
  u^{\left(\mu\right.} b^{\left.\nu\right)\kappa} \, e_{\kappa}
  - \left( e^{\mu} \, e^{\nu} + b^{\mu} \, b^{\nu} \right) \, \\
  &\phantom{:}= \frac{1}{2} \left( n^{\mu} \, n^{\nu}
  + \gamma^{\mu\nu} \right)
  \left( E^2 + B^2 \right) \, \nonumber \\
  &\phantom{:==}
  + 2 \, \sqrt{B^{2}} \, n^{\left(\mu\right.} B^{\left.\nu\right)\kappa} \,
  E_{\kappa}
  - \left( E^{\mu} \, E^{\nu} + B^{\mu} \, B^{\nu} \right) \, .
\end{align}
The fluid- and normal-frame fields are related by
contracting Eqs.~\eqref{eq.Fmunun} and \eqref{eq.sFmunun} with $u_{\nu}$,
leading to
\begin{align}
  e^{\mu} &= W \, n^{\mu} \left( E^{i} \, v_{i} \right) + W \, E^{\mu}
  + W \, \levciv^{\mu jk} \, v_{j} \, B_{k} \, ,
  \label{eq.e_comov} \\
  b^{\mu} &= W \, n^{\mu} \left( B^{i} \, v_{i} \right) + W \, B^{\mu}
  - W \, \levciv^{\mu jk} \, v_{j} \, E_{k} \, ,
  \label{eq.b_comov}
\end{align}
where $W := -n_{\mu} \, u^{\mu}$ is the Lorentz factor of the fluid
and $v^{\mu}$ is the fluid velocity, both as seen by a normal observer.

\subsection{Relativistic two-fluid theory}

Microscopically, a {two-fluid} plasma consists of two charge carriers with
charges of opposite sign,
which we will refer to as {\it electrons} and {\it ions}
without loss of generality regarding the charge ratio $Z= q_i/q_e$,
or mass ratio $m_e/m_i$.

One of the crucial differences between a Newtonian and a relativistic
dissipative approach -- as we will show -- is that one formally needs to retain
dynamical degrees of freedom of the electrons.
Taking Newtonian-like relativistic MHD reductions \cite{Palenzuela:2008sf,Bucciantini:2012sm,Ripperda:2019lsi} has been shown to lead to a loss
of mathematical well-posedness \cite{Schoepe:2018} (see also Refs. \cite{Lier:2025wfw,Lier:2026owd} for alternative approaches using BDNK theory \cite{Armas:2022wvb}).\\

We here briefly review the two-fluid approach as laid out by Ref. \cite{Most:2021uck},
before describing the particular dissipative MHD limit we will take of these equations.

\subsubsection{Dissipative 14-moment hydrodynamics}

Our plasma consists of two components, denoted by $X \in \left\{e, i\right\}$.
Before discussing the interplay of {the components in} this (two-)fluid
theory, we will describe each component individually using a 14-moment
description.
We start by introducing, for component $X$, an energy--momentum tensor
$T_{X}^{\mu\nu}$ and number current $N_{X}^\mu$.
Defining the rest frame of component $X$ as $u_{X}^{\mu}$,
we can then decompose these components according to
\begin{align}
  T_{X}^{\mu\nu}
  :=& \mathcal{E}_{X} \, u_{X}^{\mu} \, u_{X}^{\nu}
  + \mathcal{P}_{X} \, \Delta_{X}^{\mu\nu}
  + q_{X}^{\mu} \, u_{X}^{\nu} + q_{X}^{\nu} \, u_{X}^{\mu}
  + \pi_{X}^{\mu\nu} \, ,
  \label{eq:species-stress-decomposition}
\end{align}
where we have introduced the frame projector
$\Delta_{X}^{\mu\nu}= g^{\mu\nu} + u_{X}^{\mu} \, u_{X}^{\nu}$,
as well as the fluid energy density $\mathcal{E}_{X}$, heat fluxes $q_{X}^{\mu}$,
isotropic and (symmetric) anisotropic pressures $\mathcal{P}_{X}$ and
$\pi_{X}^{\mu\nu}$, respectively.
Using the orthogonality relations $q_{X} \cdot u_{X} = 0$,
$\pi_{X} \cdot u_{X} = 0$, as well as $\mathrm{Tr}\left(\pi_{X}\right) = 0$,
this decomposition is unique for {any symmetric stress-energy tensor $T_{X}^{\mu\nu}$ at a specified $u_X^\mu$}.
Similarly, we may write
\begin{align}
  N_{X}^{\mu} = n_{X} \, u_{X}^{\mu} + V_{X}^{\mu} \, , \label{eq.NX}
\end{align}
where $n_{X}$ is the particle number density and
$V_{X}^{\mu}$ is the particle diffusion current,
both of which are uniquely specified because $V_{X} \cdot u_{X} = 0$.
In total, the system comprises $14$ independent degrees of freedom for
each component $X$.
{A symmetric stress tensor has ten independent components, and a number current has four. If the three independent components of $u_X^\mu$ are also unknown, a hydrodynamic frame condition removes the redundancy.}
It is important to realize that the actual equations of motion governing
($N_{X}^{\mu}$, $T_{X}^{\mu\nu}$) {require} an effective closure that is system-specific.
Since from a kinetic theory perspective
{$N_{X}^{\mu}$ and $T_{X}^{\mu\nu}$} are computed as moments of the
{distribution function \cite{Israel:1979wp,Denicol:2012cn,Denicol:2018rbw},} these are usually called $14$ moments.\\

Usually, these descriptions assume the existence of an equilibrium thermodynamic
state, which imposes
\begin{align}
  \mathcal{E}_{X} = e_{X} + \delta e_{X} \, , \\
  \mathcal{P}_{X} = P_{X} + \Pi_{X} \, ,
\end{align}
where now $\Pi_{X}$ and $\delta e_{X}$ are the out-of-equilibrium
dynamical pressure and energy corrections,
and $P_{X}$ and $e_{X}$ are the equilibrium pressure
and energy density, respectively.
The latter are fixed via an equation of state {(EOS)},
$P_{X} = P_{X}(e_{X}, n_{X}, \ldots)$,
which also defines a temperature $T_{X} = T_{X}(e_{X}, n_{X}, \ldots)$.
We caution that because of the choice of the hydrodynamic frame above,
the definition of a temperature is fluid-frame dependent,
and the related choice has implications for causality of the resulting effective
system \cite{Bemfica:2017wps,Kovtun:2019hdm,Noronha:2021syv}.
This choice also introduces an equilibrium stress-energy tensor,
\begin{align}
  \bar{T}_{X}^{\mu\nu}
  := e_{X} \, u_{X}^{\mu} \, u_{X}^{\nu}
  + {P}_{X} \, \Delta_{X}^{\mu\nu} \, .
\end{align}

{We choose an Eckart frame for each species, $V_X^\mu=0$, and impose energy matching, $\delta e_X=0$.}
Following Ref. \cite{Denicol:2018rbw}, in the collisional limit, the leading-order terms take the following form:
\begin{align}
  &\tau_{\Pi X} \, u_{X}^{\mu} \, \nabla_{\mu} \, \Pi_{X}
    = -\zeta_{X} \, \nabla_{\mu} \, u_{X}^{\mu} - \Pi_{X} \, ,
    \label{eq.comovPi}\\
  &\tau_{\pi X} \, u_{X}^{\mu} \, \nabla_{\mu} \, \pi_{X}^{<\alpha\beta>}
    = -2\eta_{X} \, \sigma_{X}^{<\alpha\beta>}
    - {\tau_{\pi X}\omega_{gX}} \, b_{X \, \gamma}^{\left<\alpha\right.}
    \pi_{X}^{\left.\beta\right>\gamma} \notag \\
    &\hspace{10em} - \pi_{X}^{\alpha\beta} \, ,
    \label{eq.comovpi} \\
  &\tau_{q X} \, u_{X}^{\mu} \, \nabla_{\mu} \, q_{X}^{<\alpha>}
    = {-\kappa_{X} \left( \nabla^{<\alpha>} \, T_{X}
    + T_{X} \, a_{X}^{\alpha} \right)}
    - q_{X}^{\alpha} \notag \\
    &\hspace{10em}+ {\tau_{qX}\omega_{gX}} \, b^{\alpha}_{X \nu} \, q_{X}^{\nu} \, ,
    \label{eq.comovq}
\end{align}
where $\tau_{\Pi X}$, $\tau_{\pi X}$, $\tau_{q X}$ are effective (collisional)
relaxation times, $\zeta_{X}$ is the bulk viscosity, $\eta_{X}$ is the shear viscosity,
$\omega_{gX}$ is the effective cyclotron frequency,
and $\kappa_{X}$ is the heat conductivity.

We have also introduced
$a_{X}^{\alpha} := u_{X}^{\mu} \, \nabla_{\mu} \, u_{X}^{\alpha}$
as the fluid four-acceleration,
$\sigma_{X}^{\alpha\beta}
:=\Delta^{\alpha}_{X\mu}\Delta^{\beta}_{X\nu}
\nabla^{\left(\mu\right.}u_X^{\left.\nu\right)}
- (1/3) \Delta^{\alpha\beta}_{X} \, \nabla_{\gamma} \, u^{\gamma}_{X}$
as the fluid shear tensor.  Parentheses carry weight $1/2$, so this definition
is trace-free and orthogonal to $u_X^\mu$, and the {Navier--Stokes (NS)} limit is
$\pi_{\rm NS}^{\mu\nu}=-2\eta\sigma^{\mu\nu}$.
We also use the angle-bracket notation
{$q_X^{\left<\alpha\right>} := \Delta_X^\alpha{}_{\mu} \, q_X^{\mu}$}.
Also, {$\pi_X^{\left<\alpha\beta\right>}
:= \Delta_X^{\alpha\beta}{}_{\mu\nu} \, \pi_X^{\mu\nu}$},
where
{$\Delta_X^{\alpha\beta}{}_{\mu\nu}
:= \frac{1}{2}\left(
\Delta_X^\alpha{}_{\mu} \, \Delta_X^\beta{}_{\nu}
+ \Delta_X^\alpha{}_{\nu} \, \Delta_X^\beta{}_{\mu}
- \frac{2}{3} \, \Delta_X^{\alpha\beta} \, \Delta_{X\mu\nu}\right)$}
is the symmetric, traceless, {$u_X$}-orthogonal projector.
{Angle brackets project the comoving derivatives after differentiation:
$\dot q^{\langle\alpha\rangle}:=\Delta^\alpha{}_{\nu}u^\mu\nabla_\mu q^\nu$ and
$\dot\pi^{\langle\alpha\beta\rangle}:=\Delta^{\alpha\beta}{}_{\gamma\delta}u^\mu\nabla_\mu\pi^{\gamma\delta}$.}

\subsubsection{Two-fluid {19-moment} theory}

Now that we have defined evolution equations for each component separately,
we are going to describe their interaction. We then define the electric current
that couples to Maxwell's equations,
\begin{align}\label{eqn:current}
  \mathcal{J}^{\mu}
  = q_{i} \, N_{i}^{\mu} - q_{e} \, N_{e}^{\mu}
  = q_{e} (Z \, N_{i}^{\mu} - N_{e}^{\mu}) \, ,
\end{align}
where $q_{X} > 0$ is the electric {charge magnitude} of component $X$.
{Both species contribute to the current. Neglecting electron stress requires a large mass ratio and bounds on electron heating and drift. The total stress is}
\begin{align}
  T^{\mu\nu} = T_{i}^{\mu\nu} + T_{e}^{\mu\nu} + T_{\rm EM}^{\mu\nu}\, .
  \label{eq:two-fluid-total-stress}
\end{align}
We can then define an effective rest-frame observer relative to which
$T^{\mu\nu}$ can be decomposed.
This decomposition should be done in such a way as to keep effective deviations
from that frame small.
A commonly adopted choice for electron-ion plasmas is to exploit the large mass
difference between the components.
As such, we can define a new observer using mass-weighted averages,
\begin{align}
  N^{\mu}
  = {f_e} \, N_{e}^{\mu} + {f_i} \, N_{i}^{\mu} =: n \, u^{\mu} \, ,
\end{align}
where ${f_X} = m_{X} / (m_{e} + m_{i})$
is the relative mass weight of component $X$,
$u^{\mu}$ is the rest-frame observer associated with $N^{\mu}$,
and $n$ is some mixed-component number density
defined by demanding $u \cdot u = -1$.\\

The main assumption of an effective single-fluid rest frame
is then the starting point of deriving an MHD system in the next Section.
As a direct consequence of choosing an Eckart frame for the single-fluid
observer, whose rest frame coincides with neither component frame,
we find that both components as seen from the single-fluid frame acquire a
diffusion current, i.e.,
\begin{align}
  N_{e}^{\mu} &= \bar{n}_{e} \, u^\mu +\bar{V}_{e}^{\mu} \, , \\
  N_{i}^{\mu} &= \bar{n}_{i} \, u^\mu +\bar{V}_{i}^{\mu} \, ,
\end{align}
where
\begin{align}
  \bar{V}_{i}^{\mu} = -{\frac{f_e}{f_i}} \, \bar{V}_{e}^{\mu} \, ,
\end{align}
is suppressed with the mass ratio.
{These expressions for $N_{X}^{\mu}$ should be compared
with \myeqref{eq.NX} for a single component, but keeping
in mind that the system now possesses three rest frames, namely
those of each charge carrier and the mass-weighted single-fluid one.}\\

Here, we will focus primarily on a specific subset of the system,
which will not only allow us to recover an effective generalization of the
commonly used resistive Navier--Stokes equations,
but also do so in a meaningful and systematic way.

\subsubsection{Dissipative magnetohydrodynamics}
\label{sec:electron-energy-extension}

We will now take a limiting case of the resulting equations commonly used to
derive MHD systems.
A full derivation including the discussion of various limits is given in
Ref. \cite{Most:2021uck}.\\

Now, and in the following, we will make the following assumptions.
{We assume that electrons only couple to the Maxwell equations
\mbox{\eqref{eqn:current}},
and not to the energy--momentum tensor. Essentially, this means expanding the
equations in the mass ratio $\chi = {f_e/f_i} = m_{e} / m_{i} \ll 1$.
This is valid for electron-ion plasmas, but technically not for pair plasmas,
where ideally all terms should be retained.}
However, in practice we will find
that with adjusted closure models, solutions of this system can reproduce
all main features of GRPIC pair plasma simulations well (Section \ref{sec:grplasmas}).\\

As a result, we may write
\begin{align}
  N^{\mu} \approx N_{i}^{\mu}
  + \mathcal{O}\left({\chi}\right) \, ,
\end{align}
meaning that in this limit,
{$\bar V_i^\mu=O(\chi)$},
{$\bar n_i=n+O(\chi^2)$ at fixed $\bar V_e^\mu/n$},
and we are essentially modeling an ion fluid.
Hence,
\begin{align}
  \mc{J}^{\mu} \approx q_{e} \left[\left(Z \, \bar{n}_{i} -\bar{n}_{e} \right)
  u^{\mu} - \bar{V}_{e}^{\mu}\right]
  + \mathcal{O}\left({\chi}\right) \, .
\end{align}
{The exact conduction current in the mass-weighted frame is $j^\mu=-q_e(1+Z\chi)\bar V_e^\mu$. The expression above neglects ion diffusion.}
{With this ordering of the electron stress, we may approximate}
\begin{align}
  T^{\mu\nu} \approx T_{i}^{\mu\nu} + T_{\rm EM}^{\mu\nu}
  + \mathcal{O}\left({\chi}\right) \, ,
\end{align}
so that the system we model is thermodynamically just the ion part of the system.
{For coincident species frames and gamma-law equations of state with fixed $\Gamma_X>1$, define $\Theta_X:=P_X/(m_Xn_X)$. Then}
\begin{align}
 \frac{e_e}{e_i}
 &{=\chi\frac{n_e}{n_i}
 \frac{1+\Theta_e/(\Gamma_e-1)}{1+\Theta_i/(\Gamma_i-1)}}.
 \label{eq:electron-mass-thermal-ordering}
\end{align}
{The mass ordering holds for bounded dimensionless temperatures and density ratio. Heating electrons to $\Theta_e=O(\chi^{-1})$ instead gives an order-one electron energy as $\chi\to0$.}

At the same time, since the electrons still couple to the electric and
magnetic fields, we need to evolve their conduction current, and we also need to evolve the electron temperature.
This follows directly from the electron
energy--momentum balance, with effective electron--ion collisionality supplying
a net relaxation \cite{CercignaniKremer2002},
\begin{align}
  \nabla_{\mu} \, T_{e}^{\mu\nu}
  = -q_{e} \, F^{\nu}_{\phantom{\nu}\mu} \, N_{e}^{\mu}
  + \frac{1}{\tauJ} \left(T_{e}^{\mu\nu}
  - \bar{T}_{e}^{\mu\nu}\right) u_{\mu} \, .
  \label{eq:electron-momentum}
\end{align}
We further make the simplifying assumption that the electrons do not have their
own heat fluxes or anisotropic pressures in this work,
though our framework does naturally include those as well \cite{Most:2021uck}. Future work will be devoted to modeling these; see also Refs. \cite{Ressler:2015ipa,Chael:2018gzl}.\\

{To first order in the relative drift, the electron stress in the single-fluid (ion) frame is}
\begin{align}
  T_{e}^{\mu\nu} = e_{e} u^{\mu}u^{\nu} + P_{e}\Delta^{\mu\nu}
  + u^{\mu}Q_{e}^{\nu} + Q_{e}^{\mu}u^{\nu} \, ,
  \label{eq:electron-Te}
\end{align}
where we set the intrinsic electron bulk, $\Pi_e$, shear, $\pi_e^{\mu\nu}$ and heat fluxes, $q_e^\mu$, to zero, but
retain the electron enthalpy and pressure, together with the heat flux carried
by the relative drift,
\begin{align}
  Q_{e}^{\mu} := h_{e} \, \bar{V}_{e}^{\mu}
  = -\frac{h_{e}}{q_{e}} \, \ju^{\mu} \, ,
  \qquad
  H_{e} := e_{e} + P_{e} = \bar{n}_{e} \, h_{e} \, ,
  \label{eq:electron-Qe}
\end{align}
with $h_{e}$ the electron enthalpy per particle. 
{The drift expansion also requires $\bar V_e^2/\bar n_e^2\ll1$, independently of the mass ratio. We neglect the quadratic drift energy and stress of the boosted electron fluid. Proper and ion-frame number densities therefore agree only to first order in drift.} Projecting
\myeqref{eq:electron-momentum} then gives, to leading order in the drift,
\begin{align}
  \Delta_\nu^\alpha\nabla_{\mu} \, \mathcal{A}_{e}^{\mu}{}_{\alpha}
  ={}& -q_{e} \, \bar{n}_{e} \, e_{\nu}
  + q_{e} \sqrt{b^{2}} \, b^{\mu}{}_{\nu} \, \bar{V}_{e\mu}
  \notag \\
  &- \frac{Q_{e\nu}}{\tauJ}
  + \mathcal{O}\left({\chi}\right) \, ,
  \label{eq.comovj}
\end{align}
where the flux carrying the electron inertia and pressure force is
\begin{align}
  \mathcal{A}_{e}^{\mu}{}_{\nu}
  := u^{\mu} Q_{e\nu} + Q_{e}^{\mu} u_{\nu}
  + H_{e} \, u^{\mu} u_{\nu} + P_{e} \, \delta^{\mu}{}_{\nu} \, .
  \label{eq:electron-A-tensor}
\end{align}
This equation acts as a dynamical Ohm's law.
In the cold, stationary limit, $h_e\to m_e$ and $P_e\to0$, the projected momentum equation gives
\begin{align}
 j_\nu=-q_e\bar V_{e\nu}
 &=\sigma_J e_\nu+\omega_{ge}\tau_J b^\mu{}_\nu j_\mu,\notag\\
 \sigma_J&=\omega_{pe}^2\tau_J.
 \label{eq:audit-cold-ohm}
\end{align}
Without Hall rotation, we recover the usual $j_\nu=\sigma_Je_\nu$ \cite{Palenzuela:2008sf,Palenzuela:2012my}.

Here we have introduced the electron-ion collision time, $\tauJ$,
and the characteristic frequencies,
\begin{align}
  \omega_{pe} &= \sqrt{\frac{q_{e}^{2} \, \bar{n}_{e}}{m_{e}}}
  = \frac{q_{e}}{m_{e}} \, \sqrt{\frac{m_{e}}{m_{i}} \, m_{i} \,
  \bar{n}_{e}} \, , \label{eq.omegape} \\
  \omega_{ge} &= \frac{q_{e} \, b}{m_e} \, ,
\end{align}
{corresponding to the rest-frame electron plasma frequency
and cyclotron frequency, respectively.}
{These frequencies use the electron rest mass. At fixed thermodynamics and fluid velocity, retaining electron temperature instead leads to a homogeneous unmagnetized plasma coefficient $\omega_{pe}^2/w_e$ and a Hall rotation coefficient $\omega_{ge}/w_e$, where $w_e=h_e/m_e$. The rest-mass values follow in the cold limit, $w_e\to1$.}
This means that in a two-fluid system, the scale-free character of MHD is
explicitly broken, and effective kinetic scales are included,
in addition to the effective collisionality rate $1/\tauJ$.\\

Stress-energy conservation of the electrons also contains the energy.  Contracting
\myeqref{eq:electron-momentum} with $u_{\nu}$ gives, with
$\theta := \nabla_{\mu} u^{\mu}$ and
$a^{\mu} := u^{\lambda}\nabla_{\lambda}u^{\mu}$,
\begin{align}
  \nabla_{\mu}\left(e_{e} u^{\mu} + Q_{e}^{\mu}\right)
  = -P_{e} \, \theta - Q_{e}^{\mu} a_{\mu} + e_{\mu} \, \ju^{\mu} \, ,
  \label{eq:electron-energy-conservative-untruncated}
\end{align}
{The projection removes the longitudinal force $u_\nu e_\mu j^\mu$ in}
\myeqref{eq.comovj}.
{We neglect $Q_e^\mu a_\mu$ in the energy equation. This term is second order only if both drift and acceleration are first order. We adopt a gamma-law electron closure,}
\begin{align}
  P_{e} = \left(\Gamma_{e}-1\right)
  \left(e_{e} - m_{e} \, \bar{n}_{e}\right) \, ,
  \label{eq:electron-eos}
\end{align}
where $e_{e}$ includes the electron rest energy.  We stress that the electron
stress tensor enters only through \myeqref{eq.comovj} and
\myeqref{eq:electron-energy-conservative-untruncated}, {i.e., we neglect its bulk-stress contribution under the electron-stress ordering above.}
We remark that we can similarly parameterize the ion cyclotron frequency in the
ion heat fluxes and shear stresses,
\begin{align}
  \omega_{gi} &= \frac{Z \, q_{e} \, b}{m_{i}}
  = \frac{m_{e}}{m_{i}} \, \frac{q_{e}}{m_{e}} \, Z \, b \, ,
  \label{eq.omegagi}
\end{align}
where $Z$ is the charge ratio between ions and electrons,
and $m_{e}$ and $m_{i}$ are the electron and ion rest masses, respectively.\\

{For a stationary, homogeneous response without Hall rotation, at fixed fluid velocity and thermodynamics, the conductivity is}
\begin{align}
  \sigma_{E} =\frac{\omega_{pe}^{2}\tauJ}{w_e}
  =\frac{\sigma_J}{w_e} \, ,
  \label{eq.sigmaE}
\end{align}
{so, for a given electron {enthalpy}, conductivity fixes the ratio of the effective plasma coefficient to the collision frequency. The rest-mass expression follows in the cold electron limit.}
In that sense, one can easily translate between the two formalisms. \\

In order to generalize the system to different closures,
we express the electric current via
\begin{align}
  \mathcal{J}^{\mu} = \rhoq \, u^{\mu} + \ju^{\mu} \, ,
  \label{eqn:Jdecomp}
\end{align}
where $\rhoq := -u_\nu \mathcal{J}^\nu
\simeq q_{e} \, (Z \, \bar{n}_{i} - \bar{n}_{e})$
and $\ju^{\mu} := \Delta^{\mu}_{\phantom{\mu}\alpha} \, \mc{J}^{\alpha}
\simeq - q_{e} \, \bar{V}_{e}^{\mu}$.\\

{For fixed charge magnitudes and conserved ion number, charge conservation implies electron-number conservation. In the ion-frame approximation, $N_i^\mu=n_i u^\mu +\mathcal{O}(\chi)$, giving}
\begin{align}
  N_{e}^{\mu}
  = \left(\frac{q_{i}}{q_{e}} \, n_{i}
  - \frac{1}{q_{e}} \, \rhoq\right) u^{\mu} - \frac{1}{q_{e}} \, j^{\mu} \, ,
  \label{eq:electron-number-current}
\end{align}
which allows us to reconstruct the electron number density
$n_{e} = \sqrt{-N_{e} \cdot N_{e}}$
in the rest frame of the electron fluid,
as well as to compute $\bar{n}_{e} = Z \, \rhob / m_{i} - \rhoq / q_{e}$.
{These relations use the ion-frame approximation. In the exact mass-weighted frame, Eq.~\eqref{eq:electron-number-current} instead uses $\bar n_i$ and $j^\mu/[q_e(1+Z\chi)]$ in place of $n_i$ and $j^\mu/q_e$. An electron rest frame exists only for future-timelike $N_e^\mu$, which is a requirement for this set of equations.}

\subsection{Characteristic structure}
\label{sec:characteristics-main}

It is instructive to study the characteristic wave structure of the system.
In covariant four-notation, let $n^\mu$
be the hypersurface normal and $s^\mu$ a unit Eulerian normal to a
numerical face, $n_\mu s^\mu=0$ and $s_\mu s^\mu=1$.  A local plane wave
$\delta U\propto\exp[ik(s_ix^i-\lambda t)]$ has characteristic covector
\begin{align*}
 \phi_\mu&=k(yn_\mu+s_\mu),&
 \lambda&=\alpha y-\beta_s,
 \\
 v_s&=v^is_i,& \beta_s&=\beta^is_i.
\end{align*}
Thus, $y$ is its Eulerian normal speed and $\lambda$ its coordinate speed.
The wave exists when the principal symbol
$\mathbb P(\phi)=\mathbb A^\mu\phi_\mu$ is singular.  A right null vector
is its primitive perturbation, while a left null vector extracts its
amplitude.  Define the normalized fluid-frame root and normal by
\begin{align*}
 a&=\frac{u^\mu\phi_\mu}
 {(\Delta^{\alpha\beta}\phi_\alpha\phi_\beta)^{1/2}},
 \\
 \hat n^\mu&=\frac{\Delta^{\mu\nu}\phi_\nu}
 {(\Delta^{\alpha\beta}\phi_\alpha\phi_\beta)^{1/2}}.
\end{align*}
Set $D_a=a^2+(1-a^2)W^2$ and
$R_a=a^2+(1-a^2)W^2(1-v_s^2)$.  Then
\begin{align*}
 y(a)&=\frac{(1-a^2)W^2v_s-a\sqrt{R_a}}{D_a},
 \\
 \lambda(a)&=\alpha y(a)-\beta_s,
\end{align*}
Each fluid-frame root then gives the corresponding propagation speed.
On the physical constraint manifold,
\begin{align}
 \det\mathbb P\propto
 a^8(1-a^2)^4\mc C_3(a)\mc Q_8(a).
 \label{eq:main-characteristic-factor}
\end{align}
Here $\mc C_3$ is the electron contact--acoustic cubic defined in
Eq.~\eqref{eq:main-electron-cubic}.  The factor $\mc Q_8$ is the
degree-eight constrained ion--dissipative characteristic polynomial,
defined by the primitive determinant in
Eq.~\eqref{eq.ev.Q8-primitive-pencil} and factorized in the aligned case
in Eq.~\eqref{eq.Q8.resistive.factorization}.
Thus, the system has eight material modes, eight light-cone modes,
three electron modes, and eight ion modes:
\begin{itemize}
\item \emph{Light-cone modes.}
The two electromagnetic polarizations and two cleaning fields give
four left--right pairs.  Their {coordinate} speeds are
$\lambda_{\gamma,\pm}=\pm\alpha-\beta^is_i$.

\item \emph{Material modes.}
One thermodynamic, one bulk, two transverse heat, two
transverse-traceless shear, and two transverse-current perturbations
are advected with the fluid.  Hence, $a=0$ and
$\lambda_0=(\alpha v^i-\beta^i)s_i$.

\item \emph{Electron modes.}
These are an electron contact wave and an acoustic pair.  Drift mixes
the three through
\begin{align}
 \mc C_3(a)=a^3-(1-\sigma_e)v_{\rm d}a^2
 -c_{\rm se}^2a+c_{\rm se}^2v_{\rm d},
 \label{eq:main-electron-cubic}
\end{align}
where $v_{\rm d}=j_{\hat n}/(q_e\bar n_e)$,
$\sigma_e=\bar n_eP_{e,n}/H_e$, and
$c_{\rm se}^2=P_{e,e}+\sigma_e$.  At zero drift,
$a=0,\pm c_{\rm se}$. The finite-drift roots are given in
Eq.~\eqref{eq.char.speed.electron}.

\item \emph{Ion modes.}
Two viscous-shear pairs, one acoustic pair, and one causal-thermal
pair form the ion octic.  For heat parallel to a principal shear
normal, each transverse pair has
\begin{align}
 a_{\perp A,\pm}&=
 \frac{-q_{\hat n}\pm
 \sqrt{q_{\hat n}^2+4\mathcal I_Ay_\eta}}
 {2\mathcal I_A},\notag\\
 \mathcal I_A&=e+P+\Pi+\Lambda_A-\kappa T/\tauq,
 \qquad y_\eta=\eta/\tau_\pi .
 \label{eq:main-ion-shear-speed}
\end{align}
At zero projected heat this reduces to
$a_{\perp A,\pm}=\pm\sqrt{y_\eta/\mathcal I_A}$.
The aligned longitudinal channel gives
\begin{align}
 a_L&=\pm c_{L,+},\ \pm c_{L,-},&
 c_{L,\pm}^2&=
 \frac{\mathcal M_q\pm
 \sqrt{\mathcal M_q^2-4\mathcal H\mathcal F_q}}
 {2\mathcal H},
 \label{eq:main-ion-longitudinal-speed}
\end{align}
with coefficients in
Eq.~\eqref{eq.char.speed.longitudinal.equilibrium}.  Heat and shear
mix these channels for a generic normal.  The fully resistive principal
system therefore has viscous-shear and Maxwell waves, but no separate
Alfv\'en or magnetosonic branches in the general case.
\end{itemize}

{Equation~\eqref{eq.char.speed.inverse} derives the speed
map above.}  Appendices~\ref{app:characteristics} and
\ref{app:eigenvectors} give the full characteristic polynomials and
eigenvectors.  The current
needed to keep the electron and ion waves distinct is discussed below.

\subsection{Nonlinear causality and strong hyperbolicity}
\label{sec:causality}

Following Ref.~\cite{Cordeiro:2026}, and using the principal symbol of
Eq.~\eqref{eq.quasilinear}, we call the system nonlinearly causal at a
given state if every root $\phi_0(\phi_i)$ of
$\det(\mathbb A^\alpha\phi_\alpha)=0$ is real and every characteristic
covector is non-timelike, $\phi^\alpha\phi_\alpha\ge0$.
{The two requirements are independent: a light-cone bound
locates real roots but does not exclude a complex-conjugate pair.}
Strong hyperbolicity additionally requires a complete
eigenbasis with {both its diagonalizer and inverse bounded uniformly
in the wave normal.}
The material and Maxwell factors in
Eq.~\eqref{eq:main-characteristic-factor} already propagate with the
fluid and on the light cone.  Only the ion octic and electron cubic
require further restrictions.\footnote{{In their common
limits, these conditions reproduce the bulk--shear results of
Ref.~\cite{Cordeiro:2026}.  Setting $q^\mu=\kappa=0$ and removing the
electromagnetic and electron blocks gives their equilibrium shear and sound
speeds [Eq.~(40)], bulk-only speed [Eq.~(30)], and conformal shear bound
[Eq.~(41)].  On the common Maxwell--Cattaneo closure, with noncommon
second-order couplings set to zero, the nonlinear determinant agrees with
their Eqs.~(54a) and (54c), and with Eq.~(54b) after
$3\eta_{\rm eff}^2/\tau_\pi^2$ is replaced by
$3(\eta_{\rm eff}/\tau_\pi)E$.  Their strict cubic discriminant excludes
the semisimple transverse coincidence $\Lambda_2=\Lambda_3$, which our
joint eigenbasis retains.  In the heat-only limit, the physical symbol also
agrees for every normal with Ref.~\cite{Cordeiro:2025diffusion} on the
common closure $\Omega_q=\tauq/(\kappa T^2)=\mathrm{const.}$ and
$\delta_{qq}=\tauq/2$, after the printed corrections to their
Eqs.~(40a) and (42b) detailed in Appendix~\ref{app:char_causality}.}}

For arbitrary heat flux and shear stress, the exact ion criterion is
Eq.~\eqref{eq.resgeneral.complete}; the exact electron criterion is given by
Eqs.~\eqref{eq.reselectron.exact-linear} and
\eqref{eq.reselectron.exact-reality}.  These tests include repeated
roots.  The full fixed-state strong-hyperbolicity test is the resolvent
bound \eqref{eq.ev.full-resolvent-sh}.

We can generally summarize sufficient conditions for strong hyperbolicity as follows:
\begin{align*}
 \tau_\Pi,\tau_\pi,\tauq,H_e,\rhob,\bar n_e&>0,\notag\\[-2pt]
 \min_A\mathcal I_A&>0,
 \tag{SH1}\\
 \eta/\tau_\pi>0,\quad \kappa/\tauq>0,\quad c_{\rm se}^2&>0,
 \tag{SH2}\\
 \mc C_3,\mc Q_8:\quad
 &\text{all roots real with }|a|<1,
 \tag{SH3}\\
 \operatorname{Disc}_a\mc C_3>0,\quad
 \min_{|\hat n|=1}\operatorname{Disc}_a\mc Q_8&>0,\notag\\[-2pt]
 &\text{and Eq.~\eqref{eq.strong.material-gap} holds},
 \tag{SH4}\\
 \dim\ker\mathbb P(a_*)&=
 \operatorname{ord}_{a_*}\det\mathbb P,\notag\\[-2pt]
 &\text{with Eq.~\eqref{eq.ev.full-resolvent-sh} uniform in $\hat n$}.
 \tag{SH5}
\end{align*}
(SH1) makes the time symbol invertible, (SH2) keeps the restoring
channels nondegenerate, (SH3) is the strict causal spectrum, and
(SH4)--(SH5) give a uniformly complete eigenbasis.
{Here $\operatorname{ord}_{a_*}\det\mathbb P$ counts the characteristic
roots that coincide at $a_*$, including multiplicity, whereas
$\dim\ker\mathbb P(a_*)$ counts the linearly independent eigenvectors at
that speed.  Thus (SH5) requires the geometric multiplicity to equal the
algebraic multiplicity.}

While nonlinear causality was already established in the previous Section, the following provides a sufficient domain for strong hyperbolicity.  Define:
\begin{align}
 E_0&=e+P+\Pi,&Q&=\sqrt{q_\mu q^\mu},\notag\\
 S&=\sqrt{2\pi_{\mu\nu}\pi^{\mu\nu}/3},&p&=P_{,e},\notag\\
 B_\rho&=|\rhob P_{,\rhob}|,\notag\\
 M&=(1-p)E_0-B_\rho-(1+p)(S+2Q),\notag\\
 C_q&=\kappa T+|\dqq|Q+\kappa|\rhob T_{,\rhob}|,\notag\\
 C_\pi&=4\eta/3+|\dpipi|S,\notag\\
 C_\Pi&=|\zeta+\dPiPi\Pi|,\notag\\
 A_q&=(1+p)C_q+\kappa T_{,e}(2E_0-B_\rho).
 \label{eq.caus.resistive-primitive-cone-defs}
\end{align}
For $0\le p<1$, $\kappa,T,T_{,e},\eta\ge0$, and $M>0$,
\begin{align}
 \tauq&>\frac{3A_q}{M},&
 \tau_\pi&>\frac{3C_\pi}{M},&
 \tau_\Pi&>\frac{3C_\Pi}{M}
 \label{eq.caus.resistive-primitive-cone-floors}
\end{align}
These floors exclude every ion root on or outside the light cone, but
do not enforce root reality. The exact all-normal test is
Eq.~\eqref{eq.resgeneral.complete}.  For a primitive sufficient
condition, let $\Lambda$ be the fluid-frame shear matrix and set
\begin{align*}
 \bar E&=E_0-\frac{\kappa T}{\tauq},\qquad
 y=\frac{\eta}{\tau_\pi},\qquad
 c=\frac{\kappa T_{,e}}{\tauq},\\
 R&=\frac{\kappa\rhob T_{,\rhob}}{\tauq},\qquad
 d_q=\frac{\dqq}{\tauq},\qquad
 d_\pi=\frac{\dpipi}{\tau_\pi},\\
 Z&=\rhob P_{,\rhob}
 +\frac{\zeta+\dPiPi\Pi}{\tau_\Pi}+\frac{4y}{3},\\
 I_2&=\operatorname{tr}\Lambda^2,\qquad
 I_3=\operatorname{tr}\Lambda^3,\qquad
 D_\pi=\frac{I_2^3}{2}-3I_3^2,\\
 H_\pm&=\bar E\pm S,\\
 D_p&=(p\bar E-y)^3-\frac{p^2}{2}(p\bar E-y)I_2
 +\frac{p^3}{3}I_3,\\
 s_-&=\min\{\sqrt p,\sqrt{y/H_+}\},\qquad
 s_+=\max\{\sqrt p,\sqrt{y/H_-}\},\\
 g_*&=\min\!\left\{2s_-,
 \frac{\sqrt{yD_\pi}}{8S^2H_+^{3/2}},
 \frac{|D_p|}{2s_+H_+(pH_++y)^2}\right\},\\
 r_*&=\frac14\min\{g_*,1-s_+\},\qquad
 t_*=2s_--r_*,\\
 B_*&=(1+|1-d_q|)Q+|d_\pi|S+|Z-R-y|,\\
 E_*&=2|p-c|+\frac{(2+|d_q|)Q+B_*}{H_+},\\
 k_1&=\frac{1+H_+/H_-}{r_*t_*},\qquad
 k_2=\frac{H_++|R|}{H_-r_*},\\
 K_*&=k_1+\frac{k_2}{(g_*-r_*)t_*^2}.
\end{align*}
For
\begin{align*}
 p,y,H_->0,\qquad s_+<1,\qquad
 D_\pi>0,\qquad D_p\ne0,
\end{align*}
the primitive root reality condition is
\begin{align*}
 E_*K_*<1.
\end{align*}
It gives eight distinct ion roots in $(-1,1)$ for every normal and
implies the material gap below.  The reference-direction continuation
condition is Eq.~\eqref{eq.ev.ion-simple-reality-continuation}.

{For gamma-law ions on the positive-inertia branch, the six material
modes are separated from the ion block when, with }
$d_\pi=\delta_{\pi\pi}/\tau_\pi$,
$R_m=P+(\zeta+\delta_{\Pi\Pi}\Pi)/\tau_\Pi+4\eta/(3\tau_\pi)$, and
$I_r=\operatorname{tr}(\pi^r)$,
\begin{align}
 R_m&>0,\qquad 6R_m^2-d_\pi^2I_2>0,\notag\\
 6R_m^3-3R_md_\pi^2I_2+2d_\pi^3I_3&>0.
 \label{eq.strong.material-gap}
\end{align}
This condition automatically holds with the primitive simple-root condition.
{For a general EOS replace $P$ in $R_m$ by
$\rhob(\partial P/\partial\rhob)_T$; the signed criterion is
Eq.~\eqref{eq.ev.material-gap-signed}.}

For the first-order electron sector, require
$H_e>0$, $\bar n_e>0$, and $q_e\ne0$, and set
\begin{align}
 0<c_{\rm se}^2&=P_{e,e}+\frac{\bar n_eP_{e,n}}{H_e}<1,
 \label{eq.caus.cse}\\
 \sigma_e&=\frac{\bar n_eP_{e,n}}{H_e},&
 \mc V&=\frac{\sqrt{j_\mu j^\mu}}{|q_e|\bar n_e}.
 \label{eq:electron-current-normalization}
\end{align}
Its causal and simple-root ceilings are
\begin{align}
 v_{\rm caus}&=\min\!\left\{
 \frac{3-c_{\rm se}^2}{2|1-\sigma_e|},
 \frac{1-c_{\rm se}^2}{|1-\sigma_e-c_{\rm se}^2|}\right\},
 \label{eq.caus.vd}\\
 v_{\rm SH}^2&=
 \begin{cases}
 \infty,&\sigma_e<0,\\[2pt]
 \dfrac{8c_{\rm se}^2}
 {8+20\sigma_e-\sigma_e^2+\sqrt{\sigma_e(8+\sigma_e)^3}},
 &\sigma_e\ge0 .
 \end{cases}
 \label{eq:electron-sh-current-bound}
\end{align}
For $c_{\rm se}^2>0$, electron root reality is exactly
\begin{align*}
 \mc V^2\!\left[
 8+20\sigma_e-\sigma_e^2+
 \sqrt{\sigma_e(8+\sigma_e)^3}\right]
 \le8c_{\rm se}^2,\qquad \sigma_e>0 .
\end{align*}
For $\sigma_e\le0$ (as enforced by the gamma-law) no additional drift bound is required for root reality; the strict $v_{\rm SH}$ bound makes the cubic roots simple.
A vanishing denominator in $v_{\rm caus}$ imposes no bound.  For
$e_e\ge P_e\ge0$, the dominant energy condition {(DEC)} for the truncated
electron tensor gives
\begin{align}
 v_{\rm DEC}=
 \begin{cases}
 \tfrac12,&e_e\ge3P_e,\\[2pt]
 \dfrac{\sqrt{2P_e(e_e-P_e)}}{e_e+P_e},&P_e\le e_e<3P_e,
 \end{cases}
 \label{eq:linear-electron-energy-condition-bound}
\end{align}
and the electron-only interior is
\begin{align}
 \mc V<\min\{v_{\rm caus},v_{\rm SH},v_{\rm DEC}\}.
 \label{eq:electron-current-combined-bound}
\end{align}
For hot gamma-law electrons, $e_e>\rho_e>0$ implies
$\sigma_e<0$, $v_{\rm SH}=\infty$, and $v_{\rm caus}\ge1/2$.
The DEC is therefore the tighter electron-only bound.
{A sufficient strongly hyperbolic domain separates the electron and
ion speeds.} For hot gamma-law electrons on the simple-root domain of
Eqs.~\eqref{eq.resgeneral.simple-reality-inputs}--
\eqref{eq.resgeneral.simple-current-inputs},
write $x=c_{\rm se}^2$ and set
\begin{align*}
 D_x&=(x\bar E-y)^3-\tfrac12x^2(x\bar E-y)I_2
 +\tfrac13x^3I_3,\\
 s_e&=\max\{\sqrt{x},s_+\},\\
 \delta_e&=\min\!\left\{|\sqrt{x}-\sqrt{p}|,
 \frac{|D_x|}{2s_eH_+(xH_++y)^2}\right\}.
\end{align*}
For $E_*K_*<1$ and $\delta_e>r_*$, define
\begin{align}
 J_{\rm ei}&=\min\!\left\{|q_e|\bar n_e(s_--r_*),
 \frac{|q_e|\bar n_e}{1-\sigma_e}(\delta_e-r_*)\right\},
 \label{eq:electron-ion-acoustic-gap-j}\\
 \sqrt{j_\mu j^\mu}
 &<\min\{|q_e|\bar n_e v_{\rm DEC},J_{\rm ei}\}.
 \label{eq:electron-ion-acoustic-current-limit}
\end{align}
The first bound separates the electron contact; the second separates
the acoustic pair.  Unlike the DEC ceiling, $J_{\rm ei}$ depends on the
ion spectrum and can be tighter.  

\section{Numerical method}
\label{sec:numerics}

We solve the equations of general-relativistic 19-moment two-fluid
dissipative magnetohydrodynamics (GR19M) numerically,
using high-resolution shock-capturing methods with
implicit-explicit {(IMEX)} time stepping on {graphics processing units (GPUs)}.
In the following, we provide an extended overview of our approach,
including ways of implicitly integrating these equations.

\subsection{Flux-conservative form of the equations}

One numerical challenge that arises when using these equations is that in most
global simulations, {because of limited computational resources,}
dissipative scales are not always resolved everywhere on the numerical grid.
This is problematic, as either the equations become stiff and require special
integration schemes (as we will outline later in this Section),
or gradients are unresolved, leading to artificial oscillations and numerical
instability of the simulations.
The latter situation already exists in ideal (magneto-)hydrodynamics,
where unresolved shocks require some form of artificial grid viscosity to
regulate them \cite{Marti:1999wi}.
This is usually done by applying high-resolution shock-capturing schemes to the
equations, which requires the use of a flux-conservative formulation (see Ref. \cite{Font:2008fka} for a review).
In an earlier work \cite{Most:2021rhr}, we have already laid out such a reformulation of the
non-resistive sector, which we shall generalize here to the case of two-fluid
(resistive) dissipation.
In general, we are looking for a reformulation of the equations in which
\begin{align}
  \nabla_{\mu} \, \mc{F}^{\mu} =
  \frac{1}{\sqrt{-g}} \left( \p_{t} \left( \sqrt{\gamma} \, U \right)
  + \p_{i} \left( \sqrt{-g} \, F^{i} \right) \right)
  = S \, ,
  \cc{\label{eq.fluxform}}
\end{align}
where $\mc{F}^{\mu} = \left( \alpha^{-1} \, U, F^{i} \right)^{\T}$,
with $U$ a vector of conserved, or evolved, variables,
$F^{i}\left(G,U\right)$ a vector of fluxes in the $i$-th direction,
and $S$ a vector of source terms, which we allow to depend on $G$,
$\p_{j} \, G$, $U$, but not on derivatives of $U$.
{This divergence identity applies to scalar currents. 
For tensor currents we use it componentwise, placing the connection terms from the free indices in $S$.}

In order to facilitate this, we recast Eqs.~(\ref{eq.comovPi}-\ref{eq.comovq})
in flux-conservative form using the approach of Ref. \cite{Most:2021rhr}.
This approach enforces a flux-conservative nature of the equations
by breaking the algebraic constraints of the form
$\mathcal{A}_{\mu} \, u^{\mu} = 0$ and replacing them with a set of
relaxation terms akin to what is done in implicit approaches to
force-free electrodynamics \cite{Alic:2012df,Palenzuela:2012my}.
A full derivation of how to recast these equations for non-resistive relativistic MHD is provided in
Ref. \cite{Most:2021rhr}.

For signature $(-+++)$ a vector constraint is damped by
$+\omega_A(u\cdot A)u_\nu=-\omega_A[A_\nu-\Delta_\nu{}^\mu A_\mu]$.
For shear, we use the projector defect
\begin{align}
 \mathcal C_\pi^{\alpha\beta}
 &:=\pi^{\alpha\beta}
 -\Delta^{\alpha\beta}{}_{\gamma\delta}\pi^{\gamma\delta},
 \label{eq:audit-shear-defect}
\end{align}
with source $-\omega_\pi\mathcal C_\pi^{\alpha\beta}$. These additions vanish on the physical constraint surface. At fixed $u^\mu$ in its rest frame they damp the temporal and trace defects.
{The penalties relax violations of the heat and shear constraints at finite rates. As $u\cdot q\to0$ and $\mathcal C_\pi\to0$, the products $\omega_q(u\cdot q)u_\nu$ and $\omega_\pi\mathcal C_\pi^{\alpha\beta}$ can approach finite constraint forces. The constrained limit therefore follows by projecting the residual. These balances also include the nonlinear expansion terms proportional to $\delta_{qq}$, $\delta_{\pi\pi}$ and $\delta_{\Pi\Pi}$, omitted from the leading-order species equations above.}

{For the electrons, we project the weighted current tensor orthogonal to the fluid velocity.}  It is numerically preferable
to evolve an enthalpy-weighted current rather than $\ju_\mu$ itself, so with
\begin{align}
 w_e&:=\frac{h_e}{m_e}=\frac{H_e}{\rho_e}\,, &
 k_\mu&:=w_e\,\ju_\mu\,, \notag\\
 \omega_e&:=\frac{q_e}{m_e}H_e\,, &
 \chi_e&:=\frac{q_e}{m_e}P_e\,,
 \label{eq:electron-scaled-current}
\end{align}
and $\rho_e=m_e\bar n_e$, multiplying \myeqref{eq.comovj} by $-q_e/m_e$
collects the electron inertia and pressure force into
\begin{align}
 \mathcal B^\mu{}_\nu
 &:=-\frac{q_e}{m_e}\,\mathcal A_e^\mu{}_\nu \notag\\
 &=u^\mu k_\nu+k^\mu u_\nu-\omega_e\,u^\mu u_\nu
   -\chi_e\,\delta^\mu{}_\nu\,,
 \label{eq:electron-B-tensor}
\end{align}
{We combine the truncated energy equation with the projected current equation and $u^\nu k_\nu=0$. Projection acts outside the derivatives. We write}
\begin{align}
 {L_\nu\overset{\perp}{=}R_\nu}
 &:\Longleftrightarrow
 \Delta_\nu{}^\alpha(L_\alpha-R_\alpha)=0.
 \label{eq:electron-projected-equality}
\end{align}
{For shear, we use the symmetric, trace-free {(STF)} projection}
\begin{align}
 {L^{\alpha\beta}\overset{\rm STF}{=}R^{\alpha\beta}}
 &{:\Longleftrightarrow
 \Delta^{\alpha\beta}{}_{\gamma\delta}
 (L^{\gamma\delta}-R^{\gamma\delta})=0}.
 \label{eq:moment-stf-equality}
\end{align}
{The current row is then}
\begin{align}
 \nabla_\mu\mathcal B^\mu{}_\nu
 \,{\overset{\perp}{=}}{}&-\frac{k_\nu}{\tauJ}+\omega_{pe}^2\,e_\nu
   +\frac{\omega_{ge}}{w_e}\,b^{\mu}{}_{\nu}\,k_\mu \notag\\
 &-\frac{q_e}{m_e}\,u_\nu\,e_\mu \ju^\mu\,.
 \label{eq:electron-B-ohm}
\end{align}
{The projection removes the longitudinal Joule term. Combined with the truncated energy equation, the unprojected balance would need the additional term $u_\nu k^\mu a_\mu$ on its right-hand side. We solve the three projected current equations together with $u\cdot k=0$. The four stored components then obey}
\begin{align}
 {u^\nu k_{*\nu}}
 &{=\frac{q_e}{m_e}\mathcal E_{e*}}.
 \label{eq:electron-dependent-current-moment}
\end{align}
This serves as the effective Ohm's law for our system and also as the
evolution equation for the electric current.  The physical current is
recovered as $\ju_\mu=k_\mu/w_e$, and since $Q_e^\mu=-(m_e/q_e)k^\mu$ the
energy row \eqref{eq:electron-energy-summary} is a divergence of the same
variable. The kinetic coefficients are those of the cold system,
\begin{align}
 \frac{\sigma_J}{\tauJ}&=\omega_{pe}^2\,, &
 \frac{\delta_{JB}\sqrt{b^2}}{\tauJ}&=\omega_{ge}\,,
 \label{eq:electron-hot-coefficients}
\end{align}

so that the electric drive in the $k_\mu$ row is $\sigma_J$ and its Hall
coefficient $\delta_{JB}/w_e$, whereas Amp\`ere's law, which uses $\ju_\mu$,
sees $\sigma_J/w_e$.

{With flux-form predictors and a projected current balance, the evolution equations are}
\begin{widetext}
\begin{subequations}
\label{eq:full-evolution-system}
\begin{align}
  &\nabla_{\mu} \left({\rhob} \, u^{\mu}\right) = 0 \, , \\
  &\nabla_{\mu} \left( T_{i}^{\mu\nu}+T_{\rm EM}^{\mu\nu}
    \right) = 0 \, , \\
  &\nabla_{\mu} \left(      F^{\mu\nu} + g^{\mu\nu} \, \psi \right)
    = {\kappa_{\rm dc}} \, n^{\nu} \, \psi - \mc{J}^{\nu} \, , \\
  &\nabla_{\mu} \left( \starF^{\mu\nu} + g^{\mu\nu} \, \phi \right)
    = {\kappa_{\rm dc}} \, n^{\nu} \, \phi \, , \\
  &{\nabla_{\mu} \left( \rho_{e} \, u^{\mu}
    - \frac{m_{e}}{q_{e}} \, \ju^{\mu} \right) = 0} \, , \\
  &{\nabla_\mu(e_eu^\mu+Q_e^\mu)
    =-P_e\nabla_\mu u^\mu+e_\mu j^\mu
    =-P_e\nabla_\mu u^\mu+\frac{e_\mu k^\mu}{w_e}} \, ,
    \label{eq:electron-energy-summary} \\
  &\nabla_{\mu} \left( \left( \Pi + \frac{\zeta}{\tau_{\Pi}} \right) u^{\mu} \right)
    = -\frac{1}{\tau_{\Pi}} \, \Pi
    + \left( 1 - \frac{\dPiPi}{\tau_{\Pi}} \right) \Pi \,
    \nabla_{\mu} \, u^{\mu} + u^{\mu} \, \nabla_{\mu} \, \frac{\zeta}{\tau_{\Pi}}
    \, ,
    \label{eq.bigPi} \\
  & \nabla_{\mu} \left( q_{\nu} \, u^{\mu} + \frac{\kappa}{\tauq} \, T \,
    \Delta^{\mu}_{\phantom{\mu}\nu}\right)
    = - \frac{1}{\tauq} \, q_{\nu}
    + \left( 1 - \frac{\dqq}{\tauq} \right) q_{\nu} \, \nabla_{\mu} \, u^{\mu}
    + T \, \nabla_{\nu} \, \frac{\kappa}{\tauq}
    - \omega_{gi} \, b^{\mu}_{\phantom{\mu}\nu} \, q_{\mu}
    {+ \omegaq \left(q_{\mu} \, u^{\mu}\right) u_{\nu}} \, ,
    \label{eq.q} \\
  & \nabla_{\mu} \left( \pi^{\alpha\beta} \, u^{\mu}
    + 2 \, \frac{\eta}{\tau_{\pi}} \,
    g^{\mu\left(\alpha\right.} u^{\left.\beta\right)}\right)
    =  -\frac{1}{\tau_{\pi}} \, \pi^{\alpha\beta}
    + \left( 1 - \frac{\dpipi}{\tau_{\pi}} \right)
    \pi^{\alpha\beta} \, \nabla_{\mu} \, u^{\mu}
    - \omega_{gi} \,
    b^{\delta\mu} \, \pi^{\gamma}_{\phantom{\gamma}\mu} \,
    \Delta^{\alpha\beta}_{\phantom{\alpha\beta}\gamma\delta}
{-\omega_\pi\mathcal C_\pi^{\alpha\beta}} \, ,
    \label{eq.pi} \\
  &{\nabla_\mu\mathcal B^\mu{}_\nu
    \,{\overset{\perp}{=}}-\frac{k_\nu}{\tauJ}
      +\omega_{pe}^2 e_\nu
      +\frac{\omega_{ge}}{w_e}\,
       b^{\mu}_{\phantom{\mu}\nu}k_\mu
      {-\frac{q_e}{m_e}u_\nu e_\mu\ju^\mu}}\, ,
    \label{eq:electron-B-summary}
\end{align}
\end{subequations}
\end{widetext}
where we introduced constraint-enforcing relaxation rates
{$\omegaq$ and $\omega_{\pi}$}.
We provide a complete list of these equations expanded into 3+1 form in
Appendix~\ref{app:3p1}.
We now introduce the following set of conserved, or evolved, variables,

\begin{align}
  \widetilde{U} &:= \sqrt{\gamma} \, U
  := \sqrt{\gamma} \, \left( \Db, \tau, S_{j}, B^{j},
  \phi, E^{j}, \psi, {D_{e}}, \notag \right.\\
  &\left.\hspace{6em} \Pi_{*},
  Q_{\mu}, \pi_{*}^{\mu\nu}, {k_{*\mu}, \mathcal{E}_{e*}}
  \right)^{\T} \, ,
  \label{eq.consvars}
\end{align}

{Here $p=P$, $h=h_i=(e+P)/\rhob$, and
$e_{\rm int}:=e-\rhob=\rhob\epsilon$.}
{The conserved variables are} functions of the primitive variables,
\begin{align}
  V &:= \left( \rhob, \rhob \, \epsilon, W \, v^{j},
  \sqrt{\gamma} \, B^{j},
  \sqrt{\gamma} \, \phi, \sqrt{\gamma} \, E^{j}, \sqrt{\gamma} \, \psi,
  {\rho_{e}}, \right. \notag \\
  &\left.\hspace{4em} \Pi,
  q_{\mu}, \pi^{\mu\nu}, \ju_{\mu}, {e_{e}} \right)^{\T} \, ,
\end{align}
where
\begin{subequations}
\begin{align}
  \Db &:= \rhob \, W \, , \\
  \tau &:=\left( \rhob \, h + \Pi \right) W^{2} - \left( p + \Pi \right)
    - \rhob \, W \notag \\
    &\phantom{:==} + \frac{1}{2} \left( E^{2} + B^{2} \right)
    - 2 \, W \, n_{\mu} \, q^{\mu} + n_{\mu} \, n_{\nu} \, \pi^{\mu\nu} \, ,
    \label{eq.tauenergy} \\
  S_{j} &:= \left( \rhob \, h + \Pi \right) W^{2} \, v_{j}
    + \levciv_{jmn} \, E^{m} \, B^{n} \notag \\
    &\phantom{:==} + W \, q_{j} - n_{\mu} \, q^{\mu} \, W \, v_{j}
    - n_{\mu} \, \pi_{j}^{\phantom{j}\mu} \, ,
    \label{eq.Smomentum} \\
  {D_{e}} & {:= W \, \rho_{e} + \frac{m_{e}}{q_{e}} \, n_{\mu} \,
    \ju^{\mu} = W \, \rho_{e} - \frac{\kappa_{e}}{w_{e}}} \, , \\
  \Pi_{*} &:= W \left( \Pi + \frac{\zeta}{\tau_{\Pi}} \right) \, , \\
  Q_{\nu} &:= W \, q_{\nu} -  \, \frac{\kappa}{\tauq} \, T \,
  \Delta^{\mu}_{\phantom{\mu}\nu} \, n_{\mu}  \, \\
  \pi_{*}^{\alpha\beta}
    &= W \, \pi^{\alpha\beta}
    - 2 \, \frac{\eta}{\tau_{\pi}} \,
    n^{\left(\alpha\right.} u^{\left.\beta\right)} \, , \\
  {k_{*\mu}} &{:= W \left( k_{\mu} - \omega_{e} u_{\mu} \right)
    - \left( n_{\nu} k^{\nu} \right) u_{\mu} + \chi_{e} \, n_{\mu}} \, ,
    \label{eq:electron-code-kstar} \\
  {\mathcal{E}_{e*}} &{:= W e_{e} - \kappa_{e} \, ,
    \qquad \kappa_{e} := \alpha \, \frac{m_{e}}{q_{e}} \, k^{0}} \, .
    \label{eq:electron-energy-density}
\end{align}
\end{subequations}

We evolve the electron rest-mass density rather than the charge density, since it allows us to more easily ensure positive electron number density.  The charge density and its conserved
moment then follow algebraically,
\begin{align}
  \rhoq &= \frac{q_{e}}{m_{e}} \left( \mu \, Z \, \rhob
    - \rho_{e} \right) \, , &
  \Dq &= \frac{q_{e}}{m_{e}} \left( \mu \, Z \, \Db - D_{e} \right) \, ,
  \label{eq.chargefromelectron}
\end{align}
with $\mu = m_{e}/m_{i}$, and are supplied to Gauss' law and to the Amp\`ere
source in that form. {Charge conservation is thereby enforced exactly.}
We point out that in this form, there are still derivatives of the transport
coefficients appearing in the source terms.
We will handle them in a special way in the next Section.

\subsection{High-resolution shock capturing}

\subsubsection{General considerations}

Magnetohydrodynamics features a wealth of strong-gradient and shock solutions,
both in the hydrodynamical and strongly magnetized limits.
While the inclusion of dissipative effects can efficiently regulate the steep
gradients present in these solutions, astrophysical dissipation is
known to vary strongly with local conditions, as reflected by the
density and temperature dependence introduced in our subgrid model.
As a result, many of the applications we target will have dynamically important
dissipative effects in small parts of the computational domain, whereas unresolved viscous effects elsewhere will be indistinguishable from ideal flows within the limits set by our numerical resolution (i.e., numerical viscosity).
We must therefore treat these equations with the same care as {typical} compressible hydrodynamic simulations \cite{Marti:1999wi}.

These schemes normally work by replacing gradient terms in flux-conservative
forms with solutions of the Riemann problem \cite{toro2013riemann}.
In particular, our numerical scheme {reconstructs
and solves the Riemann problem one coordinate direction at a time}, and we adopt
the {Eulerian conservative high-order (ECHO)} finite difference scheme \cite{DelZanna:2007pk},
which has been shown to work well in the context of
neutron star simulations \cite{Most:2019kfe}.
As shown in {Appendix~\ref{app:char-speeds}}, even with the light-cone
families set aside, there is no state-independent
ordering that selects one of the remaining families as ``the'' second-fastest
mode that could be used directly in an approximate Riemann solver. Better Riemann solvers \cite{Mattia:2021bwh} will be investigated in future work.

{Consequently, we adopt an approximate Riemann solver, namely the {Harten--Lax--van Leer--Einfeldt (HLLE)} flux formula \cite{harten1983upstream,einfeldt1988godunov}, which requires only one speed bounding every left-going mode and one bounding every
right-going mode, which we approximate to be null. {Future work will be devoted to building better approximate solvers that can leverage the Eulerian characteristic decomposition in Appendix~\ref{app:eigenvectors}.}
For now, adopting the HLLE flux, we write}
\begin{align}
  {F = \frac{c_{+}\,F_{L} + c_{-}\,F_{R}
  - c_{+}\,c_{-}\left(U_{R} - U_{L}\right)}{c_{+}+c_{-}}},
\end{align}
{where, for a face with unit normal {$s_i$},}
\begin{align}
 {\lambda_{\pm}}&{={-\beta^i s_i
 \pm\alpha\sqrt{\gamma^{ij}s_i s_j}}},\notag\\
 {c_+}&{=\max(0,\lambda_{+,L},\lambda_{+,R})},&
 {c_-}&{=\max(0,-\lambda_{-,L},-\lambda_{-,R})}.
\end{align}

{For causal reconstructed states, these bounds contain all material, electron, ion, electromagnetic, and cleaning characteristics. The reconstructed states must themselves be checked for causality.}  The expression reduces to the Rusanov, or local
Lax--Friedrichs {(LLF)}, flux only for $c_+=c_-\equiv c$.
We use an a priori positivity limiter for both conserved number currents \cite{2013JCoPh.242..169H}.  The
high-order interface flux is mixed with a local Lax--Friedrichs flux formed
from the two adjacent cell-centered states,
$F\rightarrow\theta F+(1-\theta)F_{\rm LLF}$.
{Here $0\le\theta\le1$. The low-order forward-Euler update must preserve both number densities at the chosen timestep.}
A single $\theta$ is applied to
every evolved row and is reduced until the one-face forward-Euler contribution
keeps both $\Db\sqrt\gamma$ and $D_e\sqrt\gamma$ non-negative in both adjoining
cells.
Here, the left and right fluxes, $F_{L}$, $F_{R}$, and states, $U_{L}$, $U_{R}$,
are obtained by reconstructing the primitives $V$ using a {{piecewise-linear method (PLM)} \cite{van1979towards} or {weighted essentially non-oscillatory (WENO-Z)} algorithm \cite{2008JCoPh.227.3191B}.
{The primitive variables $V$ are reconstructed directly at the interfaces from
their {cell-centered} values; the conserved face states $U_{L}$, $U_{R}$ are then
built from the reconstructed primitives (rather than interpolating $U$ from
{centers} and recovering $V$ afterwards), and enter only through the diffusive jump
term $U_{R}-U_{L}$ in the flux above.}
We have also implemented an optional local-frame HLLE solver \cite{Pons:1998ae}, using the tetrad of Ref. \cite{White:2016Athena}.}
The characteristic speeds $\lambda_\pm$ for our system are explicitly computed in Appendix~\ref{app:characteristics}.

\subsubsection{Transport gradient treatment}

As we can see in Eqs. \eqref{eq.bigPi}-\eqref{eq:electron-B-summary},
the principal part {on the left-hand side (LHS)} of the
dissipative sector, when recast using our prior approach \cite{Most:2021rhr},
naturally assumes the flux divergence form of {\myeqref{eq.fluxform}}.
However, in rewriting the equations, we have introduced three types of gradients
in the transport sector, which need to be handled {separately:}
\begin{itemize}
  \item {Velocity divergence: $\nabla_{\mu} \, u^\mu \, ,$}
  \item {Advective derivative: $u^{\mu} \, \nabla_{\mu} \, A \, ,$}
  \item {General gradient: $\nabla_{\mu} \, B \, .$}
\end{itemize}
Rather than employing commonly used finite difference approaches,
which fundamentally cannot easily treat time derivatives and will become
inaccurate in the presence of shocks, we adopt a different treatment here.

We begin with the velocity divergence to illustrate the approach.
We note that
\begin{align}
  u^{\mu} \, \nabla_{\mu} \, \rhob = - \rhob \nabla_{\mu} \, u^{\mu} \, ,
\end{align}
meaning that $\nabla_{\mu} \, u^{\mu}$ appears as the difference between
mass conservation and density advection.
That is, if we introduce a scalar, $Y_{\theta}$,
which {at the beginning of an {IMEX} stage} we initialize to $\rho$,
then we can estimate the difference between the two equations
{(and therefore estimate $\nabla_{\mu} \, u^{\mu}$)}
by computing
\begin{align}
  u^{\mu} \, \nabla_{\mu} \, Y_{\theta} = - \rhob \, \mathcal{I}_{\theta} \, ,
\end{align}
where $\mathcal{I}_{\theta}$ can be estimated numerically by implicit time
stepping, with
$\mathcal{I}_{\theta} \rightarrow \nabla_{\mu} \, u^{\mu}$.\footnote{
The implicit update takes backward-Euler form.  The exact stage relation,
including the lapse and densitization factors, is given in
Eq.~\eqref{eq.implyrhob}.}

We can trivially generalize this to the second type of gradient,
the advective derivative.
There, we introduce a variable, $y_{A}$, which relaxes to $A(\rho, T, \ldots)$,
and obtain
\begin{align}
  \nabla_{\mu} \left( \rhob \, u^{\mu} \, y_{A} \right)
  &= \rhob \, u^{\mu} \, \nabla_{\mu} \, y_{A}
  = - \omega_{A} \, \rhob \, \left( y_{A} - A\right) \notag \\
  &= \rhob \, \mathcal{I}_{A} \, ,
\end{align}
where
\begin{align}
  \mc{I}_{A} := u^{\mu} \, \nabla_{\mu} \, y_{A}
\end{align}
will be treated as a stiff source term in the above sense of
$\mathcal{I}_{\theta}$.
{The continuum rate $\omega_A$ includes a factor $u^0$ relative to the numerical rate $\omega$ in Appendix~\ref{app:relaxation}. At finite IMEX step size, the recovered sources are stage differences. They approach spacetime derivatives only as the timestep vanishes with a consistent transport predictor, even at infinite relaxation rate.}

The third gradient type is specific to the heat-flux equation.
To this end, we introduce a new variable,
$z^{\mu}_{\phantom{\mu}\nu}$, such that,
for some scalar field $B$,
$z^{\mu}_{\phantom{\mu}\nu} \rightarrow B \, \delta^\mu_\nu$.
We further denote $\alpha \, z^{0}_{\phantom{0}\nu} =: {Z^{B}_{\nu}}$.
Then,
\begin{align}
  \mathcal{I}^{B}_{\nu} := \nabla_{\mu} \, z^{\mu}_{\phantom{\mu}\nu}
\end{align}
defines the same type of implicit gradient estimate as above.

{Using this approach, the full set of equations we solve is}

\begin{widetext}
\begin{subequations}
\label{eq:implemented-system}
\begin{align}
  &\nabla_{\mu} \, {N_b^{\mu}} = 0 \, ,
    \label{eq:implemented-mass} \\
  &\nabla_{\mu} \left( T^{\mu\nu}_{\rm hydro}
    + T^{\mu\nu}_{\rm EM}
    + T^{\mu\nu}_{\rm diss} \right) = 0 \, ,
    \label{eq:implemented-stress} \\
  &\nabla_{\mu} \left(      F^{\mu\nu} + g^{\mu\nu} \, \psi \right)
    = {\kappa_{\rm dc}} \, n^{\nu} \, \psi - \mc{J}^{\nu} \, ,
    \label{eq:implemented-electric-maxwell} \\
  &\nabla_{\mu} \left( \starF^{\mu\nu} + g^{\mu\nu} \, \phi \right)
    = {\kappa_{\rm dc}} \, n^{\nu} \, \phi \, ,
    \label{eq:implemented-magnetic-maxwell} \\
  &{\nabla_{\mu} \left( \rho_{e} \, u^{\mu}
    - \frac{m_{e}}{q_{e}} \, \ju^{\mu} \right) = 0} \, ,
    \label{eq:implemented-electron-carrier} \\
  &{\nabla_{\mu} \left( e_{e} \, u^{\mu} + Q_{e}^{\mu} \right)
    = P_{e} \, \rhob^{-1} \, \mc{I}_{\rhob}
    + e_{\mu} \, \ju^{\mu}} \, ,
    \label{eq:implemented-electron-energy} \\
  &\nabla_{\mu} \left( \left( \Pi + y_{\zeta} \right) u^{\mu} \right)
    = -\frac{1}{\tau_{\Pi}} \, \Pi
    - \left( 1 - \frac{\dPiPi}{\tau_{\Pi}} \right) \Pi \, \rhob^{-1} \,
    \mc{I}_{\rhob} + \mc{I}_{\zeta} \, ,
    \label{eq:implemented-bulk} \\
  & \nabla_{\mu} \left( q_{\nu} \, u^{\mu} + y_{\kappa} \, T \,
    \Delta^{\mu}_{\phantom{\mu}\nu}\right)
    = - \frac{1}{\tauq} \, q_{\nu}
   - \left( 1 - \frac{\dqq}{\tauq} \right) q_{\nu} \, \rhob^{-1} \,
   \mc{I}_{\rhob}
    + T \, \mc{I}^{\kappa}_{\nu}
    - \omega_{gi} \, b^{\mu}_{\phantom{\mu}\nu} \, q_{\mu}
    {+ \omegaq \left(q_{\mu} \, u^{\mu}\right) u_{\nu}} \, ,
    \label{eq:implemented-heat} \\
  & \nabla_{\mu} \left( \pi^{\alpha\beta} \, u^{\mu}
    + 2 \, y_{\eta} \,
    g^{\mu\left(\alpha\right.} u^{\left.\beta\right)}\right)
    =  -\frac{1}{\tau_{\pi}} \, \pi^{\alpha\beta}
    - \left( 1 - \frac{\dpipi}{\tau_{\pi}} \right)
    \pi^{\alpha\beta} \, \rhob^{-1} \, \mc{I}_{\rhob}
    - \omega_{gi} \,
    b^{\delta\mu} \, \pi^{\gamma}_{\phantom{\gamma}\mu} \,
    \Delta^{\alpha\beta}_{\phantom{\alpha\beta}\gamma\delta} {-\omega_\pi\mathcal C_\pi^{\alpha\beta}} \, ,
    \label{eq:implemented-shear} \\
  & {\nabla_{\mu} \, \mc{B}^{\mu}_{\phantom{\mu}\nu}
    {\overset{\perp}{=}} - \frac{1}{\tauJ} \, k_{\nu} + \omega_{pe}^{2} \, e_{\nu}
    + \frac{\omega_{ge}}{w_{e}} \,
    b^{\mu}_{\phantom{\mu}\nu} \, k_{\mu}
    - \frac{q_{e}}{m_{e}} \, u_{\nu} \, e_{\mu} \, \ju^{\mu}
    {+ \omegaJ \, \left(k_{\mu} \, u^{\mu}\right) u_{\nu}}} \, ,
    \label{eq:implemented-current} \\
  & \nabla_{\mu} \left( \rhob \, y_{\zeta} \, u^{\mu} \right)
    = \rhob \, \mc{I}_{\zeta} \, ,
    \label{eq:implemented-yzeta} \\
  & {\nabla_{\mu} \left( \rhob \, y_{\kappa} \, u^{\mu} \right)
    = \rhob \, \mc{I}_{\kappa}} \, ,
    \label{eq:implemented-ykappa} \\
  & \nabla_{\mu} \left( \rhob \, y_{\rhob} \, u^{\mu} \right)
    = \rhob \, \mc{I}_{\rhob} \, ,
    \label{eq:implemented-yrhob} \\
  & \nabla_{\mu} \, z^{\mu}_{\phantom{\mu}\nu}
      = \mc{I}^{\kappa}_{\nu} \, ,
      \label{eq:implemented-zkappa}
\end{align}
\end{subequations}
\end{widetext}
The new conserved variables are
\begin{align}
  \Pi_{*} &:= W \left( \Pi + y_{\zeta} \right){\, ,} \\
  Q_{\nu} &:= q_{\nu} \, W
    - \, y_{\kappa} \, T \, \Delta^{\mu}_{\phantom{\mu}\nu} \, n_{\mu}
    \label{eq.Qnu} \, , \\
  \pi_{*}^{\alpha\beta} &:= \pi^{\alpha\beta} \, W
    - 2 \, y_{\eta} \,
    n^{\left(\alpha\right.} u^{\left.\beta\right)} \, , \\
  Y_{\zeta} &:= \Db \, y_{\zeta} \, , \\
  {Y_{\kappa}} & {:= \Db \, y_{\kappa}} \, , \\
  Y_{\rhob} &:= \Db \, y_{\rhob} \, , \\
  Z^{\kappa}_{\nu} &:= \alpha \, {z^{0}_{\phantom{0}\nu}} \, .
\end{align}

\subsection{{Time integration}}

The equations of GR19M naturally contain a series
of different timescales.
Some of these, including the {gyration timescale $|\omega_{ge/gi}|^{-1}$ (with period $2\pi/|\omega_{ge/gi}|$)},
which {mediates} the local degree of anisotropy relative to the magnetic field,
can be much shorter than the numerical timestep; i.e.,
{$|\omega_{ge/gi}|^{-1}\ll\Delta t$ in a common time convention}.
{The relaxation time $\tau_J$ and plasma time $\omega_{pe}^{-1}$ may also be shorter than a timestep. Electric conductivity, $\sigma_J=\omega_{pe}^2\tau_J$, fixes neither time separately. With finite current inertia, large $\sigma_J$ need not give $e^\mu\to0$: rapid electric oscillations can remain in the homogeneous Maxwell--current system. Recovering ideal MHD therefore requires an asymptotic ordering of the relevant timescales.} \\
Since these regimes naturally arise in astrophysical systems, we need a
numerical scheme that is capable of smoothly capturing this behavior.
In contrast to perturbative approaches around either limit \cite{Wright:2019blb,Lier:2025wfw},
we adopt a strategy previously employed in resistive \cite{Palenzuela:2008sf,Bucciantini:2012sm,Ripperda:2019lsi},
force-free \cite{Alic:2012df,Palenzuela:2012my}, and non-resistive dissipative MHD \cite{Most:2021rhr}.
Specifically, we use implicit numerical time integrators,
which allow us to overstep these scales efficiently.
Implicit source terms are marked in color in Appendix \ref{app:3p1}.

In this work, we use {IMEX} time integration for the
{stiff and non-stiff} sectors, respectively.
While we do not wish to repeat a full discussion of the implicit numerical
integration scheme here, the scheme schematically splits
the conserved state into variables with stiff source terms, $\mathcal{V}_i$,
and variables without them, $\mathcal{U}_i$.
In particular, we have $\mc{U}\left( \bs{x}, t \right)
:= \sqrt{\gamma} \, \left( \Db, \tau, S_{j}, S, B^{j},
 {D_{e}} \right)^{\T}$, and
$\mc{V}\left( \bs{x}, t \right)
:= \sqrt{\gamma} \, \left( E^{j}, {\phi,} {\psi,} \Pi_{*}, Q_{\mu}, \pi_{*}^{\alpha\beta}, Y_{\zeta}, Y_{\rhob}, Y_{\kappa}, Z^{\kappa}_{\mu},
{k_{*\mu}}, {\mathcal{E}_{e*}} \right)^{\T}$.

We can then split the
evolution equations,
\begin{align}
  \p_{t} \, \mc{U}_{i} &= \mc{H}_{i} \, , \\
  \p_{t} \, \mc{V}_{i} &= \mc{E}_{i} + \mc{I}_{i} \,,
\end{align}
where $\mc{H}_{i}$ and $\mc{E}_{i} $ are explicitly treated, and $\mc{I}_{i}$ is potentially stiff and requires implicit integration.

Schematically, and for the implicit variables only,
the time integration scheme takes the following form
\cite{pareschi2005implicit},
\begin{align}
  &\mathcal{V}^{\left(k\right)}_{i}
    = \mathcal{V}_{i} + {\Delta t}\sum_{l=1}^{k-1}
    a_{kl} \, \mathcal{E}_{i}\left(\mathcal{U}_{i}^{\left(l\right)},
    \mathcal{V}_{i}^{\left(l\right)}\right) \nonumber \\
    &\phantom{\mathcal{V}^{\left(k\right)}_{i} = \mathcal{V}_{i}}
    + {\Delta t}\sum_{l=1}^{k}
    \tilde{a}_{kl} \, \mathcal{I}_{i}\left(\mathcal{U}_{i}^{\left(l\right)},
    \mathcal{V}_{i}^{\left(l\right)}\right) \, , \\
  &\mathcal{V}_{i}\left(t+{\Delta t}\right)
    = \mathcal{V}_{i} + {\Delta t}\sum_{l=1}^{n}
    b_{l} \, \mathcal{E}_{i}\left(\mathcal{U}_{i}^{\left(l\right)},
    \mathcal{V}_{i}^{\left(l\right)}\right) \nonumber \\
    &\phantom{\mathcal{V}_{i}\left(t+{\Delta t}\right) = \mathcal{V}_{i}}
    + {\Delta t}\sum_{l=1}^{n} \tilde{b}_{l} \, \mathcal{I}_{i}
    \left(\mathcal{U}_{i}^{\left(l\right)},
    \mathcal{V}_{i}^{\left(l\right)}\right){\, ,}
\end{align}
where $a_{kl}$ and $\tilde{a}_{kl}$ are the explicit and implicit stage
coefficients, and $b_l$ and $\tilde{b}_l$ are their corresponding weights.
{Two integrators are available: the second-order IMEX SSP2(2,2,2) scheme of
Ref.~\cite{pareschi2005implicit}, and the split scheme of
Ref.~\cite{Bucciantini:2012sm}, hereafter IMEX12.  The latter combines
the second-order explicit midpoint rule with first-order backward-Euler solves
for the stiff sector. It is therefore first order in general when the implicit
source is nonzero, and is the most robust of the three integrators.  The double Butcher tableaux of the two-stage methods are given in
Appendix~\ref{app:butcher}.}

\noindent With that, the implicit equation in every substage can be written as
\begin{align}
  &\mathcal{V}^{\left(k\right)}_{i}
  = \mathcal{V}^{\ast\,\left(k\right)}_{i}
  + {\akk} \, \Delta t \,
  \mathcal{I}_{i}\left(\mathcal{U}_{i}^{\left(k\right)},
  \mathcal{V}_{i}^{\left(k\right)}\right) \, ,
  \label{eqn:ImEx1}
\end{align}
where we have introduced the shorthand
\begin{align}
  &\mathcal{V}^{\ast\,\left(k\right)}_{i}
    = \mathcal{V}_{i} + {\Delta t}\sum_{l=1}^{k-1}
    a_{kl} \, \mathcal{E}_{i}\left(\mathcal{U}_{i}^{\left(l\right)},
    \mathcal{V}_{i}^{\left(l\right)}\right) \nonumber\\
    &\phantom{\mathcal{V}^{\ast\,\left(k\right)}_{i} = \mathcal{V}_{i}}
    + {\Delta t}\sum_{l=1}^{k-1}
    \tilde{a}_{kl} \, \mathcal{I}_{i}\left(\mathcal{U}_{i}^{\left(l\right)},
    \mathcal{V}_{i}^{\left(l\right)}\right) \label{eqn:ImEx}
\end{align}
for the explicitly updated part of the variable.
We can see from \myeqref{eqn:ImEx1} that the implicit equation by construction
always takes backward-Euler form.

\subsubsection{{Implicit equations}}\label{sec:implicit}

We now explicitly list the set of implicit equations solved in this work.
The challenge will be, here and in the following, to numerically solve these
equations.
We will show that while most of these equations can be solved analytically,
assuming the velocity $u^{\mu}$ and the ion and electron temperatures $T_{e/i}$ are known,
{these five variables will have to be determined using a nonlinear root-finding
scheme, which we will outline later in this Section.}

Defining
$\widehat{\mc{V}}_{i} := \gamma^{-1/2} \, \mc{V}_{i}^{\ast\,\left(k\right)}$
(e.g., $\widehat{\mc{V}}_{1}
= \left(E^{1}\right)^{\ast\left(k\right)}$),
\begin{align}
  &{\phi=\widehat\phi-\alpha\akk\Delta t\,\kappa_{\rm dc}\phi,}
    \label{eq.implphi}\\
  &{\psi=\widehat\psi-\alpha\akk\Delta t\,\kappa_{\rm dc}\psi,}
    \label{eq.implpsi}\\
  & E^{i} = \widehat{E}^{i}
    - \alpha \, \akk \, \Delta t
    \left( \jn^{i} - v^{i} \, v_{k} \, \jn^{k} \right) \, ,
    \label{eq.implE} \\
  & W \left( \Pi + y_{\zeta} \right) = \widehat{\Pi}_{*} + \alpha \, \akk \, \Delta t
    \left( \mc{I}_{\zeta} - \Delta_{\Pi} \, \Pi \right) \, ,
    \label{eq.implPi} \\
  & q_{\nu} \, W + \alpha \, y_{\kappa} \, T \,
    \Delta^{0}_{\phantom{0}\nu} = \widehat{Q}_{\nu}
    - \alpha \, \akk \, \Delta t \left( \Delta_{\mathrm{q}} \, q_{\nu}
    - T \, {\mc{I}^{\kappa}_{\nu}} \right. \notag \\
    & \left.\hspace{3em} + \omega_{gi} \,
    b^{\mu}_{\phantom{\mu}\nu} \, q_{\mu}
    {- \omegaq \left( q_{\mu} \, u^{\mu} \right) u_{\nu}} \right) \, ,
    \label{eq.implq} \\
  & \pi^{\alpha\beta} \, W
    - 2 \, y_{\eta} \,
    n^{\left(\alpha\right.} u^{\left.\beta\right)} = \left(\widehat{\pi}_{*}\right)^{\alpha\beta} \notag \\
    &\hspace{3em}- \alpha \,
    \akk \, \Delta t \left[ \Delta_{\pi} \, \pi^{\alpha\beta}
    + \omega_{gi} \,
    b^{\delta\mu} \, \pi^{\gamma}_{\phantom{\gamma}\mu} \,
\Delta^{\alpha\beta}_{\phantom{\alpha\beta}\gamma\delta}\right. \notag \\
    &\left.\hspace{8em}{+\omega_\pi\mathcal C_\pi^{\alpha\beta}} \right] \, ,
    \label{eq.implpi} \\
  & W \left( k_{\nu} - \omega_{e} \, u_{\nu} \right)
    - \left( n_{\mu} \, k^{\mu} \right) u_{\nu} \notag \\
  &\quad + \chi_{e} \, n_{\nu} \overset{\perp}{=} \widehat{k}_{*\nu}
    - \alpha \, \akk \, \Delta t \, \Bigl(
    \frac{1}{\tauJ} \, k_{\nu}
    - \omega_{pe}^2 \, e_{\nu} \notag \\
  &\hspace{8em}
    - \frac{\omega_{ge}}{w_{e}} \,
    b^{\mu}_{\phantom{\mu}\nu} \, k_{\mu} \notag \\
  &\hspace{8em}
    + \frac{q_{e}}{m_{e}} \, u_{\nu} \, e_{\mu} \, \ju^{\mu} \notag \\
  &\hspace{8em}
    {- \omegaJ \, \left(k_{\mu} \, u^{\mu}\right) u_{\nu}} \Bigr) \, ,
    \label{eq.implj}
\end{align}
where $\jn^{k} := \gamma^{k\mu} \, \ju_{\mu}$
and
\begin{align}
  \Delta_{\Pi}
  &:= \frac{1}{\tau_{\Pi}}
  + \left( 1 - \frac{1}{\tau_{\Pi}} \, \dPiPi \right) \frac{1}{\rhob} \, \mc{I}_{\rhob} \, , \\
  \Delta_{\mathrm{q}}
  &:= \frac{1}{\tauq}
  + \left( 1 - \frac{1}{\tauq} \, \dqq \right) \frac{1}{\rhob} \, \mc{I}_{\rhob} \, ,
  \\
  \Delta_{\pi}
  &:= \frac{1}{\tau_{\pi}}
  + \left( 1 - \frac{1}{\tau_{\pi}} \, \dpipi \right)
  \frac{1}{\rhob} \, \mc{I}_{\rhob} \, .
    \label{eq:audit-relax-diagonal}
\end{align}
{We solve the projected current equation, Eq.~\eqref{eq:electron-projected-equality}, and recover the dependent stored component from Eq.~\eqref{eq:electron-dependent-current-moment}.}
{For a scalar tracker $Y_A=\Db y_A$, define its primitive predictor by
$y_A^{*,(k)}:=\widehat{Y}_A^{(k)}/\Db^{(k)}$, where
$\widehat{Y}_A^{(k)}=\gamma^{-1/2}\widetilde{Y}_A^{*,(k)}$.  Since $\Db$ has
no implicit source, it is unchanged during the implicit substage.  The generic
stage update is therefore}
{
\begin{align}
  \mc{I}_A^{(k)}
  &= \frac{Y_A^{(k)}-\widehat{Y}_A^{(k)}}
  {\alpha\,\akk\,\Delta t\,\rhob^{(k)}}
   = \frac{u^{0(k)}}{\akk\,\Delta t}
  \left(y_A^{(k)}-y_A^{*,(k)}\right),
  \label{eq.impltracker_generic}
\end{align}
}
{where $W=\alpha u^0$.  In the strong-relaxation limit,
$y_{\rhob}=\rhob$ and the transport trackers approach their local closure
values.}

The evolved transport ratios are
\begin{align}
  y_\eta&:=\frac{\eta}{\tau_\pi}, &
  y_\zeta&:=\frac{\zeta}{\tau_\Pi}, &
  y_\kappa&:=\frac{\kappaq}{\tauq}.
  \label{eq.transport_ratios}
\end{align}
The Cattaneo tracker is initialized as
\begin{align}
  {Z^\kappa_\mu}:=\alpha y_\kappa\delta^0_{\phantom{0}\mu},
  \label{eq:transport-cattaneo-tracker}
\end{align}
so only its time component is nonzero before each time step.  The same closure evaluation
is used to update these quantities after each successful primitive recovery and after
reconstruction or another state-altering correction, after which the causal bounds of
Sec.~\ref{sec:causality} need to be reapplied.
{Suppressing the stage label, the strong-relaxation limit gives}
{
\begin{align}
  \mc{I}_{\rhob}
  &= \frac{u^0}{\akk\,\Delta t}
  \left(\rhob-y_{\rhob}^{*}\right),
  \label{eq.implyrhob}\\
  \mc{I}_{\theta}:=-\frac{\mc{I}_{\rhob}}{\rhob}
  &= \frac{u^0}{\akk\,\Delta t}
  \left(\frac{y_{\rhob}^{*}}{\rhob}-1\right),\\
  \mc{I}_{\zeta}
  &= \frac{u^0}{\akk\,\Delta t}
  \left(\frac{\zeta}{\tau_\Pi}-y_{\zeta}^{*}\right),
  \label{eq.implyzeta}\\
  \mc{I}_{\kappa}
  &= \frac{u^0}{\akk\,\Delta t}
  \left(\frac{\kappa}{\tau_q}-y_{\kappa}^{*}\right),
  \label{eq.implykappa}\\
  \mc{I}^{\kappa}_{\nu}
  &= \frac{-n_{\nu}\,\kappa/\tau_q-\widehat{Z}^{\kappa}_{\nu}}
  {\alpha\,\akk\,\Delta t} \, .
  \label{eq.implZkappanu}
\end{align}
}

Solving Eqs.~\eqref{eq.implyrhob}-\eqref{eq.implZkappanu}
for $\mc{I}_{\rhob}$, $\mc{I}_{\zeta}$, $\mc{I}_{\kappa}$, and $\mc{I}^{\kappa}_{\nu}$
and substituting them into Eqs.~\eqref{eq.implPi}-\eqref{eq.implj},
{and projecting the heat and shear equations while imposing current orthogonality, gives}
\begin{align}
  \Pi &=
  \frac{\widehat{\Pi}_{*}
  - \widehat{Y}_{\zeta} / \rhob}
    {W + \alpha \, \akk \, \Delta t \, \Delta_{\Pi}} \, ,
    \label{eq.Pibulk} \\
  q_{\nu} &= \frac{\mc{Q}^{\mu}_{\phantom{\mu}\nu}
    \left( \widehat{Q}_{\mu} - T \, \widehat{Z}^{\kappa}_{\mu} \right)}
    {W + \alpha \, \akk \, \Delta t \, \Delta_{\mathrm{q}}} \, ,
    \label{eq.qproj} \\
  \pi^{\alpha\beta}
    &= \frac{\mc{P}^{\alpha\beta}_{\phantom{\alpha\beta}\gamma\delta} \,
    \Delta^{\gamma\delta}_{\phantom{\gamma\delta}\mu\nu} \,
    \left(\widehat{\pi}_{*}\right)^{\mu\nu}}{W
    + \alpha \, \akk \, \Delta t \, \Delta_{\pi}}     \label{eq.piproj}
\, ,\\
  &\begin{aligned}
    {\ju_\nu}&{=k_\nu/w_e\,,}\\
   \phi
  &= \frac{\widehat{\phi}}
  {1 + \alpha \, \akk \, \Delta t \, \kappa_{\rm dc}} \, ,\\  \psi
  &= \frac{\widehat{\psi}}
  {1 + \alpha \, \akk \, \Delta t \, \kappa_{\rm dc}} \, .
  \end{aligned}
\end{align}
where
\begin{align}
  \mc{Q}^{\mu}_{\phantom{\mu}\nu} &:= \Delta^{\mu}_{\phantom{\mu}\nu}
    - \frac{\varepsilon_{\mathrm{q}}^2}
    {1 + \varepsilon_{\mathrm{q}}^2} \,
    \Xi^{\mu}_{\phantom{\mu}\nu} - \frac{\varepsilon_{\mathrm{q}}}
    {1 + \varepsilon_{\mathrm{q}}^2} \,
    b^{\mu}_{\phantom{\mu}\nu} \, , \label{eq.heatfluxprojector} \\
  \left(\mc{Q}_{\textsc{j}}\right)^{\mu}_{\phantom{\mu}\nu}
  &:= \Delta^{\mu}_{\phantom{\mu}\nu}
    - \frac{\varepsilon_{\textsc{j}}^2}
    {1 + \varepsilon_{\textsc{j}}^2} \,
    \Xi^{\mu}_{\phantom{\mu}\nu} + \frac{\varepsilon_{\textsc{j}}}
    {1 + \varepsilon_{\textsc{j}}^2} \,
    b^{\mu}_{\phantom{\mu}\nu} \, ,\\
  \mc{P}^{\alpha\beta}_{\phantom{\alpha\beta}\gamma\delta} &:=
    b^{-4} \, b^{\alpha} \, b^{\beta} \, b_{\gamma} \, b_{\delta}
    + 2 \, b^{\left(\alpha\right.}
    \mathfrak{p}^{\left.\beta\right)}_{\phantom{\beta}\gamma\delta}
    + \mathfrak{B}^{\alpha\beta}_{\phantom{\alpha\beta}\gamma\delta} \, ,
\end{align}
with
\begin{align}
  {b^{2}} \, \mathfrak{p}^{\alpha}_{\phantom{\alpha}\gamma\delta}
    &:= \frac{ b_{\gamma} \, \Xi^{\alpha}_{\phantom{\alpha}\delta}
    - \varepsilon_\pi \, b_{\gamma} \,
    b^{\alpha}_{\phantom{\alpha}\delta} }{1 + {\varepsilon_\pi^2} } \, , \\
  \mathfrak{B}^{\alpha\beta}_{\phantom{\alpha\beta}\gamma\delta}
    &:= \frac{1 + 2 \, \varepsilon_\pi^2}{1 + 4 \, \varepsilon_\pi^2} \,
    \Xi^{\alpha}_{\phantom{\alpha}\gamma} \,
    \Xi^{\beta}_{\phantom{\beta}\delta}
    + \frac{2 \, \varepsilon_\pi^2}{1 + 4 \, \varepsilon_\pi^2} \,
    b^{\alpha}_{\phantom{\alpha}\gamma} \,
    b^{\beta}_{\phantom{\beta}\delta} \notag \\
    &\phantom{:==}
    - \frac{\varepsilon_\pi}
    {1 + 4 \, \varepsilon_\pi^2} \left(
    b^{\alpha}_{\phantom{\alpha}\gamma} \, \Xi^{\beta}_{\phantom{\beta}\delta}
    + b^{\beta}_{\phantom{\beta}\gamma} \, \Xi^{\alpha}_{\phantom{\alpha}\delta}
    \right) \,,\\
 & {\hspace{-3em}\left[\left(\mathcal{Q}_{\textsc{j}}^{-1}
    + X'\right)^{-1}\right]^{\nu}_{\phantom{\nu}\mu}}\notag\\
 & {= \left(\mathcal{Q}_{\textsc{j}}\right)^{\nu}_{\phantom{\nu}\mu}
    + \varepsilon_{\mathcal{Q}_\textsc{j}} \frac{ n^{\alpha}
    \left( \mathcal{Q}_{\textsc{j}}\right)^{\nu}_{\phantom{\nu}\alpha} \,
    n_{\beta} \, \left(\mathcal{Q}_{\textsc{j}}\right)^{\beta}_{\phantom{\beta}\mu}}
    {{W^{2}} - \varepsilon_{\mathcal{Q}_\textsc{j}} \, n_{\beta} \,
    \left(\mathcal{Q}_{\textsc{j}}\right)^{\beta}_{\cc{\phantom{\beta}}\alpha} \,
    n^{\alpha}}{\, ,}}\notag\\
\end{align}
and
\begin{align}
  \Xi^{\mu\nu}
    &:= \Delta^{\mu\nu} - \frac{1}{b^{2}} \, b^{\mu} \, b^{\nu} \, , \\
  \varepsilon_{\mathrm{q}}
    &:= \frac{\dtI \, \omega_{gi}}{W + \dtI \, \Delta_{\rm q}} \, , \\
  \varepsilon_{\pi}
    &:= {\frac{\dtI\omega_{gi}}{2(W+\dtI\Delta_\pi)}} \, , \\
  \varepsilon_{\textsc{j}}
    &:= \frac{\dtI \, {\omega_{ge} / w_{e}}}{W + \dtI \, \Delta_{\textsc{j}} / \tauJ} \, , \\
{\varepsilon_{\mathcal{Q}_\textsc{j}}} 
 &{:=\frac{W\,\dtI^2\omega_{pe}^2}
 {w_e\left(W+\dtI\Delta_{\textsc{j}}/\tauJ\right)}}\,,\notag\\
  \Delta_{\textsc{j}}
    &:= 1 + {\sigma_J} / w_{e} \, W \, \dtI \, .
\end{align}
{All current inverses act on $u^\perp$. We define $\bar n^\mu:=\Delta^\mu{}_{\alpha}n^\alpha$ and $X'^\mu{}_{\nu}:=-\varepsilon_{\mathcal Q_J}\bar n^\mu\bar n_\nu/W^2$. Crucially, the expression for $\varepsilon_{\mathcal Q_J}$ remains valid in the limit without Hall rotation.}
The numerical closure represents the same physical gyrofrequencies by the
{Hall transport coefficients (with inverse-field units before field normalization)}
\begin{align}
  \dqB&=\frac{\omega_{gi}\tauq}{\sqrt{b^2}}, &
  \dpiB&=\frac{\omega_{gi}\tau_\pi}{\sqrt{b^2}}, &
  \dJB&=\frac{\omega_{ge}\tauJ}{\sqrt{b^2}}.
  \label{eq.transport_hall}
\end{align}
All three are zero by default, in which case the associated rotations are
skipped.

\subsubsection{Primitive inversion and implicit solver}
\label{sec:c2p}

A crucial component of every relativistic MHD scheme is the recovery of the
primitive state $V$ from the evolved conserved state $U$.  Although the
primitive-to-conserved {(P2C)} relation $U(V)$ is algebraic, the conserved-to-primitive
(C2P) inversion generally requires a nonlinear root solve.  In ideal GRMHD the
problem can be reduced to a single equation on a known physical interval
\cite{Mignone:2007fw,Palenzuela:2015dqa,Kastaun:2020uxr,Noble:2005gf,Siegel:2017sav},
and not every conserved state admits a physical primitive state
\cite{Wu:2017grhd,WuTang:2017rmhd}.  The GR19M system is more restrictive
because the dissipative ion moments, electric current, and electron energy
must be recovered simultaneously.  Since this recovery is among the most
expensive parts of the evolution, physicality corrections are made only where
required and the nonlinear system is kept low dimensional.  For the C2P
inversion we set $\dtI=0$; at an IMEX stage, $\dtI>0$, the primitive variables
must also satisfy the implicit equation.  We solve both problems using the five
unknowns $X_5=(z_i,h_iW,w_eW)$.  The individual stages are briefly summarized as follows. Further
details are given in Appendices~\ref{app:dt0_admissibility} and
\ref{app:4D}.

\begin{enumerate}
    \item \textit{(Physicality enforcement and energy preconditioning).}
    Using the admissibility conditions of
    Appendix~\ref{app:dt0_admissibility}, we first test the input conservative
    state and, when necessary, project it in several steps.  At the grid level,
    a failed cell's electron and ion number densities are replaced by the
    smallest admissible positive pair from Eq.~\eqref{eq:failed-c2p-gauss-pair}.
    Whenever a same-level discrete charge can be obtained from the Gauss
    constraint (the temporal projection of Eq.~\eqref{eq.maxwellE}), this
    construction preserves that charge; see
    Appendix~\ref{app:coupled_energy_projection}. { That analytic energy interval applies to zero ion bulk pressure, heat flux and shear stress. With nonzero ion moments, admissibility must instead be checked through the full dissipative momentum/energy map.}
    {For that nondissipative specialization, once the number densities are corrected, the ion energy density is restricted to the interval}
    \eqref{eq.energy_projection_bounds} allowed at fixed mass and momentum, and
    the electron energy is reconstructed for the same flow using
    Eq.~\eqref{eq.energy_projection_electron}.  Finally, we enforce the
    {combined current bound \eqref{eq:electron-current-combined-bound},
    which includes characteristic causality, electron square-freeness,
    and the dominant energy condition}.

    \item \textit{(Conservative-to-primitive inversion).}
    We first set $\dtI=0$ and solve the C2P residual system,
    Eqs.~\eqref{eq.4D_residual} and \eqref{eq.4D_hot_residual_rows}.  The
    coupled ion--electron energy projection in
    Eqs.~\eqref{eq.energy_projection_clamp} and
    \eqref{eq.energy_projection_electron} provides a physical initial guess;
    its electron roots are enumerated by
    Eq.~\eqref{eq.dt0_electron_quadratic}.  The solution is obtained using the
    five-dimensional root-finding method of
    Appendix~\ref{app:4D_ladder}, with the unknowns defined in
    Eq.~\eqref{eq.4D_hot_unknowns_five}.
    In case of nonconvergence, we make a second attempt using the previous primitive state as an initial guess.

    \item \textit{(Implicit equation).}
    For $\dtI>0$, the C2P solution is used to initialize the coupled
    implicit stage in Eq.~\eqref{eqn:ImEx1}.  We solve the three momentum
    equations together with the ion- and electron-enthalpy equations.  We use a
    safeguarded Broyden iteration for Eqs.~\eqref{eq.4D_residual} and
    \eqref{eq.4D_hot_residual_rows}, restricted to the physical density,
    Lorentz-factor, and enthalpy domain \eqref{eq.4D_hot_cone}, with a
    safeguarded Picard direction from the same map whenever the Broyden step
    leaves this domain or fails to reduce the residual.  Success at this step
    requires convergence of both the electron and ion sectors.

    \item \textit{(Fallback strategy for the implicit equation).}
    If the direct implicit solve does not converge, we follow the physical
    branch continuously from the C2P solution to the full IMEX stage.  We write
    the residual equations as $F(X (\lambda),\lambda)=0$, with $\lambda\in[0,1]$
    multiplying the implicit timestep, i.e., $\lambda \dtI$.  The tangent satisfies
    $D_XF\,dX/d\lambda=-F_\lambda$, where $D_XF$ is the Jacobian, and each predicted state is corrected using the original five equations.  The step size in $\lambda$ is reduced whenever the
    correction leaves the physical domain or no longer follows the local
    tangent.  If the electron enthalpy reaches an EOS floor or ceiling, it is
    held on that boundary while the remaining four equations are solved.  A full solution is only accepted upon convergence of the full residual at $\lambda=1$. {By construction, this procedure
    selects the branch connected to the C2P solution.}
    We caution that the solution to the implicit equation is not unique in the
    electron sector (see the discussion around
    Fig.~\ref{fig.wald_multiroot_branches}).

   \item \textit{(Atmosphere, velocity, and magnetization floors).}
   A recovered state is restricted to the {prescribed ion rest-mass density and velocity}
   ranges and to the maximum magnetization,
   \begin{align}
    {\rho_{\rm lo}:=\max(\rho_{\rm atmo},b^2/\sigma_{\rm limit}),}\notag\\
     {\rhob\longrightarrow\min[\rho_{\max},\max(\rho_{\rm lo},\rhob)]}
     \quad{(\rho_{\rm lo}\le\rho_{\max}).}
     \label{eq.rhofloor}
   \end{align}

{If the ion rest-mass density is raised, the electron rest-mass density is adjusted}
   so as to preserve the incoming net charge density, $D_q^{\rm in}$,
   \begin{align}
     D_e^{\rm new}=\mu ZD_b^{\rm new}-\frac{m_e}{q_e}D_q^{\rm in}.
     \label{eq:fixed-charge-density-floor}
   \end{align}
   If the resulting carrier density lies below its atmosphere value,
   $D_b^{\rm new}$ is increased by the minimum additional amount required to
   support it, subject to $\rho_{\max}$.
   For a C2P inversion there is no implicit Amp\`ere or
   electron-energy source.  We therefore preserve the recovered electron specific energy, reconstruct the current for the charge-preserving density
   pair, and reapply the {combined current bound \eqref{eq:electron-current-combined-bound}}.  The state
   is rejected if the subsequent primitive-to-conserved conversion does not
   reproduce both species densities.

   \item \textit{(Causal regulation and thermodynamic floors).}
   The transport closure is recomputed from the recovered state.
   We re-enforce the algebraic constraints
   $u_\mu\pi^{\mu\nu}=0$, $\pi^\mu{}_\mu=0$, and
   $u^\mu q_\mu=u^\mu\ju_\mu=0$, followed by the norm bounds
   \begin{align}
    &{\frac{|\Pi|}{\rhob}\leq\sigma_\Pi^{\rm max},}\\
    &{\frac{\sqrt{\pi^{\mu\nu}\pi_{\mu\nu}}}{\rhob h}\leq\sigma_\pi^{\rm max},}\\
    &{\frac{\sqrt{q^\nu q_\nu}}{\rhob h}\leq\sigma_{\rm q}^{\rm max},}\\
    &{\sqrt{\ju^\nu\ju_\nu}\leq\sigma_{\textsc{j}}^{\rm max}J_{\rm ref}.}
   \end{align}
   where $\sigma^{\rm max}_A$ are problem-dependent upper bounds.
Require the electron bound \eqref{eq:electron-current-combined-bound}.
For the coupled system, use either the primitive simple-root condition
\eqref{eq.resgeneral.simple-reality-screen} with the current gap
\eqref{eq:electron-ion-acoustic-current-limit}, or the exact causal and
uniform-eigenbasis tests \eqref{eq.resgeneral.complete} and
\eqref{eq.ev.full-resolvent-sh}.  On the $M>0$ branch, the time floors
\eqref{eq.caus.resistive-primitive-cone-floors} give a separate sufficient bound.  After any correction, recompute the closure and redo these checks.
   Whenever $\tauq$ or $\tau_\pi$ is raised, $\dqB$ or $\dpiB$ is rescaled
  proportionally so that the physical ion gyrofrequency in
   Eq.~\eqref{eq.transport_hall} remains unchanged.
   In case of failure, reduce, in order, $\pi^{\mu\nu}$, the transport ratios,
   $q^\mu$, and $\Pi$, then recompute the floors and the exact
   all-direction tests.
   Every carrier or electron-energy update changes $w_e$. We therefore reconstruct
   $j_\mu=k_\mu/w_e$ and both bounds from the same primitive state.  The
   electron energy is first varied within its EOS interval at fixed current.
   If this does not open the required electron-ion gap, we scan the full velocity segment
   \eqref{eq.precall_current_segment}, couple $E^i$ through
   Eq.~\eqref{eq.ohm_ampere}, and reclose
   Eq.~\eqref{eq.ohm_electron_energy} at every trial.  The largest certified
   $s$ is retained, noting that the segment can have disconnected solutions.  The electric field must also satisfy
   Eq.~\eqref{eq.appendix_terminal_electric}.

   We finally impose the plasma-beta floor
   \begin{align}
     e_{\rm th} \ge \frac{\beta_{\rm floor} \, b^{2}}
     {2(\Gamma-1)}
     \quad &\Longrightarrow \quad
     p \ge \frac{\beta_{\rm floor}}{2} \, b^{2}\,,
     \label{eq.betafloor}
   \end{align}
   with $\beta_{\rm floor}=10^{-5}$.  This is the last thermodynamic change,
   after which the current--field--electron-energy replay is repeated.
   Conserved variables and the stiff source are rebuilt only from that final
   primitive, which is rechecked for thermodynamic and physical consistency.
   
   \item \textit{(Fallback hierarchy).}
   If all the attempts and stages above fail, we apply a terminal atmosphere fallback.
   The final atmosphere fallback is a stationary,
   carrier-supporting state with $E^i=E^i_{\rm ideal}=0$. {For evolution of the conserved variables alone, one may instead retain an
   IMEX predictor with zero stiff source when both species densities and the
   electron energy are positive, and the C2P of that predictor converges.}
\end{enumerate}

\subsection{Numerical infrastructure and performance portability}

Our GR19M implementation is based on our earlier numerical work done
in the context of large-scale force-free electrodynamics simulations of binary
spacetimes \cite{Most:2020ami,Most:2022ojl,Most:2024eig}.
The numerical infrastructure uses the massively parallel adaptive
mesh-refinement (AMR) infrastructure AMReX \cite{Zhang2019,2020arXiv200912009Z,2024arXiv240312179M}.
As such, our code is fully able to leverage the AMR capabilities and
performance portability of AMReX, and runs on both {central processing units (CPUs)} and GPUs.

\section{Validation and numerical results}
\label{sec:grplasmas}

Having presented the formulation, well-posedness conditions, and numerical method for the GR19M scheme, we now demonstrate the correctness of the implementation and compare it against a rigorous series of problems whose full solutions are known from kinetic GRPIC simulations.

\subsection{Dissipative Bondi accretion}
\label{sec.bondiDissipative}

A classical validation test for GRMHD codes is stationary, spherically symmetric
inflow onto a Schwarzschild black hole with a radial magnetic field.  The analytic ideal solution \cite{Bondi:1952ni,Michel:1972} can easily be computed (see, e.g., Ref.~\cite{White:2016Athena}).  Here, we generalize that problem to the GR19M system, including ion bulk pressure, heat flux, shear stress, electron number and energy, electric current, and Maxwell
equations.  We construct a transcritical stationary solution of this coupled
system and use it as a reference solution to assess the validity of our numerical implementation. A full derivation is provided in Appendix~\ref{app:bondi-reference}.

{We construct the stationary reference in Schwarzschild coordinates
$(t,r,\theta,\phi)$, with
${\rm d}s^2=-\mathcal A\,\mathrm dt^2+\mathcal A^{-1}\mathrm dr^2
+r^2(\mathrm d\theta^2+\sin^2\theta\,\mathrm d\phi^2)$,
as in Eq.~\eqref{eq:bondi-schwarzschild-metric}.}
Define $v:=u^r<0$, $R:=\sqrt{\mathcal A+v^2}$,
$\mathcal A=1-2M/r$, and let $s^\mu$ be the outward radial unit vector in the
fluid frame.  Spherical symmetry and the constraints
$u_\mu q^\mu=0$, $u_\mu\pi^{\mu\nu}=0$,
$\pi^{\mu\nu}=\pi^{\nu\mu}$, and $\pi^\mu{}_\mu=0$ give
\begin{align}
 q^\mu&=q_\parallel s^\mu,&
 \ju^\mu&=j s^\mu,\notag\\
 e^\mu&=E s^\mu,&
 b^\mu&=B s^\mu,
 \label{eq:bondi-main-heat}\\
 \pi^{\mu\nu}
 &=\varpi\left(s^\mu s^\nu
   -\frac{1}{2}\mathcal P_\perp^{\mu\nu}\right),
 &\mathcal P_\perp^{\mu\nu}
 &:=\Delta^{\mu\nu}-s^\mu s^\nu .
 \label{eq:bondi-main-shear}
\end{align}
Thus, the shear eigenvalues are
$(\varpi,-\varpi/2,-\varpi/2)$ and
\begin{align}
 p_\parallel=P+\Pi+\varpi,
 p_\perp&=P+\Pi-\frac{\varpi}{2},\notag\\
 \Delta P:=p_\perp-p_\parallel&=-\frac{3}{2}\varpi .&&
 \label{eq:bondi-main-pressures}
\end{align}
Here $q_\parallel=q^\mu s_\mu$ and $j=\ju^\mu s_\mu$ are the physical scalar
heat flux and conduction current. {The signed comoving radial magnetic
field is $B:=b^\mu s_\mu$, with $b_\mu b^\mu=B^2$.}
The source-free monopole has $B\propto r^{-2}$ and
exerts no net force on the aligned ideal solution.

{The ideal critical radius is used to fix the initial variables, but the full dissipative problem must be solved to determine the dissipative critical radii and accretion rate.} The infalling matter then develops nonideal contributions primarily through compression. We therefore review the ideal solution first. Following Ref.~\cite{White:2016Athena}, the ideal seed may be written with
\begin{align}
 n=\frac{1}{\Gamma-1},\qquad T=\frac{P}{\rhob}.
\end{align}
{Here $n$ is the polytropic index and $K_{\rm p}$ below is the polytropic constant.}
{For $M>0$, $K_{\rm p}>0$, and $1<\Gamma\le2$, a chosen ideal critical radius $r_c>(n+3)M/2$ fixes}
\begin{align}
 u_c^r&=-\sqrt{\frac{M}{2r_c}},\notag\\
 T_c&=\frac{n}{n+1}
 \frac{(u_c^r)^2}{1-(n+3)(u_c^r)^2},\notag\\
 C_1&=T_c^n u_c^r r_c^2,\notag\\
 C_2&=\left[1+(n+1)T_c\right]^2
 \left[\mathcal A(r_c)+(u_c^r)^2\right].
\end{align}
At any radius, $T$ is a positive root of
\begin{align}
 \left[1+(n+1)T\right]^2
 \left[\mathcal A(r)+\frac{C_1^2}{r^4T^{2n}}\right]=C_2,
 \label{eq:bondi-white-temperature}
\end{align}
where the transonic branch uses the lower-temperature root inside $r_c$ and
the higher-temperature root outside it.  The remaining fields are
\begin{align}
 v=\frac{C_1}{r^2T^n},\qquad
 \rhob=\left(\frac{T}{K_{\rm p}}\right)^n,\qquad
 \dot M_{\rm ideal}=-\frac{4\pi C_1}{K_{\rm p}^n}.
 \label{eq:bondi-white-state}
\end{align}
This ideal solution supplies the starting point for the dissipative continuation.  The
dissipative problem is posed on a finite radial interval.  At its outer
boundary we fix the ion density and temperature and place the heat flux and
bulk pressure at their local Navier--Stokes targets.  The heat condition
determines the constant Killing-energy flux
$\dot{\mathcal E}=4\pi r^2T^r{}_t$, from which $q_\parallel$ is reconstructed
throughout the domain {using Eq.~\eqref{eq:bondi-heat-reconstruction}}.

All three ion dissipative channels use the same local collision time and
dimensionless amplitude:
\begin{align}
 \kappa&=fP\tau_i,& \eta&=fP\tau_i,& \zeta&=fP\tau_i,\notag\\
 \tau_\Pi&=\tau_q=\tau_\pi=\tau_i.
 \label{eq:bondi-main-transport}
\end{align}
To demonstrate that the code can handle complex nonlinear relaxation times,
{we use a Coulomb collision-inspired test closure, writing the ion collision time for $T=P/\rhob$} \cite{kulsrud2020plasma},
\begin{align}
 \tau_i=\frac{12\pi^{3/2}}{\ln\Lambda}
 \left(\frac{q_e}{m_e}\right)^{-4}\mu^{-4}\,{m_i^{-1}}
 \frac{T^{3/2}}{Z^4\rhob}\,,
 \label{eq:bondi-main-taui}
\end{align}
{where $\ln \Lambda \simeq 10$ is the Spitzer logarithm and $\mu=m_e/m_i$.}
The electron thermodynamics and the enthalpy-weighted current are
\begin{align}
 P_e&=\bar n_eT_e,&
 e_e&=m_e\bar n_e+\frac{\bar n_eT_e}{\Gamma_e-1},\notag\\
 H_e&=e_e+P_e,&
 w_e&=\frac{H_e}{m_e\bar n_e},\notag\\
 k&=w_ej,& K&=\frac{m_e}{q_e}k.
 \label{eq:bondi-main-electron-eos}
\end{align}
{The collision time follows the same collision model, while the electric response is rescaled for this test problem:}
\begin{align}
 \tauJ&=\tau_e=Z^2\sqrt{\frac{\mu}{2}}\,\tau_i,&
 \omega_{pe}&=\frac{q_e/m_e}{d_0}
   \sqrt{\frac{\bar n_e}{n_0}},\notag\\
 \sigma_J&=\omega_{pe}^2\tauJ,
 \label{eq:bondi-main-electron-transport}\\
 \delta_{qB}=\delta_{\pi B}
 &=\frac{Zq_e}{m_i}\tau_i,&
 \delta_{JB}&=\frac{q_e}{m_e}\tauJ.
 \label{eq:bondi-main-gyro-parameters}
\end{align}
{Here $d_0$ is the prescribed electric-response length and $n_0$ is its reference number density; their numerical values are specified below.}

The stationary spherical reduction of the coupled system follows directly
from the flux-divergent equations.  With
$\delta_{\Pi\Pi}=\tau_\Pi$, $\delta_{qq}=\tau_q$, and
$\delta_{\pi\pi}=\tau_\pi$, its dissipative, current, carrier, temperature,
and Maxwell rows are
\begin{align}
 0&=\tau_\Pi v\Pi'+\Pi+\zeta\theta+\tau_\Pi\Pi\theta,
 \label{eq:bondi-main-bulk-row}\\
 0&=\tau_qvq_\parallel'+q_\parallel+\kappa(RT)'
       +\tau_qq_\parallel\theta,
 \label{eq:bondi-main-heat-row}\\
 0&=\tau_\pi v\varpi'+\varpi+2\eta\sigma_\parallel
       +\tau_\pi\varpi\theta,
 \label{eq:bondi-main-shear-row}\\
 0&=vk'+\theta k+kv'
 -\frac{q_e}{m_e}\left(H_eR'+RP_e'\right)
 +\frac{k}{\tauJ}-\omega_{pe}^2E,
 \label{eq:bondi-main-current-row}\\
 0&=v\bar n_e'+\bar n_e\theta
 -\frac{1}{q_e}\left[(jR)'+\frac{2jR}{r}\right],
 \label{eq:bondi-main-number-row}\\
 0&=\left(v\bar n_e-\frac{\Gamma_ejR}{q_e}\right)T_e'
 -(\Gamma_e-1)vT_e\bar n_e'
 -(\Gamma_e-1)Ej,
 \label{eq:bondi-main-temperature-row}\\
 0&=E'+\frac{2E}{r}+\frac{j}{v}.
 \label{eq:bondi-main-gauss-row}
\end{align}
where
\begin{align}
 \theta=v'+\frac{2v}{r},\qquad
 \sigma_\parallel=\frac{2}{3}\left(v'-\frac{v}{r}\right),\notag\\
 R'=\frac{M/r^2+vv'}{R}.
\end{align}

Equivalently, for $X\in\{\Pi,q_\parallel,\Delta P\}$,
\begin{align}
 \frac{dX}{dr}=-\frac{1}{v}
 \left[\frac{X-X_{\rm NS}}{\tau_X}+X\theta\right],
 \label{eq:bondi-moment-derivatives}
\end{align}
with
\begin{align}
 \Pi_{\rm NS}&=-\zeta\theta,&
 q_{\rm NS}&=-\kappa(RT)',&
 \Delta P_{\rm NS}&=3\eta\sigma_\parallel.
\end{align}
{The subscript ${\rm NS}$ denotes the corresponding Navier--Stokes target.}
These equations remain coupled rather than forming independent {ordinary differential equations (ODEs)}; for example, compression introduces the velocity derivative into the system.
We can apply some simplifications before the system is solved.
First, strict stationarity of the radial electric
field requires $\mathcal J^r=0$.  With $n_Z:=Z\rhob/m_i$ and
$\rhoq=q_e(n_Z-\bar n_e)$,
\begin{align}
 0&=\rhoq v+jR,&
 \bar n_e&=n_Z+\frac{jR}{q_ev}
 =n_Z+\frac{kR}{q_ev w_e}.
 \label{eq:bondi-main-carrier-closure}
\end{align}
Consequently, the radial electron-number flux is
$r^2(\bar n_ev-jR/q_e)=r^2n_Zv$, already constant by ion mass conservation,
so Eq.~\eqref{eq:bondi-main-number-row} is imposed algebraically.
The last expression in
Eq.~\eqref{eq:bondi-main-carrier-closure} is implicit because $w_e$ depends on
$\bar n_e$; {the gamma-law closure gives the explicit charge-neutral branch
\mbox{\eqref{eq:bondi-ee-carrier-root}}, while the other quadratic root approaches
$H_e=0$ and is inadmissible.}

Together with the Killing-energy and radial-momentum equations, the system can then be solved with suitable boundary conditions. Analogously to the ideal Bondi solution, the setup features several critical points. {At a critical point, the stationary derivative matrix is singular. Its determinant gives the candidate critical speeds. A finite crossing also requires the source to be orthogonal to the left null vector.} {For $r_c>2M$ and an inward longitudinal characteristic root $-1<\hat z<0$ evaluated on the local state,}
\begin{align}
 \hat z=\frac{v_c}{R_c},\qquad
 v_c=\hat z\sqrt{\frac{\mathcal A(r_c)}{1-\hat z^2}}\,.
 \label{eq:bondi-main-critical-speed}
\end{align}
The ion quartic in Appendix~\ref{app:char_causality} supplies the ion acoustic
and second-sound candidates, while the electron cubic supplies the
electron-acoustic candidates. {Simple roots are locally analytic for analytic coefficients, but this can fail at degeneracies. The critical radii are eigenparameters of the transcritical boundary-value problem.}
The system is solved numerically as a differential-algebraic equation using the approach outlined in Appendix~\ref{app:bondi-dae}.

\paragraph{Numerical parameters and results.}
For the validation problem considered here, we now adopt a specific set of parameters.  {We write $r_g:=GM/c^2$ for the gravitational radius.} We take
$\Gamma=4/3$, $\Gamma_e=5/3$, $K_{\rm p}=6^{-1/3}$, and $f=0.1$, on
$r_{\rm in}=2.01r_g\le r\le r_{\rm out}=160r_g$.  The ideal seed has
$r_c=5.152976303711r_g$ and
$\dot M_{\rm ideal}=\,{3.29540949070849}$. {At fixed temperature and velocity, this seed increases the $K_{\rm p}=1$ density and mass flux by six and the monopole flux by $\sqrt6$.} The monopole is normalized to
$\beta=2P/B^2=0.1$ at the ideal critical point. For the closure, we adopt
$m_i=\mu=q_e/m_e=Z=1$ and $\ln\Lambda=10$. {The specified effective response length is $d_0=5\times10^{-3}r_g$, independently normalized at the critical-radius number density $n_0=6.0927516\times10^{-2}r_g^{-3}$.}

At the outer boundary of the coupled reference, we set
$\rhob=7.8551347579\times10^{-3}$ and
$T_i=1.09395837263986\times10^{-1}$, impose $T_e=T_i$, place $\Pi$ and
$q_\parallel$ at their local Navier--Stokes targets, and set $k=0$.

As a reference solution, we adopt the following parameters:
\begin{align}
 \dot M&=2.7885883176,&
 \dot{\mathcal E}&=3.9751794909,\notag\\
 \frac{r_{c,e}}{r_g}&=3.3405183,&
 \frac{r_{c,i}}{r_g}&=4.4364586,\notag\\
 \frac{r_{c,q}}{r_g}&=18.0554749.
 \label{eq:bondi-main-hot-reference}
\end{align}
From small to large radii, these are the electron acoustic, the ion acoustic and the second sound points.  Although the electric sector is
weak for these parameters, it is not identically neutral:
$\max|E|=1.88\times10^{-7}$,
$\max|k|=4.35\times10^{-8}$, and
$\max|\rhoq|=1.90\times10^{-8}$.
With these values, the solution both remains causal and satisfies the dominant energy condition throughout the domain.
Appendix~\ref{app:bondi-reference} gives the boundary
conditions, critical regularity construction, and numerical audit.

For the chosen normalization $\tauJ=\tau_i/\sqrt{2}$; at $r=3r_g$,
$\tau_i=14.61r_g/c$, $\tauJ=10.33r_g/c$, and
$\sigma_J=3.53\times10^5$.  At the same radius,
$\omega_{gi}\tau_i=14.3$ and $\omega_{ge}\tauJ=10.1$, with the latter reduced
to $5.1$ by the enthalpy factor.  The far outer profile is not uniformly in
the $\omega_g\tau\gg1$ limit, so the closure there is more precisely
Braginskii-like.  Although all gyro-rotation terms vanish on the aligned
stationary profile, transverse perturbations retain these coefficients.

{We evolve this reference solution in axisymmetry on the $(\varpi,z)$
half-plane,}
\begin{align}
 0\leq\varpi/r_g\leq40,\qquad -40\leq z/r_g\leq40,
\end{align}
using one uniform $256\times512$ level
($\Delta\varpi=\Delta z=0.15625r_g$), piecewise-linear reconstruction, the
{IMEX12} integrator, and {Courant--Friedrichs--Lewy (CFL) number} $\cfl=0.2$, through $t=600r_g/c$.  The ion relaxation
time is $\tau_i\simeq16$--$27r_g/c$ over $4\le r/r_g\le30$.
{Because the physical electric and current sectors are smaller than the
truncation residual of the analytic profile on this grid, we adopt a simple
well-balanced prescription \cite{2022LRCA....8....2K}.  The complete discrete residual of the
initial exact state is stored and subtracted from the explicit stages at every timestep.
The fixed radial monopole is likewise removed from the Amp\`ere background flux.}

\begin{figure*}[!tp]
 \centering
 \includegraphics[width=0.96\textwidth]{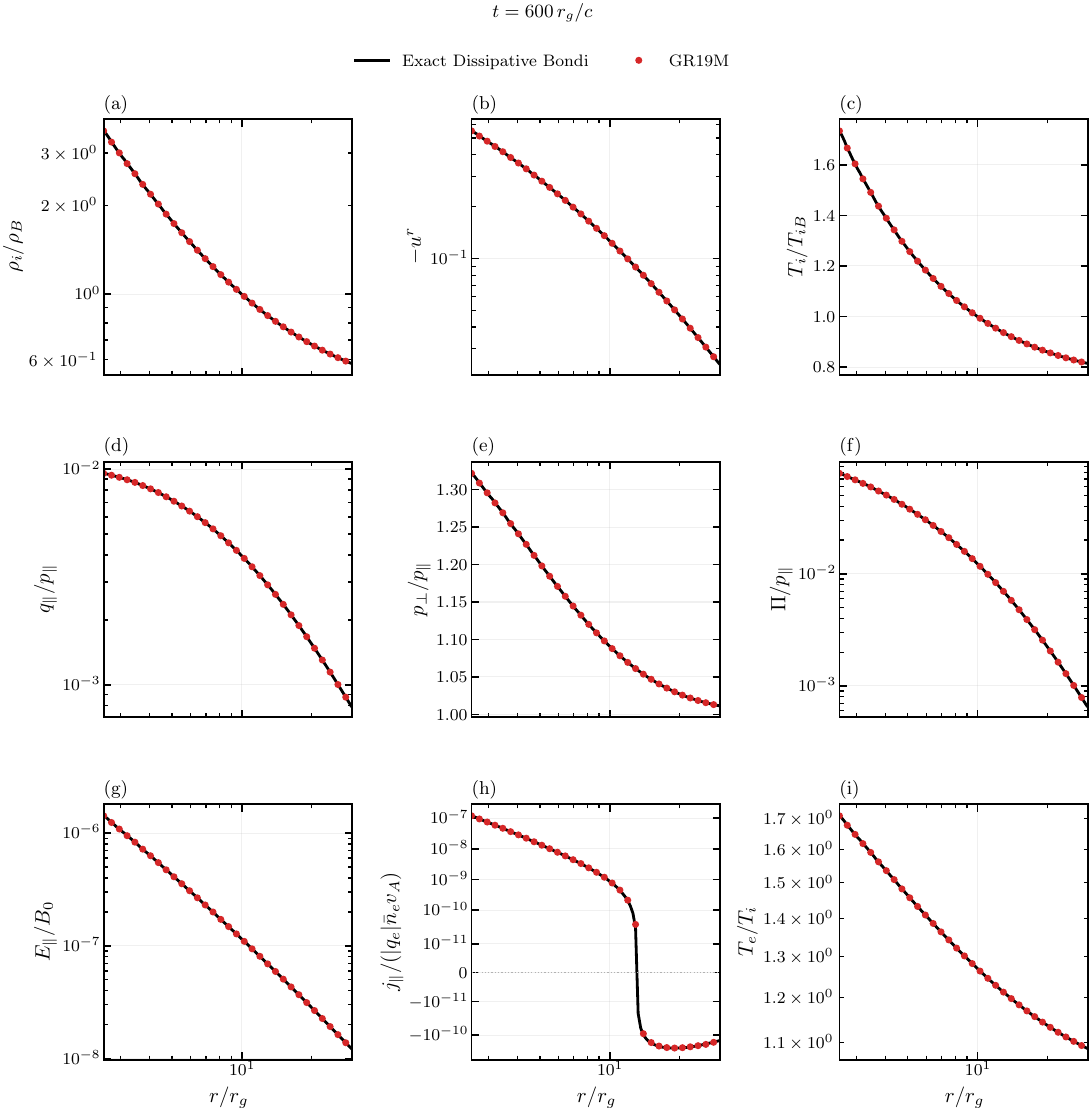}
 \caption{{Dissipative Bondi solution for $f=0.1$ at
 $t=600r_g/c$.  {Solid black curves are the exact Bondi reference solution and red
 circles are proper-volume-weighted shell averages from the full {GR19M} simulation.}
 The panels show ion density, {$\rhob/\rho_B$}, radial velocity $-u^r$, ion temperature $T_i/T_{iB}$,
 parallel heat flux $q_\parallel/p_\parallel$, {perpendicular-to-parallel matter-pressure ratio} $p_\perp/p_\parallel$,
 bulk scalar pressure $\Pi/p_\parallel$, parallel electric field $E_\parallel/B_0$,
 parallel dissipative electric current $j_\parallel/(|q_e|\bar n_e {v_{\rm A}})$, and electron temperature $T_e/T_i$.
 Here the {Bondi radius} is $r_B=10r_g$, $\rho_B:={\rhob}(r_B)$ is the reference ion rest-mass density,
 $T_{iB}:=T_i(r_B)$, $B_0:=B(r_B)$ is the initial magnetic field strength, and {${v_{\rm A}}=[B^2/(w_i+B^2)]^{1/2}$, with $w_i:={\rhob}+e_{\rm th}+p={\rhob} h$ the ion thermal enthalpy density, is the nominal Alfv\'en speed. {Here $e_{\rm th}$ and $p$ are the ion thermal internal-energy density and pressure.}}}}
 \label{fig.bondiDissipative}
\end{figure*}

Figure~\ref{fig.bondiDissipative} shows that the evolved solution reproduces
the reference outside the excised region.  On $2.5\le r/r_g\le30$, the
proper-volume-weighted relative $L_2$ errors in
$({\rhob},-u^r,T_i,q_\parallel,\varpi,\Pi)$ are
$(2.90\times10^{-8},6.74\times10^{-9},6.83\times10^{-8},
4.76\times10^{-7},9.44\times10^{-8},7.81\times10^{-8})$.
For $(E_\parallel,k_\parallel,e_e/(m_e\bar n_e)-1)$ they are
$(6.07\times10^{-7},1.42\times10^{-3},2.24\times10^{-8})$. Overall, we find excellent agreement in all quantities, confirming that our numerical evolution code fully reproduces two-fluid GR19M solutions.

\subsubsection{Bondi-like accretion onto a Kerr black hole}
\label{sec:collisionless-accretion-surrogate}

We next investigate Bondi-like accretion onto a spinning Kerr black hole.
This scenario has been extensively investigated using GRPIC simulations by
\citet{Galishnikova:2023} and provides a useful comparison point for
our assessment of the {GR19M} code.

\begin{figure}[!tbp]
 \centering
 \includegraphics[width=0.88\linewidth,height=0.72\textheight,keepaspectratio]
 {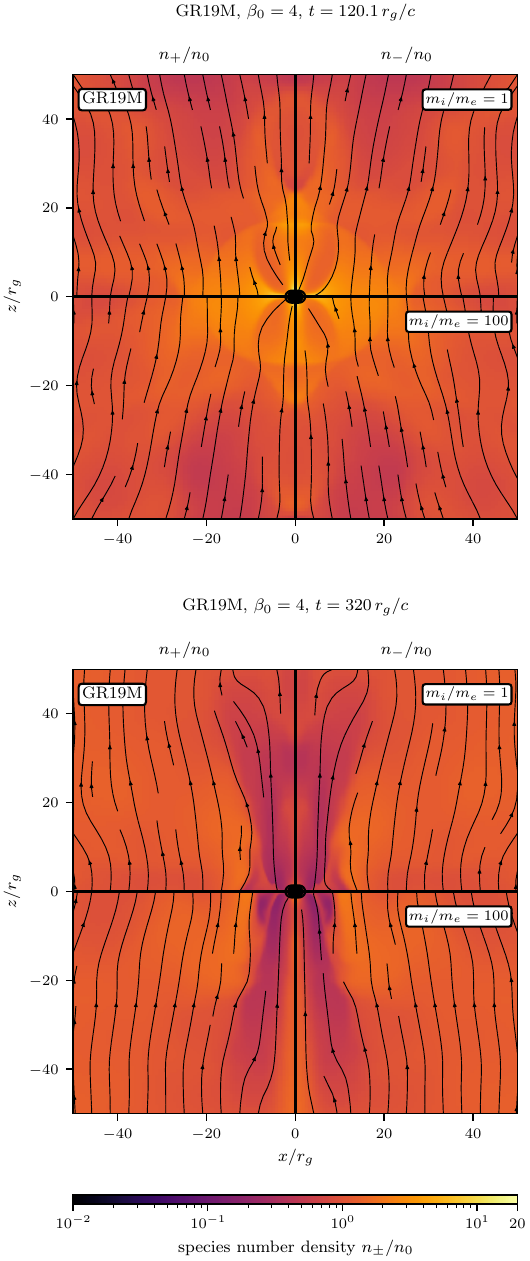}
 \caption{{Weakly collisional Bondi-like accretion flow onto a Kerr black hole
 at {initial total plasma beta} $\beta_0=4$ and $t=120.1\,r_g/c$ (top) and $t=320\,r_g/c$ (bottom), with
 $r_g=GM/c^2$.
 {Each panel has four quadrants.  Across $x=0$: the ion number density
 $n_+/n_0$ on the left and the electron number density $n_-/n_0$ on the right,
 normalized to the {homogeneous initial density $n_0=1$}.  Across $z=0$:
 the equal-mass pair run, $m_i/m_e=1$, above and an otherwise identical
 unequal-mass run, {$m_i/m_e=100$}, below, as labeled.  Both evolve the
 half-plane
 $\varpi\geq0$, so only $x<0$ is mirrored, and each half shows the corresponding
 run's actual hemisphere; all four quadrants share one color scale.}
 Black curves are poloidal magnetic-field lines of the run in that half, and
 the filled black region is the event-horizon interior.}}
 \label{fig:collisionless-beta1-morphology}
\end{figure}

The fixed background is now a Kerr black hole with $M=1$ and spin $a=0.95$,
written in ingoing Kerr--Schild coordinates
{\cite{Gammie:2003rj}, with the cylindrical metric given in
Appendix~\ref{app:fidotetrad}}.  We evolve the complete
meridional half-plane in axisymmetric cylindrical coordinates,
\begin{equation}
 0\leq\varpi/r_g\leq100,\qquad -100\leq z/r_g\leq100 .
\end{equation}
The homogeneous reference plasma has
\begin{align}
 n_i&=1,&m_i&=1,&m_i/m_e&=1,\notag\\
 T_i&=T_e=0.02,&P_i&=P_e=0.02,&\Gamma&=5/3 ,
\end{align}
so the initialized total isotropic pressure is $P_0=0.04$. In the {GR19M}
state, $\rho_b=n_i m_i=1$ is the ion rest-mass density used by the carrier
closure.  The bulk stress tensor does not separately add the conserved
electron rest mass.  Thus, for this equal-mass choice, the evolved bulk
rest-mass term is one half of the intended two-species rest mass, and
$P_0/\rho_b=0.04$ rather than $P_0/[n_i(m_i+m_e)]=0.02$. 

The initially static state is understood in the normal-observer frame,
$Wv^i=0$.  Inside $r=6M$, density, pressure, and the vector potential are
multiplied by
\begin{equation}
 f_{\rm h}(r)=\exp\left[5\left(1-\frac{6M}{r}\right)\right],
\end{equation}
while the initial floors are
$\rho_b\geq10^{-6}\rho_{b,0}r^{-3/2}$ and
$P\geq3.33\times10^{-9}P_0r^{-5/2}$, with the radius clamped at the
horizon for excised cells. {Here $\rho_{b,0}=1$ is the initial homogeneous ion rest-mass density.} A pressure floor additionally imposes
$P/(b^2/2)\geq10^{-3}$ on the initialized state.  The magnetic field is the
curl of
\begin{equation}
 A_\phi=B_0 f_{\rm h}(r)
 \left[\frac{r^2\sin^2\theta}{2}
 +k_{\rm loop}\sum_{n=1}^{1000}A_n(\varpi,z)\right],
 \label{eq:collisionless-accretion-aphi}
\end{equation}
where {$B_0=\sqrt{2P_0/\beta_0}=0.1414214$} and
$k_{\rm loop}=0.08473782435$. {Here $k_{\rm loop}$ is the loop-amplitude coefficient.} The compact, same-sign $A_n$ form a
deterministic catalogue of 1000 loops with radii in $[1,20) r_g$ and centers
outside $r=20 r_g$; the normalization makes the loop and background proxy
magnetic energies equal on the base-grid quadrature.  
{In the meridional coordinates $(x,z)=(r\sin\theta,r\cos\theta)$, used here as
transformed diagnostic coordinates while the native cylindrical grid radius is
$\varpi=\sqrt{r^2+a^2}\sin\theta$, each loop has
$A_n=r(L_n-d_n)\exp[1-L_n^2/(L_n^2-d_n^2)]$ for $d_n<L_n$ and vanishes
otherwise, where $d_n=[(x-x_n)^2+(z-z_n)^2]^{1/2}$.  We draw $L_n$ uniformly
from $[1,20)r_g$, $x_n$ uniformly from $[20r_g,100r_g-L_n]$, and $z_n$
uniformly over the interval that keeps the full support inside $r=100r_g$.
The catalogue is fixed and common to both mass ratios; all vector-potential
amplitudes have the same sign.}
Charge, conduction
current, heat flux, and pressure anisotropy initially vanish, and the
electric field is initialized from the ideal condition.

We retain the ion Larmor radius $\rho_{L,i}=0.018 r_g$ and derive the carrier
parameters as follows.  For the present
{$m_i/m_e=1$, $\beta_0=4$ case, we have
\begin{equation}
 \frac{|q_e|}{m_e}=\frac{|q_i|}{m_i}=55.6,\qquad
 d_e=d_i=0.018 r_g.
\end{equation}}
This choice self-consistently gives the conductivity
\begin{equation}
 \sigma_J=\min\left(\omega_{pe}^2\tau_J,10^5\right) c/r_g.
 \label{eq:collisionless-accretion-conductivity}
\end{equation}
We cap its maximum value at $10^{5}$ for numerical stability; the
initial value is {$\sigma_J=1543.2 c/r_g$}.
 All four dissipative sectors use the
fixed effective relaxation time
$\tau_J=\tau_\Pi=\tau_q=\tau_\pi=0.5 r_g/c$, {and the shear and heat-conduction
drives carry the multiplicative prefactors $f_\pi=6.75$ and $f_q=1$}.
  Hard firehose, mirror, and free-streaming
limiters constrain the recovered pressure anisotropy and heat flux \cite{Foucart:2015cws},
\begin{equation}
 -b^2\leq\Delta P\leq\Delta P_{\rm mirror},\qquad
 |q|\leq\rho_b c_s^3 .
\end{equation}
{For the implemented mirror bound,
$\Delta P_{\rm mirror}=-(b^2+3P)/2+
\left[(b^2+3P)^2/4+3b^2P/2\right]^{1/2}$, and $c_s$ is the ion sound speed.}

We adopt a $64\times128$ base grid, with five additional fixed mesh
refinement levels centered on the black hole, leading to the finest resolution
of $\Delta x=0.049\,r_g$.  We use the IMEX12
integrator with a fixed step $\Delta t=0.1\,r_g/c$. {The equal-mass run continues to $t=520\,r_g/c$ and the unequal-mass run to $t=320\,r_g/c$. The snapshot comparisons and decay fits use the common interval ending at $320\,r_g/c$.}

We begin by describing the overall flow morphology of the Bondi-like accretion flow onto a Kerr black hole (Fig.~\ref{fig:collisionless-beta1-morphology}). As magnetic flux is advected inward, a magnetized funnel forms and the system transitions into a {magnetically arrested disk (MAD)} regime at late times \cite{Galishnikova:2023}.
The two charge carriers track one another closely, as they must for the
equal-mass pair configuration adopted here.

\begin{figure}[!tbp]
 \centering
 \includegraphics[width=\linewidth,height=0.72\textheight,keepaspectratio]
 {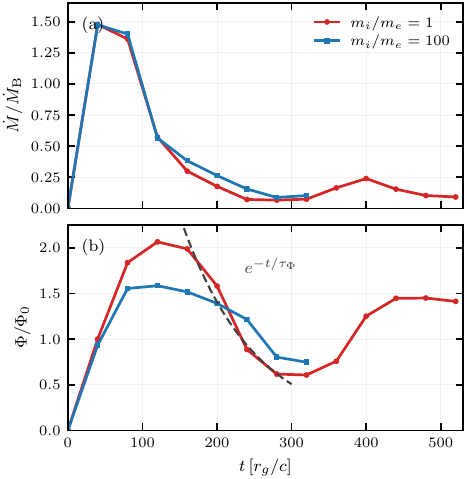}
 \caption{Horizon diagnostics for the $\beta_0=4$ model, for two species mass
 ratios: the equal-mass pair run, $m_i/m_e=1$ (red circles), and the otherwise
 identical unequal-mass run, {$m_i/m_e=100$} (blue squares).
 Panel (a) shows the mass accretion rate, $\dot{M}$, through the
 outer event horizon, normalized to the Bondi rate of the initial
 state, $\dot M_{\rm B}$.  Panel (b) shows the magnetic
 flux threading the horizon,
 $\Phi=\tfrac12\oint|B^r|\sqrt{-g}\,{\rm d}\theta\,{\rm d}\phi$.  Both curves are normalized to
 the same $\Phi_0=44.5$ so that the
 panel compares absolute horizon fluxes. The black hole features a flux decay episode, which we fit with $\Phi\propto e^{-t/\tau_\Phi}$, giving $\tau_\Phi=98\pm11\,r_g/c$ (dashed line, pair run only), in remarkable agreement with collisionless reconnection-mediated flux decay \cite{Bransgrove:2021heo,Galishnikova:2023}.  The
 {unequal-mass run accumulates less flux initially and decays more slowly,
 retaining $0.75\,\Phi_0$ at $t=320\,r_g/c$.}}
 \label{fig:collisionless-beta4-averaged}
\end{figure}

We can more quantitatively analyze and confirm the behavior of our {GR19M} implementation in two important ways. First, we can analyze a proxy for resistive flux decay on the horizon (a more complete current-sheet dissipation analysis will be presented in Sec. \ref{sec:waldMagnetosphere}), as well as anisotropic pressures and heat fluxes characteristic of this accretion regime.
Figure~\ref{fig:collisionless-beta4-averaged} shows the horizon accretion rate
and the magnetic flux threading the horizon as functions of time.  Starting from the static initial state, the accretion rate rises to $\dot M\simeq1.5\,\dot M_{\rm B}$ by
$t\simeq40\,r_g/c$ as the inner region drains, then decays by more than an
order of magnitude to $0.067\,\dot M_{\rm B}$ at $t\simeq280\,r_g/c$.  In the equal-mass case, the
horizon magnetic flux builds up over the same interval, reaching
$2.1\,\Phi_0$ at $t\simeq120\,r_g/c$, and then falls back to $0.61\,\Phi_0$ by
$t\simeq320\,r_g/c$.  That exponential decay agrees well with GRPIC simulations \cite{Bransgrove:2021heo} (see also Sec. \ref{sec:waldMagnetosphere}), where fitting $\Phi\propto e^{-t/\tau_\Phi}$ over $150\leq t\,c/r_g\leq300$ gives a decay time
$\tau_\Phi=98\pm11\,r_g/c$.  This establishes that resistive dissipation in this problem is well captured.

For $m_i/m_e=100$, the same fit gives $\tau_\Phi=195\pm52\,r_g/c$.
{Applying the flux-decay estimate of Ref.~\mbox{\cite{Bransgrove:2021heo}} to the
measured geometry and equatorial electric field gives $205\,r_g/c$ for the pair and unequal-mass runs,
respectively. The geometric factor is nearly unchanged, while the effective
speed of flux transport is reduced by about a factor of two.} 

\begin{figure*}[!tp]
 \centering
 \includegraphics[width=\textwidth,height=0.72\textheight,keepaspectratio]
 {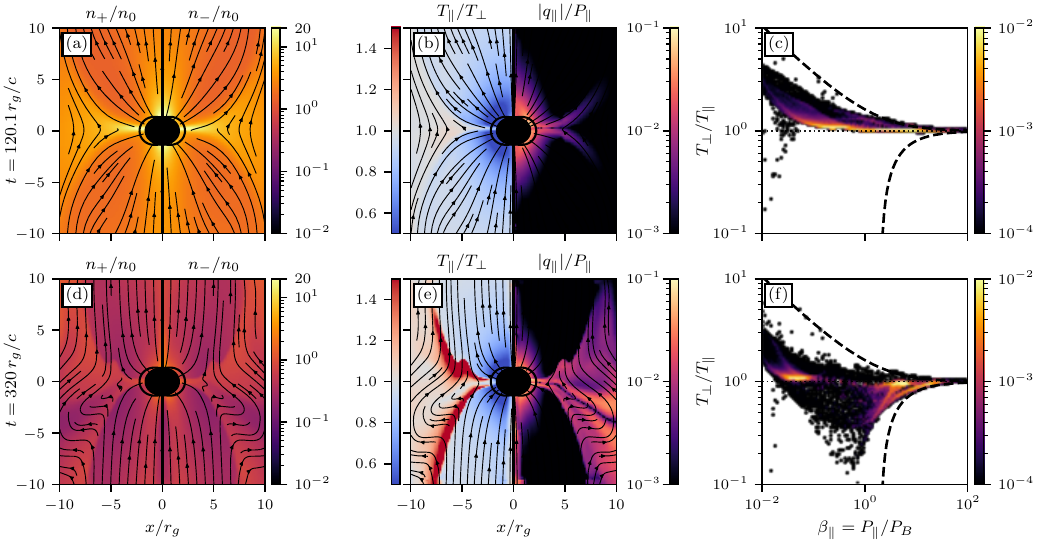}
 \caption{{Accretion properties of the near-horizon flow for the equal-mass
 pair run, $m_i/m_e=1$, of the $\beta_0=4$ model. The panels are shown at different times $t=120.1\,r_g/c$ (top row) and
 $t=320\,r_g/c$ (bottom row), where $r_g=GM/c^2$ is the gravitational radius.
 Panels (a) and (d) show the ion number density $n_+/n_0$ on the left
 half-plane and the electron number density $n_-/n_0$ on the right.  Panels
 (b) and (e) are split in the same way, with the field-aligned temperature
 anisotropy $T_\parallel/T_\perp$ on the left and the ratio of the
 field-aligned heat flux to the parallel pressure, $|q_\parallel|/P_\parallel$,
 on the right.  Panels (c) and (f) show the rest-mass-weighted probability
 distribution of the flow in the $(\beta_\parallel,T_\perp/T_\parallel)$
 plane, where $\beta_\parallel=2P_\parallel/b^2$ is the parallel plasma beta
 and $b^2$ the comoving magnetic energy density. The dashed curves are the
{hard mirror and firehose bounds imposed by the closure.} 
  Black lines denote the magnetic field, filled circles the outer
 event horizon, and open circles the ergosphere of the Kerr black hole.}}
 \label{fig:collisionless-beta4-kinetic}
\end{figure*}

\begin{figure*}
 \centering
 \includegraphics[width=\textwidth,height=0.72\textheight,keepaspectratio]
 {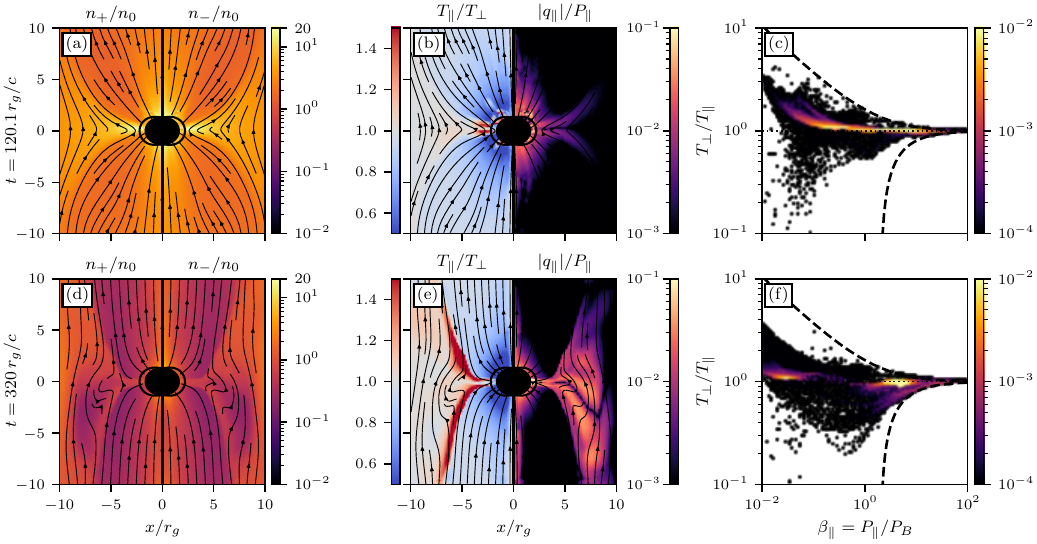}
 \caption{{Same as Fig. \ref{fig:collisionless-beta4-kinetic} but for $m_i/m_e = 100$.}}
 \label{fig:collisionless-massratio-kinetic}
\end{figure*}

Next, we investigate the buildup of pressure anisotropy $\Delta P$ and the emergence of parallel heat flux. These features are characteristic of a Braginskii-like weakly collisional plasma and are mediated in nature by the mirror instability \cite{Kunz:2014qha}, which locally enhances collisionality between the particles and the field. Consequently, we impose hard mirror cutoffs. The firehose instability is fully captured by our equations in this limit {(see Sec.~\ref{sec.firehose} and Appendix~\ref{app:characteristics})}, but we must still impose hard firehose bounds for the system to maintain causality \cite{Cordeiro:2023ljz}.
As demonstrated in the spherical Bondi case, compression of the plasma during infall produces pressure anisotropy (Fig.~\ref{fig.bondiDissipative}). Similarly, for a rotating black hole, we expect deviations from perfect-fluid stresses to be largest near the event horizon. Consequently, we analyze near-horizon dynamics at two different times (Fig.~\ref{fig:collisionless-beta4-kinetic}). We find that the anisotropic temperatures, $T_{\parallel/\perp} \sim p_{\parallel/\perp}/\rho$, are indeed enhanced near the black hole, with $T_\perp > T_\parallel$ initially. Heat flux is also present, but initially only near the event horizon.
At later times ($t=320 r_g/c$), the plasma supply wanes because our initial conditions do not prescribe a fixed inflow, and the plasma approaches the firehose threshold near the magnetized funnel. Heat fluxes, $q_\parallel$, become active throughout the inner flow but remain $<0.1 P_\parallel$. Overall, our results agree very well with similar GRPIC simulations \cite{Galishnikova:2023}.

\subsection{Wald pair magnetosphere}
\label{sec:waldMagnetosphere}

Finally, we perform a detailed assessment of reconnection in a black hole magnetosphere by evolving the pair-loaded magnetosphere of a black hole embedded in a uniform magnetic field.
This is a controlled problem, as the case of vacuum (no plasma) can be solved analytically \cite{Wald:1974}, and variants of this problem have been studied extensively in GRMHD \cite{Komissarov:2007rc} and {general-relativistic force-free electrodynamics (GRFFE)} approaches \cite{Komissarov:2004ms,Palenzuela:2010xn,Nathanail:2014aua,Mahlmann:2018ukr,Kim:2024mau}. Given that we model reconnection physics beyond what is captured by numerical resistivity in ideal GRMHD or ad hoc prescriptions in resistive GRMHD and GRFFE (although see Ref. \cite{Ripperda:2026EffectiveResistivity}), our solution here is closest to the GRPIC work of Ref. \cite{Parfrey:2019}, which we will use as a direct comparison and validation reference.

Initially, the black hole is immersed in the stationary, axis-aligned vacuum field \cite{Wald:1974}.  In standard Kerr coordinates, its four-potential is
\begin{equation}
 A_{\mu}=\frac{B_{0}}{2}
 \left(g_{\mu\phi}+2a g_{\mu t}\right),
 \label{eq:waldPotential}
\end{equation}

Throughout the magnetosphere tests, $\boldsymbol D$ denotes the Eulerian electric field $\boldsymbol E$ of the evolution system. The coordinate Maxwell auxiliaries are
\begin{align}
 \boldsymbol{\mathcal E}&=\alpha\boldsymbol D+\boldsymbol\beta\times\boldsymbol B,\notag\\
 \boldsymbol H&=\alpha\boldsymbol B-\boldsymbol\beta\times\boldsymbol D.
\end{align}
Here, $H_\phi$ is the covariant azimuthal component of $\boldsymbol H$, and cross products use the spatial metric. The asymptotic magnetic-field strength is $B_{0}$, and $a$ is the dimensionless spin for $M=1$. The full vacuum magnetic-field solution in Kerr--Schild coordinates is given in Appendix C of Ref.~\cite{Kim:2024mau}. Direct computation shows that Eq.~\eqref{eq:waldPotential} includes an electric field induced by frame dragging, so the plasma starts far from the screened condition $\boldsymbol{D}\mathbin{\cdot}\boldsymbol{B}=0$.
We use ingoing Kerr--Schild coordinates and a black-hole spin $a/M=0.999$, giving a horizon radius $r_{\rm H}/r_g=1+\sqrt{1-a^{2}}\approx 1.045$ and a horizon angular frequency $\Omega_{\rm H}={a}/{(r_{\rm H}^{2}+a^{2})}$.

The positive and negative charge carriers, denoted by $i$ and $e$, have
$m_i=m_e=1$, $Z=1$, and $|q_e|/m_e=10^{5}$.  We set $B_{0}=10^{-2}$, so that
the asymptotic cyclotron frequency, Larmor radius, reference magnetization,
and skin depth are
\begin{align}
 \Omega_{B0} &={10^{3}\,c/r_g}\,\\
 r_{L0} &=10^{-3} r_g,\\
 \sigma_{0} &=\frac{\Omega_{B0}}{\Omega_{\rm H}}=2091.51,\\
 d_e &=\left(\Omega_{\rm H}\Omega_{B0}\right)^{-1/2}=0.0457331 r_g .
\end{align}
Both fluids adopt a $\Gamma=4/3$ equation of state.
{The ion-frame carrier densities are $\bar n_i=\rhob/m_i$ and $\bar n_e=Z\rhob/m_i-\rhoq/q_e$. The density panels instead use the {fiducial-observer (FIDO)} densities $n_\pm^{\rm FIDO}=\alpha N_\pm^0$ defined below.}

\begin{figure}[!tbp]
 \centering
 \includegraphics[width=\columnwidth,height=0.72\textheight,keepaspectratio]{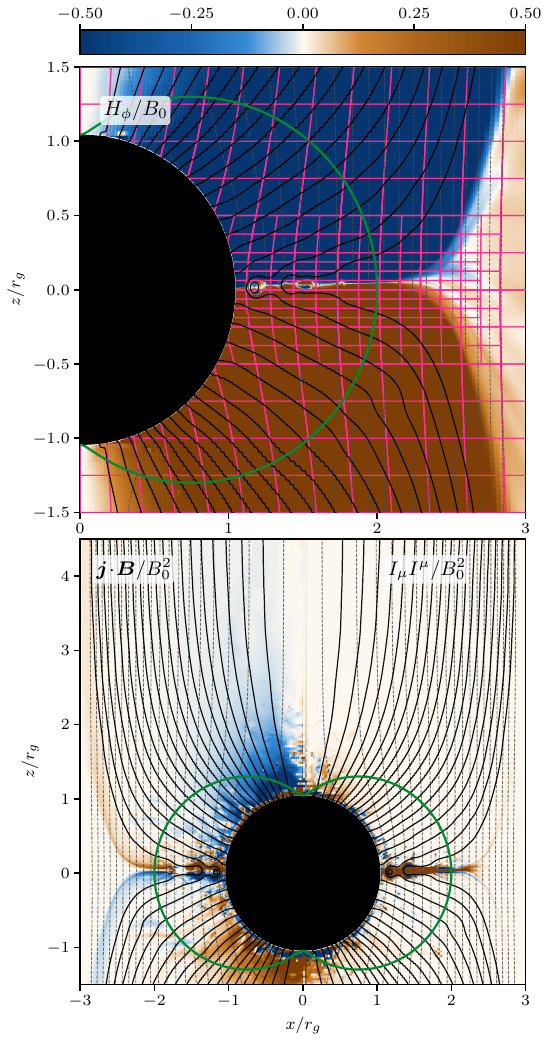}
 \caption{Wald magnetosphere at time $t=80\, r_g/c$ modeled using a {GR19M} pair closure.
 The upper panel shows the relative toroidal magnetic
 field $H_\phi/B_0$ normalized to the {field strength $B_0$ at infinity.}
 The lower panel shows the parallel current
{$\boldsymbol{j}\mathbin{\cdot}\boldsymbol{B}/B_0^2$ on the left and the
 four-current norm ${\mathcal J_\mu\mathcal J^\mu}/B_0^2$ on the right.} Solid black curves denote magnetic field lines and dashed gray curves are the same
 lines in the initial Wald state.  The event-horizon interior is black and
 the ergosphere is green.  {Pink lines in the upper panel denote the patch boundaries of the adaptively refined mesh tracking the equatorial current sheet.}}
 \label{fig:waldParfreyMorphology}
\end{figure}

\begin{figure*}[!tp]
 \centering
 \includegraphics[width=\textwidth,height=0.72\textheight,keepaspectratio]{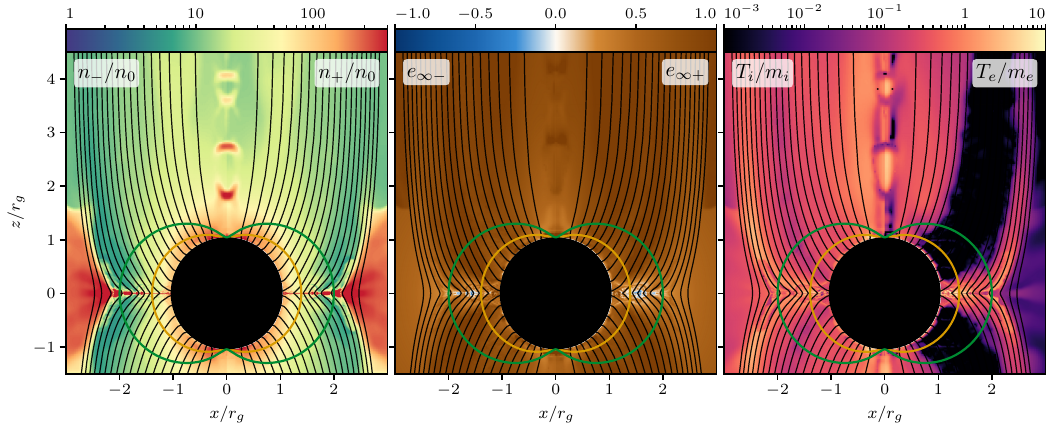}
 \caption{Overall morphology of the jet region in the Wald problem at $t=60 r_g/c$, where $r_g=GM/c^2$ is the gravitational radius.
 In the first two panels, the left and right half-planes show the negative and positive carriers, respectively.  From left
 to right the panels display {FIDO number densities,} $n_\pm$, normalized to the Goldreich--Julian density, ${n_{\rm GJ}}$,
 energy-at-infinity, ${e_{\infty\pm}}$, and the ion and electron temperatures, $T_i/m_i$ and $T_e/m_e$.
 Black curves are evolved poloidal magnetic-flux surfaces, the event-horizon
 interior is black, the ergosphere is green, and the gold curve is the
 analytic inner light surface for {field-line angular frequency} $\Omega_F=\Omega_{\rm H}/2$.  }
 \label{fig:waldParfreyCarriers}
\end{figure*}

\begin{figure*}[!tp]
 \centering
 \includegraphics[width=\textwidth,height=0.72\textheight,keepaspectratio]{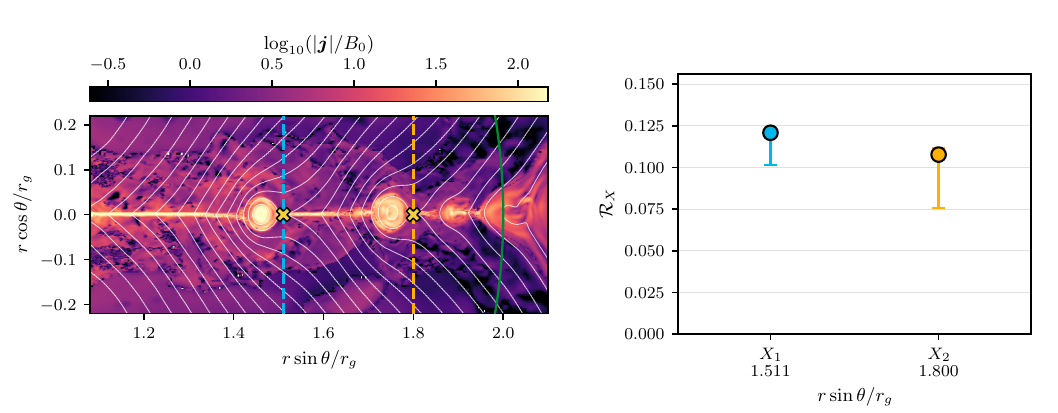}
 \caption{{Wald current-sheet zoom-in at $t=62\,r_g/c$.} \textit{(Left)} {Magnitude of the dissipative current},
 {$|\boldsymbol{j}|=(j_\mu j^\mu)^{1/2}$ with $j_\mu u^\mu=0$}, shown as
 $\log_{10}(|\boldsymbol{j}|/B_0)$.  White curves are evolved poloidal flux surfaces and the
 green curve is the ergosphere.  {The light-blue and yellow dashed lines and gold crosses mark two X-points, at $r\sin\theta/r_g=1.511$ and $1.800$.}
 \textit{(Right)} {Instantaneous reconnection rates $\mathcal{R}_X$}.}
 \label{fig:waldReconnectionRate}
\end{figure*}

Although the corresponding
particle-in-cell problem begins in vacuum, GRMHD formulations require a nonzero density to define a reference velocity and frame.
We use $T/m=2.5\times10^{-4}$ and limit the one-species floor
magnetization to $200$ by injecting density locally.

One challenge in MHD simulations of magnetospheres is preserving a highly magnetized state. Various approaches have been developed, including prescribed magnetization profiles \cite{Tchekhovskoy:2012hm} and tracer-based flooring approaches \cite{Parfrey:2019,Most:2024qgc,Chatterjee:2026tdp}. We find that, because positive and negative charge carriers are independent fluids, simple magnetization-based injections are not always stable in this regime. Instead,
pairs are supplied for $r_{\rm H}<r<6 r_g$ using the pair injection scheme of
Ref.~\cite{Parfrey:2019}, extended to AMR and arbitrary positive charge number
$Z$.  A neutral parcel contains one positive carrier and $Z$ negative
carriers.  On level $\ell$, its requested positive-species load is
\begin{align}
 \Delta n_+^{(0)}
 &=w(r)\frac{\Delta t_\ell}{\Delta t_{\rm ref}}\,
 \mathcal R\frac{|\boldsymbol D\mathbin{\cdot}\boldsymbol B|}
 {Z|q_e|B},\notag\\
 \Delta n_-^{(0)}&=Z\Delta n_+^{(0)} .
 \label{eq:waldPairLoading}
\end{align}
The committed load is the bounded parcel
\begin{align}
 \Delta n_+&=\min\!\Bigg\{
 \Delta n_+^{(0)},\notag\\[-2pt]
 &\hspace{2em}
 \frac{[B^2/\sigma_{\rm inj}-\rho_{\rm pair}]_+}
 {m_++Zm_-},
 \frac{[n_{-,\max}-n_-]_+}{Z}
 \Bigg\},\notag\\
 \Delta n_-&=Z\Delta n_+,
 \label{eq:waldPairLoadingRoom}\\
 n_{-,\max}
 &=\frac{\sigma_{J,\max}}
 {(|q_e|/m_-)^2m_-\max(\tau_J,10^{-10})},\notag\\
 \rho_{\rm pair}&=\frac{D_+^*+D_-^*}{\sqrt\gamma}.
\end{align}
where $[x]_+=\max(x,0)$.  The second entry prevents one discrete parcel from
crossing the magnetization surface that activated it, so the source vanishes
continuously as $B^2\rightarrow\sigma_{\rm inj}\rho_{\rm pair}$.  The third
reserves room below the same conductivity ceiling used by the closure.  For
the present pair plasma $Z=1$, and Eq.~\eqref{eq:waldPairLoading} reduces to
equal positive and negative loads.
{Here $w(r)$ is the radial window specified below; $\mathcal R$ is the
dimensionless pair-loading gain; $\Delta t_{\rm ref}$ is its reference step;
$\sigma_{\rm inj}$ is the activation threshold in total-pair magnetization;
$m_+=m_i$ and $m_-=m_e$;
$D_+^*$ and $D_-^*$ are the pre-injection conserved carrier rest-mass densities;
$n_-$ is the negative-carrier density entering the conductivity closure; and
$\sigma_{J,\max}=10^5 c/r_g$ is the prescribed conductivity ceiling.}

We use $\mathcal R=0.5$ and
$\Delta t_{\rm ref}=0.00625\, r_g/c$.  The factor
$\Delta t_\ell/\Delta t_{\rm ref}$ gives every subcycled AMR level the same
physical supply per unit coordinate time.  The radial window is unity through
$r=5.9 r_g$ and decreases to zero with a quintic smoothstep over
$5.9<r/r_g<6$.  Loading occurs only where
$|\boldsymbol D\mathbin{\cdot}\boldsymbol B|/B^{2}>10^{-3}$ and the local
total-pair magnetization exceeds $\sigma_{\rm inj}=50$.  The injected
temperatures are $T_i/m_i=0.5$ and $T_e/m_e=5\times10^{-4}$, and the
carrier-aware current closure uses $\tau_J=0.25 r_g/c$.
Since we focus exclusively on resistive dissipation in the current sheet, bulk
viscosity, shear viscosity, and heat conduction are disabled in this test.

The evolution is axisymmetric in cylindrical Kerr--Schild coordinates
$(\varpi,z)$ on $0\leq\varpi/r_g\leq64$ and $-64\leq z/r_g\leq64$.
We use the reference-metric flux formulation, IMEX12 time integration,
$\cfl=0.25$, and divergence-cleaning rate $\kappa_{\rm dc}=16c/r_g$.
{We use a total of eight refinement levels with the finest resolution}
$\Delta x_8/r_g= 1/256$ (the grid structure is shown in the top panel of Fig.
\ref{fig:waldParfreyMorphology}). The simulation is evolved until
$t=80\, r_g/c$.
 The general morphology we discover resembles that of the force-free Wald magnetosphere \cite{Nathanail:2014aua}, as expected.

We begin by outlining the basic features of a Wald magnetosphere before presenting a more detailed analysis of the reconnection and two-fluid physics captured by our scheme (Fig.~\ref{fig:waldParfreyMorphology}).
Initially, the Wald solution consists of vertical field lines. {The initial Wald magnetic field has an unscreened component $\boldsymbol D\cdot\boldsymbol B$, which triggers pair injection inside the magnetosphere \cite{Parfrey:2019}.} The plasma falls along magnetic field lines toward the black hole, enhancing the connected flux through the hole. An equatorial current sheet forms, and the magnetic field lines bend toward the ergosphere (compare the solid and dashed initial field lines). This spacetime rotation strongly twists the magnetic field, producing a substantial toroidal field inside the magnetic funnel, as is characteristic of force-free solutions of this type \cite{1996MNRAS.279..389L}.
{Inside the current sheet and along the sides of the funnel, we observe that the {total current becomes spacelike, $\mathcal J^\mu\mathcal J_\mu>0$}, meaning that ${\boldsymbol j}^2 > {|q_e|^2}(n_+ -n_-)^2$, and the numbers of negative and positive charge carriers are almost equal.}
The current sheet itself becomes tearing unstable and forms plasmoids \cite{2007PhPl...14j0703L,Samtaney:2009bq,2009PhPl...16k2102B}.

For each timelike current we define the representative carrier velocity
$u_\pm^\mu=N_\pm^\mu/\sqrt{-N_{\pm\nu}N_\pm^\nu}$, 
where in our preceding notation, $N_i^\mu = N_+^\mu$ and $N_e^\mu = N_-^\mu$.
Figure~\ref{fig:waldParfreyCarriers}
then shows the FIDO densities $n_\pm^{\rm FIDO}=\alpha N_\pm^0$ (see Appendix~\ref{app:fidotetrad}), conserved
energy per carrier $e_{\infty\pm}=-u_{\pm t}$, and the ion and electron temperatures.
For $\Gamma_i=\Gamma_e=4/3$, the plotted temperatures are $T_i/m_i=P_i/\rho_i$ and $T_e/m_e=P_e/\rho_e$, {where $\rho_i:=m_i\bar n_i=\rhob$ and $\rho_e:=m_e\bar n_e$}.
We normalize number densities relative to the Goldreich--Julian density \cite{Goldreich:1969sb},
 ${n_{\rm GJ}}=\Omega_{\rm H}B_0/{|q_e|}=4.78\times10^{-8}$.
 We can see that plasmoids formed in the current sheet develop negative energies, allowing energy extraction from the black hole consistent with the Penrose process in the special case of magnetic reconnection \cite{Comisso:2020ykg}.
 In this test, the electrons are deliberately injected at a lower temperature than the ions, allowing us to identify where electron heating occurs directly. As expected, electrons are heated by compression near the jet sheath and by resistive dissipation inside the current sheet.

\begin{figure*}[!tp]
 \centering
 \includegraphics[width=0.8\textwidth,keepaspectratio]{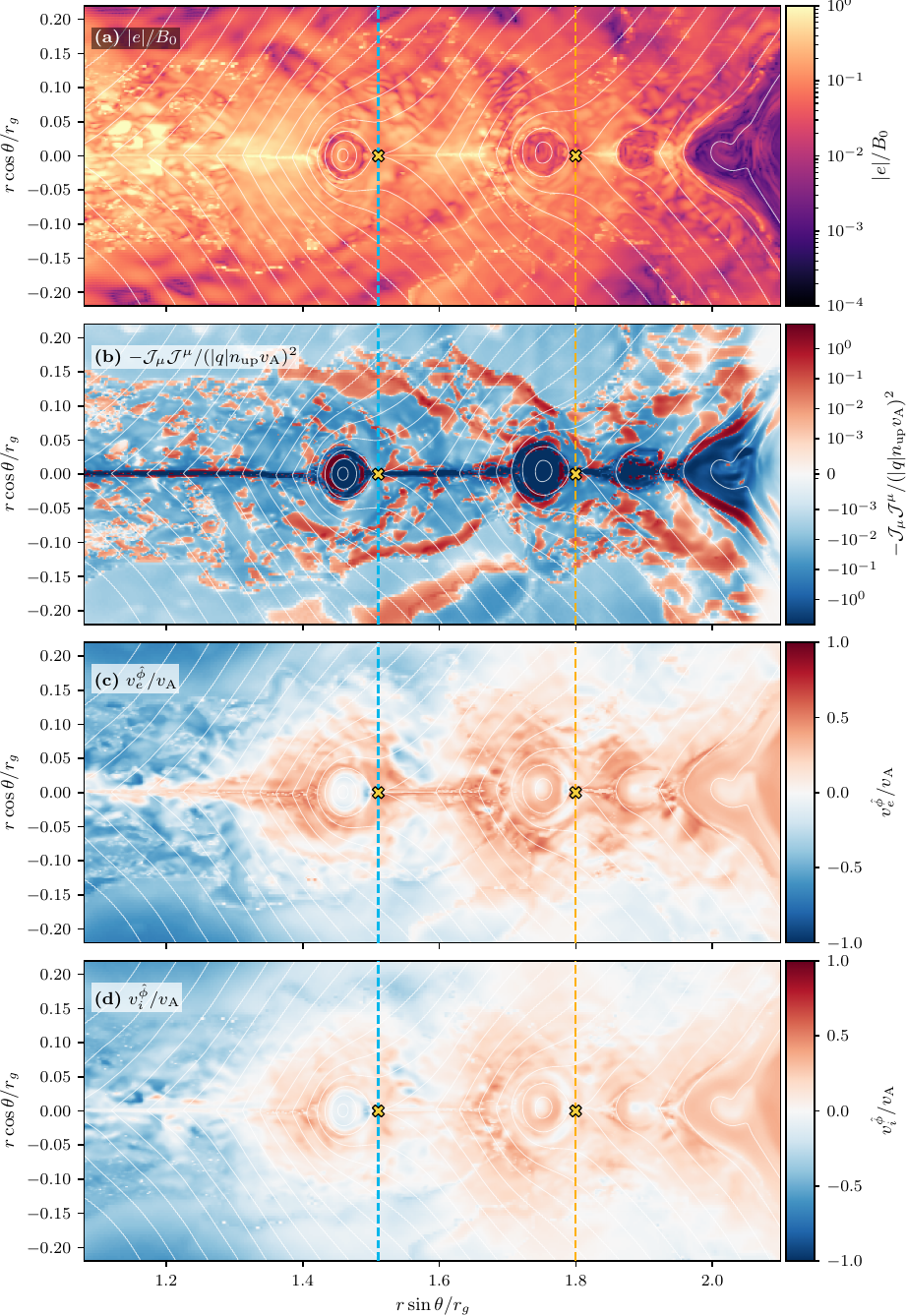}
 \caption{{Current and nonideal-field diagnostics in the Wald current sheet at}
 {$t=62 r_g/c$, with the two X-points from Fig.~\ref{fig:waldReconnectionRate} highlighted.}  From top to bottom: the comoving {nonideal} electric-field
 magnitude $|e|/{B_0}$, the total four-current invariant
 $-{\mathcal J_\mu\mathcal J^\mu}/(|q_e|n_{\rm up}{v_{\rm A}})^2$ ({negative values identify spacelike total current}), and the FIDO-frame
 toroidal velocities $v_e^{\hat\phi}/{v_{\rm A}}$ and $v_i^{\hat\phi}/{v_{\rm A}}$.  White curves are evolved poloidal magnetic-flux
 surfaces. {The light-blue and yellow dashed lines and gold crosses mark the selected X-points.} {Here $n_{\rm up}$ and $v_{\rm A}$ are the total FIDO carrier density and nominal Alfv\'en speed in the upstream.}}
 \label{fig:waldChargeStarvation}
\end{figure*}

As shown in Fig.~\ref{fig:waldParfreyMorphology}, reconnection occurs in the plasmoid regime \cite{2007PhPl...14j0703L,2009PhPl...16k2102B,Uzdensky:2010ts}, which should produce fast reconnection in a collisionless, highly magnetized plasma \cite{Sironi:2014jfa,Guo:2015cua,Goodbred:2022}. As an important assessment of the {19-moment} dissipative GRMHD system, we now demonstrate that our closure captures this behavior.

To this end, we quantify the instantaneous reconnection rate, $\mathcal{R}_X$, at X-points in the current sheet.  In the local orthonormal frame of the equatorial sheet
{defined by Eqs.~\eqref{eq.fidonormal} and \eqref{eq.fidotetrad} in
Appendix~\ref{app:fidotetrad}}, we define
\begin{equation}
 \mathcal{R}_X=\frac{|E_{\hat{\phi}}(X)|}{B_{p,\mathrm{up}}}.
 \label{eq:waldReconnectionRate}
\end{equation}{Here $\mathcal R_X$ is dimensionless in $c=1$ units. Under ideal upstream inflow it estimates $v_{\rm in}/c$; the Alfv\'en-normalized rate is $\mathcal R_X/({v_{\rm A}}/c)$.}
Here $B_{p,\mathrm{up}}$ is the mean of the upper- and lower-side medians of
the sheet-tangent magnetic field over
$0.10\leq |r\cos\theta|/M\leq0.15$.

{Figure~\ref{fig:waldReconnectionRate} shows the magnitude of the evolved dissipative current density, $j^\mu$, at $t=62 r_g/c$.  The sheet contains four X-points.  The two displayed X-points lie at $r\sin\theta/r_g=1.511$ and $1.800$, with instantaneous reconnection rates $\mathcal{R}_X=0.121$ and $0.108$, respectively.  Both are of order $0.1c$ and exceed the characteristic $0.01c$ rate of MHD reconnection \cite{2007PhPl...14j0703L,2009PhPl...16k2102B}.  We use the outer displayed X-point below because its rate is close to $0.1c$ and its systematic envelope, $0.076\leq\mathcal{R}_X\leq0.112$, is comparatively narrow.} This controlled setup allows us to investigate the origin of fast reconnection further and determine whether it is consistent with alternative approaches recently proposed in the literature \cite{Moran:2025aqb,Ripperda:2026EffectiveResistivity} (see also Refs.~\cite{Selvi:2022uup,Bugli:2024fby}).
{In highly magnetized pair plasmas, the energy required to sustain the
current suppresses the pressure at the X-line.  The upstream magnetic field
then bends inward, opening the exhaust and enabling fast reconnection
\cite{Goodbred:2022}.  Kinetic simulations also find currents close to the
charge-carrying limit, $|\boldsymbol J|\simeq |q_e|n_t c$, where {$\boldsymbol J$ is the total spatial current and $n_t$ is the total carrier density} in the measurement frame; this motivates the effective
resistivity prescription of Ref.~\cite{Moran:2025aqb}.}

To test whether the {GR19M} closure proposed here correctly captures this regime,
we now perform a detailed analysis of the X-point (Fig.~\ref{fig:waldChargeStarvation}).

We proceed in three steps. {First, we confirm that the selected X-point features substantial resistive dissipation: the magnetic-field topology is X-like, and a substantial nonideal field $\left|e\right|/B \gg 0$ is present as $B$ approaches zero at the X-point.}
{Having established the presence of a dissipating X-point, we now establish that it is charge-starved.
The four-norm of the total electric current can be written as}
\begin{align*}
{{\mathcal J_\mu\mathcal J^\mu}=\ju_\mu\ju^\mu-{\rhoq^2}\,,}
\end{align*}
{implying that $-{\mathcal J_\mu\mathcal J^\mu} < 0$ means that the comoving charge density ${\rhoq}$ becomes negligible compared to the dissipative current. This is consistent with the fact that positive and negative charge carriers are evacuating the X-point, as can be seen by considering the electron and ion velocities.}

\begin{figure*}[!tp]
 \centering
 \includegraphics[width=\textwidth,height=0.72\textheight,keepaspectratio]{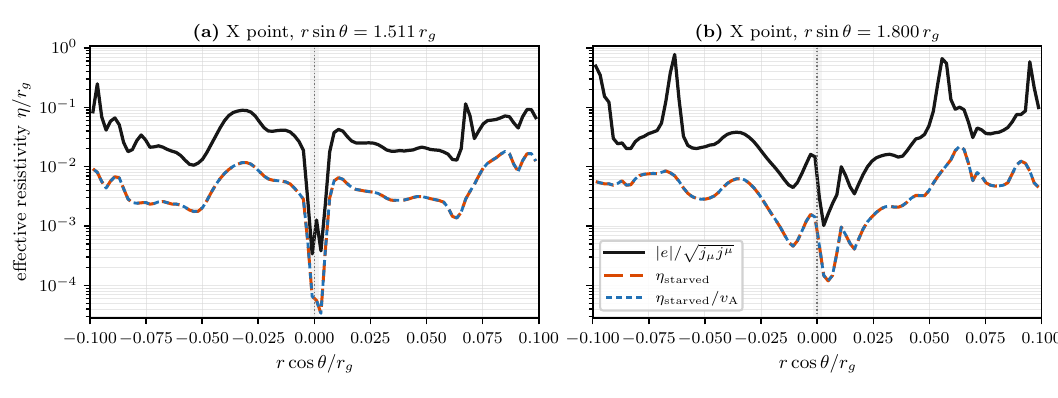}
 \caption{Effective resistivity along vertical cuts through the
 {two X-points in Fig. \ref{fig:waldReconnectionRate}. The solid black
 curve is the effective scalar resistivity,
 $\eta_{\rm eff}=|e|/(\ju_\mu\ju^\mu)^{1/2}$, inferred from the {19-moment} evolution. The dashed orange curve is the approximate charge-starved value from Eq.~\eqref{eq:waldRipperdaResistivity} using the local upstream quantities
and charge-starvation thickness at each X-point; the dashed blue curve shows $\eta_{\rm starved}/v_{\rm A}$.  The two nearly coincide because $v_{\rm A}\simeq1$.  The dotted vertical
 line marks the equatorial plane. }}
 \label{fig:waldEffectiveResistivity}
\end{figure*}

As a final diagnostic, we compare the effective resistivity probed at the X-points of the Wald current sheet with that predicted by a nonuniform resistivity derived under a charge-starvation assumption and validated against PIC \cite{Moran:2025aqb}.
This comparison can be summarized as follows \cite{Ripperda:2026EffectiveResistivity}. If the entire Ohm's law were represented by a scalar closure \cite{Palenzuela:2008sf}, $\ju^\mu=\eta_{\rm eff}^{-1}e^\mu$, the effective scalar resistivity in our simulation would be
\begin{equation}
 \eta_{\rm eff}=\frac{|e|}{(\ju_\mu\ju^\mu)^{1/2}}.
 \label{eq:waldInferredResistivity}
\end{equation}
The prediction for the charge-starved regime is \cite{Ripperda:2026EffectiveResistivity}:
\begin{equation}
 \eta_{\rm starved}=4\pi t_c\frac{\Delta_0}{L}
 \frac{|e|}{{B_{\rm up}}}\frac{{n_{\rm up}}}{n}\,.
 \label{eq:waldRipperdaResistivity}
\end{equation}
{Here $t_c$ and $L$ are the characteristic current-sheet time and length scales.} We set $t_c=L$, with ${B_{\rm up}}$ and ${n_{\rm up}}$ the local
two-sided upstream reconnecting field and total FIDO carrier density, and use
the accompanying charge-starvation estimate
$\Delta_0={B_{\rm up}}/(4\pi|q_e|{n_{\rm up}})$.  Equation~\eqref{eq:waldRipperdaResistivity}
then reduces to ${\eta_{\rm starved}}=|e|/(|q_e|n)$, {where $n$ is the local total carrier density} \cite{Moran:2025aqb,Ripperda:2026EffectiveResistivity}. This has a particularly straightforward interpretation: the effective resistivity replaces the {nonideal} electric field contribution with that expected for a charge-starved current.

{Figure~\ref{fig:waldEffectiveResistivity} evaluates both expressions at the two X-points; the right cut is also used for the reconnection-rate measurement in Fig.~\ref{fig:waldReconnectionRate}.  Over $|r\cos\theta|\leq0.04\,r_g$, the median ratios $\eta_{\rm eff}/\eta_{\rm starved}$ are $6.69$ and $8.04$.  The charge-starved prescription, hence, can capture the local structure of the effective resistivity but differs in amplitude, despite similar reconnection rates.  The {GR19M} scheme nevertheless captures fast reconnection without imposing a non-uniform scalar resistivity prescription.}

\subsection{Balding black holes}
\label{sec:baldingBlackHoles}

{A second, more challenging problem that tests nonideal (effective resistive) dissipation in a black-hole magnetosphere is black-hole balding. The no-hair theorem states that a black hole is characterized only by its mass $M$, spin parameter $a$, and electric charge $Q$, as in the Kerr--Newman solution \mbox{\cite{Newman:1965my}}, which has both electric and magnetic fields. In the presence of plasma, this solution changes, allowing the black hole to discharge and the magnetic field to bald. Balding occurs primarily on the effective resistive timescale of the plasma \mbox{\cite{2011PhRvD..84h4019L}} (though see Ref. \mbox{\cite{Most:2024qgc}} for balding on the dynamical timescale of an oscillating spacetime). Because black holes cannot support closed magnetic-field lines \mbox{\cite{MacDonald:1982zz}}, the magnetic field stretches, opens, and forms a split-monopole configuration with an equatorial current sheet \mbox{\cite{2011PhRvD..84h4019L,Selvi:2025dur}}. If the magnetic moment of the progenitor field is not aligned with the black-hole spin axis, a black-hole pulsar state forms, resembling a pulsar wind \mbox{\cite{Selvi:2024lsh,Kim:2024fuy}}. Dissipation in this current sheet has been studied using resistive GRMHD \mbox{\cite{Bransgrove:2021heo,Ripperda:2026EffectiveResistivity}} and GRPIC approaches \mbox{\cite{Bransgrove:2021heo}}. We compare against the latter.}

\begin{figure*}[!tp]
 \centering
 \includegraphics[width=\textwidth,height=0.72\textheight,keepaspectratio]{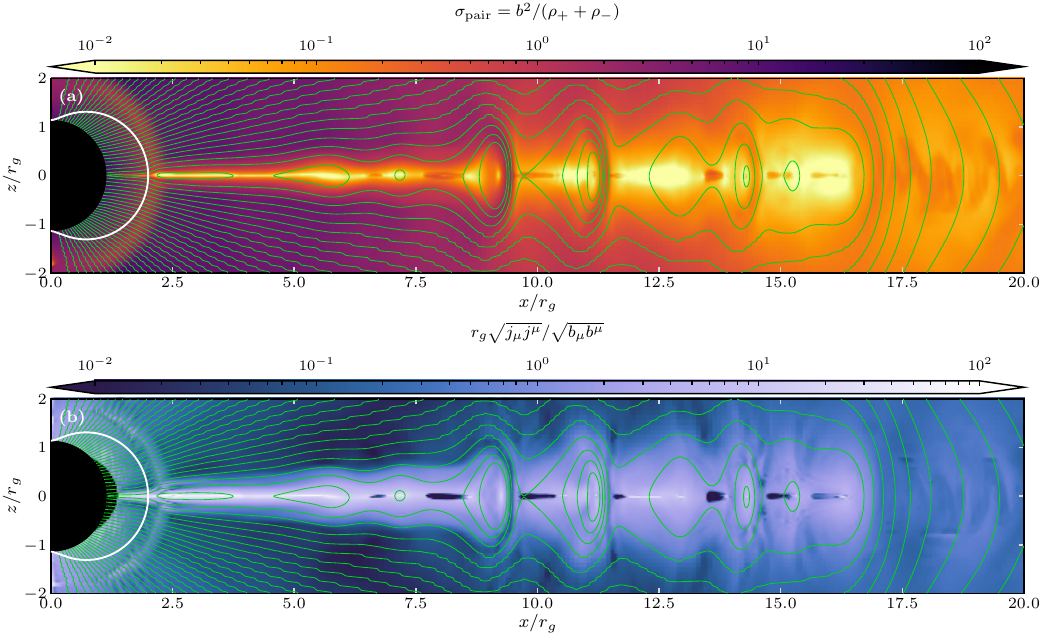}
 \caption{{Balding black hole magnetosphere at $t=60 \, r_g/c$.
 {Panel (a) shows the cold total-pair magnetization
 $\sigma_{\rm pair}=b_\mu b^\mu/(\rho_++\rho_-)$.  Panel (b) shows
 $r_g(j_\mu j^\mu)^{1/2}/(b_\mu b^\mu)^{1/2}$, with
 $j^\mu=\Delta^\mu{}_{\nu}\mathcal J^\nu$ the conduction current,
 $j^\mu u_\mu=0$, as in Fig.~\ref{fig:waldReconnectionRate}.
 Both norms are spacetime invariants, and distances are measured in $r_g$.}
 Green curves denote poloidal magnetic-flux surfaces, the white curve is the
 ergosphere, and the event-horizon interior is black and masked.}}
 \label{fig:baldingMagnetization}
\end{figure*}

 As in the preceding Wald magnetosphere and other MHD reconnection studies in this context, the reconnection rate measured in GRPIC \cite{Bransgrove:2021heo} matches the collisionless rate of order $0.1$, whereas GRMHD yields a considerably slower rate of $0.01$--$0.02$ \cite{Komissarov:2007wk,Ripperda:2019lsi}.
 The problem is considerably more demanding than the Wald problem, as the open spinning black hole magnetosphere features a net outflow of plasma that needs to be continuously replenished. The wind also exerts a net pressure on the sheet.

\begin{figure*}[!tp]
 \centering
 \includegraphics[width=\textwidth,height=0.72\textheight,keepaspectratio]{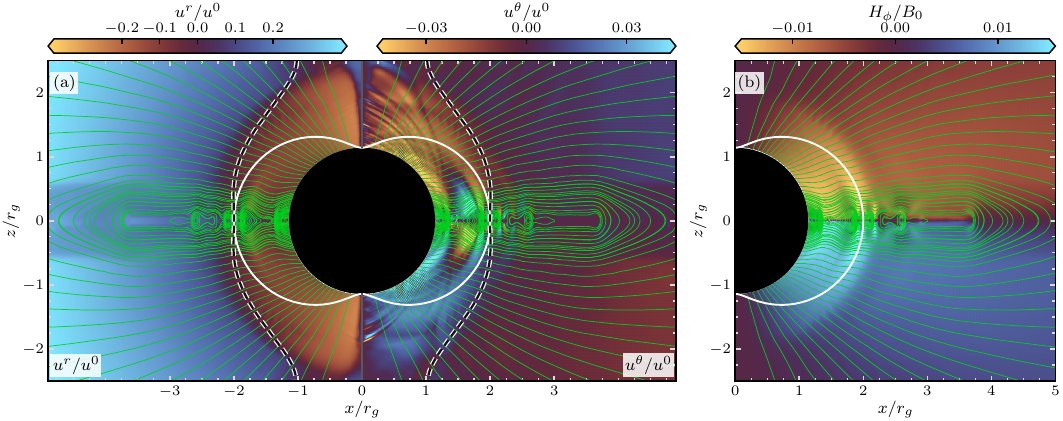}
 \caption{{Near-horizon flow and toroidal field during the onset of balding at {$t=35\, r_g/c$.}
Green lines represent poloidal magnetic-flux surfaces. White solid lines correspond to the ergosphere; white dashed lines indicate the stagnation surface, where the radial transport velocity, $u^r/u^0$, changes sign. Also shown are {the angular coordinate rate, $d\theta/dt=u^\theta/u^0$}, and the toroidal magnetic field $H_\phi$ relative to the initial magnetic-field scale $B_0$.
 }}
 \label{fig:baldingVelocities}
\end{figure*} 

In our setup, the fixed background is a Kerr black hole with $M=1$ and spin $a=0.99$, which we model in axisymmetry. The magnetosphere is initialized with an aligned vacuum dipole expressed in terms of the magnetic vector potential,
\begin{equation}
 A_\phi=B_0\frac{\sin^2\theta}{r}
       =B_0\frac{\varpi^2}{r\left(r^2+a^2\right)}\,,
\end{equation}
so that the toroidal field vanishes identically at $t=0$.  The horizon radius and angular velocity are
\begin{align}
 r_{\rm H}&=1+\sqrt{1-a^{2}}\approx 1.14 r_g,\\
 \Omega_{\rm H}&=\frac{a}{r_{\rm H}^{2}+a^{2}}\approx 0.43\, r_g^{-1}\, .
\end{align}
We measure how much magnetic field threads the black hole using the absolute horizon flux, $\Phi_{\rm BH}=2\pi B_0/r_{\rm H}$.

We again adopt a pair-plasma composition, $m_e=m_i=1$, $Z=1$, and ${|q_e|/m_e}=1$, and set $B_0=10^{4}$, giving the reference cyclotron frequency and Larmor radius $\Omega_{B0}=10^{4} c/r_g$ and ${r_{L0}}=10^{-4} r_g$.  The plasma is initially static in the normal-observer frame, $u_i=0$, with vanishing electric field, charge, and dissipative moments, a $\Gamma=4/3$ equation of state, and a reference total pressure $10^{-4}$.  We impose a cold magnetization limit of $\sigma<200$ and require the plasma beta to be $>10^{-5}$.  We choose an effective collisional time $\tau_J=0.25$ and limit the conductivity to $\sigma_J < 10^5$, {bounding the effective current-response length $d_\sigma=(\tau_J/\sigma_J)^{1/2}$ away from zero.}
 As before, we focus on reconnection dynamics: bulk viscosity, shear viscosity, and heat conduction are disabled, and both divergence-cleaning fields use a damping rate $\kappa_{\rm dc}=4$.

As discussed, a rotating split monopole features a net outflow in this scenario. We must therefore inject pairs continuously through the pair-creation module, which critically controls the plasma parameters in the current sheet.
Pairs are supplied for $r_{\rm H}<r<6$ by the scheme of Eq.~\eqref{eq:waldPairLoading}, now with gain $\mathcal{R}=1.5$, and reference step $\Delta t_{\rm ref}=0.00625\, r_g/c$. Loading is triggered where $|\boldsymbol{D}\mathbin{\cdot}\boldsymbol{B}|/B^{2}>10^{-3}$, with $\boldsymbol{D}$ the normal-observer electric field, and where the total magnetization exceeds $50$. Injected pair matter carries $T/m=0.5$.

\begin{figure}[!tbp]
 \centering
 \includegraphics[width=\columnwidth,height=0.72\textheight,keepaspectratio]{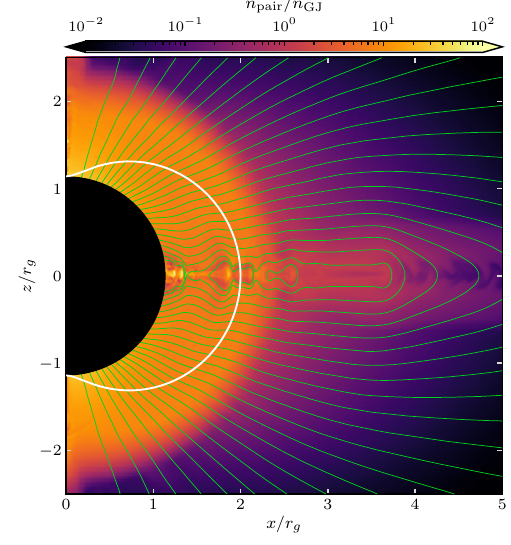}
 \caption{{Pair multiplicity relative to the Goldreich--Julian number density,
 $n_{\rm pair}/n_{\rm GJ}$, at $t=35 \, r_g/c$.  Here
 {$n_{\rm pair}=(\rho_{+}+\rho_{-})/m$ is the total number density of the equal-mass pair sector and $n_{\rm GJ}=\Omega_{\rm H}B_{0}/{|q_e|}$}, where
 $\Omega_{\rm H}=a/(r_{\rm H}^{2}+a^{2})$ is the horizon angular frequency
 and $B_{0}$ is the initial reference field strength. {Here $m=m_i=m_e$}, and the charge-to-mass ratio is ${|q_e|/m_e}=1$.
}}
 \label{fig:baldingPairMultiplicity}
\end{figure}

\begin{figure}[!tbp]
 \centering
 \includegraphics[width=\columnwidth,height=0.72\textheight,keepaspectratio]{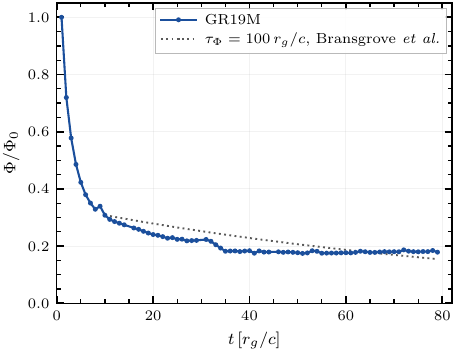}
 \caption{{Unsigned horizon flux of the balding black hole, normalized
 to its first recorded value $\Phi_0$ at times $t$, where $r_g = GM/c^2$ is the gravitational radius.  The blue curve shows
 {GR19M}.  The dotted curve is an exponential guide with
 $\tau_\Phi=100\,r_g/c$ from Ref.~\cite{Bransgrove:2021heo}, matched to
 the measured flux at $t=10\,r_g/c$.  The initial flux loss slows markedly
 after $t\simeq50\,r_g/c$.}}
 \label{fig:baldingHorizonFlux}
\end{figure}

\begin{figure*}[!tp]
 \centering
 \includegraphics[width=\textwidth,height=0.72\textheight,keepaspectratio]{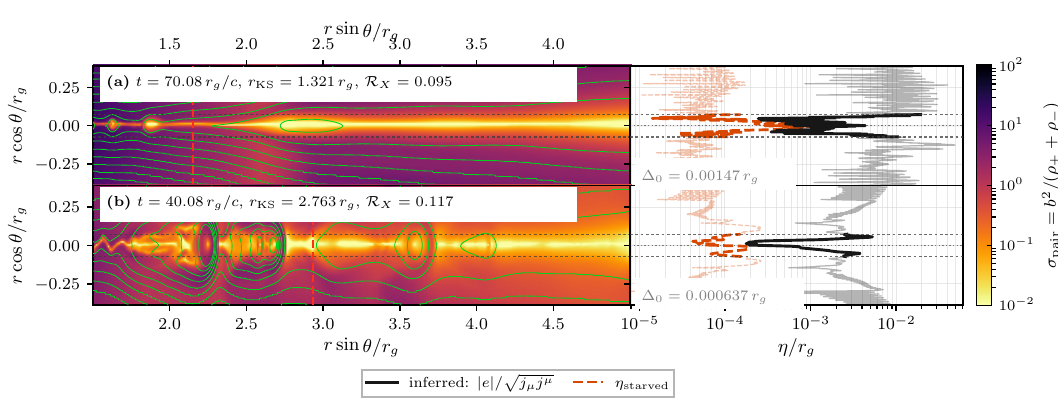}
 \caption{X-point reconnection properties of the balding magnetosphere. Here, we focus on two representative X-points at different times, one inside and one outside the stagnation radius.
 We report both the reconnection rate $\mathcal{R}_X$ at the X-point and the inferred effective scalar resistivity, $\eta_{\rm eff}$.
 Left: the cold total-pair magnetization $\sigma_{\rm pair}$.
 Green curves are poloidal magnetic-flux surfaces. The
 dashed red vertical line is the column along which the resistivity is
 extracted.
 Right: the resistivity $\eta$, sharing the left panels' vertical axis. {The dashed horizontal lines indicating the layer appear in both columns.}
 The solid black curve is the scalar-Ohm-law inference
 $\eta_{\rm eff}=|e|/(\ju_\mu\ju^\mu)^{1/2}$ and the dashed orange curve is
 $\eta_{\rm starved}=|e|/(|q_e|n)$, the charge-starved resistivity prediction in Eq.~(4) of
 \citet{Ripperda:2026EffectiveResistivity}.  For completeness, we also report the
 charge-starvation thickness $\Delta_0=B_{\rm up}/(4\pi|q_e|n_{\rm up})$ 
 in each panel. {Here $B_{\rm up}$ and $n_{\rm up}$ are the upstream reconnecting-field magnitude and total FIDO carrier density, respectively.}}
 \label{fig:baldingResistivityCuts}
\end{figure*}

We adopt an axisymmetric domain, $0\leq\varpi\leq75M$ and $-75M\leq z\leq75M$, using the reference-metric flux formulation, PLM reconstruction, an HLL solver, IMEX12 time integration, and $\cfl=0.3$. We use nested box-in-box fixed mesh refinement with a coarse resolution of $\Delta x_0=0.78 r_g$. In total, we use eight additional refinement levels: the seventh follows the sheet out to $\varpi=20.5M$ with $|z|\leq0.25M$, and the eighth covers the inner reconnection region out to $\varpi=6M$ with $|z|\leq0.125M$, giving the finest spacing $\Delta x_8=0.003 r_g$.

Figure~\ref{fig:baldingMagnetization} shows the global structure of the balding black-hole magnetosphere. The initially closed dipolar magnetic field has opened and stretched outward, while the return flux is compressed into a thin, weakly magnetized equatorial layer that extends across the displayed domain.
The upstream plasma near the current sheet has $\sigma \simeq 10$--$20$ and is thermally supported, as expected. The layer is tearing unstable and exhibits a plasmoid chain \cite{2011PhRvD..84h4019L,Bransgrove:2021heo}. We can also identify plasmoid mergers. Initial reconnection and plasmoid formation occur mainly near the stagnation point \cite{Bransgrove:2021heo}, where the radial velocity $u^r/u^0$ changes sign (see the white dashed lines in Fig.~\ref{fig:baldingVelocities}). As in Ref.~\cite{Bransgrove:2021heo}, plasmoids are rapidly accelerated outward beyond the stagnation point but move only slowly inward within it. In that region, our pair-injection scheme produces pair multiplicities $\mathcal{M} \simeq 10$ relative to the Goldreich--Julian number density $n_{\rm GJ}$, indicating consistently that enough charge carriers are supplied to screen the parallel electric field locally (Fig.~\ref{fig:baldingPairMultiplicity}). In fact, $\mathcal{M}>1$ throughout most of the current sheet, confirming that our pair-injection scheme operates as intended.

{The stagnation point is important because it gives rise to the primary X-point, where most magnetic-flux dissipation occurs. Assuming essentially ideal inflow into the stagnation point and a collisionless reconnection rate $\mathcal{R}\simeq 0.1$, Ref. \mbox{\cite{Bransgrove:2021heo}} argued that the black-hole magnetic flux should decay as $\Phi(t) \simeq \exp(-t/\tau_\Phi)$, where $\tau_\Phi \simeq 0.3 r_g/\left<u^\theta/u^0\right>$. Consistent with their GRPIC results, we find $\left<u^\theta/u^0\right>\simeq 0.03c$.
We can also estimate this more carefully by considering the inflow drift speed into the current sheet, $\boldsymbol{v}_{\rm in}= {\boldsymbol{D}\times\boldsymbol{B}}/{B^{2}}$,
which we average over an interval $|r-r_X|\leq0.04\, r_g$, $0.06\leq|z|/r_g\leq0.12$, {where $r_X$ is the X-point radius}.
At $t=50\, r_g/c$ the upstream inflow has median $0.0330$, with the local upstream magnetization being $\sigma_{\rm up}=7.98$, so the nominal Alfv\'en speed ${v_{\rm A}}=[\sigma_{\rm up}/(1+\sigma_{\rm up})]^{1/2}=0.9427$ is close to unity. This estimate is consistent with $\left<u^\theta/u^0\right>$ inferred above.
Such an inflow speed predicts a decay time $\tau_\Phi \simeq 100\, r_g/c$. We can then compare this explicitly with the measured decay of magnetic flux on the horizon (Fig. \mbox{\ref{fig:baldingHorizonFlux}}). We can see that overall the measured flux decay matches the GRPIC prediction and result, though in line with Ref. \mbox{\cite{Ripperda:2026EffectiveResistivity}}, we observe small oscillations in the decaying flux associated with plasmoids falling into the horizon.}

Finally, having established the basic properties of the balding magnetosphere, we confirm the collisionless-like reconnection properties of its current sheet.
We repeat the local X-point analysis performed in Sec.~\ref{sec:waldMagnetosphere}.
We measure the reconnection rate in the local orthonormal FIDO tetrad (Appendix~\ref{app:fidotetrad}) of the equatorial current sheet (Fig.~\ref{fig:baldingResistivityCuts}).
We focus on $t=40\, r_g/c$ and $t=70\, r_g/c$, when we identify pronounced X-points inside and outside the stagnation surface. The sheet has undergone strong plasmoid formation at both times. We then extract the effective reconnection rates at the X-points, finding $\mathcal{R}_X=0.117$ and $\mathcal{R}_X=0.095$, respectively; both are consistent with collisionless relativistic reconnection \cite{Sironi:2014jfa,Guo:2015cua}.
{We then analyze the effective resistivity, $\eta_{\rm eff}$, computed along a vertical cut. At both X-points, we find ${\eta_{\rm eff}} \simeq 2$--$4\times 10^{-4} r_g c$. Comparison with the effective charge-starved resistivity in Eq. \mbox{\eqref{eq:waldRipperdaResistivity}} shows good agreement at the X-point itself, although the agreement is poorer for the earlier X-point and the shape of the resistivity profile does not match perfectly. We further observe that the reconnection rate decreases with distance from the black hole, as also observed in the GRPIC simulation of Ref. \mbox{\cite{Meringolo:2025bdu}}.}

While this is overall a more challenging simulation than the Wald magnetosphere, our simulations demonstrate that the {GR19M} scheme is well equipped to capture effective collisionless dissipation in a full GR multi-fluid setup.

\section{Conclusions}
\label{sec:conclusions}

We have developed a general-relativistic {19-moment} formulation for dissipative
plasmas, {derived its nonlinear causality conditions and necessary and sufficient conditions for strong hyperbolicity}, and implemented a numerical scheme for
global simulations. This scheme explicitly enforces physical admissibility.
The formulation treats anisotropic transport and electron
dynamics together. In black-hole accretion and magnetospheres, the scheme
shows excellent agreement with the macroscopic dynamics and fast reconnection
found in general-relativistic particle-in-cell (GRPIC) calculations.

The equations retain bulk and shear viscosity, heat conduction, resistivity,
and Hall terms, together with electron number, momentum, and energy. Electron
inertia and pressure enter the dynamical Ohm law, while compression and Joule
heating evolve the electron thermodynamics. Our characteristic analysis bounds
the heat flux, anisotropic stresses, electron drift, and transport coefficients
on finite non-equilibrium backgrounds. We construct the left and right
eigenvectors and {identify conditions for a uniform eigenbasis,
including coincident characteristic speeds}. In
particular, electron pressure supplies the restoring force that removes the
defective electron modes of the pressureless, only weakly hyperbolic reduction.

Writing the equations in flux-divergence form allows us to evolve the complete
system with high-resolution shock-capturing methods. Implicit integration of
the relaxation and electromagnetic source terms permits timesteps longer than
the microscopic plasma and gyration times in regions where these scales need not be
resolved. Physical admissibility conditions \cite{Wu:2017grhd} guide the coupled primitive
recovery and implicit solve, constraining the densities, thermodynamics, and
dissipative currents. The implementation supports adaptive mesh refinement
and runs on CPUs and GPUs. We also constructed a stationary, transcritical
dissipative Bondi solution containing all three ion transport channels and the
electron sector, and recovered this reference in a well-balanced evolution.

The black-hole calculations test whether this description captures the
macroscopic consequences of collisionless plasma physics. Bondi-like accretion
onto a Kerr black hole develops a magnetically arrested flow, pressure
anisotropy, and field-aligned heat flux in close agreement with GRPIC
\cite{Galishnikova:2023}. The pair magnetospheres reproduce the field geometry
and plasmoid formation seen in kinetic calculations
\mbox{\cite{Parfrey:2019,Bransgrove:2021heo}}, with measured X-point reconnection rates
$\mathcal{R}_{X}=0.095$--$0.121$. These rates are characteristic of fast
collisionless reconnection. The horizon-flux decay matches the GRPIC results well \mbox{\cite{Bransgrove:2021heo}}. The inferred
X-point resistivity is also consistent with charge starvation \mbox{\cite{Goodbred:2022,Moran:2025aqb}}. Thus, the
evolved carrier supply and current reproduce both the local dissipation and
its effect on the large-scale magnetic field.

The next step is systematic calibration against GRPIC across magnetization,
plasma beta, species mass ratio, and guide-field strength \cite{Galishnikova:2023,Moran:2025aqb}.
{Independent scans of conductivity and current-relaxation time should
separate the effects of resistive dissipation and current inertia.} Measurements of
anisotropy relaxation, heat transport, electron heating, and current-sheet
structure can constrain the transport closure and replace the imposed mirror
and firehose bounds with dynamical prescriptions. Three-dimensional
comparisons should also test the influence of pair supply and scale
separation. For quantitative pair thermodynamics, the present ion-frame
reduction must be extended to retain electron stresses in the total
energy--momentum tensor and intrinsic electron dissipation.

The framework can now be applied to systems in which transport through the
bulk plasma and dissipation in current sheets determine the same global
evolution. In low-luminosity accretion flows, it will allow us to study how
anisotropic stresses and heat conduction affect accretion while reconnection
releases magnetic energy in flares. The evolution of magnetic flux near a
spinning black hole can likewise be followed together with jet launching
and dissipation in the jet boundary layer \cite{Sironi:2020mzd}.
Further applications include dissipation in pulsar and magnetar
magnetospheres, and the electromagnetic transients produced during
neutron-star collapse and compact-object mergers. Connecting these
calculations to observed emission will require radiation and nonthermal
particle models calibrated to the underlying kinetic physics.

\begin{acknowledgments}
ERM gratefully acknowledges discussions with J. Beattie, M. Disconzi, A. Galishnikova, C. Gammie, A. Hegade, P. Hopkins, J. Noronha, J. Noronha-Hostler, J.-F. Paquet, A. Philippov, B. Ripperda, L. Sironi, A. Spitkovsky, and S. Teukolsky.
ERM and SD acknowledge support by the U.S. National Science Foundation {(NSF)} under Grant Nos. AST-2307394, PHY-2541792, and PHY-2309210. ERM is also supported by a Research Fellowship from the Sloan Foundation and a William H. Hurt Scholarship at Caltech.
The simulations were performed on the NSF Vista supercomputer under grant AST21006. The authors also acknowledge the use of Delta at the National Center for Supercomputing Applications (NCSA) through allocation PHY210074 from the Advanced Cyberinfrastructure Coordination Ecosystem: Services \& Support (ACCESS) program, which is supported by National Science Foundation grants \#2138259, \#2138286, \#2138307, \#2137603, and \#2138296. Support also comes from the Resnick High Performance Computing Center, a facility supported by Resnick Sustainability Institute at the California Institute of Technology. We acknowledge support from OpenAI’s Researcher Access Program for providing access to ChatGPT models to support this work. Work in this paper primarily used OpenAI's ChatGPT 5.6-sol, 5.5, and 6-astra, with additional tasks carried out by Anthropic's Opus 4.8 and 5, as well as Fable 5 and 5.1.
\end{acknowledgments}

\section*{Use of large language models}
We used large language models (LLMs), identified in the acknowledgements, to assist with algebraic derivations and symbolic checks using Mathematica and with the development,
debugging, and optimization of numerical and analysis code based on the initial version developed by the authors. We also used them for consistency checks between the manuscript and the implementation and for assistance in writing technical appendices that match the Mathematica derivations and code implementation. Verification of this work included explicit computer-algebra
calculations in Mathematica, comparisons with the implemented equations, and extensive numerical tests and comparisons with reference results in the literature beyond what is presented in the main text.
\newpage
\clearpage

\appendix

\section{Double-adiabatic matching of the shear coefficient in the Braginskii limit}
\label{app:double-adiabatic-shear}

Here we briefly connect our closure prescription to recent calculations of
relativistic double-adiabatic closures, which determine the collisionless
production of a small pressure anisotropy from an initially isotropic,
gyrotropic distribution \cite{Ley:2026DoubleAdiabatic,Wierzchucka:2026DoubleAdiabatic}.
As we will see, this can be straightforwardly incorporated into our approach.

First, we need to match the notation of Refs. \cite{Ley:2026DoubleAdiabatic,Wierzchucka:2026DoubleAdiabatic}.

Let $D:=u^\mu\nabla_\mu$,
$B:=\sqrt{b^2}$, and
$\sigma_\parallel:=\hat b^\mu\hat b^\nu\sigma_{\mu\nu}$.
Then, conservation of particle number and ideal induction yield
\begin{align}
 D\ln n&=-\theta,
 &D\ln B&=\sigma_\parallel-\frac{2}{3}\theta,
 &D\ln\!\left(Bn^{-2/3}\right)&=\sigma_\parallel .
 \label{eq:double-adiabatic-kinematics}
\end{align}

For an ultrarelativistic isotropic reference state $(n_0, B_0)$, define
$n':=n/n_0$, $B':=B/B_0$, and $A:=B'^3/n'^2$, which measures the anisotropy.  The exact angular moments of Ref.~\cite{Wierzchucka:2026DoubleAdiabatic} have the near-isotropic expansion
\begin{align}
 P_\perp'&\approx B'^2 \left[1-\frac{2}{5}(A-1)\right]\,,\\
 P_\parallel'&\approx \frac{n'^2}{B'} \left[1-\frac{1}{5}(A-1) \right]\,.
 \label{eq:double-adiabatic-moments}
\end{align}
Consequently,
\begin{align}
 D\ln P_\perp
 &=\frac{4}{5}\left(D\ln n+D\ln B\right),
 \notag\\
 D\ln P_\parallel
 &=\frac{4}{5}\left(3D\ln n-2D\ln B\right),
 \label{eq:double-adiabatic-pressure-laws}
\end{align}
and, at isotropy,
\begin{align}
 D\Delta P
 =\frac{12}{5}P\sigma_\parallel,
 \qquad \Delta P:=P_\perp-P_\parallel .
 \label{eq:double-adiabatic-drive-ur}
\end{align}

The small-amplitude kinetic integral of
Ref.~\cite{Ley:2026DoubleAdiabatic} gives the corresponding result at arbitrary
particle temperature,
\begin{align}
 D\Delta P&=3\varepsilon P\sigma_\parallel,
 &\varepsilon&:=1-\frac{1}{5}\left\langle v^2\right\rangle_P,
 \notag\\
 \left\langle v^2\right\rangle_P
 &:=\frac{\int_0^\infty dp\,p^3v^3f_0(p)}
 {\int_0^\infty dp\,p^3vf_0(p)},
 &\frac45&\leq\varepsilon\leq1.
 \label{eq:double-adiabatic-epsilon}
\end{align}
This follows by integrating their small-anisotropy moment by parts, using
$p\,dv/dp=v(1-v^2)$ and a vanishing boundary term.  Thus,
$\varepsilon\to1$ in the nonrelativistic limit (recovering the original Braginskii-type expression) and
$\varepsilon\to4/5$ in the ultrarelativistic limit.

In our convention the gyrotropic shear stress is
\begin{align}
 \pi^{\mu\nu}
 =-\Delta P\left(\hat b^\mu\hat b^\nu
 -\frac13\Delta^{\mu\nu}\right),
 \qquad
 \hat b_\mu\hat b_\nu\pi^{\mu\nu}=-\frac23\Delta P.
 \label{eq:double-adiabatic-stress-map}
\end{align}
Projecting Eq.~\eqref{eq.comovpi} twice along $\hat b^\mu$ and using
$\hat b_\mu D\hat b^\mu=0$ yields, to linear order in the anisotropy,
\begin{align}
 D\Delta P
 =\frac{3\eta}{\tau_\pi}\sigma_\parallel
 -\frac{\Delta P}{\tau_\pi}.
 \label{eq:double-adiabatic-gr14m}
\end{align}
Identifying $1/\tau_\pi$ with the isotropization rate and matching the
collisionless driving term therefore fixes
\begin{align}
 {\frac{\eta}{\tau_\pi}=\varepsilon P},
 \qquad
 \frac{\eta}{\tau_\pi}\xrightarrow[T/m\gg1]{}\frac45P.
 \label{eq:double-adiabatic-match}
\end{align}
We stress that this holds formally only in the Braginskii-like non-resistive limit.

\section{{Characteristic eigenvectors and hyperbolicity}}
\label{app:eigenvectors}

Appendix~\ref{app:characteristics} provides the characteristic speeds and
the causal domain of the system.  Here we construct the primitive right and
left eigenvectors in the Eulerian frame for future use in simulation codes such as ours,
following the method of Ref.~\cite{Teukolsky:2026}. At a repeated root or a
coincidence between characteristic families, the individual columns are
replaced by a basis of the joint eigenspace.  Section~\ref{app:ev_causality} states
the strong-hyperbolicity conditions and the compatibility requirement at
such coincidences. 

For a characteristic covector $\xi_\mu$, let
$\mathbb A(\xi):=\mathbb A^\mu\xi_\mu$ be the principal symbol in primitive
variables.  A right eigenvector is a nonzero column $R_I$ satisfying
$\mathbb A(\xi_I)R_I=0$; it gives the primitive perturbation carried by that
mode.  A left eigenvector is a nonzero row $L_I$ satisfying
{$L_I\mathbb A(\xi_I)=0$. The modal projector of the evolution problem is the time-weighted row $\ell_I=L_I\mathbb A^0$. For simple roots normalize $\ell_IR_J=\delta_{IJ}$.}
  At a repeated
root we instead choose bases of the full right and left kernels, as required
for strong hyperbolicity.

{As in Sec. \ref{app:characteristics}, we consider only perturbations tangent to}
\begin{align}
 {u_\mu u^\mu}&{=-1,} &
 {u_\mu q^\mu}&{=0,} &
 {u_\mu k^\mu}&{=0,}\notag\\
 {u_\mu\pi^{\mu\nu}}&{=0,} &
 {\pi^{\mu\nu}}&{=\pi^{\nu\mu},} &
 {\pi^\mu{}_\mu}&{=0.}
 \label{eq.ev.constraints}
\end{align}

\subsection{{Eulerian representation}}

{Let $n^\mu$ be the hypersurface normal and let
$(s^\mu,\tau_1^\mu,\tau_2^\mu)$ be an orthonormal Eulerian triad, with
$s^\mu$ normal to the numerical face, and set
$v_s:=v^\mu s_\mu=v^is_i$.  Following
Ref.~\cite{Teukolsky:2026}, write the characteristic covector in terms of the
Eulerian speed $y$ as}
\begin{align}
 {\xi_\mu(y)}&{:=y n_\mu+s_\mu,} &
 {u^\mu}&{=W(n^\mu+v^\mu),}\notag\\
 {z(y)}&{:=u^\mu\xi_\mu=W(v_s-y),} &
 {\ell(y)}&{:=\sqrt{1-y^2+z^2}>0,}\notag\\
 {a(y)}&{:=\frac{z}{\ell},} &
 {\widehat n^\mu(y)}&{:=\frac{\xi^\mu+z u^\mu}{\ell}.}
 \label{eq.ev.eulerian-covector}
\end{align}
{Thus, $a$ is the normalized fluid-frame root $\hat z$ used in
Appendix~\ref{app:characteristics}, not the Eulerian speed.  Material modes
have $y=v_s$, light-cone modes have $y=\pm1$, and every other $y_I$ follows by
inserting $a(y_I)$ and $\widehat n^\mu(y_I)$ into the corresponding
 polynomial in Appendix~\ref{app:characteristics}.  The coordinate speed is}
\begin{align}
{\lambda_I=\alpha y_I-\beta^i s_i.}
 \label{eq.ev.coordinate-speed}
\end{align}
{The Eulerian and coordinate eigenvectors are identical.  Equation
\eqref{eq.ev.eulerian-covector} also keeps the aberration of the wave normal.}

\subsubsection{{Independent amplitudes and frame map}}

{We here choose the fluid tetrad obtained by the canonical boost
of the Eulerian orthonormal frame.  In local normal coordinates, with
$U^i:=Wv^i$,}
\begin{align}
 {u^\mu}&{=(W,U^i),} &
 {e_{\hat a}^{\,0}}&{=U_{\hat a},} &
 {e_{\hat a}^{\,i}}&{=\delta^i{}_{\hat a}
 +\frac{U^iU_{\hat a}}{W+1}.}
 \label{eq.ev.boost-tetrad}
\end{align}
{All background tensors are expressed in this fixed tetrad, which is not
rotated separately for different modes.  For mode $I$, set
$\widehat n_I^{\hat a}:=e^{\hat a}{}_{\mu}\widehat n_I^\mu$ and evaluate the
fluid-frame symbol at $(a_I,\widehat n_I^{\hat a})$.  The constrained
fluid-frame amplitudes are}
\begin{widetext}
\begin{align}
 {\widehat V_{\rm R}=\bigl(}
 &{\delta e,\delta\rhob,\delta\Pi,\delta u^{\hat a},
 \delta q^{\hat a},
 \delta\pi^{\hat1\hat1},\delta\pi^{\hat1\hat2},
 \delta\pi^{\hat1\hat3},\delta\pi^{\hat2\hat2},
 \delta\pi^{\hat2\hat3},}\notag\\[-0.2ex]
 &{\delta k^{\hat a},\delta\rhoq,\delta e_e,
 \delta\mathcal E^{\hat a},\delta\mathcal B^{\hat a},
 \delta\psi,\delta\phi\bigr)^{\rm T},}
 \label{eq.ev.comoving-state-r}
\end{align}
\end{widetext}
{Here
$\delta\pi^{\hat3\hat3}=-\delta\pi^{\hat1\hat1}
-\delta\pi^{\hat2\hat2}$.  The variables
$\delta\mathcal E^{\hat a}$ and $\delta\mathcal B^{\hat a}$ are the six
components of the independent perturbation $\delta F^{\mu\nu}$ in this fixed
tetrad.}

\begin{widetext}
\begin{samepage}
{Set ${e_{\rm int}}:=e-\rhob=\rhob\epsilon$.  The $27$ independent resistive
Eulerian amplitudes, in public primitive order, are}
\begin{align}
 {V_{\rm R}^{\rm E}=\bigl(}
 &{\delta\rhob,{\delta e_{\rm int}},
 \delta U^s,\delta U^1,\delta U^2,
 \delta B^s,\delta B^1,\delta B^2,\delta\phi,}
 \notag\\[-0.2ex]
 &{\delta E^s,\delta E^1,\delta E^2,\delta\psi,
 \delta\rho_e,\delta\Pi,
 \delta Q^s,\delta Q^1,\delta Q^2,}
 \notag\\[-0.2ex]
 &{\delta\Sigma^{ss},\delta\Sigma^{s1},\delta\Sigma^{s2},
 \delta\Sigma^{11},\delta\Sigma^{12},
 \delta J^s,\delta J^1,\delta J^2,\delta e_e\bigr)^{\rm T}.}
 \label{eq.ev.state}
\end{align}
\end{samepage}
\end{widetext}
{We use
$Q^i:=\gamma^i{}_\mu q^\mu$,
$J^i:=\gamma^i{}_\mu\ju^\mu$, and
$\Sigma^{ij}:=\gamma^i{}_\mu\gamma^j{}_\nu\pi^{\mu\nu}$.
The factors of $\sqrt\gamma$ multiplying the public field primitives are
fixed coefficients in the principal symbol and have been suppressed.  The
five displayed shear components are independent. }

{Before boosting a fluid-frame column, its dependent components must be
restored.  For $X^\mu=q^\mu,k^\mu$,}
\begin{align}
 {\delta u^\mu}
 &{=e_{\hat a}^{\,\mu}\delta u^{\hat a},}\notag\\
 {\delta X^\mu}
 &{=e_{\hat a}^{\,\mu}\delta X^{\hat a}
 +u^\mu X_{\hat a}\delta u^{\hat a},}\notag\\
 {\delta\pi^{\mu\nu}}
 &{=e_{\hat a}^{\,\mu}e_{\hat b}^{\,\nu}
 \delta\pi^{\hat a\hat b}
 +2u^{(\mu}e_{\hat a}^{\,\nu)}
 \pi^{\hat a}{}_{\hat b}\delta u^{\hat b}.}
 \label{eq.ev.constraint-lift}
\end{align}
{The Eulerian velocity, heat flux, current flux, and shear entries are}
\begin{align}
 {\delta U^i}&{=\gamma^i{}_\mu\delta u^\mu,} &
 {\delta Q^i}&{=\gamma^i{}_\mu\delta q^\mu,}\notag\\
 {\delta K^i}&{=\gamma^i{}_\mu\delta k^\mu,} &
 {\delta\Sigma^{ij}}&{=\gamma^i{}_\mu\gamma^j{}_\nu
 \delta\pi^{\mu\nu}.}
 \label{eq.ev.vector-boost}
\end{align}
{Writing $h^{\rm E}_{ij}:=\gamma_{ij}-v_i v_j$, the missing shear
component follows from}
\begin{align}
 {h^{\rm E}_{ij}\Sigma^{ij}}&{=0,}\notag\\
 {h^{\rm E}_{ij}\delta\Sigma^{ij}}
 &{=2\Sigma^{ij}v_i\delta v_j,\qquad
 \delta v^i=\frac{\gamma^i{}_j-v^iv_j}{W}\,\delta U^j.}
 \label{eq.ev.eulerian-shear-constraint}
\end{align}
{This equation fixes $\delta\Sigma^{22}$; its coefficient
$h^{\rm E}_{22}=1-v_2^2$ is positive for a subluminal fluid velocity.}

{The electron variables used by the fluid-frame symbol and by the code
are related by}
\begin{align}
 {{\delta e_{\rm int}}}
 &{=\delta e-\delta\rhob,}\notag\\
 {\delta\rho_e}
 &{=\frac{m_eZ}{m_i}\,\delta\rhob
 -\frac{m_e}{q_e}\delta\rhoq,}\notag\\
 {\delta J^i}
 &{=\frac{\delta K^i}{w_e}
 -\frac{K^i}{w_e^2}\delta w_e,}\notag\\
 {\delta w_e}
 &{=\frac{1+P_{e,e}}{\rho_e}\delta e_e
 +\left(\frac{P_{e,n}}{m_e\rho_e}
 -\frac{H_e}{\rho_e^2}\right)\delta\rho_e.}
 \label{eq.ev.electron-jacobian}
\end{align}

{For the resistive field entries,}
\begin{align}
 {\delta F^{\mu\nu}}
 &{=2u^{[\mu}e_{\hat a}^{\,\nu]}
 \delta\mathcal E^{\hat a}
 +\levciv^{\mu\nu\rho\sigma}
 e_{\hat a\rho}u_\sigma\delta\mathcal B^{\hat a},}\notag\\
 {\delta E^i}
 &{=\gamma^i{}_\mu\delta F^{\mu\nu}n_\nu,\qquad
 \delta B^i=\gamma^i{}_\mu\delta{\star F}^{\mu\nu}n_\nu.}
 \label{eq.ev.resistive-field-jacobian}
\end{align}
{No $\delta u^\mu$ term is added here because $F^{\mu\nu}$ is an
independent resistive primitive.}

{Equations~\eqref{eq.ev.constraint-lift} and
\eqref{eq.ev.resistive-field-jacobian}, together with the scalar maps, define
the square tangent Jacobian}
\begin{align}
 {\mathcal J_{\rm R}}
 &{:=\frac{\partial V_{\rm R}^{\rm E}}
 {\partial\widehat V_{\rm R}}.}
 \label{eq.ev.forward-jacobians}
\end{align}
{Its tangent inverse, denoted by $\mathcal K_{\rm R}$, follows from
$\delta u^\mu=\gamma^\mu{}_i\delta U^i+n^\mu v_i\delta U^i$,
$X^0=v_iX^i$, Eq.~\eqref{eq.ev.eulerian-shear-constraint}, and}
\begin{align}
 {\delta F^{\mu\nu}}
 &{=2n^{[\mu}\delta E^{\nu]}
 +\levciv^{\mu\nu\rho\sigma}\delta B_\rho n_\sigma,}
 \label{eq.ev.inverse-field-r}
\end{align}
{These identities are inverse only on the constrained tangent space,
which is precisely where the characteristic symbol is defined.}

\subsubsection{{Explicit Eulerian components}}

{We now give one right column and one left row for every resistive characteristic
family.  We label the irreducible roots by $a_I$
and complete each $\widehat n_I$ by an oriented pair
$(t_{I1},t_{I2})$.  The entries are rational functions of $a_I$ and the
background, and we express each wave frame in the fixed tetrad of
Eq.~\eqref{eq.ev.boost-tetrad}.  We suppress the label $I$.}

{Put $U^i=Wv^i$ and define, for $i=(s,1,2)$,}
\begin{align}
 {L^i{}_a}&{:=\delta^i{}_a+\frac{U^iU_a}{W+1},}&
 {A^i{}_a}&{:=W\delta^i{}_a-\frac{U^iU_a}{W+1},}\notag\\
 {C^i{}_a}&{:=\epsilon^i{}_{ja}U^j.}
 \label{eq.ev.explicit-boost-blocks}
\end{align}
{For a resistive fluid-frame amplitude}
\begin{align}
 {\widehat r_{\rm R}=
 (\varepsilon,r,\varpi,w^a,h^a,S^{ab},c^a,
 \varrho,x_e,\epsilon^a,\beta^a,\psi,\phi)^{\rm T},}
 \label{eq.ev.rtuple}
\end{align}
{These are, in order, the energy, baryon density, bulk pressure, velocity,
heat, shear, current, charge, electron-energy, electric, magnetic, and two
cleaning perturbations.  The tensor $S^{ab}$ is symmetric and trace-free.
Define}
\begin{align}
 {U_r^i}&{:=L^i{}_aw^a,}\notag\\
 {Q_r^i}&{:=L^i{}_ah^a+U^iq_aw^a,}\notag\\
 {K_r^i}&{:=L^i{}_ac^a+U^ik_aw^a,}\notag\\
 {\Sigma_r^{ij}}&{:=L^i{}_aL^j{}_bS^{ab}
 +(U^iL^j{}_a+U^jL^i{}_a)\pi^a{}_bw^b,}\notag\\
 {\rho_{e,r}}&{:=\frac{m_eZ}{m_i}r-\frac{m_e}{q_e}\varrho,}\notag\\
 {w_{e,r}}&{:=\frac{1+P_{e,e}}{\rho_e}x_e
 +\left(\frac{P_{e,n}}{m_e\rho_e}-\frac{H_e}{\rho_e^2}\right)
 \rho_{e,r},}\notag\\
 {J_r^i}&{:=\frac{K_r^i}{w_e}-\frac{K^i}{w_e^2}w_{e,r},}\notag\\
 {E_r^i}&{:=A^i{}_a\epsilon^a-C^i{}_a\beta^a,}\notag\\
 {B_r^i}&{:=A^i{}_a\beta^a+C^i{}_a\epsilon^a.}
 \label{eq.ev.explicit-res-blocks}
\end{align}
\begin{widetext}
{Thus, the complete Eulerian column is the following ordered
$27$-tuple:}
\begin{align}
 {R_{\rm R}^{\rm E}[\widehat r_{\rm R}]=
 \bigl(r,\varepsilon-r,U_r^s,U_r^1,U_r^2,
 B_r^s,B_r^1,B_r^2,\phi,
 E_r^s,E_r^1,E_r^2,\psi,\rho_{e,r},\varpi,
 Q_r^s,Q_r^1,Q_r^2,
 \Sigma_r^{ss},\Sigma_r^{s1},\Sigma_r^{s2},
 \Sigma_r^{11},\Sigma_r^{12},J_r^s,J_r^1,J_r^2,x_e\bigr)^{\rm T}.}
 \label{eq.ev.explicit-R27}
\end{align}
\end{widetext}
{Each mode below specifies every nonzero
amplitude.}

{The left rows require both the change of primitive variables and the
Eulerian time symbol.  We give both explicitly.  Define}
\begin{align}
 {F_a{}^i}&{:=\frac{A_a{}^i}{W}
 =\delta_a{}^i-\frac{U_aU^i}{W(W+1)}.}
 \label{eq.ev.explicit-inverse-blocks}
\end{align}
\begin{widetext}
{For a resistive differential $dV_{\rm R}^{\rm E}$, the
fluid-frame amplitudes are the following Eulerian one-forms:}
\begin{align}
 {r}&{=d\rhob,}&
 {\varepsilon}&{={d e_{\rm int}}+d\rhob,}&
 {\varpi}&{=d\Pi,}&
 {w^a}&{=F^a{}_i\,dU^i,}\notag\\
 {h^a}&{=F^a{}_i\,dQ^i-\frac{U^a}{W}q_bw^b,}&
 {\varrho}&{=\frac{q_eZ}{m_i}\,d\rhob
 -\frac{q_e}{m_e}\,d\rho_e,}&
 {x_e}&{=d e_e,}\notag\\
 {d w_e}&{=\frac{1+P_{e,e}}{\rho_e}\,d e_e
 +\left(\frac{P_{e,n}}{m_e\rho_e}-\frac{H_e}{\rho_e^2}\right)d\rho_e,}&
 {c^a}&{=F^a{}_i\left(w_e\,dJ^i+\frac{K^i}{w_e}\,d w_e\right)
 -\frac{U^a}{W}k_bw^b,}\notag\\
 {\epsilon^a}&{=A^a{}_i\,dE^i+C^a{}_i\,dB^i,}&
 {\beta^a}&{=A^a{}_i\,dB^i-C^a{}_i\,dE^i.}
 \label{eq.ev.explicit-coframe}
\end{align}
\end{widetext}
\begin{widetext}
{To recover the shear form, put $h_{ij}^{\rm E}:=\delta_{ij}-v_iv_j$,
restore the omitted component through}
\begin{align}
 {d\Sigma^{22}=\frac{1}{h_{22}^{\rm E}}\bigl[
 2\Sigma^{ij}v_i\,dv_j-h_{ss}^{\rm E}d\Sigma^{ss}
 -h_{11}^{\rm E}d\Sigma^{11}-2h_{s1}^{\rm E}d\Sigma^{s1}
 -2h_{s2}^{\rm E}d\Sigma^{s2}-2h_{12}^{\rm E}d\Sigma^{12}\bigr],}
 \label{eq.ev.explicit-shear-completion}\\
 {dv^i=\frac{\delta^i{}_j-v^iv_j}{W}\,dU^j,\qquad
 Z^{ij}:=(U^iL^j{}_a+U^jL^i{}_a)\pi^a{}_bw^b,\qquad
 S^{ab}=F^a{}_iF^b{}_j(d\Sigma^{ij}-Z^{ij}).}
 \label{eq.ev.explicit-shear-coframe}
\end{align}
\end{widetext}

\begin{widetext}
{These amplitudes determine all constrained spacetime variations:}
\begin{align}
 {d u^\mu}&{=e_a{}^\mu w^a,}&
 {d q^\mu}&{=e_a{}^\mu h^a+u^\mu q_aw^a,}&
 {d k^\mu}&{=e_a{}^\mu c^a+u^\mu k_aw^a,}\notag\\
 {d\pi^{\mu\nu}}&{=e_a{}^\mu e_b{}^\nu S^{ab}
 +2u^{(\mu}e_a{}^{\nu)}\pi^a{}_bw^b,}&
 {f^{\mu\nu}}&{=2n^{[\mu}dE^{\nu]}
 +\levciv^{\mu\nu\rho\sigma}dB_\rho n_\sigma.}
 \label{eq.ev.explicit-lifts}
\end{align}
\end{widetext}
{Here $f^{\mu\nu}=dF^{\mu\nu}$.  Set}
\begin{align}
 {{d\bar n_e}}&{:=\frac{Z}{m_i}r-\frac{\varrho}{q_e},}&
 {dP_e}&{:=P_{e,e}x_e+P_{e,n} {d\bar n_e},}\notag\\
 {d\chi_e}&{:=\frac{q_e}{m_e}dP_e,}\notag\\
 {dP}&{:=P_{,e}\varepsilon+P_{,\rhob}r,}\notag\\
 {dT}&{:=T_{,e}\varepsilon+T_{,\rhob}r.}
 \label{eq.ev.explicit-thermo-forms}
\end{align}
\begin{widetext}
{With $E:=e+P+\Pi$, the stress-tensor variations are}
\begin{align}
 {dT_{\rm fl}^{\mu\nu}}
 &{=\varepsilon u^\mu u^\nu+2E u^{(\mu}d u^{\nu)}
 +(dP+\varpi)\Delta^{\mu\nu}+2d q^{(\mu}u^{\nu)}
 +2q^{(\mu}d u^{\nu)}+d\pi^{\mu\nu},}\notag\\
 {dT_{{\rm EM},{\rm R}}^{\mu\nu}}
 &{=f^{\mu\alpha}F^\nu{}_\alpha+F^{\mu\alpha}f^\nu{}_\alpha
 -\frac12g^{\mu\nu}F^{\alpha\beta}f_{\alpha\beta}.}
 \label{eq.ev.explicit-stress-forms}
\end{align}
\end{widetext}

\clearpage
\begin{widetext}
{Let $z_0:=u^\mu n_\mu=-W$, $p_a:=e_a{}^\mu n_\mu=-U_a$, and
$\Theta_0:=p_aw^a$.  The Eulerian time rows of the equations are}
\begin{align}
 {\mathcal T_{\rm R}^\nu}&{:=n_\mu
 (dT_{\rm fl}^{\mu\nu}+dT_{{\rm EM},{\rm R}}^{\mu\nu}),}&
 {\mathcal R}&{:=z_0r+\rhob\Theta_0,}&
 {\mathcal P}&{:=\tau_\Pi z_0\varpi+(\zeta+\dPiPi\Pi)\Theta_0,}\notag\\
 {\mathcal Q_a}&{:=\tauq z_0h_a+\dqq q_a\Theta_0
 +\kappa(p_a dT+Tz_0w_a),}\notag\\
 {\mathcal S_{ab}}&{:=\tau_\pi z_0S_{ab}+\dpipi\pi_{ab}\Theta_0
 +\eta\left(p_aw_b+p_bw_a-\frac23\delta_{ab}\Theta_0\right).}
 \label{eq.ev.eulerian-time-hydro}
\end{align}
\begin{align}
 {\mathcal K_a}&{:=e_{a\nu}\Delta^\nu{}_\lambda
 \left[z_0d k^\lambda+k^\lambda\Theta_0+(n_\mu k^\mu)d u^\lambda
 -\omega_e z_0d u^\lambda-\Delta^{\lambda\mu}n_\mu d\chi_e\right],}\notag\\
 {\mathcal C_q}&{:=z_0\varrho+\rhoq\Theta_0
 +\frac{n_\mu d k^\mu}{w_e}-\frac{n_\mu k^\mu}{w_e^2}d w_e,}&
 {\mathcal E_e}&{:=z_0x_e+H_e\Theta_0-\frac{m_e}{q_e}n_\mu d k^\mu,}\notag\\
 {\mathcal M^\nu}&{:=n_\mu f^{\mu\nu}+n^\nu d\psi,}&
 {\widetilde{\mathcal M}^{\nu}}&{:=n_\mu {\star f}^{\mu\nu}+n^\nu d\phi.}
 \label{eq.ev.eulerian-time-res}
\end{align}
\end{widetext}
{For any vector-valued equation row $\mathcal Y^\nu$ in these formulas,
we use the fluid-tetrad components}
\begin{align}
 {\mathcal Y_0:=-u_\nu\mathcal Y^\nu,\qquad
 \mathcal Y_a:=e_{a\nu}\mathcal Y^\nu,}\notag\\
 {\mathcal Y_n:=\widehat n^a\mathcal Y_a,\qquad
 \mathcal Y_A:=t_A^a\mathcal Y_a.}
 \label{eq.ev.eulerian-time-projections}
\end{align}
{Each object in Eqs.~\eqref{eq.ev.eulerian-time-hydro} and
\eqref{eq.ev.eulerian-time-res} is therefore a displayed row in Eulerian variables.  The left eigenvectors below are written directly as
linear combinations of these rows.}

\subsubsection{{The explicit three-row chart}}

{The eight dissipative roots can be written compactly.  Define}
\begin{align}
 {\mathcal K}&{:=\frac{\kappa T}{\tauq},}&
 {c_T}&{:=\frac{\kappa T_{,e}}{\tauq},}&
 {c_\rho}&{:=\frac{\kappa\rhob T_{,\rhob}}{\tauq},}\notag\\
 {d_q}&{:=\frac{\dqq}{\tauq},}&
 {d_\pi}&{:=\frac{\dpipi}{\tau_\pi},}\notag\\
 {y_\eta}&{:=\frac{\eta}{\tau_\pi},}&
 {Y_\Pi}&{:=\frac{\zeta+\dPiPi\Pi}{\tau_\Pi},}\notag\\
 {Y_P}&{:=\rhob P_{,\rhob}+Y_\Pi,}&
 {q_n}&{:=q_i\widehat n^i.}
 \label{eq.ev.reduced-defs}
\end{align}
{After solving the baryon, bulk, heat, and shear rows component by
component, the four conservation rows are}
\begin{align}
 {A\varepsilon+B_jw^j}&{=g_0,}&
 {C_i\varepsilon+D_{ij}^{\rm R}w^j}&{=g_i,}
 \label{eq.ev.reduced-system}
\end{align}
\begin{widetext}
{where}
\begin{align}
 {A}&{:=a-\frac{c_T}{a},}\notag\\
 {B_j}&{:=2aq_j+(\pi\widehat n)_j
 +\left(E-\mathcal K-\frac{d_qq_n}{a}
 +\frac{c_\rho}{a^2}\right)\widehat n_j,}\notag\\
 {C_i}&{:=(P_{,e}-c_T)\widehat n_i,}
 \label{eq.ev.ABC}\\
 {D_{ij}^{\rm R}}&{:=a[(E-\mathcal K)\delta_{ij}+\pi_{ij}]
 +q_n\delta_{ij}+(1-d_q)q_i\widehat n_j}\notag\\
 &{\quad-\frac{d_\pi}{a}(\pi\widehat n)_i\widehat n_j
 -\frac{y_\eta}{a}\left(\delta_{ij}+\frac13\widehat n_i\widehat n_j\right)
 +\frac{c_\rho-Y_P}{a}\widehat n_i\widehat n_j.}
 \label{eq.ev.DR}
\end{align}
\end{widetext}
\begin{widetext}
{Put $c_C:=P_{,e}-c_T$ and define the primitive coefficients}
\begin{align}
 \kappa_0&:=A\left[a(E-\mathcal K)+q_n-\frac{y_\eta}{a}\right],&
 \kappa_\pi&:=Aa,\notag\\
 \kappa_q&:=A(1-d_q),&
 \kappa_{\pi n}&:=-\frac{Ad_\pi}{a},\notag\\
 \kappa_{nq}&:=-2ac_C,&
 \kappa_{n\pi}&:=-c_C,\notag\\
 \kappa_{nn}&:=\frac{A}{a}\left(c_\rho-Y_P-\frac{y_\eta}{3}\right)
 -c_C\left(E-\mathcal K-\frac{d_qq_n}{a}+\frac{c_\rho}{a^2}\right).
 \label{eq.ev.K-coefficients}
\end{align}
{Then}
\begin{align}
 K_{ij}^{\rm R}&=\kappa_0\delta_{ij}+\kappa_\pi\pi_{ij}
 +\kappa_qq_i\widehat n_j+\kappa_{\pi n}(\pi\widehat n)_i\widehat n_j
 \notag\\[-2pt]
 &\quad+\kappa_{nq}\widehat n_iq_j
 +\kappa_{n\pi}\widehat n_i(\pi\widehat n)_j
 +\kappa_{nn}\widehat n_i\widehat n_j.
 \label{eq.ev.K3}
\end{align}
{In the oriented wave frame
$(e_1,e_2,e_3)=(\widehat n,t_1,t_2)$, so that
$q_1=q_n$ and $\pi_{11}=\pi_{nn}$, all nine entries are}
\begin{align}
 \mathsf k_{11}&:=\kappa_0+(\kappa_\pi+\kappa_{\pi n}+\kappa_{n\pi})\pi_{11}
 +(\kappa_q+\kappa_{nq})q_1+\kappa_{nn},\notag\\
 \mathsf k_{12}&:=(\kappa_\pi+\kappa_{n\pi})\pi_{12}+\kappa_{nq}q_2,&
 \mathsf k_{13}&:=(\kappa_\pi+\kappa_{n\pi})\pi_{13}+\kappa_{nq}q_3,\notag\\
 \mathsf k_{21}&:=(\kappa_\pi+\kappa_{\pi n})\pi_{12}+\kappa_qq_2,&
 \mathsf k_{22}&:=\kappa_0+\kappa_\pi\pi_{22},&
 \mathsf k_{23}&:=\kappa_\pi\pi_{23},\notag\\
 \mathsf k_{31}&:=(\kappa_\pi+\kappa_{\pi n})\pi_{13}+\kappa_qq_3,&
 \mathsf k_{32}&:=\kappa_\pi\pi_{23},&
 \mathsf k_{33}&:=\kappa_0+\kappa_\pi\pi_{33},
 \label{eq.ev.K-nine}
\end{align}
{so that $K_{ij}^{\rm R}=\mathsf k_{ij}$ and
$\boldsymbol k_i^{\rm R}=(\mathsf k_{i1},\mathsf k_{i2},\mathsf k_{i3})$.}
\end{widetext}
{The same reduction gives a compact primitive-only construction of the
ion--dissipative octic.  To retain zero-speed multiplicities when specializing the state, clear the fixed fourth-order pole and extend the polynomial to $a=0$:}
\begin{align}
 \widetilde{\mathcal Q}_8^{\,\rm prim}(a,\widehat n;\mathcal P)
 &:=-a^4\det
 \begin{pmatrix}
  A&B_j\\ C_i&D_{ij}^{\rm R}
 \end{pmatrix},\notag\\
 \overline{\mathcal Q}_8^{\,\rm prim}
 &:=\frac{\widetilde{\mathcal Q}_8^{\,\rm prim}}
 {[a^8]\widetilde{\mathcal Q}_8^{\,\rm prim}}.
 \label{eq.ev.Q8-primitive-pencil}
\end{align}
{Equations~\eqref{eq.ev.reduced-defs}--\eqref{eq.ev.DR} make every entry of
this $4\times4$ determinant an explicit contraction of the reconstructed
primitives with $\widehat n$.  Eliminating the ten ion equations gives the determinant factor $\tau_\Pi\tauq^3\tau_\pi^5a^{10}$, leaving the material factor $a^6$ times this octic. Cancelling $a$ after specializing the state can remove physical roots.
{This polynomial has the roots and multiplicities of the physical
factor $\mathcal Q_8$; its monic form is $F_8$ of
Eq.~\eqref{eq.resgeneral.polynomial}.} The leading coefficient is nonzero under
(SH1). }
\begin{widetext}
{At a simple octic root define the three polarization representatives}
\begin{align}
 {w^{(12)}}&{:=
 (\mathsf k_{12}\mathsf k_{23}-\mathsf k_{13}\mathsf k_{22},
 \mathsf k_{13}\mathsf k_{21}-\mathsf k_{11}\mathsf k_{23},
 \mathsf k_{11}\mathsf k_{22}-\mathsf k_{12}\mathsf k_{21}),}\notag\\
 {w^{(23)}}&{:=
 (\mathsf k_{22}\mathsf k_{33}-\mathsf k_{23}\mathsf k_{32},
 \mathsf k_{23}\mathsf k_{31}-\mathsf k_{21}\mathsf k_{33},
 \mathsf k_{21}\mathsf k_{32}-\mathsf k_{22}\mathsf k_{31}),}\notag\\
 {w^{(31)}}&{:=
 (\mathsf k_{32}\mathsf k_{13}-\mathsf k_{33}\mathsf k_{12},
 \mathsf k_{33}\mathsf k_{11}-\mathsf k_{31}\mathsf k_{13},
 \mathsf k_{31}\mathsf k_{12}-\mathsf k_{32}\mathsf k_{11}).}
 \label{eq.ev.right-cross-charts}
\end{align}
\end{widetext}
{Every $\mathsf k_{ij}$ in Eq.~\eqref{eq.ev.right-cross-charts} is the
primitive rational expression printed in Eqs.~\eqref{eq.ev.K-coefficients} and
\eqref{eq.ev.K-nine}.  Two reduced rows
annihilate each triple identically, while the third gives $\det K^{\rm R}$.
For the reduced $4\times4$ conservation block,
$\det K^{\rm R}=A^2\det\bigl(\begin{smallmatrix}A&B\\ C&D^{\rm R}\end{smallmatrix}\bigr)$.
After the nonzero factor $A^2$ is removed, multiplication by $-a^4$ gives
Eq.~\eqref{eq.ev.Q8-primitive-pencil}.  Thus, for $A\ne0$,
Eq.~\eqref{eq.ev.right-cross-charts} gives three explicit charts for the same
one-dimensional eigenspace; at rank two at least one is nonzero.}
\begin{widetext}
{The three corresponding left-polarization charts are}
\begin{align}
 {\lambda^{(12)}}&{:=
 (\mathsf k_{21}\mathsf k_{32}-\mathsf k_{31}\mathsf k_{22},
 \mathsf k_{31}\mathsf k_{12}-\mathsf k_{11}\mathsf k_{32},
 \mathsf k_{11}\mathsf k_{22}-\mathsf k_{21}\mathsf k_{12}),}\notag\\
 {\lambda^{(23)}}&{:=
 (\mathsf k_{22}\mathsf k_{33}-\mathsf k_{32}\mathsf k_{23},
 \mathsf k_{32}\mathsf k_{13}-\mathsf k_{12}\mathsf k_{33},
 \mathsf k_{12}\mathsf k_{23}-\mathsf k_{22}\mathsf k_{13}),}\notag\\
 {\lambda^{(31)}}&{:=
 (\mathsf k_{23}\mathsf k_{31}-\mathsf k_{21}\mathsf k_{33},
 \mathsf k_{33}\mathsf k_{11}-\mathsf k_{31}\mathsf k_{13},
 \mathsf k_{21}\mathsf k_{13}-\mathsf k_{23}\mathsf k_{11}),}\notag\\
 {\lambda_0}&{:=-\frac{\lambda_iC_i}{A},\qquad
 \lambda_n:=\lambda_i\widehat n^i.}
 \label{eq.ev.left-cross-charts}
\end{align}
\end{widetext}
{Any nonzero chart may be used, with $A\ne0$ in the expression for
$\lambda_0$.  If one cofactor chart vanishes at rank two, another nonzero chart supplies the polarization. At $A=0$ one instead pivots
directly on a nonzero entry of the original $4\times4$ block.}

\subsection{{Resistive system}}
\label{app:ev_res}

{The restored current flux and its spatial symbol are}
\begin{align}
 {\mathcal B^\mu{}_\nu}
 &{=u^\mu k_\nu+k^\mu u_\nu-\omega_e u^\mu u_\nu
 -\chi_e\delta^\mu{}_\nu,}\notag\\
 {0}&{=a\,\delta k^i+k^i\Theta+k_n\delta u^i
 -\omega_e a\,\delta u^i-\widehat n^i\delta\chi_e,}\notag\\
 {\omega_e}&{:=\frac{q_eH_e}{m_e},\qquad
 \chi_e:=\frac{q_eP_e}{m_e},\qquad
 \Theta:=\widehat n_i\delta u^i.}
 \label{eq.ev.Btensor}
\end{align}
{In particular, the $k_n\delta u^i$ term is retained.  The physical
determinant and mode count are}
\begin{align}
 {\det\widehat{\mathbb A}_{\rm R}}
 &{\propto a^8(1-a^2)^4\mathcal C_3(a)\mathcal Q_8(a),}
 \label{eq.ev.det}\\
 {27}&{=8_{\rm mat}+4_{\gamma,+}+4_{\gamma,-}
 +3_e+8_h.}
 \label{eq.ev.res-count}
\end{align}

\subsubsection{{All right columns}}

{Define the trace-free wave-frame tensors}
\begin{align}
 {S_0}&{:=\widehat n\widehat n-t_2t_2,}&
 {S_+}&{:=t_1t_1-t_2t_2,}\notag\\
 {S_\times}&{:=t_1t_2+t_2t_1,}&
 {S_{nA}}&{:=\widehat nt_A+t_A\widehat n,}
 \label{eq.ev.wave-tensors}\\
 {\mathcal X}&{:=P_{,\rhob}T_{,e}-P_{,e}T_{,\rhob}.}\notag
\end{align}
\begin{widetext}
{The eight material columns are
$R_X^{\rm E}=R_{\rm R}^{\rm E}[\widehat r_X]$, with the following complete
nonzero entries:}
\begin{align}
 {\begin{array}{c|l}
 T&\varepsilon=T_{,\rhob},\quad r=-T_{,e},\quad
 S=\mathcal XS_0,\quad\varrho=-q_e(Z/m_i)T_{,e},\\
 \Pi&\varpi=1,\quad S=-S_0,\\
 q_A&h=t_A,\qquad A=1,2,\\
 \pi_+&S=S_+,\\
 \pi_\times&S=S_\times,\\
 k_A&c=t_A,\qquad A=1,2.
 \end{array}}
 \label{eq.ev.res-material-right}
\end{align}
\end{widetext}
{Every unlisted entry is zero.  Equation
\eqref{eq.ev.explicit-R27} shows, in particular, that the $T$ column has
$\rho_{e,r}=0$.  At zero drift the electron contact joins this eigenspace:}
\begin{align}
 {\widehat r_{P_e}:\qquad
 \varrho=q_eP_{e,e},\qquad x_e=P_{e,n}.}
 \label{eq.ev.XPe}
\end{align}
{Its only nonzero entries are
$\delta\rho_e=-m_eP_{e,e}$ and $\delta e_e=P_{e,n}$.}

{At each $a=\pm1$, the two radiative columns have}
\begin{align}
 {\widehat r_{\gamma1}:}&\quad
 {\epsilon=t_1,\qquad\beta=-a t_2,}\notag\\
 {\widehat r_{\gamma2}:}&\quad
 {\epsilon=t_2,\qquad\beta=a t_1.}
 \label{eq.ev.radiative-seeds}
\end{align}
{They contain no fluid or electron perturbation; their six public field
components are printed by Eq.~\eqref{eq.ev.explicit-res-blocks}.}

{The electron polynomial is}
\begin{align}
 {\mathcal C_3(a)}
 &{=a^3-(1-\sigma)v_{\rm d}a^2-c_{\rm se}^2a
 +c_{\rm se}^2v_{\rm d},}\notag\\
 {c_{\rm se}^2}&{:=P_{e,e}+\frac{\bar n_e}{H_e}P_{e,n},\qquad
 \sigma:=\frac{\bar n_e}{H_e}P_{e,n}.}
 \label{eq.ev.cubic}
\end{align}
{For $P_{e,n}\ne0$, at each of its three roots the complete right column is
$R_{\rm R}^{\rm E}[\widehat r_e(a)]$, with}
\begin{align}
 {c^i}&{=\frac{q_e}{m_e}aP_{e,n}\widehat n^i,}&
 {\varrho}&{=-q_e(a^2-P_{e,e}),}&
 {x_e}&{=P_{e,n}.}
 \label{eq.ev.electron-right}
\end{align}
{All ion and field entries vanish.  Explicitly, its electron
components are}
\begin{align}
 {\rho_{e,r}}&{=m_e(a^2-P_{e,e}),}\notag\\
 {w_{e,r}}&{=\frac{1+P_{e,e}}{\rho_e}P_{e,n}
 +\left(\frac{P_{e,n}}{m_e\rho_e}-\frac{H_e}{\rho_e^2}\right)
 m_e(a^2-P_{e,e}),}\notag\\
 {J_r^i}&{=\frac{q_eaP_{e,n}}{m_ew_e}L^i{}_a\widehat n^a
 -\frac{K^i}{w_e^2}w_{e,r},\qquad e_{e,r}=P_{e,n}.}
 \label{eq.ev.electron-public-right}
\end{align}

\begin{widetext}
{For each of the eight roots of $\mathcal Q_8$, take any nonzero $w$ from
Eq.~\eqref{eq.ev.right-cross-charts} with $K=K^{\rm R}$ and set}
\begin{align}
 {\varepsilon}&{=-\frac{B_iw^i}{A},}&
 {\Theta}&{=\widehat n_iw^i,}&
 {r}&{=-\frac{\rhob\Theta}{a},}&
 {\varpi}&{=-\frac{Y_\Pi\Theta}{a},}\notag\\
 {\delta T}&{:=T_{,e}\varepsilon
 -\frac{\rhob T_{,\rhob}\Theta}{a},}\notag\\
 {h^i}&{=-\frac{\dqq q^i\Theta
 +\kappa(\widehat n^i\delta T+Ta w^i)}{\tauq a},}\notag\\
 {S^{ij}}&{=-\frac{\dpipi\pi^{ij}\Theta
 +\eta(\widehat n^iw^j+\widehat n^jw^i
 -\frac23\delta^{ij}\Theta)}{\tau_\pi a}.}
 \label{eq.ev.res-Q8-slaves}
\end{align}
\end{widetext}
\begin{widetext}
{The field entries vanish.  The finite-drift electron entries are also
literal.  Put $\nu:=Z/m_i$, $k_n:=k_i\widehat n^i$, and}
\begin{align}
 {N_0}&{:=\nu r,}&
 {s_0}&{:=a k_iw^i+\omega_e\Theta-\frac{2k_n\Theta}{a},}\notag\\
 {B_x}&{:=\frac{1+P_{e,e}}{\rho_e},}&
 {B_q}&{:=\frac{m_ew_e-P_{e,n}}{q_e\rho_e},}&
 {B_0}&{:=\frac{(P_{e,n}-m_ew_e)N_0}{\rho_e},}\notag\\
 {E_{11}}&{:=a-\frac{P_{e,e}}a,}&
 {E_{12}}&{:=\frac{P_{e,n}}{q_ea},}\notag\\
 {E_{21}}&{:=\frac{q_eP_{e,e}}{m_eaw_e}
 -\frac{k_nB_x}{w_e^2},}&
 {E_{22}}&{:=a-\frac{P_{e,n}}{m_eaw_e}
 -\frac{k_nB_q}{w_e^2},}\notag\\
 {f_1}&{:=-H_e\Theta+\frac{m_e}{q_e}s_0
 +\frac{P_{e,n}N_0}{a},}\notag\\
 {f_2}&{:=-\rhoq\Theta-\frac{s_0}{w_e}
 -\frac{q_eP_{e,n}N_0}{m_eaw_e}+\frac{k_nB_0}{w_e^2},}\notag\\
 {\Delta_e}&{:=E_{11}E_{22}-E_{12}E_{21}.}
 \label{eq.ev.res-electron-lift-defs}
\end{align}
\begin{align}
 {x_e}&{=\frac{f_1E_{22}-E_{12}f_2}{\Delta_e},}&
 {\varrho}&{=\frac{E_{11}f_2-E_{21}f_1}{\Delta_e},}\notag\\
 {\delta P_e}&{:=P_{e,e}x_e
 +P_{e,n}\left(N_0-\frac{\varrho}{q_e}\right),}\notag\\
 {c^i}&{=\omega_e w^i-\frac{k_n}{a}w^i
 -\frac{k^i}{a}\Theta+\frac{q_e}{m_ea}\widehat n^i\delta P_e.}
 \label{eq.ev.res-electron-lift}
\end{align}
\end{widetext}
{Equations~\eqref{eq.ev.res-Q8-slaves}--
\eqref{eq.ev.res-electron-lift}, inserted into
Eq.~\eqref{eq.ev.explicit-R27}, are all eight Eulerian ion--dissipative
columns. This reconstruction requires $aA\Delta_e\ne0$ and a nonzero cofactor chart. For vanishing denominators or repeated roots, use the joint kernel of the full symbol. At zero drift they reduce to}
\begin{align}
 {c^i=\omega_e w^i,\qquad
 \varrho=-\frac{(q_e\bar n_e+\rhoq)\Theta}{a},\qquad x_e=0.}
 \label{eq.ev.ion-electron-lift}
\end{align}

\begin{widetext}
{It remains to print the two cleaning columns at each light-cone root.
Their field seeds and conservation targets are}
\begin{align}
 {\widehat r_\psi:}&\quad
 {\epsilon=\widehat n,\quad\psi=-a,\qquad
 (g_0,g_i)=(-a\mathcal E_n,
 \mathcal E_i-a(\widehat n\times\mathcal B)_i),}\notag\\
 {\widehat r_\phi:}&\quad
 {\beta=\widehat n,\quad\phi=-a,\qquad
 (g_0,g_i)=(-a\mathcal B_n,
 \mathcal B_i+a(\widehat n\times\mathcal E)_i).}
 \label{eq.ev.res-cleaning-targets}
\end{align}
\end{widetext}
\begin{widetext}
{In the wave frame of Eq.~\eqref{eq.ev.K-nine}, the two primitive right-hand
sides $h_i^{\rm cl}:=Ag_i-C_ig_0$ are}
\begin{align}
 (h_1,h_2,h_3)_\psi
 &=\bigl((A+ac_C)\mathcal E_1,
 A(\mathcal E_2+a\mathcal B_3),
 A(\mathcal E_3-a\mathcal B_2)\bigr),\notag\\
 (h_1,h_2,h_3)_\phi
 &=\bigl((A+ac_C)\mathcal B_1,
 A(\mathcal B_2-a\mathcal E_3),
 A(\mathcal B_3+a\mathcal E_2)\bigr).
 \label{eq.ev.cleaning-h-primitive}
\end{align}
{For either row, write its entries as $h_i$.  Cramer's rule then gives}
\begin{align}
 {\Delta_K}&{:=\mathsf k_{11}(\mathsf k_{22}\mathsf k_{33}-\mathsf k_{23}\mathsf k_{32})
 -\mathsf k_{12}(\mathsf k_{21}\mathsf k_{33}-\mathsf k_{23}\mathsf k_{31})}\notag\\
 &{\quad+\mathsf k_{13}(\mathsf k_{21}\mathsf k_{32}-\mathsf k_{22}\mathsf k_{31}),}
 \label{eq.ev.cleaning-delta}\\
 {w_1}&{:=\frac{h_1(\mathsf k_{22}\mathsf k_{33}-\mathsf k_{23}\mathsf k_{32})
 -\mathsf k_{12}(h_2\mathsf k_{33}-\mathsf k_{23}h_3)
 +\mathsf k_{13}(h_2\mathsf k_{32}-\mathsf k_{22}h_3)}{\Delta_K},}\notag\\
 {w_2}&{:=\frac{\mathsf k_{11}(h_2\mathsf k_{33}-\mathsf k_{23}h_3)
 -h_1(\mathsf k_{21}\mathsf k_{33}-\mathsf k_{23}\mathsf k_{31})
 +\mathsf k_{13}(\mathsf k_{21}h_3-h_2\mathsf k_{31})}{\Delta_K},}\notag\\
 {w_3}&{:=\frac{\mathsf k_{11}(\mathsf k_{22}h_3-h_2\mathsf k_{32})
 -\mathsf k_{12}(\mathsf k_{21}h_3-h_2\mathsf k_{31})
 +h_1(\mathsf k_{21}\mathsf k_{32}-\mathsf k_{22}\mathsf k_{31})}{\Delta_K}.}
 \label{eq.ev.cleaning-velocity}
\end{align}
\end{widetext}
{All $\mathsf k_{ij}$ are the primitive expressions in
Eq.~\eqref{eq.ev.K-nine}.  Finally set
$\varepsilon=(g_0-B_iw^i)/A$ and use
Eqs.~\eqref{eq.ev.res-Q8-slaves}--\eqref{eq.ev.res-electron-lift} for
$(r,\varpi,h,S,c,\varrho,x_e)$.  Together with the appropriate seed in
Eq.~\eqref{eq.ev.res-cleaning-targets}, Eq.~\eqref{eq.ev.explicit-R27}
then prints every one of the $27$ components of all four cleaning columns.
These expressions use the chart $A\Delta_K\Delta_e\ne0$.  For $A\ne0$,
$\Delta_K=0$ signals a light--octic coincidence; $\Delta_e=0$ similarly
signals a light--electron coincidence.  Such points require another pivot or
a joint rank calculation.}

\subsubsection{{All left rows}}

{We give the left eigenvectors as equation weights applied to the
Eulerian time rows above.  All projections in this subsection refer to the
mode-dependent wave frame $(\widehat n,t_1,t_2)$; the Eulerian boost remains
arbitrary.  Put}
\begin{align}
 {Y_n}&{:=\frac{\dpipi\pi_{nn}+4\eta/3}{\tau_\pi},}\notag\\
 {Y_1}&{:=\frac{\dpipi\pi_{11}-2\eta/3}{\tau_\pi},}\notag\\
 {Y_\times}&{:=\frac{\dpipi\pi_{12}}{\tau_\pi},}\notag\\
 {c_{qA}}&{:=\frac{\dqq q_A}{\tauq\rhob},}\notag\\
 {c_{kA}}&{:=\frac{1}{\rhob}
 \left(k_A-\frac{k_n\dpipi}{\eta}\pi_{nA}\right),}\notag\\
 {\Delta_{\mathrm R}}&{:=\mathcal X
 +\frac{T_{,e}}{\rhob}(Y_\Pi+Y_n).}
 \label{eq.ev.res-material-left-defs}
\end{align}
{Define the equation-space rows}
\begin{align}
 {r_n}&{:=-\frac{Y_n}{\rhob}\mathcal R
 +\frac{1}{\tau_\pi}\mathcal S_{nn},}\notag\\
 {r_\Pi}&{:=-\frac{Y_\Pi}{\rhob}\mathcal R
 +\frac{1}{\tau_\Pi}\mathcal P,}\notag\\
 {r_1}&{:=-\frac{Y_1}{\rhob}\mathcal R
 +\frac{1}{\tau_\pi}\mathcal S_{11},}\notag\\
 {r_\times}&{:=-\frac{Y_\times}{\rhob}\mathcal R
 +\frac{1}{\tau_\pi}\mathcal S_{12},}\notag\\
 {r_{qA}}&{:=\frac{1}{\tauq}\mathcal Q_A-c_{qA}\mathcal R,}\notag\\
 {r_{kA}}&{:=\mathcal K_A-\frac{k_n}{\eta}\mathcal S_{nA}
 -c_{kA}\mathcal R,\qquad A=1,2.}
 \label{eq.ev.res-material-left-bases}
\end{align}
{This material chart requires $\Delta_{\mathrm R}\ne0$ and the nonzero transport and density denominators in (SH1)--(SH2). At singular points, use the full material kernel.
The eight material left eigenvectors are}
\begin{align}
 {L_T^{\rm E}}&{:=\frac{r_n+r_\Pi}{\Delta_{\mathrm R}},}\notag\\
 {L_\Pi^{\rm E}}&{:=r_\Pi
 -\frac{T_{,e}Y_\Pi}{\rhob}L_T^{\rm E},}\notag\\
 {L_{qA}^{\rm E}}&{:=r_{qA}-T_{,e}c_{qA}L_T^{\rm E},}\notag\\
 {L_{\pi+}^{\rm E}}&{:=r_1
 -\frac{T_{,e}Y_1}{\rhob}L_T^{\rm E},}\notag\\
 {L_{\pi\times}^{\rm E}}&{:=r_\times
 -\frac{T_{,e}Y_\times}{\rhob}L_T^{\rm E},}\notag\\
 {L_{kA}^{\rm E}}&{:=r_{kA}-T_{,e}c_{kA}L_T^{\rm E}.}
 \label{eq.ev.res-material-left}
\end{align}
{At zero drift the ninth (contact) row is}
\begin{align}
 {L_{P_e}^{\rm E}:=\frac{-(q_e\bar n_e+\rhoq)\mathcal R/\rhob
 +\mathcal C_q+(q_e\bar n_e/H_e)\mathcal E_e}
 {q_ec_{\rm se}^2}.}
 \label{eq.ev.res-contact-left}
\end{align}

{The eight light-cone rows are particularly simple.  At each
$a=\varsigma=\pm1$,}
\begin{align}
 {L_{\gamma1,\varsigma}^{\rm E}}&{:=\frac12
 (\mathcal M_1-\varsigma\widetilde{\mathcal M}_2),}&
 {L_{\gamma2,\varsigma}^{\rm E}}&{:=\frac12
 (\mathcal M_2+\varsigma\widetilde{\mathcal M}_1),}\notag\\
 {L_{\psi,\varsigma}^{\rm E}}&{:=\frac12
 (\varsigma\mathcal M_0+\mathcal M_n),}&
 {L_{\phi,\varsigma}^{\rm E}}&{:=\frac12
 (\varsigma\widetilde{\mathcal M}_0+\widetilde{\mathcal M}_n).}
 \label{eq.ev.res-light-left}
\end{align}
{Subscripts $0,n,1,2$ denote projections on
$(-u_\nu,e_{n\nu},e_{1\nu},e_{2\nu})$.  These rows remain pure field rows,
although the cleaning columns carry a fluid response.}

\begin{widetext}
{For the finite-drift electron rows, write}
\begin{align}
 {v}&{:=\frac{k_nm_e}{q_eH_e},}&
 {\sigma_e}&{:=\frac{\bar n_eP_{e,n}}{H_e},}&
 {p_e}&{:=P_{e,e},}&
 {c_e}&{:=q_e\bar n_e,}\notag\\
 {d_e(a)}&{:=a-(1-\sigma_e)v,}&
 {y_e(a)}&{:=-\frac{P_{e,n}}{m_ed_e(a)},}\notag\\
 {z_e(a)}&{:=\frac{q_ep_e/m_e
 +v c_e(1+p_e)y_e/H_e}{a}.}
 \label{eq.ev.electron-left-defs}
\end{align}
\end{widetext}
\begin{widetext}
{The electron equation weights are $(1,y_e,z_e)$ on the current,
charge, and electron-energy equations.  Their ion forcing is}
\begin{align}
 {S_\rho}&{:=-\frac{q_e(Z/m_i)P_{e,n}}{m_e}
 +y_ev q_e(Z/m_i)(1-\sigma_e),}\notag\\
 {S_i}&{:=(2k_n-\omega_e a)\widehat n_i
 +y_e\left(\rhoq\widehat n_i+\frac{a}{w_e}k_i\right)
 +z_e\left(H_e\widehat n_i-\frac{m_e}{q_e}ak_i\right),}\notag\\
 {F_i}&{:=S_i-\frac{\rhob S_\rho}{a}\widehat n_i.}
 \label{eq.ev.electron-left-source}
\end{align}
\end{widetext}
\begin{widetext}
{Let $\boldsymbol c_j:=(K_{1j}^{\rm R},K_{2j}^{\rm R},K_{3j}^{\rm R})^{\rm T}$
be the three columns of $K^{\rm R}$.  The conservation weights are explicitly}
\begin{align}
 {\Delta_K}&{:=\boldsymbol c_1\mathbin\cdot
 (\boldsymbol c_2\mathbin\times\boldsymbol c_3),}\notag\\
 {\boldsymbol\lambda}&{:=-\frac{A}{\Delta_K}\left[
 F_1(\boldsymbol c_2\mathbin\times\boldsymbol c_3)
 +F_2(\boldsymbol c_3\mathbin\times\boldsymbol c_1)
 +F_3(\boldsymbol c_1\mathbin\times\boldsymbol c_2)\right],}\notag\\
 {\lambda_0}&{:=-\frac{C_i\lambda_i}{A},}&
 {b_q^i}&{:=-\frac{\lambda_0\widehat n^i+a\lambda^i}{\tauq a},}\notag\\
 {b_\rho}&{:=-\frac{S_\rho+P_{,\rhob}\lambda_n
 +\kappa T_{,\rhob}\widehat n_i b_q^i}{a},}&
 {b_\Pi}&{:=-\frac{\lambda_n}{\tau_\Pi a},}&
 {b_\pi^{ij}}&{:=-\frac{\operatorname{STF}
 (\widehat n^i\lambda^j)}{\tau_\pi a}.}
 \label{eq.ev.electron-left-hyd}
\end{align}
\end{widetext}
\begin{widetext}
{For conservation weights $(\ell_0,\boldsymbol\ell)$, define}
\begin{align}
 {X_E}&{:=\ell_n\mathcal E_n+\ell_1\mathcal E_1+\ell_2\mathcal E_2
 -a\ell_2\mathcal B_1+a\ell_1\mathcal B_2-a\ell_0\mathcal E_n,}\notag\\
 {X_B}&{:=\ell_n\mathcal B_n+\ell_1\mathcal B_1+\ell_2\mathcal B_2
 -a\ell_0\mathcal B_n+a\ell_2\mathcal E_1-a\ell_1\mathcal E_2,}\notag\\
 {m_0}&{:=\frac{X_E}{a^2-1},}&
 {m_n}&{:=\frac{aX_E}{a^2-1},}&
 {m_1}&{:=\ell_2\mathcal B_n-\ell_n\mathcal B_2-\ell_0\mathcal E_1,}\notag\\
 {m_2}&{:=-\ell_1\mathcal B_n+\ell_n\mathcal B_1-\ell_0\mathcal E_2,}\notag\\
 {\widetilde m_0}&{:=\frac{X_B}{a^2-1},}&
 {\widetilde m_n}&{:=\frac{aX_B}{a^2-1},}&
 {\widetilde m_1}&{:=-\ell_0\mathcal B_1-\ell_2\mathcal E_n+\ell_n\mathcal E_2,}\notag\\
 {\widetilde m_2}&{:=-\ell_0\mathcal B_2+\ell_1\mathcal E_n-\ell_n\mathcal E_1,}\notag\\
 {\mathcal W_a[\ell_0,\boldsymbol\ell]}
 &{:=m_0\mathcal M_0+m_i\mathcal M_i
 +\widetilde m_0\widetilde{\mathcal M}_0
 +\widetilde m_i\widetilde{\mathcal M}_i.}
 \label{eq.ev.res-maxwell-left-weights}
\end{align}
\end{widetext}
\begin{widetext}
{The background $\mathcal E_a$ and $\mathcal B_a$ are measured in the
fluid tetrad.  For the electron row, Eq.~\eqref{eq.ev.res-maxwell-left-weights}
reduces to the two primitive scalars}
\begin{align}
 \Xi_E&:=\lambda_n\mathcal E_n+\lambda_1\mathcal E_1
 +\lambda_2\mathcal E_2-a\lambda_2\mathcal B_1
 +a\lambda_1\mathcal B_2-a\lambda_0\mathcal E_n,\notag\\
 \Xi_B&:=\lambda_n\mathcal B_n+\lambda_1\mathcal B_1
 +\lambda_2\mathcal B_2-a\lambda_0\mathcal B_n
 +a\lambda_2\mathcal E_1-a\lambda_1\mathcal E_2.
 \label{eq.ev.res-electron-maxwell-scalars}
\end{align}
{At a root of $\mathcal C_3$, the complete Eulerian electron row is}
\begin{align}
 {L_e^{\rm E}(a)}&{:=\lambda_0\mathcal T_{{\rm R},0}
 +\lambda_i\mathcal T_{{\rm R},i}+b_\rho\mathcal R+b_\Pi\mathcal P
 +b_q^i\mathcal Q_i+b_\pi^{ij}\mathcal S_{ij}}\notag\\
 &{\quad+\mathcal K_n+y_e\mathcal C_q+z_e\mathcal E_e
 +\frac{\Xi_E}{a^2-1}(\mathcal M_0+a\mathcal M_n)}\notag\\
 &{\quad+(\lambda_2\mathcal B_n-\lambda_n\mathcal B_2
 -\lambda_0\mathcal E_1)\mathcal M_1}\notag\\
 &{\quad+(-\lambda_1\mathcal B_n+\lambda_n\mathcal B_1
 -\lambda_0\mathcal E_2)\mathcal M_2}\notag\\
 &{\quad+\frac{\Xi_B}{a^2-1}
 (\widetilde{\mathcal M}_0+a\widetilde{\mathcal M}_n)}\notag\\
 &{\quad+(-\lambda_0\mathcal B_1-\lambda_2\mathcal E_n
 +\lambda_n\mathcal E_2)\widetilde{\mathcal M}_1}\notag\\
 &{\quad+(-\lambda_0\mathcal B_2+\lambda_1\mathcal E_n
 -\lambda_n\mathcal E_1)\widetilde{\mathcal M}_2.}
 \label{eq.ev.res-electron-left}
\end{align}
\end{widetext}
{This chart applies away from $aA\Delta_Kd_e(a)=0$ and from coincidences with
the light cone.  At zero drift the two acoustic rows reduce to}
\begin{align}
 {L_{\rm se}^{\rm E}(a)=
 \frac{q_e(Z/m_i)P_{e,n}}{m_ea}\mathcal R+\mathcal K_n
 -\frac{P_{e,n}}{m_ea}\mathcal C_q
 +\frac{q_eP_{e,e}}{m_ea}\mathcal E_e.}
 \label{eq.ev.res-electron-left-zero}
\end{align}
{All three electron rows are unnormalized.}

{Finally consider a root $a_{h,r}$, $r=1,\ldots,8$, of $\mathcal Q_8$.
Take $\lambda$ from Eq.~\eqref{eq.ev.left-cross-charts} with
$K=K^{\rm R}$ and define the equation weights}
\begin{align}
 {b_q^i}&{:=-\frac{\lambda_0\widehat n^i+a\lambda^i}{\tauq a},}
 &
 {b_\rho}&{:=-\frac{P_{,\rhob}\lambda_n
 +\kappa T_{,\rhob}\widehat n_ib_q^i}{a},}\notag\\
 {b_\pi^{ij}}&{:=-\frac{\operatorname{STF}
 (\widehat n^i\lambda^j)}{\tau_\pi a}.}
 \label{eq.ev.res-Q8-left-weights}
\end{align}
\begin{widetext}
{The complete Eulerian row is}
\begin{align}
 {L_h^{\rm E}(a):=\lambda_0\mathcal T_{{\rm R},0}
 +\lambda_i\mathcal T_{{\rm R},i}+b_\rho\mathcal R
 -\frac{\lambda_n}{\tau_\Pi a}\mathcal P
 +b_q^i\mathcal Q_i+b_\pi^{ij}\mathcal S_{ij}
 +\mathcal W_a[\lambda_0,\boldsymbol\lambda].}
 \label{eq.ev.res-Q8-left}
\end{align}
\end{widetext}
{It has no current, carrier, or electron-energy equation weight.  A
nonzero scalar multiple represents the same left eigenvector.}

\begin{widetext}
{The complete resistive basis is therefore}
\begin{align}
 {\mathbb R_{\rm R}^{\rm E}}
 &{=\bigl(R_T,R_\Pi,R_{q1},R_{q2},R_{\pi+},R_{\pi\times},R_{k1},R_{k2},
 R_{\gamma1,+},R_{\gamma2,+},R_{\psi,+},R_{\phi,+},}\notag\\[-0.3ex]
 &{\qquad R_{\gamma1,-},R_{\gamma2,-},R_{\psi,-},R_{\phi,-},
 R_{e,1},R_{e,2},R_{e,3},}\notag\\[-0.3ex]
 &{\qquad R_{h,1},R_{h,2},R_{h,3},R_{h,4},R_{h,5},R_{h,6},R_{h,7},R_{h,8}\bigr),}
 \label{eq.ev.res-Rcomplete}\\
 {\widetilde{\mathbb L}_{\rm R}^{\rm E}}
 &{=\bigl(L_T,L_\Pi,L_{q1},L_{q2},L_{\pi+},L_{\pi\times},L_{k1},L_{k2},
 L_{\gamma1,+},L_{\gamma2,+},L_{\psi,+},L_{\phi,+},}\notag\\[-0.3ex]
 &{\qquad L_{\gamma1,-},L_{\gamma2,-},L_{\psi,-},L_{\phi,-},
 L_{e,1},L_{e,2},L_{e,3},}\notag\\[-0.3ex]
 &{\qquad L_{h,1},L_{h,2},L_{h,3},L_{h,4},L_{h,5},L_{h,6},L_{h,7},L_{h,8}\bigr)^{\rm T}.}
 \label{eq.ev.res-Lcomplete}
\end{align}
\end{widetext}
{At zero drift replace the eight material members and the three electron
members by the nine material members including $P_e$ and the two nonzero
acoustic members.  The displayed left rows are unnormalized.  Within a
degenerate family they span the left eigenspace.  A biorthogonal dual basis
generally requires a change of basis within that family, rather than separate
scalar rescalings.}

\subsubsection{{Strong hyperbolicity}}
\label{app:ev_causality}

Strong hyperbolicity requires an invertible time symbol, real speeds,
and a complete eigenbasis whose {diagonalizer and inverse are uniformly bounded in}
the wave normal.  At a repeated speed $a_*$ this requires
\begin{align}
 \dim\ker\widehat{\mathbb A}(a_*)
 =\operatorname{ord}_{a_*}\det\widehat{\mathbb A},
 \label{eq.ev.sh-semisimple}
\end{align}
and the bound must remain locally uniform on an open state domain.
{Here $\operatorname{ord}_{a_*}\det\widehat{\mathbb A}$ counts the
characteristic roots that coincide at $a_*$, including multiplicity, while
$\dim\ker\widehat{\mathbb A}(a_*)$ counts the independent zero modes of the
pencil at that speed.  Equality therefore says that the geometric and
algebraic multiplicities agree, so the repeated characteristic has no
defective Jordan block.}

For the constrained $27\times27$ resistive pencil, put
$P_z=\widehat{\mathbb A}_{\rm R}(t+i\epsilon,\hat n)$.
On the real-root branch, with invertible time symbol, strong
hyperbolicity is equivalent to the existence of a finite $K$ such that
\begin{align}
 \epsilon^2\operatorname{tr}\!\left[
 (\operatorname{adj}P_z)(\operatorname{adj}P_z)^\dagger\right]
 &\le K|\det P_z|^2,\notag\\
 &t\in\mathbb R,\quad\epsilon>0,\quad|\hat n|=1.
 \label{eq.ev.full-resolvent-sh}
\end{align}
This resolvent bound is necessary and sufficient, including at
material, ion, electron, and light-wave coincidences.  On an open
domain $K$ must remain locally bounded in state.

The ion--material contact has a closed primitive variable condition for
$\rhob,y,c>0$ and $\Delta_H\ne0$.
Let $y=\eta/\tau_\pi$, $c=\kappa T_{,e}/\tauq$,
$d_\pi=\delta_{\pi\pi}/\tau_\pi$, and
\begin{align}
 P_{\rm mat}&=\rhob P_{,\rhob}
 -p\frac{\rhob T_{,\rhob}}{T_{,e}},\notag\\
 R_{\rm mat}&=Z-\frac{pR}{c}
 =P_{\rm mat}
 +\frac{\zeta+\delta_{\Pi\Pi}\Pi}{\tau_\Pi}+\frac{4y}{3}.
 \label{eq.ev.material-restoring-general}
\end{align}
$P_{\rm mat}=\rhob(\partial P/\partial\rhob)_T$ is the isothermal
pressure stiffness.
For a gamma law, $P_{\rm mat}=P$ and $R_{\rm mat}=R_m$.
Equation~\eqref{eq.resgeneral.polynomial}
gives, for every unit normal,
\begin{align}
 F_8(0,\hat n)&=\frac{y^2c}{\Delta_H}
 \bigl(R_{\rm mat}+d_\pi\pi_{\hat n\hat n}\bigr).
 \label{eq.ev.material-octic-zero}
\end{align}
On the real-octic branch the material contact is semisimple for every
normal exactly when $\Delta_H(R_{\rm mat}I+d_\pi\pi)$ is positive
definite, or equivalently
\begin{align}
 \Delta_H R_{\rm mat}&>0,\notag\\
 6R_{\rm mat}^2-d_\pi^2I_2&>0,\notag\\
 \Delta_H(6R_{\rm mat}^3-3R_{\rm mat}d_\pi^2I_2
 +2d_\pi^3I_3)&>0.
 \label{eq.ev.material-gap-signed}
\end{align}
{Here semisimple means that the repeated characteristic has
as many independent eigenvectors as its algebraic multiplicity,
Eq.~\eqref{eq.ev.sh-semisimple}.}
For a gamma law with $\Delta_H>0$, this is
Eq.~\eqref{eq.strong.material-gap}.

{For hot gamma-law electrons, $e_e>\rho_e>0$ and $1<\Gamma_e\leq2$,
the exact fixed-state electron--ion test also controls approach to a
retained contact.} Let $H_8(\hat n)$ be the material-separated
eight-wave ion speed matrix, $P_u$ its velocity projection, and put
$v=j/(q_e\bar n_e)$, $t=v\cdot\hat n$,
$x=c_{\rm se}^2$, {$\beta_e=1-\sigma_e>1$}. Define
\begin{align}
 L&=H_8^3-\beta_etH_8^2-xH_8+xtI_8,\notag\\
 b&=x(v^{\rm T}P_u)(H_8-tI_8)
 -2t(\hat n^{\rm T}P_u)(H_8-\beta_etI_8).
 \label{eq.ev.ei-graph}
\end{align}
For $v\ne0$, provided the ion block has a uniform eigenbasis and the
electron bound \eqref{eq:electron-sh-current-bound} holds, their
first-order coupling has a uniformly bounded
eigenbasis exactly when, for some finite state-dependent $K$,
\begin{align}
 \|b\operatorname{adj}L\|^2
 &\le K(\det L)^2\qquad\text{for every }|\hat n|=1.
 \label{eq.ev.ei-uniform}
\end{align}
At zero drift the coupling vanishes at that state, though an acoustic
match need not be interior.  Equation~\eqref{eq.ev.ei-uniform}
retains fixed-state compatible contacts; the primitive bound
\eqref{eq.resgeneral.simple-current-gap} instead gives a sufficient
open strict gap.
{For other EOS branches or a simultaneous light-wave contact, use
the full test \eqref{eq.ev.full-resolvent-sh}.}
The strict-gap current bound in the main text is the compact sufficient
choice used for admission.

{For $\Delta_H\ne0$, establish eight distinct real ion roots
along one reference direction and require a positive discriminant as the
propagation direction varies.  Choose a fluid-frame unit normal
$\hat n_0$ and set}
\begin{align}
 \mathfrak H_k&:=\det\bigl[M_{i+j}(\hat n_0)\bigr]_{i,j=0}^{k-1}>0,
 \quad k=2,\ldots,7,\notag\\
 &\min_{x\cdot x=1}\mathcal R_8(x)>0.
 \label{eq.ev.ion-simple-reality-continuation}
\end{align}
Here $M_0=8$ and $\mathcal R_8$ is the octic discriminant.
{{Together with $M_0=8$ and $\mathcal R_8(\hat n_0)>0$, the six minors
make the Hermite matrix positive definite and establish eight real,
simple roots at $\hat n_0$.} The strict discriminant prevents collisions on the connected
normal sphere.}  Conversely, eight simple real roots in
every direction imply \eqref{eq.ev.ion-simple-reality-continuation}.

For hot gamma-law ions and electrons with
$y=\eta/\tau_\pi>0$ and $c=\kappa T_{,e}/\tauq>0$,
a compact sufficient domain follows by combining the primitive screen
\eqref{eq.resgeneral.simple-reality-screen}, the strict cone floors, and
the current gap \eqref{eq.resgeneral.simple-current-gap}.  Alternatively,
combine the reference-direction continuation test
\eqref{eq.ev.ion-simple-reality-continuation}, the
strict cone floors, the material gap \eqref{eq.strong.material-gap}, the
electron bound \eqref{eq:electron-current-combined-bound}, and the exact
graph test \eqref{eq.ev.ei-uniform}.  In either case the ion and electron
waves have a uniformly bounded eigenbasis on the compact normal sphere.

\subsection{Validation}

We provide a companion \texttt{Mathematica} script, which constructs the constrained
resistive symbol from the covariant equations.  It checks the $27\times27$
determinant, the displayed sparse families and slaving relations, the
Eulerian maps, and the rank defect obtained by deleting the pressure force.
Independent numerical right and left bases verify completeness and all
residuals for a generic simple-root state,
\begin{align}
 {\mathbb A_{\rm R}^{\rm E}(\xi_I)R_I^{\rm E}}&{=0,}&
 {L_I^{\rm E}\mathbb C_{\rm R}^{\rm E}}&{=y_IL_I^{\rm E}.}
 \label{eq.ev.validation-identities}
\end{align}

\section{Characteristic structure and nonlinear causality}
\label{app:characteristics}

{We compute the characteristics of the system with a projected electron-current balance summarized
in Sec.~\ref{sec:methods} and Appendix~\ref{app:3p1}.  The electron sector
contains charge conservation, the energy equation
\eqref{eq:electron-energy-summary}, and the three fluid-projected current equations represented by
\eqref{eq:electron-B-summary}.  The electron stress is not included in the total stress-energy tensor $T^{\mu\nu}$,
and the characteristic symbol is restricted to the physical constraint
manifold. Heat and shear also use the projected constrained limit. The physical symbol excludes finite-rate modes of the constraint violations. {This is the resistive electron sector.
}}

{We recall that $P_e$ is the electron pressure and use $P_{,e}$, $P_{,\rhob}$, $T_{,e}$, and
$T_{,\rhob}$ for ion equation-of-state derivatives.  For the electron sector
we define}
\begin{align}
  {P_{e,e}}&{:=\left(\frac{\p P_e}{\p e_e}\right)_{\bar n_e} \, ,} &
  {P_{e,n}}&{:=\left(\frac{\p P_e}{\p\bar n_e}\right)_{e_e} \, ,}
  \notag\\
  {h_e}&{:=\frac{H_e}{\bar n_e} \, ,} &
  {c_e}&{:=q_e\bar n_e \, .}
  \label{eq.char.eosderivs}
\end{align}

\subsection{Method and set-up}

{We follow Ref.~\cite{Cordeiro:2026} in our discussion and derivation of nonlinear causality constraints. 
{We begin by writing the system in quasi-linear form}
\begin{align}
  {\left(\mathbb A^\alpha\p_\alpha+\mathbb B\right)U=0 \, ,}
  \label{eq.quasilinear}
\end{align}
{where $\mathbb A^\alpha$ contains the principal part and $\mathbb B$ the
derivative-free terms.  A characteristic surface $\varphi(x)=0$, with normal
$\phi_\mu:=\nabla_\mu\varphi$, satisfies}
\begin{align}
  {\det\left(\mathbb A^\alpha\phi_\alpha\right)=0 \, .}
\end{align}
{Causality requires all roots $\phi_0(\phi_i)$ to be real and every
characteristic covector to be non-timelike,
$\phi_\alpha\phi^\alpha\geq0$ \cite{Cordeiro:2026}.  We introduce}
\begin{align}
  {z}& {:=u^\alpha\phi_\alpha \, ,} &
  {v^\mu}& {:=\Delta^{\mu\alpha}\phi_\alpha \, ,} &
  {w^\mu}& {:=\pi^{\mu\alpha}\phi_\alpha \, ,}
\end{align}
{so that $\phi_\alpha\phi^\alpha=v^2-z^2$, with
$v^2:=v^\mu v_\mu$.  Defining}
\begin{align}
  {\hat z}& {:=\frac{z}{\sqrt{v^2}} \, ,} &
  {\hat n^\mu}& {:=\frac{v^\mu}{\sqrt{v^2}} \, ,}
\end{align}
{reduces the causal bound to $|\hat z|\leq1$.  We evaluate the symbol in a
local orthonormal fluid frame aligned with the eigenvectors of
$\pi^{\mu\nu}$,}
\begin{align}
  {u^\mu}&{=(1,0,0,0) \, ,} &
  {\pi^{\mu\nu}}& {=\operatorname{diag}
  (0,\Lambda_1,\Lambda_2,\Lambda_3) \, ,} &
  {\sum_{a=1}^{3}\Lambda_a}& {=0 \, .}
  \notag
\end{align}

{We recall that causality constraints depend on the physical constraint manifold they are evaluated on. 
We restrict the symbol to perturbations tangent to the physical constraint
manifold,
\begin{align}
 u_{\mu}u^{\mu}&=-1,&
 u_{\mu}q^{\mu}&=0,&
 u_{\mu}k^{\mu}&=0,\notag\\
 u_{\mu}\pi^{\mu\nu}&=0,&
 \pi^{\mu\nu}&=\pi^{\nu\mu},&
 \pi^{\mu}{}_{\mu}&=0.
 \label{eq.char.constraints}
\end{align}
Thus, $\delta u^0=0$, $q^\mu$ and $k^\mu$ each have three independent
components, and $\pi^{\mu\nu}$ has five. }

\subsection{Block structure and reduction}

The Maxwell equations replace the electromagnetic-stress divergence by the Lorentz force together with derivatives of cleaning fields. The row operation preserves the determinant. Because $\mathcal J^\mu=\rho_q u^\mu+j^\mu$ enters Maxwell's equations algebraically, the field/cleaning block has no principal forcing from matter, but cleaning perturbations can force fluid rows. Thus, the principal symbol is block triangular, not generally block diagonal. Its determinant factorizes while its eigenvectors retain the cleaning-induced fluid response.

{Second, $k_\mu$, $\rhoq$, and $e_e$ do not enter the
principal part of the hydrodynamic equations, because the electron stress tensor is not included in stress-energy conservation.  The three independent
components of $k_\mu$, together with $\rhoq$ and $e_e$, therefore form a
five-dimensional electron block. The electron pressure flux
$-\chi_e\delta^\mu{}_\nu$ couples $(\delta\rhoq,\delta e_e)$ to the current
rows, while $Q_e^\mu=-(m_e/q_e)k^\mu$ couples $\delta k^\mu$ to the energy
row.  After the homogeneous covector
normalization, its determinant is $\hat z^2\mc C_3$, where $\mc C_3$ is given in
Eq.~\eqref{eq.Qe}.  The symmetric $k^\mu u_\nu$ flux adds only a
velocity-to-current coupling below this block and does not change its
determinant.}

{On the constrained state space, the resistive determinant therefore factorizes into the product of the individual block determinants,}
\begin{align}
  {\det\left(\mathbb A^\alpha\phi_\alpha\right)
  \propto {\hat z^8}\left(1-\hat z^2\right)^4
  \mc C_3(\hat z)\mc Q_8(\hat z) \, ,}
  \label{eq.charfactor}
\end{align}
{The $27$ physical primitives are listed in Eq.~\eqref{eq.ev.state}.  Here $\mc Q_8$ is the degree-eight
characteristic polynomial of the constrained ion--dissipative block.}

\subsection{The characteristic modes}
\label{app:char_modes}

{\it Electromagnetic modes.} The factor $(1-\hat z^2)^4$ contains the two electromagnetic polarizations
and two cleaning scalars.  In $3+1$ form their speeds are
\begin{align}
 \lambda_\pm=-\beta^i n_i
 \pm\alpha\sqrt{\gamma^{ij}n_i n_j}.
 \label{eq.lightcone}
\end{align}
{The factor $\hat z^8$ contains the modes advected with the fluid: one
isothermal thermodynamic mode, one bulk mode, two transverse heat-flux
modes, two transverse-traceless shear modes, two transverse-current modes,
and no independent entropy mode.  They propagate at}
\begin{align}
 \lambda_0=(\alpha v^i-\beta^i)n_i.
 \label{eq.char.speed.material}
\end{align}
{At zero electron drift $\mc C_3$ supplies one additional material mode,
{whose fluid-frame seed is Eq.~\eqref{eq.ev.XPe} and whose Eulerian
column is given by Eq.~\eqref{eq.ev.explicit-R27}.}  It rearranges the electron carrier and
energy at fixed $P_e$, and is therefore the electron analogue of the entropy
or contact wave of standard hydrodynamics.  The eight roots of $\mc Q_8$
form two left--right transverse shear pairs, one longitudinal sound pair,
and one thermal, or second-sound, pair.  Bulk viscosity and longitudinal
shear stress dress the compressive modes rather than adding new branches.
Away from a principal direction, heat flux and anisotropic stress mix the
transverse and longitudinal channels.}  In the resistive system there are no
Alfv\'en or magnetosonic branches because the block-triangular determinant is independent of the background electromagnetic field at fixed thermodynamic and transport coefficients. The Lorentz force is algebraic. Cleaning perturbations still enter the fluid principal rows and change the eigenvectors.}

\paragraph*{Electron-acoustic modes.}
{These are pressure waves of the electron component relative to the ion
frame, with the electron enthalpy providing their inertia.  At zero drift
they consist of the left--right acoustic pair and the isobaric contact mode
described above.  A finite drift Doppler-shifts and mixes all three modes.}
{With $u\cdot k=0$ imposed, the
$\left(k_{\mu},\rhoq,e_e\right)$ block gives the cubic
\myeqref{eq.ev.cubic}, and the branch is the electron-acoustic pair
\begin{align}
  \mc{C}_{3}\left(\hat{z}\right)
  = \hat{z}^{3} - \left(1-\sigma_e\right)v_{\rm d}\,\hat{z}^{2}
  - c_{\rm se}^{2}\,\hat{z}
  + c_{\rm se}^{2}\,v_{\rm d} = 0 \, ,\notag\\
  \mc{C}_{3}\big|_{v_{\rm d}=0}
  = \hat{z}\left(\hat{z}^{2}
    - c_{\rm se}^{2}\right) \, ,
  \label{eq.Qe}
\end{align}
with $c_{\rm se}^{2}=P_{e,e}+\sigma_e$, $\sigma_e=\bar{n}_{e}P_{e,n}/H_{e}$ and
$v_{\rm d}=\ju_{\hat{n}}/(q_e\bar n_e)$ the electron drift.  This is the true electron
sound speed.
}

\subsection{Nonlinear causality}
\label{app:char_limits}
\label{app:char_causality}

All causality statements in this paper refer to the nonlinear principal
symbol evaluated on a finite, generally nonequilibrium background.  Following
Ref.~\cite{Cordeiro:2023ljz}, we say that a first-order quasilinear system is
\emph{nonlinearly causal} at such a state if
\begin{enumerate}
 \item[(CI)] every root $\phi_0(\phi_i)$ of
 $\det(\mathbb A^\alpha\phi_\alpha)=0$ is real, and
 \item[(CII)] every characteristic covector is non-timelike,
 $\phi^\alpha\phi_\alpha\ge0$.
\end{enumerate}
In our normalization, (CII) is $|\hat z|\le1$;
consequently every result
below fulfills this definition precisely by placing all roots of each
nontrivial characteristic factor on the real interval $[-1,1]$.  ``Nonlinear'' means that $\Pi$, $q^{\mu}$,
$\pi^{\mu\nu}$, the electron drift, and the state-dependent transport
coefficients are not linearized about zero. 

{The electron energy is evolved, and the dynamical current contributes
the cubic $\mc C_3$ to the resistive characteristic polynomial.}

\subsubsection{Resistive system}
\label{app:char_constrained}

In the resistive system $E^i$ and $B^i$ are independent Maxwell variables.
The Maxwell and cleaning modes lie on the light cone.  The material modes
have $\hat z=0$, and causality of the remaining modes is decided by the
constrained ion--dissipative octic $\mc Q_8$ and the electron cubic
$\mc C_3$.

For a wave normal along a principal axis of $\pi^{\mu\nu}$ and
$q^\mu\parallel\hat n^\mu$, the octic separates into two transverse
quadratics and one longitudinal quartic.  {If $\Lambda_{(1)}$ and
$\Lambda_{(2)}$ are the shear eigenvalues along the two transverse
polarizations, we normalize the factors so that}
\begin{align}
 {\mc Q_8(\hat z)
 =\mc Q_\parallel^c(\hat z)
 \mc Q_\perp^c(\Lambda_{(1)};\hat z)
 \mc Q_\perp^c(\Lambda_{(2)};\hat z).}
 \label{eq.Q8.resistive.factorization}
\end{align}
With
$E=e+P+\Pi$, $y_\eta=\eta/\tau_\pi$, and
$\mathcal K=\kappa T/\tauq$, the transverse factors are
\begin{align}
 \mc Q_\perp^c(\Lambda_a)
 =\left(E+\Lambda_a-\mathcal K\right)\hat z^2
 +q_{\hat n}\hat z-y_\eta=0 ,
 \label{eq.Qperpc}
\end{align}
where $\Lambda_a$ is the shear eigenvalue in the polarization direction.
For $\eta\ge0$, on the positive-inertia branch, their two roots are real and causal precisely when
\begin{align}
 y_\eta&\ge0,&
 E+\Lambda_a-\mathcal K&>0,\notag\\
 E+\Lambda_a-\mathcal K-y_\eta
 &\ge |q_{\hat n}|.
 \label{eq.causperpc}
\end{align}
For a direction-independent screen one imposes the last inequality for all
three shear eigenvalues and replaces $|q_{\hat n}|$ by
$\sqrt{q^\mu q_\mu}$.  With $\kappa=0$ and $q^\mu=0$, on that branch, this is simply
$0\le y_\eta/(E+\Lambda_a)\le1$.

The longitudinal factor is
\begin{align}
 \mc Q_\parallel^c
 &=\left(E+\Lambda_{\hat n}-\mathcal K\right)\hat z^4
 -q_{\hat n}\left(\frac{\dqq}{\tauq}-2+2P_{,e}\right)\hat z^3\notag\\
 &\quad+2q_{\hat n}T_{,e}\frac{\kappa}{\tauq}\hat z^3\notag\\
 &\quad-\mathcal C_2^c\hat z^2
 +q_{\hat n}\left(P_{,e}\frac{\dqq}{\tauq}
 -2T_{,e}\frac{\kappa}{\tauq}\right)\hat z\notag\\
 &\quad+\frac{\kappa}{\tauq}T_{,e}\mathcal Z^c
 -\frac{\kappa}{\tauq}P_{,e}\rhob T_{,\rhob}\notag\\
 &=0,
 \label{eq.Qparc}
\end{align}
where
\begin{align}
 \mathcal Z^c&:=\rhob P_{,\rhob}
 +\frac{\zeta+\dPiPi\Pi}{\tau_\Pi}
 +\frac{\dpipi}{\tau_\pi}\Lambda_{\hat n}
 +\frac{4}{3}y_\eta,\notag\\
 \mathcal C_2^c&:=P_{,e}
 \left(E+\Lambda_{\hat n}-\mathcal K\right)
 +\mathcal Z^c-\rhob T_{,\rhob}\frac{\kappa}{\tauq}.
 \label{eq.char.longdefs}
\end{align}
All four roots of Eq.~\eqref{eq.Qparc} must be real and lie in $[-1,1]$.
This requirement can be written as a finite set of exact algebraic
inequalities.  To do so, set
\begin{align}
 r_q&:=\frac{\kappa}{\tauq},&
 X_T&:=\rhob T_{,\rhob},\notag\\
 \delta_q&:=\frac{\dqq}{\tauq},&
 \mathcal H&:=E+\Lambda_{\hat n}-\mathcal K,
 \label{eq.Qpar.compactdefs}
\end{align}
and introduce the following physical combinations:
\begin{align}
 \ell_q&:=\delta_q-2+2P_{,e}-2r_qT_{,e},\notag\\
 \mathcal M_q&:=P_{,e}\mathcal H+\mathcal Z^c-r_qX_T,\notag\\
 \mathcal N_q&:=P_{,e}\delta_q-2r_qT_{,e},\notag\\
 \mathcal F_q&:=r_q(T_{,e}\mathcal Z^c-P_{,e}X_T).
 \label{eq.Qpar.physicaldefs}
\end{align}
The quartic and its coefficient identification are then
\begin{align}
 \mc Q_\parallel^c(\hat z)
 &=\mathcal H\hat z^4-q_{\hat n}\ell_q\hat z^3
 -\mathcal M_q\hat z^2+q_{\hat n}\mathcal N_q\hat z+\mathcal F_q,
 \notag\\
 (A_4,A_3,A_2)
 &=(\mathcal H,-q_{\hat n}\ell_q,-\mathcal M_q),\notag\\
 (A_1,A_0)&=(q_{\hat n}\mathcal N_q,\mathcal F_q).
 \label{eq.Qpar.coefficients}
\end{align}
Together with Eq.~\eqref{eq.char.longdefs},
Eqs.~\eqref{eq.Qpar.compactdefs} and \eqref{eq.Qpar.physicaldefs} are the
direct primitive map and yield the primitive quartic used below.

{The aligned real-root condition uses the primitive coefficient ratios
above.  The transverse bound \eqref{eq.causperpc} gives
$\mathcal H>0$ at $\Lambda_a=\Lambda_{\hat n}$.}
\begin{align}
 (m,\ell,r,f)
 &:=(\mathcal M_q,\ell_q,\mathcal N_q,\mathcal F_q)/\mathcal H,
 \notag\\
 u_q&:=8m+3q_{\hat n}^2\ell^2,\notag\\
 v_q&:=m^2+3q_{\hat n}^2\ell r+12f,\notag\\
 w_q&:=2m(m^2-36f)
 +9q_{\hat n}^2(m\ell r-3r^2-3\ell^2f).
 \label{eq.Qpar.primitive-reality-combinations}
\end{align}
The three classical quartic margins reduce respectively to
$u_q$, $(u_q^2-16v_q)/3$, and $(4v_q^3-w_q^2)/27$.  Thus, all four roots are
real if and only if
\begin{align}
 v_q&\ge0,\notag\\[-2pt]
 \mathfrak R_q^{\rm prim}
 &:=\min\left\{u_q-4\sqrt{v_q},
 2v_q^{3/2}-|w_q|\right\}\ge0.
 \label{eq.Qpar.real}
\end{align}
This is the only reduction supplied by the other causality conditions: the
transverse bound removes the repeated denominator constraint, whereas the
longitudinal interval bounds constrain root location and do not imply
Eq.~\eqref{eq.Qpar.real}.
The strict condition $\mathfrak R_q^{\rm prim}>0$ gives four distinct real
roots; a repeated real root lies on $\mathfrak R_q^{\rm prim}=0$.  These
expressions contain only reconstructed primitives, $\mathcal Z^c$, equation-of-state derivatives, and transport coefficients.

For completeness, the exact Ferrari roots are reasonably compact in terms
of these coefficients.  Define
\begin{align}
 \alpha&:=-\frac{q_{\hat n}\ell_q}{\mathcal H},&
 \beta&:=-\frac{\mathcal M_q}{\mathcal H},\notag\\
 \gamma&:=\frac{q_{\hat n}\mathcal N_q}{\mathcal H},&
 \epsilon&:=\frac{\mathcal F_q}{\mathcal H},\notag\\
 p_F&:=-\frac{\mathcal M_q}{\mathcal H}
 -\frac{3q_{\hat n}^2\ell_q^2}{8\mathcal H^2},\notag\\
 q_F&:=\frac{q_{\hat n}\mathcal N_q}{\mathcal H}
 -\frac{q_{\hat n}\ell_q\mathcal M_q}{2\mathcal H^2}
 -\frac{q_{\hat n}^3\ell_q^3}{8\mathcal H^3},\notag\\
 r_F&:=\frac{\mathcal F_q}{\mathcal H}
 +\frac{q_{\hat n}^2\ell_q\mathcal N_q}{4\mathcal H^2}
 -\frac{q_{\hat n}^2\ell_q^2\mathcal M_q}{16\mathcal H^3}
 -\frac{3q_{\hat n}^4\ell_q^4}{256\mathcal H^4}.
 \label{eq.Qpar.ferrari-defs}
\end{align}
Let $m_F\ne0$ solve the resolvent cubic
\begin{align}
 m_F^3+2p_Fm_F^2+(p_F^2-4r_F)m_F-q_F^2=0.
 \label{eq.Qpar.resolvent}
\end{align}
Then the four roots are
\begin{align}
 \hat z_{1,2}
 &=-\frac{\alpha}{4}
 +\frac{-\sqrt{m_F}\mathbin{\pm}
 \sqrt{-2p_F-m_F+2q_F/\sqrt{m_F}}}{2},\notag\\
 \hat z_{3,4}
 &=-\frac{\alpha}{4}
 +\frac{ \sqrt{m_F}\mathbin{\pm}
 \sqrt{-2p_F-m_F-2q_F/\sqrt{m_F}}}{2}.
 \label{eq.Qpar.ferrari-roots}
\end{align}
On the four-real-root locus one may choose the largest real resolvent
root.  When $S_F>0$, one has $|\chi_F|\le1$, and a manifestly real exact
expression for it is
\begin{align}
 S_F&:=p_F^2+12r_F,\notag\\
 \chi_F&:=\frac{2p_F^3-72p_Fr_F+27q_F^2}{2S_F^{3/2}},\notag\\
 m_F&=-\frac{2p_F}{3}+\frac{2\sqrt{S_F}}{3}
 \cos\!\left[\frac13\arccos(\chi_F)\right].
 \label{eq.Qpar.cardano}
\end{align}
The boundary $S_F=0$ consists of a triple root and a simple root, except
at the quadruple-root point.  On its non-quadruple part the resolvent has the
triple root $m_F=-2p_F/3>0$, which can be used directly in
Eq.~\eqref{eq.Qpar.ferrari-roots}.  The remaining case $m_F=0$ is the
quadruple-root limit and is obtained continuously.

The radicals need not be used in a numerical causality screen.  Direct
coefficient elimination gives precisely the three margins stated above:
$u_q$, $(u_q^2-16v_q)/3$, and $(4v_q^3-w_q^2)/27$.  The last is the
discriminant of the monic quartic, while the first two carry the remaining
real-root sign information.  Thus, Eq.~\eqref{eq.Qpar.real} is the complete
quartic classification, including repeated roots, without constructing a
root or expanding the discriminant.

It remains to locate those real roots.  The M\"obius transform
\begin{align}
 G(x):=(1+x)^4\mc Q_\parallel^c
 \!\left(\frac{x-1}{x+1}\right)=\sum_{k=0}^4g_kx^k
 \label{eq.Qpar.mobius}
\end{align}
maps $[-1,1]$ to the nonnegative real projective line.  Hence, once
Eq.~\eqref{eq.Qpar.real} holds, Descartes' rule of signs gives the exact
equivalence
\begin{align}
 &\hat z_i\in[-1,1]\quad(i=1,\ldots,4)\notag\\
 &\quad\Longleftrightarrow\quad
 A_4(-1)^kg_k\ge0\quad(k=0,\ldots,4).
 \label{eq.Qpar.signs}
\end{align}
Here
\begin{align}
 g_{0,4}&=A_4+A_2+A_0\mp(A_3+A_1),\notag\\
 g_{1,3}&=-4A_4+4A_0\pm2(A_3-A_1),\notag\\
 g_2&=6A_4-2A_2+6A_0.
 \label{eq.Qpar.gcoefficients}
\end{align}

In the physical sector $\mathcal H=A_4>0$, the five linear sign tests in
Eq.~\eqref{eq.Qpar.signs} pair into the following three physical
inequalities:
\begin{align}
 \mathcal H-\mathcal M_q+\mathcal F_q
 &\ge\left|q_{\hat n}\left(\ell_q-\mathcal N_q\right)\right|,\notag\\
 2\left(\mathcal H-\mathcal F_q\right)
 &\ge\left|q_{\hat n}\left(\ell_q+\mathcal N_q\right)\right|,\notag\\
 3\left(\mathcal H+\mathcal F_q\right)+\mathcal M_q&\ge0.
 \label{eq.causpar}
\end{align}
Equations~\eqref{eq.Qpar.real} and \eqref{eq.causpar} are therefore jointly
necessary and sufficient for longitudinal resistive causality.  In terms of
the coefficients \eqref{eq.Qpar.coefficients} the three lines read simply
$A_4+A_2+A_0\ge|A_3+A_1|$, $2(A_4-A_0)\ge|A_3-A_1|$ and
$3A_4-A_2+3A_0\ge0$; the first is the pair of endpoint conditions
$\mc Q_\parallel^c(\pm1)\ge0$, and the second is the corresponding pair of
endpoint-slope conditions, since
$g_{1}=4\mc Q_\parallel^c(-1)+2\mc Q_\parallel^{c\prime}(-1)$ and
$g_{3}=4\mc Q_\parallel^c(1)-2\mc Q_\parallel^{c\prime}(1)$. 

\paragraph*{Relaxation-time limits.}
For direct numerical use, define the local primitive combinations
\begin{align}
 h_{\hat n}&:=e+P+\Pi+\Lambda_{\hat n},&
 L_e&:=1-P_{,e},\notag\\
 Q_{\hat n}&:=|q_{\hat n}|,&
 X_T&:=\rhob T_{,\rhob},\notag\\
 B_\Pi&:=\zeta+\dPiPi\Pi,&
 B_\pi&:=\dpipi\Lambda_{\hat n}+\frac{4\eta}{3},\notag\\
 R_P&:=\rhob P_{,\rhob},&
 \mathcal Z^c&=R_P+\frac{B_\Pi}{\tau_\Pi}
 +\frac{B_\pi}{\tau_\pi}.
 \label{eq.caus.resistive-floor-notation-app}
\end{align}
These are only abbreviations for reconstructed primitives and local
equation-of-state or transport evaluations.  For $s=\pm1$,
define the first candidate directly by
\begingroup\footnotesize
\begin{align}
 \tau_{q,1s}:={}&
 \frac{\left[\begin{aligned}
 &s|q_{\hat n}|(1-P_{,e})\dqq\\[-2pt]
 &-\kappa\Biggl\{(1-P_{,e})(\rhob T_{,\rhob}-T)\\[-2pt]
 &\quad+T_{,e}\Biggl[\rhob P_{,\rhob}
 +\frac{\zeta+\dPiPi\Pi}{\tau_\Pi}\\[-2pt]
 &\hspace{7.6em}
 +\frac{\dpipi\Lambda_{\hat n}+4\eta/3}{\tau_\pi}
 \Biggr]\Biggr\}
 \end{aligned}\right]_+}
 {\begin{aligned}
 &(1-P_{,e})(e+P+\Pi+\Lambda_{\hat n})\\[-2pt]
 &\quad-\rhob P_{,\rhob}
 -\frac{\zeta+\dPiPi\Pi}{\tau_\Pi}\\[-2pt]
 &\quad-\frac{\dpipi\Lambda_{\hat n}+4\eta/3}{\tau_\pi}
 +2s|q_{\hat n}|(1-P_{,e})
 \end{aligned}}.
 \label{eq.caus.resistive-tauq-candidate1-app}
\end{align}
\endgroup
The second pair of half-spaces gives
\begingroup\footnotesize
\begin{align}
 \tau_{q,2s}:={}&
 \frac{\left[\begin{aligned}
 &2\kappa\Biggl\{T+T_{,e}\Biggl[\rhob P_{,\rhob}
 +\frac{\zeta+\dPiPi\Pi}{\tau_\Pi}\\[-2pt]
 &\hspace{7.6em}
 +\frac{\dpipi\Lambda_{\hat n}+4\eta/3}{\tau_\pi}
 \Biggr]-P_{,e}\rhob T_{,\rhob}\Biggr\}\\[-2pt]
 &\quad+s|q_{\hat n}|\left[(1+P_{,e})\dqq
 -4\kappa T_{,e}\right]
 \end{aligned}\right]_+}
 {2\left[e+P+\Pi+\Lambda_{\hat n}
 +s|q_{\hat n}|(1-P_{,e})\right]}.
 \label{eq.caus.resistive-tauq-candidate2-app}
\end{align}
\endgroup
Finally, the third interval condition gives
\begingroup\footnotesize
\begin{align}
 \tau_{q,3}:={}&
 \frac{\left[\begin{aligned}
 \kappa\Biggl\{&(3+P_{,e})T
 +(1+3P_{,e})\rhob T_{,\rhob}\\[-2pt]
 &-3T_{,e}\Biggl[\rhob P_{,\rhob}
 +\frac{\zeta+\dPiPi\Pi}{\tau_\Pi}\\[-2pt]
 &\hspace{7.1em}
 +\frac{\dpipi\Lambda_{\hat n}+4\eta/3}{\tau_\pi}
 \Biggr]\Biggr\}
 \end{aligned}\right]_+}
 {\begin{aligned}
 &(3+P_{,e})(e+P+\Pi+\Lambda_{\hat n})\\[-2pt]
 &\quad+\rhob P_{,\rhob}
 +\frac{\zeta+\dPiPi\Pi}{\tau_\Pi}
 +\frac{\dpipi\Lambda_{\hat n}+4\eta/3}{\tau_\pi}
 \end{aligned}}.
 \label{eq.caus.resistive-tauq-candidate3-app}
\end{align}
\endgroup
{Here $[x]_+:=\max(x,0)$; the large-$\tauq$ branch requires
every displayed denominator to be positive.}
Multiplying the two absolute-value inequalities in
Eq.~\eqref{eq.causpar} by the positive $\tauq$ resolves them into four
affine half-spaces, so the desired primitive limit is
\begin{align}
 \tau_{q,\min}
 &:=\max\left\{
 \max_{s=\pm1}\tau_{q,1s},
 \max_{s=\pm1}\tau_{q,2s},
 \tau_{q,3}\right\},\notag\\[-2pt]
 \tauq&>\max\{0,\kappa T_{,e}\},\qquad
 \tauq\ge\tau_{q,\min}.\label{eq.caus.resistive-tauq-min-app}
\end{align}
The additional
$\tauq>\kappa T_{,e}$ selects the branch on which the coefficient of
$\mathcal Z^c$ in the first inequality has the sign reached as
$\tauq\rightarrow\infty$.  {Conditional on this branch and
its positive denominators,} the heat-flux floor
is necessary and sufficient for the three interval conditions
\eqref{eq.causpar} at fixed $\tau_\Pi$ and $\tau_\pi$.  This is an equivalence
for root \emph{location}, not root reality; it does not imply any of the
quartic conditions \eqref{eq.Qpar.real}.

These viscous limits require $\kappa T_{,e}>0$. For zero or negative values, use Eq.~\eqref{eq.causpar} before dividing by this product.
For a fixed $\tauq$ on this branch, the first two lines of
Eq.~\eqref{eq.causpar} give the upper limits
\begin{align}
 U_1(\tauq)
 &:=\frac{L_e\left[h_{\hat n}
 +\frac{\kappa}{\tauq}(X_T-T)\right]}
 {1-\frac{\kappa}{\tauq}T_{,e}}\notag\\[-2pt]
 &\quad-\frac{Q_{\hat n}
 \left|L_e\left(\frac{\dqq}{\tauq}-2\right)\right|}
 {1-\frac{\kappa}{\tauq}T_{,e}},\notag\\
 U_2(\tauq)
 &:=\frac{\tauq}{\kappa T_{,e}}\Biggl[
 h_{\hat n}-\frac{\kappa}{\tauq}(T-P_{,e}X_T)\notag\\[-2pt]
 &\hspace{5.2em}-\frac{Q_{\hat n}}{2}
 \left|(1+P_{,e})\frac{\dqq}{\tauq}
 -2L_e-\frac{4\kappa}{\tauq}T_{,e}\right|\Biggr],\notag\\
 \mathcal Z_+(\tauq)&:=\min\{U_1(\tauq),U_2(\tauq)\},\notag\\
 \mathcal Z_-(\tauq)
 &:=\frac{(3+P_{,e})
 \left(\frac{\kappa T}{\tauq}-h_{\hat n}\right)
 +\frac{\kappa}{\tauq}(1+3P_{,e})X_T}
 {1+\frac{3\kappa}{\tauq}T_{,e}}.
 \label{eq.caus.resistive-zlimits}
\end{align}
The third line of Eq.~\eqref{eq.causpar} supplies the lower limit, so all
three interval conditions are exactly equivalent to
\begin{align}
 \mathcal Z_-(\tauq)
 \le R_P+\frac{B_\Pi}{\tau_\Pi}+\frac{B_\pi}{\tau_\pi}
 \le\mathcal Z_+(\tauq)&.\label{eq.caus.resistive-viscous-interval}
\end{align}
On the usual regulator branch,
$B_\Pi\ge0$, $B_\pi\ge0$, and
$\mathcal Z_-(\tauq)\le R_P<\mathcal Z_+(\tauq)$.  The lower limit is then
automatic, while the upper limit gives the two equivalent conditional forms
\begin{align}
 \tau_\pi&\ge\frac{B_\pi}
 {\mathcal Z_+(\tauq)-R_P-B_\Pi/\tau_\Pi},\notag\\
 \tau_\Pi&\ge\frac{B_\Pi}
 {\mathcal Z_+(\tauq)-R_P-B_\pi/\tau_\pi},
 \label{eq.caus.resistive-viscous-floors-app}
\end{align}
with positive denominators.  These are the same coupled upper constraint
solved for opposite variables, not two independent floors.  Allocating a
fraction $0<\vartheta<1$ of the available margin to the bulk channel gives
{the independent sufficient pair}
\begin{align}
 \tau_\Pi&>\frac{B_\Pi}
 {\vartheta[\mathcal Z_+(\tauq)-R_P]},\notag\\
 \tau_\pi&>\frac{B_\pi}
 {(1-\vartheta)[\mathcal Z_+(\tauq)-R_P]}.
 \label{eq.caus.resistive-viscous-allocation-app}
\end{align}
If $\mathcal Z_-(\tauq)>R_P$, the lower side of
Eq.~\eqref{eq.caus.resistive-viscous-interval} instead supplies an upper
relaxation-time constraint, and increasing all relaxation times cannot
restore the interval condition.  If either $B_\Pi$ or $B_\pi$ is negative,
the exact interval, rather than a lower-floor form, must likewise be used.
The
equivalent bounds include the causal endpoints. Strict inequalities may be
used to impose a finite inward margin in a numerical screen.

Consequently, the heat-flux floor
\eqref{eq.caus.resistive-tauq-min-app} and the exact viscous interval
\eqref{eq.caus.resistive-viscous-interval} are two equivalent rearrangements
of the sign tests \eqref{eq.causpar} on this branch; they are not separate
constraints that must both be imposed.  Each is necessary and sufficient for
those sign tests under its stated denominator and sign assumptions.  The
conditional floors
\eqref{eq.caus.resistive-viscous-floors-app} are equivalent rearrangements of
the upper side of that interval when their denominators are positive, whereas
the allocated pair \eqref{eq.caus.resistive-viscous-allocation-app} is
sufficient but not necessary. Full longitudinal causality requires their conjunction
with the independent reality conditions \eqref{eq.Qpar.real}.

For reference, the simplifications connecting these limits to
Eq.~\eqref{eq.causpar} are
\begin{align}
 \mathcal H-\mathcal M_q+\mathcal F_q
 &=(1-P_{,e})(\mathcal H+r_qX_T)
 -(1-r_qT_{,e})\mathcal Z^c,\notag\\
 \ell_q-\mathcal N_q
 &=(1-P_{,e})(\delta_q-2),\notag\\
 \ell_q+\mathcal N_q
 &=(1+P_{,e})\delta_q-2(1-P_{,e})-4r_qT_{,e}.
 \label{eq.Qpar.linear-primitive-identities}
\end{align}
Thus, the interval part is genuinely affine in
$(\Pi,\Lambda_{\hat n},q_{\hat n})$ once the local thermodynamic and
transport coefficients have been evaluated.

There is no analogous finite linear reduction of the real-root conditions.
Already on the slice $q_{\hat n}=0$ one has
\begin{align}
 \mc Q_\parallel^c
 &=\mathcal H\hat z^4-\mathcal M_q\hat z^2+\mathcal F_q,\notag\\
 y_\pm&=\frac{\mathcal M_q\pm
 \sqrt{\mathcal M_q^2-4\mathcal H\mathcal F_q}}{2\mathcal H},
 \qquad y:=\hat z^2.
 \label{eq.Qpar.qzero-roots}
\end{align}
Reality alone is therefore exactly
\begin{align}
 \mathcal M_q\ge0,\qquad \mathcal F_q\ge0,
 \qquad 4\mathcal H\mathcal F_q\le\mathcal M_q^2.
 \label{eq.Qpar.qzero-reality}
\end{align}

Both roots belong to $[0,1]$ precisely when
\begin{align}
 0\le\mathcal M_q\le2\mathcal H,\qquad
 \max(0,\mathcal M_q-\mathcal H)\le\mathcal F_q
 \le\frac{\mathcal M_q^2}{4\mathcal H}.
 \label{eq.Qpar.qzero-conditions}
\end{align}
Indeed, the first two linear lower bounds say that the product
$y_-y_+$ and $(1-y_-)(1-y_+)$ are nonnegative, while
$0\le\mathcal M_q\le2\mathcal H$ fixes the signs of the corresponding
sums.  The upper bound is the same irreducible quadratic discriminant
condition as in Eq.~\eqref{eq.Qpar.qzero-reality}.

\paragraph*{Primitive ray bounds.}
The aligned conditions give the scalar bounds below when thermodynamic derivatives and transport coefficients remain fixed along each ray. Full ion causality requires real, causal roots of the octic in every direction.
First, the transverse condition itself has a direct primitive form.  Put
\begin{align}
 \Delta_\perp&:=e+P-\frac{\kappa T}{\tauq}
 -\frac{\eta}{\tau_\pi},\notag\\
 Q&:=\sqrt{q^\mu q_\mu},&
 \Lambda_{\min}&:=\min_a\Lambda_a.
 \label{eq.caus.primitive-ray-transverse-defs}
\end{align}
Together with $\eta/\tau_\pi\ge0$ and the strict inertia condition
$e+P+\Pi+\Lambda_{\min}-\kappa T/\tauq>0$, the remaining inequalities in
the direction-independent transverse bound from Eq.~\eqref{eq.causperpc}, with $|q_{\hat n}|$ replaced by $Q$, reduce to
\begin{align}
 Q&\le \Delta_\perp+\Pi+\Lambda_{\min},\notag\\
 \Pi&\ge Q-\Delta_\perp-\Lambda_{\min},\qquad
 \Lambda_{\min}\ge Q-\Delta_\perp-\Pi .
 \label{eq.caus.primitive-ray-transverse}
\end{align}
Thus, the condition bounds a negative bulk pressure, rather than a positive
one, and its exact shear bound is one-sided.  If a rotationally invariant
sufficient shear ceiling is preferable, tracelessness gives
\begin{align}
 \sqrt{\pi^{\mu\nu}\pi_{\mu\nu}}
 \le \sqrt{\frac32}\,
 \left(\Delta_\perp+\Pi-Q\right),\qquad
 \Delta_\perp+\Pi-Q\ge0.
 \label{eq.caus.primitive-ray-shear-norm}
\end{align}
The factor $\sqrt{3/2}$ is sharp for a bound stated only in terms of this
norm.  It ensures that no eigenvalue can be smaller than
$-(\Delta_\perp+\Pi-Q)$.

For the longitudinal condition, a particularly simple limiter first sets
$q^\mu=0$ and scales the viscous state along its current ray,
$(\Pi,\pi^{\mu\nu})\mapsto
\alpha_{\rm v}(\Pi,\pi^{\mu\nu})$.  For every principal direction define
\begin{align}
 \mathcal H(\alpha_{\rm v})
 &=\mathcal H_0+\alpha_{\rm v}\mathcal H_1,\notag\\
{\mathcal H_0}&:=e+P-\frac{\kappa T}{\tauq},\notag\\
 \mathcal H_1&:=\Pi+\Lambda_{\hat n},\notag\\
 \mathcal Z(\alpha_{\rm v})
 &=\mathcal Z_0+\alpha_{\rm v}\mathcal Z_1,\notag\\
 \mathcal Z_0&:=\rhob P_{,\rhob}+\frac{\zeta}{\tau_\Pi}
 +\frac{4\eta}{3\tau_\pi},\notag\\
 \mathcal Z_1&:=\frac{\dPiPi\Pi}{\tau_\Pi}
 +\frac{\dpipi\Lambda_{\hat n}}{\tau_\pi},\notag\\
 \mathcal M(\alpha_{\rm v})
 &:=P_{,e}\mathcal H(\alpha_{\rm v})
 +\mathcal Z(\alpha_{\rm v})
 -\frac{\kappa}{\tauq}\rhob T_{,\rhob},\notag\\
 \mathcal F(\alpha_{\rm v})
 &:=\frac{\kappa}{\tauq}
 \left[T_{,e}\mathcal Z(\alpha_{\rm v})
 -P_{,e}\rhob T_{,\rhob}\right].
 \label{eq.caus.primitive-viscous-ray}
\end{align}
All four quantities are affine in $\alpha_{\rm v}$.  At zero heat flux the
complete longitudinal conditions reduce to
\begin{align}
 \mathcal H&>0,&
 0\le\mathcal M&\le2\mathcal H,\notag\\
 \max(0,\mathcal M-\mathcal H)&\le\mathcal F
 \le\frac{\mathcal M^2}{4\mathcal H}.
 \label{eq.caus.primitive-viscous-ray-bounds}
\end{align}

Together with
$\Delta_\perp+\alpha_{\rm v}(\Pi+\Lambda_{\min})\ge0$,
these contain only linear boundaries in $\alpha_{\rm v}$ and the single
quadratic boundary
$\mathcal M^2-4\mathcal H\mathcal F=0$.  The admissible viscous factor is
chosen in the admissible component connected to $\alpha_{\rm v}=0$, provided that state passes these bounds. Linear and quadratic boundaries give its supremum in $[0,1]$; strict boundaries require an inward margin. Passing all three principal-axis tests does not ensure oblique causality.

At fixed viscous state, the aligned heat-flux step is also explicit. For a principal wave normal with $q^\mu\parallel\hat n^\mu$, set
\begin{align}
 D_{\hat n}&:=\tauq(e+P+\Pi+\Lambda_{\hat n})
 -\kappa T>0,\notag\\
 m&:=P_{,e}+\frac{\tauq\mathcal Z^c
 -\kappa\rhob T_{,\rhob}}{D_{\hat n}},\notag\\
 b&:=\frac{\dqq-2\kappa T_{,e}
 -2(1-P_{,e})\tauq}{D_{\hat n}},\notag\\
 c&:=\frac{P_{,e}\dqq-2\kappa T_{,e}}{D_{\hat n}},\notag\\
 f&:=\frac{\kappa(T_{,e}\mathcal Z^c
 -P_{,e}\rhob T_{,\rhob})}{D_{\hat n}}.
 \label{eq.caus.primitive-heat-ray-defs}
\end{align}
Under $q^\mu\mapsto\alpha_q q^\mu$, let
$x:=\alpha_q^2q_{\hat n}^2$.  Equation~\eqref{eq.Qpar.real} is then
equivalent to
\begin{align}
 U(x)&:=8m+3b^2x,\notag\\
 V(x)&:=m^2+12f+3bcx,\notag\\
 W(x)&:=2m(m^2-36f)\notag\\&\quad
 +9(mbc-3c^2-3fb^2)x,\notag\\
 V(x)&\ge0,\qquad U(x)\ge0,\notag\\
 U(x)^2-16V(x)&\ge0,\notag\\
 4V(x)^3-W(x)^2&\ge0.
 \label{eq.caus.primitive-heat-ray-reality}
\end{align}

The four left-hand sides have degrees at most $1$, $1$, $2$, and $3$ in
$x$.  Assuming the zero-flux state has strict margins, let $x_{\rm real}$
denote the first boundary of its connected component.  The exact reality
ceiling on that component is simply $x\le x_{\rm real}$.  No quartic
characteristic roots are needed.

The root-location conditions decouple even further.  In the same variables
they are
\begin{align}
 1-m+f&\ge \sqrt{x}\,|b-c|,&
 2(1-f)&\ge \sqrt{x}\,|b+c|,\notag\\
 3(1+f)+m&\ge0.
 \label{eq.caus.primitive-heat-ray-location}
\end{align}
Thus, provided the three zero-flux left-hand sides have the required signs,
\begin{align}
 x_{\rm loc}&:=\min\Biggl\{
 \left(\frac{1-m+f}{|b-c|}\right)^2,\notag\\[-2pt]
 &\hspace{4.3em}
 \left(\frac{2(1-f)}{|b+c|}\right)^2
 \Biggr\},\notag\\
 x_*&:=\min(x_{\rm real},x_{\rm loc}) .
 \label{eq.caus.primitive-heat-ray-ceiling}
\end{align}
where a ratio with zero denominator is omitted.

{An aligned trial heat-flux factor}
is the smallest of the transverse factor and
$\sqrt{x_*/q_{\hat n}^2}$ over the principal directions.  A factor whose
denominator vanishes is again omitted.

{Within the aligned reduction, at fixed thermodynamics and transport,
this construction selects the causal component connected to the
nondissipative state.  For general heat-flux directions, retain the transverse
couplings in $\mc Q_8$ and recheck Eq.~\eqref{eq.resgeneral.complete}
after reclosure.}
  Such a branch choice is essential:
the cubic discriminant in
Eq.~\eqref{eq.caus.primitive-heat-ray-reality} can produce disconnected
admissible intervals in $x$.
Hence, no globally equivalent collection of
independent one-sided ceilings on $\Pi$, $\pi^{\mu\nu}$, and $q^\mu$ exists.
Along the two rays above, however, all boundaries are at most quadratic in
the viscous factor and cubic in the squared heat-flux factor, so the required
rescaling is unambiguous and inexpensive.

When $\kappa=q_{\hat n}=0$ in the aligned reduction, the exact conditions reduce to
\begin{align}
 0\le P_{,e}+\frac{\mathcal Z^c}{E+\Lambda_{\hat n}}\le1.
 \label{eq.causpar-noheat}
\end{align}
After setting $q^\mu=\kappa=0$ and removing the electromagnetic and
electron blocks, Eqs.~\eqref{eq.causperpc} and
\eqref{eq.causpar-noheat} reduce to the common constrained bulk--shear
system of Ref.~\cite{Cordeiro:2026}.
Our determinant reproduces their
Eqs.~(32), (33), (39), (54a), and (54c).
\footnote{The discrepancy lies in
the printed coefficient $\mathcal A_1$ in their Eq.~(54b):
$3\eta_{\rm eff}^2/\tau_\pi^2$ must be replaced by
$3(\eta_{\rm eff}/\tau_\pi)E$.  With this replacement the two determinants
agree exactly. The difference does not appear to be due to a different physical
constraint.}
The heat-flux-only comparison with Ref.~\cite{Cordeiro:2025diffusion}
uses $\Pi=\pi^{\mu\nu}=\zeta=\eta=0$ and omits the corresponding
viscous variables, electromagnetic fields, and electron block.
For constant $\Omega_q=\tauq/(\kappa T^2)$ and $\dqq=\tauq/2$, their
Eq.~(33) gives precisely our constrained heat equation.  For every wave
normal their physical principal symbol therefore agrees with ours.
Their extended determinant adds only a material root and a copy of the
physical transverse root.  In the aligned case, with
$E_q=e+P-\kappa T/\tauq>0$, the monic quintic is
\begin{align}
 P_5^{(E)}(\hat z)
 =\left(\hat z+\frac{q_{\hat n}}{E_q}\right)
 \frac{\mc Q_\parallel^c(\hat z)}{E_q}.
 \label{eq.heat-diffusion-comparison}
\end{align}
Direct evaluation of their Eq.~(36), however, requires replacing
$2[T_{,e}/(\Omega_qT^2)-P_{,e}]$ by
$4[T_{,e}/(\Omega_qT^2)-P_{,e}]$ in their printed Eq.~(40a).
The other quintic coefficients agree.  Their Eq.~(42b) also requires
$4|\mathcal A_4^{(E)}|\le10+\mathcal A_3^{(E)}$, alongside
$|\mathcal A_4^{(E)}|\le5$, as stated correctly in their Appendix~A.
With these corrections and real roots, their interval conditions and ours
describe the same causal characteristics.

For a generic propagation direction $\mc Q_8$ need not factor, and the exact
statement is correspondingly that all eight roots of $\mc Q_8$ are real and
lie in $[-1,1]$; this is exactly (CI) and (CII) for the resistive ion sector,
while its material and Maxwell roots satisfy them identically.
{Section~\ref{app:resistive-general-direction-free} gives the exact
coefficient tests over all normals, including arbitrary anisotropic stress
and heat flux, and a separate primitive sufficient domain.}

\subsubsection{Nonlinear causality of the resistive electron sector}

In the resistive system, the cubic \eqref{eq.Qe} must pass its own
causality test.  Put
$A_e=1-\sigma_e$ and
\begin{align}
 d_e&:=\sqrt{A_e^2v_{\rm d}^2+3c_{\rm se}^2},\notag\\
 \mathcal D_e&:=c_{\rm se}^2\left[
 4A_e^3v_{\rm d}^4
 +c_{\rm se}^2(A_e^2+18A_e-27)v_{\rm d}^2
 +4c_{\rm se}^4\right].
 \label{eq.causcubicdisc}
\end{align}
{Using
$c_{\rm se}^2=P_{e,e}+\bar n_eP_{e,n}/H_e$,
$A_e=1-\bar n_eP_{e,n}/H_e$, and
$v_{\rm d}=\ju_{\hat n}/(q_e\bar n_e)$ makes
Eq.~\eqref{eq.causcubicdisc} fully primitive.  Thus, $\mathcal D_e>0$ requires only
$(\bar n_e,H_e,P_{e,e},P_{e,n},\ju_{\hat n})$ and fixed species constants;
no characteristic root or eigenvector is needed.}
For $c_{\rm se}^2>0$, set $y=v_{\rm d}^2/c_{\rm se}^2$.  The
normalized discriminant is the quadratic
\begin{align}
 \frac{\mathcal D_e}{c_{\rm se}^6}
 &=4(1-\sigma_e)^3y^2
 -(8+20\sigma_e-\sigma_e^2)y+4\notag\\
 &:=f_e(y).
 \label{eq:electron-sh-discriminant-quadratic}
\end{align}
For $\sigma_e\ne1$ the quadratic discriminant is
$\sigma_e(8+\sigma_e)^3$.  For $-8<\sigma_e<0$ it is negative and the leading
coefficient is positive; for $\sigma_e\le-8$ all coefficients of
$f_e$ are positive.  Hence, $f_e(y)>0$ for every $y\ge0$ when
$\sigma_e<0$.  At $\sigma_e=1$, $f_e(y)=4-27y$.
For $\sigma_e\ge0$ the first positive zero is
\begin{align}
 y_*&=\frac{8}
 {8+20\sigma_e-\sigma_e^2+\sqrt{\sigma_e(8+\sigma_e)^3}}.
 \label{eq:electron-sh-first-zero}
\end{align}
Every $0\le y\le\mc V^2/c_{\rm se}^2$ is attained by a propagation
direction.  Positivity in all directions is therefore equivalent to
$\mc V^2<c_{\rm se}^2y_*$, yielding
Eq.~\eqref{eq:electron-sh-current-bound}.  At $\sigma_e=0$,
$f_e(y)=4(y-1)^2$; checking the discriminant only at the direction
parallel to the current would miss the oblique degeneracy when
$\mc V>c_{\rm se}$.

{At zero drift, causality alone permits $0\le c_{\rm se}^2\le1$.
On the strongly hyperbolic branch the exact condition is instead}
\begin{align}
 {\quad
 0<c_{\rm se}^2
 =P_{e,e}+\frac{\bar n_e}{H_e}P_{e,n}\le1 .
 \quad}
 \label{eq.causcse}
\end{align}
{At the lower endpoint the three roots coalesce at the material speed and the
electron block is defective; see Sec.~\ref{app:ev_causality}.
At finite drift, on this branch all three cubic roots are real and causal if and only if}
\begin{align}
 0<c_{\rm se}^2&\le1,&
 \mathcal D_e&\ge0,\notag\\
 d_e+|A_ev_{\rm d}|&\le3,&
 |(A_e-c_{\rm se}^2)v_{\rm d}|&\le1-c_{\rm se}^2.
 \label{eq.causcubic}
\end{align}
{Strong hyperbolicity further requires $\mathcal D_e>0$; at
$\mathcal D_e=0$ the repeated electron root is defective.  This is the
electron square-freeness condition
{\eqref{eq:electron-sh-current-bound}.}  Given that condition,
putting $t=|A_ev_{\rm d}|$ in the remaining inequalities and squaring
$\sqrt{t^2+3c_{\rm se}^2}\le3-t$ gives
$2t\le3-c_{\rm se}^2$.  Maximizing $|v_{\rm d}|$ over directions then
gives the causal ceiling \eqref{eq.caus.vd}.}
For the
gamma-law electron closure, $1<\Gamma_e\le2$ and $e_e>\rho_e$, these
conditions reduce to
\begin{align}
 {\quad
 |v_{\rm d}|\le
 \frac{1-c_{\rm se}^2}{1-\sigma_e-c_{\rm se}^2}.
 \quad}
 \label{eq.causvd}
\end{align}
{For the gamma law,
$c_{\rm se}^2=(\Gamma_e-1)(1-\rho_e/H_e)$ gives}
\begin{align}
 \frac{1-c_{\rm se}^2}{1-\sigma_e-c_{\rm se}^2}
 &=\frac{1-c_{\rm se}^2}
 {\Gamma_e-2c_{\rm se}^2}.
 \label{eq.causvd.primitive}
\end{align}
{For this gamma law $\sigma_e<0$, so the discriminant is positive
for every drift, including the causal boundary.  The strict invariant
screen in Eq.~\eqref{eq:electron-current-combined-bound} additionally
keeps the electron roots inside the light cone and enforces the dominant
energy condition.  These are electron spectral conditions; the full
system also requires the ion-sector and joint-kernel compatibility tests
of Sec.~\ref{app:ev_causality}.}
Equations~\eqref{eq.causcse}--\eqref{eq.causcubic} therefore fulfill
(CI) by enforcing three real roots of $\mc C_3$ and (CII) by placing
all three in $[-1,1]$.

\subsubsection{\texorpdfstring{{General propagation directions}}{General propagation directions}}
\label{app:resistive-general-direction-free}

{For each wave normal, finite coefficient tests characterize (CI)--(CII)
from the local primitives, EOS derivatives and transport coefficients.
They must hold for every unit normal.  The sufficient primitive condition
below avoids this angular minimization.  The exact tests retain arbitrary
finite anisotropic stress, bulk pressure, heat flux and electron drift.}
Assume the physical constraints \eqref{eq.char.constraints}, positive
relaxation times $\tau_\Pi,\tau_\pi,\tauq$, and $H_e>0$,
$\bar n_e>0$, $q_e\ne0$.
\paragraph*{The ion criterion.}
In any fluid-rest orthonormal triad let $q$ be the three components of
$q^\mu$ and let $\Lambda$ be the symmetric trace-free array of
$\pi^{\mu\nu}$.  With $y:=\eta/\tau_\pi$, define
\begin{align}
 \bar E&:=e+P+\Pi-\frac{\kappa T}{\tauq},&
 \mathsf H&:=\bar E\,\mathsf I+\Lambda,\notag\\
 \Delta_H&:=\det\mathsf H,&p&:=P_{,e},\notag\\
 c&:=\frac{\kappa T_{,e}}{\tauq},&
 R&:=\frac{\kappa\rhob T_{,\rhob}}{\tauq},\notag\\
 Z&:=\rhob P_{,\rhob}
 +\frac{\zeta+\delta_{\Pi\Pi}\Pi}{\tau_\Pi}+\frac{4y}{3},
 \notag\\
 d_q&:=\frac{\delta_{qq}}{\tauq},&
 d_\pi&:=\frac{\delta_{\pi\pi}}{\tau_\pi}.
 \label{eq.resgeneral.inputs}
\end{align}
  A necessary first
condition is $\Delta_H\ne0$, since otherwise the rest-frame
time covector is characteristic and timelike.

{Write the unit wave normal as $x=(x_1,x_2,x_3)$ and the speed as
$\hat z$.  Coefficient operations remove $\hat z$; the normal remains
under the minima in Eq.~\eqref{eq.resgeneral.complete}:}
\begin{align}
 s&:=q\cdot x,\notag\\
 \mathsf D&:=\hat z^2\mathsf H+(\hat z s-y)\mathsf I\notag\\
 &\quad+\bigl[(1-d_q)\hat z q-d_\pi\Lambda x
 -(Z-R-y)x\bigr]x^T,\notag\\
 b&:=2\hat z^3q+\hat z^2\mathsf Hx
 -d_q\hat z s\,x+Rx,\notag\\
 F_8(\hat z,x)&:=
 \frac{(\hat z^2-c)\det\mathsf D}{\Delta_H}\notag\\
 &\quad-\frac{(p-c)b^T\operatorname{adj}(\mathsf D)x}{\Delta_H},\notag\\
 F_8(\hat z,x)&=\sum_{i=0}^{8}a_i(x)\hat z^{8-i},
 \qquad a_0=1 .
 \label{eq.resgeneral.polynomial}
\end{align}
On $x\cdot x=1$, eliminating the baryon and nine relaxation
amplitudes from the source equations gives the retained
energy--velocity kernel
\begin{align}
 \mathsf K_4=
 \begin{pmatrix}
 (\hat z^2-c)/\hat z&b^T/\hat z^2\\
 (p-c)x&\mathsf D/\hat z
 \end{pmatrix},\notag\\
 \hat z^4\det\mathsf K_4&=\Delta_H F_8.
\end{align}
The adjugate is a finite cofactor sum, so this expression does
not divide by $\det\mathsf D$.  In particular, it retains zero
characteristic roots.  The full ion determinant is
$\tau_\Pi\tauq^3\tau_\pi^5\hat z^6\Delta_H F_8$.

Set $M_0=8$ and compute $M_1,\ldots,M_{14}$ by
\begin{align}
 M_j&=
 \begin{cases}
 -j a_j-\displaystyle\sum_{i=1}^{j-1}a_iM_{j-i},&1\le j\le8,\\
 -\displaystyle\sum_{i=1}^{8}a_iM_{j-i},&9\le j\le14.
 \end{cases}
 \label{eq.resgeneral.newton}
\end{align}
The eight reality polynomials and seven location polynomials are
the following finite scalar expressions:
\begin{align}
 \mathcal R_k(x)&:=
 \sum_{\substack{I\subset\{0,\ldots,7\}\\|I|=k}}
 \det [M_{i+j}(x)]_{i,j\in I},
 \quad 1\le k\le8,\notag\\
 \mathcal L_j(x)&:=
 \sum_{i=0}^{8-j}\binom{8-i}{j}a_i(x),\qquad 0\le j\le6 .
 \label{eq.resgeneral.scalar-polynomials}
\end{align}
Their degrees in $x$ are at most $56$ and $8-j$, respectively.
Put
\begin{align}
 I_2&:=\operatorname{tr}\Lambda^2,&
 I_3&:=\operatorname{tr}\Lambda^3,\notag\\
 \mathsf A&:=\Lambda^2-\bar E\Lambda+
 \left(\bar E^2-\tfrac12I_2\right)\mathsf I,\notag\\
 N&:=\left(3\bar E^2-\tfrac12I_2\right)q\notag\\
 &\quad +(1-d_q-2p+2c)\mathsf A q,\notag\\
 \Delta_H&=\bar E^3-\tfrac12\bar E I_2+\tfrac13I_3 .
 \label{eq.resgeneral.linear-heat-row}
\end{align}
The complete ion causality conditions are
\begin{align}
 \Delta_H&\ne0,\qquad N\cdot N\le64\Delta_H^2,\notag\\
 \min_{x\cdot x=1}\mathcal R_k(x)&\ge0,\quad 1\le k\le8,\notag\\
 \min_{x\cdot x=1}\mathcal L_j(x)&\ge0,\quad 0\le j\le6.
 \label{eq.resgeneral.complete}
\end{align}
Together with \eqref{eq.reselectron.exact-linear} and
\eqref{eq.reselectron.exact-reality}, this is a
necessary-and-sufficient criterion for nonlinear resistive
causality in the sense of (CI)--(CII).  Maxwell and cleaning
roots are already on the light cone, and material roots are zero.
For strict ion containment require $N\cdot N<64\Delta_H^2$
and make the seven $\mathcal L_j$ minima strict; retain the weak
$\mathcal R_k$ inequalities.

The sufficient floors \eqref{eq.caus.resistive-primitive-cone-floors}
also follow directly from the octic matrix.  With
$D_*=E_0-S-2Q-B_\rho$,
$N_*=E_0+S+2Q$,
$X=C_q/\tauq$, and
$Y=C_\pi/\tau_\pi+C_\Pi/\tau_\Pi$,
every unit normal and complex $|\hat z|\ge1$ satisfy, when
$D_*-X-Y>0$,
\begin{align}
 \sigma_{\min}(\mathsf D/\hat z^2)&\ge D_*-X-Y,\notag\\
 \frac{|b^T\mathsf D^{-1}x|}{|\hat z|^2}
 &\le\frac{N_*+X}{D_*-X-Y}.
 \label{eq.resgeneral.cone-proof}
\end{align}
Here $\|\Lambda\|_{\rm op}\le S$.  The sign-aware primitive condition is
\begin{align}
 0\le c<1,\qquad D_*-X-Y&>0,\notag\\
 (1-c)(D_*-X-Y)&>|p-c|(N_*+X).
 \label{eq.resgeneral.sign-aware-cone-screen}
\end{align}
It alone excludes all ion roots with $|\hat z|\ge1$, including complex
roots.  With $M=D_*-pN_*>0$, the independent floors imply the stronger
comparison
\begin{align}
 &(1-c)(D_*-X-Y)-(p+c)(N_*+X)\notag\\
 &=M-\frac{A_q}{\tauq}-Y+cY>0.
 \label{eq.resgeneral.cone-margin}
\end{align}

\paragraph*{A primitive simple-root reality condition.}
Use $Q,S$ from Eq.~\eqref{eq.caus.resistive-primitive-cone-defs} and set
\begin{align}
 I_2&=\operatorname{tr}\Lambda^2,&
 I_3&=\operatorname{tr}\Lambda^3,\notag\\
 D_\pi&=I_2^3/2-3I_3^2,\notag\\
 H_\pm&=\bar E\pm S,\notag\\
 D_p&=(p\bar E-y)^3
 -\tfrac12p^2(p\bar E-y)I_2
 +\tfrac13p^3I_3,\notag\\
 s_-&=\min\{\sqrt p,\sqrt{y/H_+}\},\notag\\
 s_+&=\max\{\sqrt p,\sqrt{y/H_-}\}.
 \label{eq.resgeneral.simple-reality-inputs}
\end{align}
For $p,y,H_->0$, $s_+<1$, $D_\pi>0$, and $D_p\ne0$, define
\begin{align}
 g_*&=\min\!\left\{2s_-,
 \frac{\sqrt{yD_\pi}}{8S^2H_+^{3/2}},\right.\notag\\[-2pt]
 &\hspace{5em}\left.
 \frac{|D_p|}{2s_+H_+(pH_++y)^2}\right\},\notag\\
 r_*&=\tfrac14\min\{g_*,1-s_+\},\notag\\
 t_*&=2s_--r_*,\notag\\
 B_*&=(1+|1-d_q|)Q+|d_\pi|S
 +|Z-R-y|,\notag\\
 E_*&=2|p-c|
 +\frac{(2+|d_q|)Q+B_*}{H_+},\notag\\
 k_1&=\frac{1+H_+/H_-}{r_*t_*},\notag\\
 k_2&=\frac{H_++|R|}{H_-r_*},\notag\\
 K_*&=k_1+\frac{k_2}{(g_*-r_*)t_*^2}.
 \label{eq.resgeneral.simple-reality-margin}
\end{align}
The direction-independent primitive condition
\begin{align}
 E_*K_*<1
 \label{eq.resgeneral.simple-reality-screen}
\end{align}
gives eight distinct real ion roots in $(-1,1)$ for every normal.
The comparison circles remain disjoint under $E_*K_*<1$; conjugation
then makes every continued root real.  Their nonzero signs also secure
the material gap.

\paragraph*{Current gap on this branch.}
For hot gamma-law electrons put $x=c_{\rm se}^2$,
$\beta_e=1-\sigma_e$, and
\begin{align}
 D_x&=(x\bar E-y)^3-\tfrac12x^2(x\bar E-y)I_2
 +\tfrac13x^3I_3,\notag\\
 s_e&=\max\{\sqrt x,s_+\},\notag\\
 \delta_e&=\min\!\left\{|\sqrt x-\sqrt p|,
 \frac{|D_x|}{2s_eH_+(xH_++y)^2}\right\}.
 \label{eq.resgeneral.simple-current-inputs}
\end{align}
If $E_*K_*<1$ and $\delta_e>r_*$, the acoustic-only current ceiling is
\begin{align}
 \sqrt{j_\mu j^\mu}&<J_{\rm ei,ac}^{(*)},\notag\\
 J_{\rm ei,ac}^{(*)}&:=
 \frac{|q_e|\bar n_e}{\beta_e}(\delta_e-r_*).
 \label{eq.resgeneral.simple-acoustic-j}
\end{align}
The sufficient acoustic, contact, and DEC limit is
\begin{align}
 \sqrt{j_\mu j^\mu}<\min\!\left\{|q_e|\bar n_e v_{\rm DEC},
 |q_e|\bar n_e(s_--r_*),J_{\rm ei,ac}^{(*)}\right\}.
 \label{eq.resgeneral.simple-current-gap}
\end{align}
The determinant $D_x$ bounds the separation of electron acoustic
speeds from the comparison shear speeds.  The ion roots lie within
$r_*$ of those speeds; the electron contact moves by at most
$\mc V$ and each electron acoustic root by at most $\beta_e\mc V$.
This condition needs no gamma-law ion equation of state and is the
primitive current bound used below.

\paragraph*{The electron criterion, including degeneracies.}
Use the primitive invariants
\begin{align}
 X&:=P_{e,e}+\frac{\bar n_eP_{e,n}}{H_e},\notag\\
 \sigma&:=\frac{\bar n_eP_{e,n}}{H_e},&
 A&:=1-\sigma,\notag\\
 V&:=\frac{\sqrt{j_\mu j^\mu}}{|q_e|\bar n_e},
 \label{eq.reselectron.exact-inputs}
\end{align}
with $H_e>0$ and $\bar n_e>0$.
For the full electron cubic, nonlinear causality in every direction is
exactly
\begin{align}
 0\le X&\le1,\qquad 2|A|V\le3-X,\notag\\
 |A-X|V&\le1-X,
 \label{eq.reselectron.exact-linear}\\
 G_\sigma&:=8+20\sigma-\sigma^2
 +\sqrt{\sigma(8+\sigma)^3},\notag\\
 X>0,\ \sigma>0
 &\Longrightarrow V^2G_\sigma\le8X .
 \label{eq.reselectron.exact-reality}
\end{align}
For root reality alone require $X\ge0$ and
\eqref{eq.reselectron.exact-reality}.  No drift restriction is needed
for reality when $\sigma\le0$, or when $X=0$.
For strict light-cone containment require $X<1$ and make the
two drift inequalities in \eqref{eq.reselectron.exact-linear}
strict; retain the weak reality comparison.

 For $X>0$, strict discriminant
positivity gives Eq.~\eqref{eq:electron-sh-current-bound}; for
$\sigma<0$ the electron cubic is simple at every drift.

\paragraph*{Zero-heat relaxation-time domain.}
\label{app:resistive-large-tau-bound}
For $q^\mu=\pi^{\mu\nu}=0$, the causal heat-time domain is elementary.
With $\tau_\Pi$ and $\tau_\pi$ fixed, define
\begin{align}
 h&=e+P+\Pi,& K&=\kappa T,& a&=\kappa T_{,e},\notag\\
 b&=\kappa\rhob T_{,\rhob},& p&=P_{,e},&
 y&=\eta/\tau_\pi,\notag\\
 Z&=\rhob P_{,\rhob}
 +\frac{\zeta+\delta_{\Pi\Pi}\Pi}{\tau_\Pi}
 +\frac{4\eta}{3\tau_\pi}.
 \label{eq.restaufloor.primitive-inputs}
\end{align}
Put
\begin{align}
 A&=ph+Z,& B&=pK+b,& F&=aZ-pb.
 \label{eq.restaufloor.zeroheat-inputs}
\end{align}
On the positive-inertia branch
$h>y\ge0$, $K\ge0$, $0<A<h$, and $F\ge0$, the cone floor is
\begin{align}
 \tau_0=\max\left\{
 0,\frac{K}{h-y},\frac{B}{A},
 \frac{2K-B}{2h-A},\frac{K-B-F}{h-A}\right\}.
 \label{eq.restaufloor.zeroheat-expanded}
\end{align}
Root reality adds
\begin{align}
 \mathcal D_\tau&=F(ABh+Fh^2-A^2K),
 \label{eq.restaufloor.zeroheat-discriminant}\\
 \tau_\pm&=\frac{AB+2Fh\pm2\sqrt{\mathcal D_\tau}}{A^2},
 \qquad \mathcal D_\tau>0.
 \label{eq.restaufloor.zeroheat-endpoints}
\end{align}
The exact ion domain is
\begin{align}
 \tauq&\in
 \begin{cases}
 [\tau_0,\tau_-]\cup
 [\max\{\tau_0,\tau_+\},\infty),&\mathcal D_\tau>0,\\
 [\tau_0,\infty),&\mathcal D_\tau\le0,
 \end{cases}\notag\\[-2pt]
 \tauq&>\frac{\kappa T}{e+P+\Pi}.
 \label{eq.restaufloor.zeroheat}
\end{align}
The first line retains both causal components.  The last inequality
excludes a singular time symbol.  At finite heat the exact
all-direction condition is Eq.~\eqref{eq.resgeneral.complete}; the
primitive sufficient floors are
Eq.~\eqref{eq.caus.resistive-primitive-cone-floors}.

\subsection{Exact characteristic speeds}
\label{app:char-speeds}

We now collect the speeds implied by the preceding characteristic
polynomials.  Let $s_i$ be a unit Eulerian face normal,
$v_s=v^is_i$, and $\beta_s=\beta^is_i$.  Solving the definition of
$\hat z$ for the Eulerian speed gives
\begin{align}
 D(\hat z)&:=\hat z^2+(1-\hat z^2)W^2,\notag\\
 R(\hat z)&:=\hat z^2+(1-\hat z^2)W^2(1-v_s^2),\notag\\
 y(\hat z)&=
 \frac{(1-\hat z^2)W^2v_s-\hat z\sqrt{R(\hat z)}}{D(\hat z)},&
 \lambda(\hat z)&=\alpha y(\hat z)-\beta_s.
 \label{eq.char.speed.inverse}
\end{align}
The sign reversal $y(1)=-1$ and $y(-1)=1$ follows from the covector
convention.  Equation~\eqref{eq.char.speed.inverse} gives $y(0)=v_s$ and,
when the tangential velocity vanishes,
$y=(v_s-\hat z)/(1-v_s\hat z)$.
{Use this map together with the aberrated fluid-frame normal in
Eq.~\eqref{eq.ev.eulerian-covector}: each root must be evaluated at that
same normal.  A principal-axis root cannot generally be converted
independently of the wave normal.}

For the aligned resistive system, define
\begin{align}
 \mathcal I_a&:=E+\Lambda_a-\mathcal K .
\end{align}
The four transverse ion roots are the two polarizations of the causal
viscoelastic, or shear, wave,
\begin{align}
 \hat z_{\perp a,\pm}
 =\frac{-q_{\hat n}\pm
 \sqrt{q_{\hat n}^2+4\mathcal I_a y_\eta}}
 {2\mathcal I_a},\qquad a=1,2.
 \label{eq.char.speed.shear}
\end{align}
At $q_{\hat n}=\mathcal K=\Lambda_a=0$ this reduces to
$\hat z_{\perp a,\pm}=\pm c_\pi$, with
\begin{align}
 c_\pi^2=\frac{\eta}{\tau_\pi(e+P+\Pi)} .
 \label{eq.char.speed.shear.equilibrium}
\end{align}
Thus, the resistive transverse roots are viscous shear waves, not Alfv\'en
waves; magnetic disturbances belong to the independent light-cone block.

The three electron roots can also be written explicitly.  Set
\begin{align}
 A_e&:={1-\sigma_e},\notag\\
 \mathfrak p_e&:=-c_{\rm se}^2-\frac{A_e^2v_{\rm d}^2}{3},\notag\\
 \mathfrak q_e&:=c_{\rm se}^2v_{\rm d}
 \left(1-\frac{A_e}{3}\right)
 -\frac{2A_e^3v_{\rm d}^3}{27},\notag\\
 \theta_e&:=\frac13\arccos\!\left[
 -\frac{\mathfrak q_e}
 {2(-\mathfrak p_e/3)^{3/2}}\right].
 \label{eq.char.speed.electron.defs}
\end{align}
On the three-real-root branch selected by Eq.~\eqref{eq.causcubic},
\begin{align}
 \hat z_{e,k}&=\frac{A_ev_{\rm d}}{3}
 +2\sqrt{-\frac{\mathfrak p_e}{3}}
 \cos\!\left(\theta_e-\frac{2\pi k}{3}\right),\notag\\
 &\hspace{10em}k=0,1,2.
 \label{eq.char.speed.electron}
\end{align}
These are the drift-mixed electron contact and acoustic characteristics;
their zero-drift values are $0$ and $\pm c_{\rm se}$.

In particular, when $q_{\hat n}=0$ the first- and second-sound speeds are
\begin{align}
 c_{L,\pm}^2&=
 \frac{\mathcal M_q\pm
 \sqrt{\mathcal M_q^2-4\mathcal H\mathcal F_q}}
 {2\mathcal H},&
 \hat z_L&=\pm c_{L,+},\ \pm c_{L,-}.
 \label{eq.char.speed.longitudinal.equilibrium}
\end{align}
{The $\pm$ labels order the speeds by magnitude.  Sound and the causal
thermal, or second-sound, branch are identified by continuation from
vanishing conductivity and can exchange ordering at a semisimple crossing.}
If $\kappa=0$, then
$\mathcal F_q=0$ and
\begin{align}
 c_{L,+}^2=P_{,e}+\frac{\mathcal Z^c}{E+\Lambda_{\hat n}},\qquad
 c_{L,-}^2=0;
 \label{eq.char.speed.longitudinal.noheat}
\end{align}
the first expression becomes
$c_s^2=P_{,e}+\rhob P_{,\rhob}/(e+P)$ in
the ideal-fluid limit, with vanishing viscous transport ratios.
On the primitive simple-root domain
\eqref{eq.resgeneral.simple-reality-inputs}--
\eqref{eq.resgeneral.simple-reality-screen}, the current bound
\eqref{eq.resgeneral.simple-current-gap} separates the electron contact
and acoustic waves from all ion waves.  Compatible coincidences may instead
be retained by the exact condition \eqref{eq.ev.ei-uniform}.

{\subsection{{Gyrotropic specialization}}
\label{app:char_brag}

{The gyrotropic limit restricts the background heat flux and shear stress
to}
\begin{align}
 {q^\mu}& {=q_\parallel\hat b^\mu,}&
 {\pi^{\mu\nu}}& {=\pi_0
 \left(-b^\mu b^\nu+\frac{b^2}{3}\Delta^{\mu\nu}\right).}
 \label{eq.brag.background}
\end{align}
{Writing $\Delta p=p_\perp-p_\parallel=\pi_0b^2$ and
$\mu=\hat b_\mu\hat n^\mu$, the projections entering the characteristic
polynomials are}
\begin{align}
 {\Lambda_{\hat n}}& {=\Delta p\left(\frac13-\mu^2\right),}&
 {q_{\hat n}}& {=q_\parallel\mu,}\notag\\
 {\Lambda_\parallel}& {=-\frac23\Delta p,}&
 {\Lambda_\perp}& {=\frac13\Delta p.}
 \label{eq.bragsubs}
\end{align}
{{These relations are substituted directly into the constrained ion octic.}
The algebraic constraints remain those of Eq.~\eqref{eq.char.constraints}; in
particular, no additional trace constraint is introduced.}

{\emph{Resistive system.}  The Maxwell and material roots remain causal
identically, while the electron cubic is subject to
Eqs.~\eqref{eq.causcse}--\eqref{eq.causcubic}.  The ion sector requires all
eight roots of $\mc Q_8$ to be real and in $[-1,1]$ for every propagation
direction.  A direction-independent screen for its transverse factors follows
from Eq.~\eqref{eq.causperpc}:}
\begin{align}
 {y_\eta}&{\ge0,}&
 {E+\Lambda_s-\mathcal K}& {>0,}\notag\\
 {E+\Lambda_s-\mathcal K-y_\eta}
 &{\ge |q_\parallel|,}&
 {\Lambda_s}&{\in\left\{-\frac23\Delta p,
 \frac13\Delta p\right\}.}
 \label{eq.brag.resistive-screen}
\end{align}
{For propagation along the magnetic field,
$\Lambda_{\hat n}=-2\Delta p/3$, both transverse eigenvalues equal
$\Delta p/3$, and $q_{\hat n}=\pm q_\parallel$.  The longitudinal factor is
then the quartic \eqref{eq.Qpar.coefficients}, with reality and causality
enforced by Eqs.~\eqref{eq.Qpar.real} and \eqref{eq.causpar}.  For oblique
propagation the channels generally mix, and the full octic root test remains
necessary.}

{{Equation~\eqref{eq.brag.resistive-screen} shows how the anisotropy changes the}
available causal margin.  Positive $\Delta p$ lowers the parallel shear
eigenvalue, whereas negative $\Delta p$ lowers the perpendicular one; heat
flux and dissipative transport reduce the remaining margin.}

}

\section{Dominant energy condition for the electrons}
\label{app:electron-dominant-energy-condition}
Since we have truncated the electron fluid to first order, we must impose necessary and sufficient conditions for the electron energy fluxes to remain physically meaningful.

The first-order-drift electron tensor used in the reduced formulation is
\begin{align}
 T_e^{\mu\nu}={}&e_eu^\mu u^\nu+P_e\Delta^{\mu\nu}
 +u^\mu Q_e^\nu+Q_e^\mu u^\nu,
 \label{eq:electron-dec-tensor}
\end{align}
where $u_\mu u^\mu=-1$, $\Delta^{\mu\nu}=g^{\mu\nu}+u^\mu u^\nu$,
and $u_\mu Q_e^\mu=0$.  Define the invariant magnitude
$\mathcal Q_e:=\sqrt{Q_{e\mu}Q_e^\mu}\ge0$.
The dominant energy condition requires the energy flux
\begin{align}
 \mathcal J_e^\mu(w):=-T_e^{\mu\nu}w_\nu
 \label{eq:electron-dec-flux-definition}
\end{align}
to be future-directed and causal, or zero, for every future-timelike
$w^\mu$.  With signature $(-,+,+,+)$, choose a local orthonormal
frame in which $u^{\hat\mu}=(1,0,0,0)$ and rotate the spatial basis so that
$Q_e^{\hat\mu}=(0,\mathcal Q_e,0,0)$; at $\mathcal Q_e=0$ the axis is
arbitrary.  Write $w^{\hat\mu}=\gamma(1,\boldsymbol v)$, where
$|\boldsymbol v|<1$ and $\gamma=(1-|\boldsymbol v|^2)^{-1/2}$.
Direct contraction gives
\begin{align}
 \gamma^{-1}\mathcal J_e^{\hat\mu}
 =\left(e_e-\mathcal Q_ev_\parallel,\,
 \mathcal Q_e-P_ev_\parallel,\,-P_e\boldsymbol v_\perp\right).
 \label{eq:electron-dec-observer-flux}
\end{align}
The condition is homogeneous in $w^\mu$.  By continuity, it also holds
for future-null vectors; conversely, every future-timelike vector is a
positive sum of future-null vectors, and the future causal cone is convex.
It therefore suffices to test $k^{\hat\mu}=(1,\boldsymbol n)$ with
$|\boldsymbol n|=1$.  Writing $x=n_\parallel$, we obtain
\begin{align}
 -\mathcal J_{e\mu}(k)\mathcal J_e^\mu(k)
 ={}& e_e^2-P_e^2-\mathcal Q_e^2
 -2\mathcal Q_e(e_e-P_e)x\notag\\
 &+\mathcal Q_e^2x^2=:F_e(x),\qquad -1\le x\le1.
 \label{eq:electron-dec-null-quadratic}
\end{align}
Future orientation requires $e_e\ge\mathcal Q_e$.  In particular, the
two longitudinal directions $x=\pm1$ require
\begin{align}
 e_e-P_e\ge0,\qquad e_e+P_e\ge2\mathcal Q_e,
 \label{eq:electron-dec-longitudinal}
\end{align}
which already imply $e_e\ge\mathcal Q_e$.

Set $A_e:=e_e-P_e\ge0$.  For $\mathcal Q_e>0$, the quadratic
\eqref{eq:electron-dec-null-quadratic} has its stationary point at
$x_*=A_e/\mathcal Q_e$.  Its exact minimum is
\begin{align}
 \min_{-1\le x\le1}F_e(x)=
 \begin{cases}
 A_e(e_e+P_e-2\mathcal Q_e),&\mathcal Q_e\le A_e,\\
 2P_eA_e-\mathcal Q_e^2,&\mathcal Q_e>A_e.
 \end{cases}
 \label{eq:electron-dec-minimum}
\end{align}
\begin{samepage}
The first branch also applies at $\mathcal Q_e=0$.  Combining this
minimum with Eq.~\eqref{eq:electron-dec-longitudinal} gives the necessary
and sufficient invariant conditions
\begin{subequations}
\label{eq:electron-dec-iff}
\begin{align}
 e_e+P_e&\ge2\mathcal Q_e,\\*
 e_e-P_e&\ge0,\\*
 e_e-P_e+\sqrt{(e_e+P_e)^2-4\mathcal Q_e^2}&\ge2|P_e|.
\end{align}
\end{subequations}
\end{samepage}
These inequalities include the degenerate boundary
$e_e+P_e=2\mathcal Q_e$.
At $\mathcal Q_e=0$ they reduce to the perfect-fluid condition
$e_e\ge|P_e|$.

For $P_e\ge0$, Eq.~\eqref{eq:electron-dec-iff} can be written as
\begin{align}
 e_e\ge P_e,\qquad
 \mathcal Q_e^2\le
 \begin{cases}
 (e_e+P_e)^2/4,&e_e\ge3P_e,\\
 2P_e(e_e-P_e),&P_e\le e_e<3P_e.
 \end{cases}
 \label{eq:electron-dec-positive-pressure}
\end{align}
For $e_e\ge3P_e$, the final inequality in
Eq.~\eqref{eq:electron-dec-iff} is automatic.  For
$P_e\le e_e<3P_e$, isolating the square root and squaring the
nonnegative sides gives $\mathcal Q_e^2\le2P_e(e_e-P_e)$.

\begin{samepage}
Using $\mathcal Q_e=h_e\sqrt{\bar V_{e\mu}\bar V_e^\mu}$ and
$e_e+P_e=\bar n_eh_e$, with $\bar n_e>0$ and $h_e>0$, gives
\begin{align}
 \frac{\sqrt{\bar V_{e\mu}\bar V_e^\mu}}{\bar n_e}
 &\simeq\frac{\sqrt{\ju_\mu\ju^\mu}}{|q_e|\bar n_e}\notag\\
 &\le
 \begin{cases}
 \dfrac12,&e_e\ge3P_e,\\[3pt]
 \dfrac{\sqrt{2P_e(e_e-P_e)}}{e_e+P_e},
 &P_e\le e_e<3P_e.
 \end{cases}
 \label{eq:electron-dec-drift-bound}
\end{align}
\end{samepage}
For $P_e\le e_e<3P_e$, the pressure-dependent ceiling is strictly
below one half.  For the gamma law
\eqref{eq:electron-eos}, $\rho_e=m_e\bar n_e>0$ and
$c_{\rm se}^2=\Gamma_eP_e/H_e$ give
\begin{align}
 \frac{P_e}{H_e}=\frac{c_{\rm se}^2}{\Gamma_e},\qquad
 \frac{e_e-P_e}{H_e}=1-\frac{2c_{\rm se}^2}{\Gamma_e}.
 \label{eq:electron-dec-gamma-substitution}
\end{align}
{Thus, $e_e\ge3P_e$ is equivalent to
$c_{\rm se}^2\le\Gamma_e/4$; substitution into
Eq.~\eqref{eq:electron-dec-drift-bound} gives the gamma-law form used in
the main text.}  Since
$0\le c_{\rm se}^2<\Gamma_e-1$, the one-half ceiling applies to all
states with $1<\Gamma_e\le4/3$.

\section{3+1 decomposition of the dissipative MHD system}
\label{app:3p1}

Here we summarize the explicit set of {GR19M} dissipative MHD equations in commonly adopted 3+1 Valencia form.
Throughout this Appendix, source terms advanced by the implicit stage solve
of Sec.~\ref{sec:implicit} are printed in \imp{dark blue}; everything else on a
right-hand side is assembled by the explicit source operator.
The moment balances below are evaluated on the algebraic
moment-constraint surface, where the constraint-penalty terms vanish.
We project the current row using $\overset{\perp}{=}$ from Eq.~\eqref{eq:electron-projected-equality} and fix its longitudinal stored component through Eq.~\eqref{eq:electron-dependent-current-moment}.

\begin{align}
  {\rhob h:=\rhob+\rhob\epsilon+p,\qquad \rhob h^*:=\rhob h+\Pi} \, .
\end{align}
\begin{widetext}
\begin{subequations}
\begin{align}
  \p_{t} \, \widetilde{D}_{\rm b}
    &+ \p_{i}\left(\sqrt{\gamma} \, \rhob
    \left(W \, \alpha \, v^{i} - W \, \beta^{i}\right)\right) = 0 \, , \\
  \p_{t} \, \widetilde{S}_{j}
    &+ \p_{i}
    \left\{\sqrt{\gamma} \, \rhob \, h^* \, W \, v_{j}
    \left( W \, \alpha \, v^{i} - W \, \beta^{i} \right)\right.
    - \alpha \left( E_{j} \, \widetilde{E}^{i}
    + B_{j} \, \widetilde{B}^{i} \right)
    - \sqrt{\gamma} \, \beta^{i} \, \levciv_{j\ell m} \, E^{\ell} \, B^{m} \notag \\
    &\phantom{\hspace{3em}}
    + \alpha \left[ \sqrt{\gamma} \, \left(p + \Pi\right)
    + \sqrt{\gamma} \, \frac{1}{2}
    \left( E^{2} + B^{2} \right) \right] \delta^{i}_{\phantom{i}j}
    + \sqrt{\gamma} \, \alpha \, \beta_{j} \, \pi^{i0}
    + \sqrt{\gamma} \, \alpha \, \gamma_{kj} \, \pi^{ik} \notag \\
    &\phantom{\hspace{3em}}
    + \sqrt{\gamma} \left[
    \left( W \, \alpha \, v^{i} - W \, \beta^{i} \right) q_{j} \right.
    \left.\left.
    + W \, v_{j} \, \alpha \, q^{i} \right] \right\} \notag \\
    &=\sqrt{\gamma} \left(
    \frac{1}{2} \, \alpha \, S^{ik} \, \p_{j} \, \gamma_{ik}
    + S_{i} \, \p_{j} \, \beta^{i}
    - \left( \tau + \Db \right) \, \p_{j} \, \alpha\right) \, , \\
  \p_{t} \, \widetilde{\tau}
    &+ \p_{i} \left\{\sqrt{\gamma}\left(\rhob \, h^* \, W - \rhob \right)
    \left( W \, \alpha \, v^{i} - W \, \beta^{i} \right)\right.
    + \sqrt{\gamma} \, \alpha \, \levciv^{ijk} \, E_{j} \, B_{k}
    + \left( \sqrt{\gamma}\left(p + \Pi\right)
    - \sqrt{\gamma} \, \frac{1}{2} \left( E^{2} + B^{2} \right) \right) \beta^{i}
    \notag \\
    &\phantom{\hspace{3em}}
    + \sqrt{\gamma} \, \alpha^{2} \, \pi^{i0}
    + \sqrt{\gamma}\left[
    \left( W \, \alpha \, v^{i} - W \, \beta^{i} \right) \alpha \, q^{0}
    \left.
    + W \, \alpha \, q^{i}\right]\right\} \notag \\
    &= \sqrt{\gamma} \left(\alpha \, K_{ij} \, S^{ij}
      - S^{i} \, \p_{i} \, \alpha\right) \, , \\
  \p_{t} \, \widetilde{B}^{i}
    &+ \p_{j}
    \left(-\beta^{j} \, \widetilde{B}^{i}
    + \alpha \, \eta^{0ijk} \, E_{k}
    + \alpha \, \gamma^{ij} \, \widetilde{\phi} \right)
    = -\widetilde{B}^{j} \, \p_{j} \, \beta^{i}
    + \widetilde{\phi}\left(\gamma^{ij} \, \p_{j} \, \alpha
    - \alpha \, \gamma^{jk} \,
    ^{\left(3\right)}\Gamma^{i}_{\phantom{i}jk}\right) \, , \\
  \p_{t} \, \widetilde{E}^{i}
    &+ \p_{j}
    \left(-\beta^{j} \, \widetilde{E}^{i}
    - \alpha \, \eta^{0ijk} \, B_{k}
    + \alpha \, \gamma^{ij} \, \widetilde{\psi} \right)
    = -\widetilde{E}^{j} \, \p_{j} \, \beta^{i}
    + \widetilde{\psi}\left(\gamma^{ij} \, \p_{j} \, \alpha
    - \alpha \, \gamma^{jk} \,
    ^{\left(3\right)}\Gamma^{i}_{\phantom{i}jk}\right) \notag \\
    &\hspace{20em}- \alpha \, \widetilde{D}_{\rm q} \, v^{i}
    \imp{{}- \alpha \, \sqrt{\gamma}
    \left(\delta^{i}_{\phantom{i}k}
    - v^{i} \, v_{k}\right) \gamma^{k\mu} \, j_{\mu}} \, , \\
  \p_{t} \, \widetilde{\phi}
    &+ \p_{i} \left(\alpha \, \widetilde{B}^{i}
    - \beta^{i} \, \widetilde{\phi}\right)
    = \widetilde{B}^{j} \, \p_{j} \, \alpha - \alpha \, \widetilde{\phi}
    \left(\imp{\kappa_{\rm dc}} + K\right) \, , \\
  \p_{t} \, \widetilde{\psi}
    &+ \p_{i} \left(\alpha \, \widetilde{E}^{i}
    - \beta^{i} \, \widetilde{\psi}\right)
    = \widetilde{E}^{j} \, \p_{j} \, \alpha - \alpha \, \widetilde{\psi}
    \left(\imp{\kappa_{\rm dc}} + K\right) + \alpha \, \widetilde{D}_{\rm q} \, , \\
  {\p_{t} \, \widetilde{D}_{e}}
    &{+ \p_{i}\left(\sqrt{\gamma} \, \rho_{e}
    \left(W \, \alpha \, v^{i} - W \, \beta^{i}\right)
    - \sqrt{\gamma} \, \alpha \, \frac{m_{e}}{q_{e}} \, j^{i}\right)
    = 0} \, , \\
     {\p_{t} \, \widetilde{k}_{*\nu}}
    & {+ \p_{i}\Bigl\{\sqrt{\gamma}\bigl[
    W\left(\alpha v^{i}-\beta^{i}\right)
    \left(k_{\nu}-\omega_{e}u_{\nu}\right)}
    +\alpha k^{i}u_{\nu}-\alpha\chi_{e}\delta^{i}_{\phantom{i}\nu}
    \bigr]\Bigr\}
     {\overset{\perp}{=}\alpha\sqrt{\gamma}\,
    \Gamma^{\rho}_{\phantom{\rho}\mu\nu}
    \mathcal B^{\mu}_{\phantom{\mu}\rho}} \notag \\
    &\qquad
    \imp{{}-\frac{\alpha\sqrt{\gamma}}{\tauJ}k_{\nu}
    +\alpha\sqrt{\gamma}\,\omega_{pe}^{2}e_{\nu}
    +\alpha\sqrt{\gamma}\,\frac{\omega_{ge}}{w_{e}}
    b^{\mu}_{\phantom{\mu}\nu}k_{\mu}
    -\alpha\sqrt{\gamma}\,\frac{q_{e}}{m_{e}}
    u_{\nu}e_{\mu}\ju^{\mu}} \, , \\
  \p_{t} \, \widetilde{\Pi}_{*}
    &+ \p_{i} \left[\sqrt{\gamma} \left(\Pi + y_{\zeta}\right)
    \left(W \, \alpha \, v^{i} - W \, \beta^{i}\right)\right]
    = \imp{-{\alpha\sqrt\gamma\Delta_\Pi\Pi}
    + \alpha \, \sqrt{\gamma} \, \mc{I}_{\zeta}} \, , \\
  \p_{t} \, \widetilde{Q}_{\nu}
    &+ \p_{i} \left[\sqrt{\gamma} \left(q_{\nu} + y_{\kappa} \, T \, u_{\nu}\right)
    \left(W \, \alpha \, v^{i} - W \, \beta^{i}\right)
    + \sqrt{\gamma} \, y_{\kappa} \, T \,
    \alpha \, \delta^{i}_{\phantom{i}\nu}\right] \notag \\
    &\overset{\perp}{=}
    \alpha \, \sqrt{\gamma} \, \Gamma^{\rho}_{\phantom{\rho}\mu\nu}
    \left( q_{\rho} \, u^{\mu}
    + y_{\kappa} \, T \, \Delta^{\mu}_{\phantom{\mu}\rho} \right)
    \imp{{}- {\alpha\sqrt\gamma\Delta_{\rm q}q_\nu}
    + \alpha \, \sqrt{\gamma} \, T \, \mc{I}^{\kappa}_{\nu}
    - \alpha \, \sqrt{\gamma} \, \omega_{gi} \,
    b^{\mu}_{\phantom{\mu}\nu} \, q_{\mu}} \, , \\
  \p_{t} \, \widetilde{\pi}_{*}^{\alpha\beta}
    &+ \p_{i}\left[
    \sqrt{\gamma} \left(W \, \alpha \, v^{i} - W \, \beta^{i}\right)
    \pi^{\alpha\beta}
    + \sqrt{\gamma} \, y_{\eta}
    \left(g^{i\alpha} \, W \, \frac{u^{\beta}}{u^{0}}
    + g^{i\beta} \, W \, \frac{u^{\alpha}}{u^{0}}\right) \right] \notag \\
    &\overset{\rm STF}{=}
    - \alpha \, \sqrt{\gamma} \left(
    u^{\mu} \, \Gamma^{\alpha}_{\phantom{\alpha}\mu\rho} \, \pi^{\rho\beta}
    + u^{\mu} \, \Gamma^{\beta}_{\phantom{\beta}\mu\rho} \, \pi^{\alpha\rho}
    + y_{\eta} \, u^{\rho} \,
    \Gamma^{\alpha}_{\phantom{\alpha}\mu\rho} \, g^{\mu\beta}
    + y_{\eta} \, u^{\rho} \,
    \Gamma^{\beta}_{\phantom{\beta}\mu\rho} \, g^{\mu\alpha}
   {+y_\eta u^\beta\Gamma^\alpha_{\mu\rho}g^{\mu\rho}
    +y_\eta u^\alpha\Gamma^\beta_{\mu\rho}g^{\mu\rho}}\right)
    \notag \\
    &\qquad
    \imp{{}-{\alpha\sqrt\gamma\Delta_\pi\,}
    \pi^{\alpha\beta}
    -\alpha\sqrt{\gamma}\,\omega_{gi}\,
    b^{\delta\mu}\pi^{\gamma}_{\phantom{\gamma}\mu}
    \Delta^{\alpha\beta}_{\phantom{\alpha\beta}\gamma\delta}} \, , \\
  {\p_{t} \, \widetilde{\mathcal E}_{e*}}
    &{+\p_{i}\left\{\sqrt{\gamma}\left[
    e_{e}\left(W\alpha v^{i}-W\beta^{i}\right)
    -\alpha\frac{m_{e}}{q_{e}}k^{i}\right]\right\}}
    {=}\imp{\alpha\sqrt{\gamma}
    \left(-P_{e}\theta+e_{\mu}\ju^{\mu}\right)} \, , \\
  \p_{t} \, \widetilde{Y}_{\zeta}
    &+ \p_{i} \left(F_{\Db}^{i} \, y_{\zeta}\right)
    = \imp{\alpha \, \sqrt{\gamma} \, \rhob \, \mc{I}_{\zeta}} \, , \\
  \p_{t} \, \widetilde{Y}_{\kappa}
    &+ \p_{i} \left(F_{\Db}^{i} \, y_{\kappa}\right)
    = \imp{\alpha \, \sqrt{\gamma} \, \rhob \, \mc{I}_{\kappa}} \, , \\
  \p_{t} \, \widetilde{Y}_{\rhob}
    &+ \p_{i} \left(F_{\Db}^{i} \, y_{\rhob}\right)
    = \imp{\alpha \, \sqrt{\gamma} \, \rhob \, \mc{I}_{\rhob}} \, , \\
  \p_{t} \, \widetilde{Y}_{\eta}
    &+ \p_{i} \left(F_{\Db}^{i} \, y_{\eta}\right)
    = {\alpha \, \sqrt{\gamma} \, \rhob \, \mc{I}_{\eta}} \, , \\
  \p_{t} \, \widetilde{Z}^{\kappa}_{\nu}
    &+ \p_{i} \left(\sqrt{\gamma} \, \alpha \, y_{\kappa} \,
    \delta^{i}_{\phantom{i}\nu}\right)
    = \alpha \, \sqrt{\gamma} \,
    \Gamma^{\rho}_{\phantom{\rho}\mu\nu} \, z^{\mu}_{\phantom{\mu}\rho}
    \imp{{}+ \alpha \, \sqrt{\gamma} \, \mc{I}^{\kappa}_{\nu}} \, ,
\end{align}
\end{subequations}
where
\begin{align}
  F_{\Db}^{i} := \sqrt{\gamma} \, \rhob \left(W \, \alpha \, v^{i}
  - W \, \beta^{i}\right) \, .
\end{align}
\end{widetext}
\begin{subequations}
\begin{align}
  \Db &:= -n_{\mu} \, {N_b^\mu} \\
    &\phantom{:}= \rhob \, W \, , \\
  \tau &:= n_{\mu} \, n_{\nu}
    \left( T_{\mathrm{hydro}}^{\mu\nu}
    + T_{\mathrm{diss}}^{\mu\nu}
    + T_{\mathrm{EM}}^{\mu\nu}
    - n^{\mu} \, {N_b^\nu} \right) \\
    &\phantom{:}=
    \left( \rhob + \rhob \, \epsilon
    + p + \Pi \right) W^{2} - \left( p + \Pi \right)
    - \rhob \, W \notag \\
    &\phantom{:===} + \frac{1}{2} \left( E^{2} + B^{2} \right)
    + 2 \, \alpha \, W \, q^{0} + \alpha^{2} \, \pi^{00} \, , \\
  S_{\rho} &:= -\gamma_{\rho\mu} \, n_{\nu}
    \left( T_{\mathrm{hydro}}^{\mu\nu}
    + T_{\mathrm{diss}}^{\mu\nu}
    + T_{\mathrm{EM}}^{\mu\nu} \right) \\
    &\phantom{:}= \left( \rhob + \rhob \, \epsilon
    + p + \Pi \right) W^{2} \, v_{\rho} \notag \\
    &\phantom{:===}+ \gamma_{\rho}^{\phantom{\rho}j} \,
    \levciv_{jmn} \, E^{m} \, B^{n}
    + W \, \gamma_{\rho\mu} \, q^{\mu} \notag \\
    &\phantom{:===}+ \alpha \, q^{0} \, W \, v_{\rho}
    + \alpha \, \gamma_{\rho\mu} \, \pi^{\mu0} \, , \\
  S^{\rho\sigma} &:=
    \gamma^{\rho}_{\phantom{i}\mu} \, \gamma^{\sigma}_{\phantom{j}\nu}
    \left( T_{\mathrm{hydro}}^{\mu\nu}
    + T_{\mathrm{diss}}^{\mu\nu}
    + T_{\mathrm{EM}}^{\mu\nu} \right) \notag \\
    &\phantom{:}= \left( \rhob + \rhob \, \epsilon
    + p + \Pi \right) W^{2} \, v^{\rho} \, v^{\sigma}
    + \left( p + \Pi \right) \gamma^{\rho\sigma} \notag \\
    &\phantom{:==}
    + 2 \, W \, q_{k} \, \gamma^{k\left(\rho\right.} v^{\left.\sigma\right)}
    + \gamma^{\rho}_{\phantom{\rho}\mu} \,
    \gamma^{\sigma}_{\phantom{\sigma}\nu} \, \pi^{\mu\nu}
    \notag \\
    &\phantom{:==}
    + \frac{1}{2} \, \gamma^{\rho\sigma} \left( E^{2} + B^{2} \right)
    - \left( E^{\rho} \, E^{\sigma} + B^{\rho} \, B^{\sigma} \right) \, , \\
    {D_{e}}
    & {:= -n_{\mu}\left(\rho_{e}u^{\mu}
    -\frac{m_{e}}{q_{e}}\ju^{\mu}\right)
    =W\rho_{e}-\frac{\kappa_{e}}{w_{e}} \, ,} \\
   {k_{*\nu}}
    & {:= -n_{\mu}\mathcal B^{\mu}_{\phantom{\mu}\nu}
    =W\left(k_{\nu}-\omega_{e}u_{\nu}\right)
    -\left(n_{\mu}k^{\mu}\right)u_{\nu}+\chi_{e}n_{\nu} \, ,} \\
  {\mathcal E_{e*}}
    & {:= -n_{\mu}\left(e_{e}u^{\mu}+Q_{e}^{\mu}\right)
    =We_{e}-\kappa_{e} \, ,} \notag \\
    & {\hspace{2em}\kappa_{e}:=
    \alpha\frac{m_{e}}{q_{e}}k^{0} \, .}
\end{align}
\end{subequations}

\section{Butcher tableaux}
\label{app:butcher}

{The IMEX12 method of Ref.~\cite{Bucciantini:2012sm} can likewise
be written as an additive Runge--Kutta pair by including the initial state as
the first stage. Its explicit (left) and implicit (right)
tableaux are}
\begin{equation}
\renewcommand{\arraystretch}{1.25}
\begin{array}{c|ccc}
\tfrac12 & \tfrac12 & 0 & 0 \\
1 & 0 & 1 & 0 \\ \hline
 & 0 & 1 & 0
\end{array}
\qquad
\begin{array}{c|ccc}
\tfrac12 & 0 & \tfrac12 & 0 \\
1 & 0 & 0 & 1 \\ \hline
 & 0 & 0 & 1
\end{array} \, .
\label{eq:split12-tableaux}
\end{equation}
{The explicit tableau is second-order accurate, whereas the implicit tableau
is first-order accurate.}

{The second two-stage option is the $L$-stable IMEX SSP2(2,2,2) pair of
Ref.~\cite{pareschi2005implicit}.  With
$\gamma_2=1-1/\sqrt{2}$, its explicit (left) and implicit (right) tableaux are}
\begin{equation}
\renewcommand{\arraystretch}{1.25}
\begin{array}{c|cc}
0 & 0 & 0 \\
1 & 1 & 0 \\ \hline
 & \tfrac12 & \tfrac12
\end{array}
\qquad
\begin{array}{c|cc}
\gamma_2 & \gamma_2 & 0 \\
1-\gamma_2 & 1-2\gamma_2 & \gamma_2 \\ \hline
 & \tfrac12 & \tfrac12
\end{array} \, .
\label{eq:ssp2-tableaux}
\end{equation}
{For SSP2(2,2,2), the constant implicit diagonal
$\tilde a_{ii}=\gamma_2$ is the coefficient $\akk$ in
Sec.~\ref{sec:implicit}.}

{IMEX12 is the default and is used for all tests below unless stated
otherwise. {The black-hole magnetosphere runs of
Sec.~\ref{sec:waldMagnetosphere} and Sec.~\ref{sec:baldingBlackHoles} use
IMEX12.} Note finally that the implicit
step size entering the algebraic solves,
$\dtI=\akk \, \alpha \, \Delta t$, is built from the \emph{stage} step rather
than from the full timestep: for IMEX12 the implemented diagonal is
$\akk=1$ with stage steps $\Delta t/2$ and $\Delta t$, which reproduces the
$\left(\tilde{a}_{22},\tilde{a}_{33}\right)=\left(1/2,1\right)$ of the tableau
above, with the initial state counted as stage one.}

\section{Physical domain of {GR19M} conservatives}
\label{app:dt0_admissibility}

Primitive recovery maps the evolved conservative state to density,
temperature, velocity, and dissipative moments.  This map is nonlinear in
relativistic fluids, and not every conservative state admits a physical
{primitive \mbox{\cite{WuTang:2017rmhd}}}.

The full {GR19M} system is more restrictive because the implicit electron map
{can admit multiple algebraic roots, whose physical admissibility must be checked separately.}
{The accepted branch must continue from an admissible $\dtI=0$ state;
a nonsingular residual Jacobian ensures local uniqueness
(Appendix~\ref{app:4D_ladder} and Fig.~\ref{fig.wald_multiroot_branches}).}
Stable evolution consequently requires explicit admissibility checks.

\subsection{{Necessary conditions for physical admissibility}}

We use undensitized quantities in an orthonormal Eulerian tetrad and suppress hats on tetrad indices. Spatial norms, dot products and cross products are Euclidean. Time components enter through normal projections, such as $-n_\mu j^\mu=\alpha j^0$. These bounds require transforming the full four-vectors and stress tensor, in addition to removing the density weight.

{As before, we use the four-velocity $u^\mu=(W,z^i)$, Lorentz factor
$W=\sqrt{1+z^2}$, and ion rest-mass density $\rhob=\Db/W$.  After subtracting the independently
evolved electromagnetic part, we recall the definitions}
\begin{align*}
 {\Db}& {=-n_\mu N_b^\mu,\qquad N_b^\mu:=\rhob u^\mu,}&
 {\tau+\Db}& {=n_\mu n_\nu T^{\mu\nu},}\\
 {S_i}& {=-\gamma_{i\mu}n_\nu T^{\mu\nu},}&
 {E^\mu}& {=F^{\mu\nu}n_\nu},\\
 B^\mu & =\starF^{\mu\nu}n_\nu.
\end{align*}
{Here $T^{\mu\nu}$ is the evolved ion--dissipative plus electromagnetic
stress tensor, and $n_\mu E^\mu=n_\mu B^\mu=0$.  The local matter remainder is}
\begin{align}
 {E_0}& {:=\tau+\Db-\frac{E^2+B^2}{2},}&
 {m_i}& {:=S_i-(E\mathbin{\times}B)_i,}\notag\\
 {m}& {:=\sqrt{m_i m_i}.}
 \label{eq.dt0_raw_matter}
\end{align}
{We write $D_e=N_e^*/\sqrt\gamma$ for the undensitized conservative
{electron rest-mass density (the corresponding Eulerian number density is $D_e/m_e$).} To avoid confusing the star in the moment name with
densitization, define
$\widetilde{\mathcal E}_{e*}=\sqrt\gamma\mathcal E_{e*}$ and
$\widetilde k_{*\mu}=\sqrt\gamma k_{*\mu}$.  Below
$U_e:=\widetilde{\mathcal E}_{e*}/\sqrt\gamma=\mathcal E_{e*}$ and
$M_\mu:=\widetilde k_{*\mu}/\sqrt\gamma=k_{*\mu}$.}

{Every physical state has {positive rest-frame mass densities, $\rho_b>0$ and $\rho_e>0$}.}
\begin{align}
 {{\Db>0,\qquad D_e>0.}}
 \label{eq.dt0_structural_density_gate}
\end{align}

{The first inequality follows from $\Db=W\rhob$.  As in the main text,
$q_e>0$ denotes the magnitude of the electron charge.  Put
$q_m=q_e/m_e>0$ and write
$D_e=W\rho_e-\alpha j^0/q_m$.  Orthogonality $u_\mu j^\mu=0$ and the causal
drift bound
$\xi=\sqrt{j_\mu j^\mu}/(q_m\rho_e)<1$ give}
\begin{align}
 {D_e\geq W\rho_e(1-|v|\xi)>0.}
 \label{eq.dt0_carrier_positive}
\end{align}
{A causal drift can nevertheless make $D_e$ smaller than its zero-current
atmosphere value, so no stronger lower bound on $D_e$ is used as a
conservative gate.}
{In this context, we adopt $[\rho_-,\rho_+]$ as the admissible interval of rest-mass density, with
$\rho_-=\rho_{\rm atmo}$ the atmosphere floor and $\rho_+=\rho_{\max}$ the
matter ceiling of Eq.~\eqref{eq.rhofloor}, and let
$W_{\max}=\sqrt{1+z_{\max}^2}$ be the Lorentz factor of the velocity ceiling.}
{The density and Lorentz-factor bounds restrict the flow to}
\begin{align}
 {\mathcal W_D}& {=[W_D^-,W_D^+],}\notag\\
 {W_D^-}& {=\max\!\left(1,\frac{\Db}{\rho_+}\right),}&
 {W_D^+}& {=\min\!\left(W_{\max},\frac{\Db}{\rho_-}\right).}
 \label{eq.dt0_WD}
\end{align}
{Thus, $\Db>0$ and $W_D^-\leq W_D^+$ are necessary and sufficient for
the baryon density and velocity windows.}

\subsubsection{Primitive inversion}

For pure primitive inversion $(\Delta t_I=0)$, we proceed as follows.
{The system must admit a four-velocity, $u^\mu$, such that}
\begin{align}
 {u^\mu Q_\mu}&{=0,}&
 {g_{\mu\nu}\pi_*^{\mu\nu}}&{=2y_\eta W,}\notag\\
 {U_e}& {=\frac{m_e}{q_e}u^\mu M_\mu.}
 \label{eq.dt0_moment_compatibility}
\end{align}
{The heat and electron moments therefore restrict $W$ to}
\begin{align}
 {\mathcal W_Q={}}&
 {\left\{W\in\mathcal W_D:
 W^2Q_0^2\leq(W^2-1)Q_iQ^i\right\},}\notag\\
 {\mathcal W_M={}}&
 {\left\{W\in\mathcal W_D:
 \left|U_e-\frac{m_e}{q_e}WM_0\right|\right.}\notag\\[-2pt]
 &{\left.\hspace{3em}\leq\left|\frac{m_e}{q_e}\right|
 \sqrt{W^2-1}\sqrt{M_iM^i}\right\}.}
 \label{eq.dt0_common_W}
\end{align}
For $y_\eta>0$, the shear trace fixes
$W=W_\pi:=g_{\mu\nu}\pi_*^{\mu\nu}/(2y_\eta)$; for $y_\eta=0$, that trace
must vanish.  A necessary condition is a nonempty common set
$\mathcal W_D\cap\mathcal W_Q\cap\mathcal W_M$, including $W_\pi$ when it is
 fixed.

The ion dissipative primitives obey the following relations:
\begin{align}
 \Pi&=\frac{\Pi_*}{W}-y_\zeta,&
 q_0&=\frac{Q_0+y_\kappa Tz^2}{W},\notag\\
 q_i&=\frac{Q_i}{W}-y_\kappa Tz_i,&
 \pi^{00}&=\frac{\pi_*^{00}}{W}+2y_\eta,\notag\\
 \pi^{0i}&=\frac{\pi_*^{0i}}{W}+y_\eta\frac{z_i}{W}.
 \label{eq.dt0_component_maps}
\end{align}
{Let $Z_+=\sqrt{(W_D^+)^2-1}$, $v_+=Z_+/W_D^+$, and let $T_+$ bound the ion temperature.} Define the absolute bulk envelope by
\begin{align}
 \Pi_+&:=\sup_{W\in\mathcal W_D}|\Pi_*/W-y_\zeta|\notag\\
 &=\max\{|\Pi_*/W_D^--y_\zeta|,|\Pi_*/W_D^+-y_\zeta|\}.
 \label{eq:audit-bulk-envelope}
\end{align}
The corresponding nonnegative energy and momentum budgets may be chosen as

\begin{align}
 \delta_E={}&Z_+^2\Pi_+
 +2\bigl(|Q_0|+|y_\kappa|T_+Z_+^2\bigr)\notag\\
 &+\frac{|\pi_*^{00}|}{W_D^-}+2|y_\eta|,
 \label{eq.dt0_deltaE}\\
 \delta_S={}&W_D^+Z_+\Pi_++\sqrt{Q_iQ_i}
 +|y_\kappa|T_+W_D^+Z_+\notag\\
 &+v_+\bigl(|Q_0|+|y_\kappa|T_+Z_+^2\bigr)
 +\frac{\sqrt{\pi_*^{0i}\pi_*^{0i}}}{W_D^-}+|y_\eta|v_+.
 \label{eq.dt0_deltaS}
\end{align}
{For an EOS whose ideal-fluid admissible set obeys the energy cone
(in particular a $\Gamma$ law with $1<\Gamma\leq2$ and nonnegative
thermal energy), the remaining perfect-fluid tensor requires}
\begin{align}
 {E_0+\delta_E\geq
 \sqrt{\Db^2+\bigl[\max(m-\delta_S,0)\bigr]^2}.}
 \label{eq.dt0_widened_cone}
\end{align}
{This is a one-sided test and a necessary but not sufficient condition.
If it fails, no primitive state in the declared
box can reproduce the conserved state, and the cell is inadmissible.  If it
passes, nothing has been proven: $\delta_E$ and $\delta_S$ were built by
pushing each dissipative contribution to its own extreme, and no single
primitive has to realize all of those extremes at once, so the widened cone
encloses more than the true admissible set.  The test is sharp only when the
dissipative fields vanish, $\Pi=q^\mu=\pi^{\mu\nu}=0$,
and Eq.~\eqref{eq.dt0_widened_cone} becomes the ordinary ideal-fluid
cone.}

\subsubsection{Implicit equation}

{Hereafter write $W_-=W_D^-$ and $W_+=W_D^+$.}
At positive time the structural density and velocity conditions still
apply.  The source-free moment equalities do not apply to the known IMEX
predictor, denoted by a hat, because the implicit maps project constraint
errors while relaxing the moments.  Before evaluating the closure, each
relaxation block must be regular.  {Every dissipative equation of the implicit
{map has denominator $D_A=W+\Delta t_I\Delta_A$, with the expansion-dependent $\Delta_A$ of Eq.~\eqref{eq:audit-relax-diagonal}. On the trial box, let $\Delta_{A,-}=\inf\Delta_A$. A sufficient regularity condition is}}

\begin{align}
 D_A&\ge W_-+\Delta t_I\Delta_{A,-}>0,
 \qquad A\in\{\Pi,{\rm q},\pi\}.
 \label{eq.imex_relaxation_lower}
\end{align}
For $\delta_{AA}=\tau_A$ or $\theta=0$, this reduces to $D_A\ge W_-+\Delta t_I/\tau_{A,+}>0$.

The Hall--Amp\`ere block is regular for either Hall sign.  With
$d_0=\tau_JW+\dtI$, $\lambda=\dtI^2\sigma_JW/w_e$,
$d=d_0+\lambda$, $a=(\dtI\delta_{JB}/w_e)/d$, $x=\lambda/d$, and
$F^2=-b^2\Xi$, {where $F^\mu{}_\nu=\sqrt{b^2}\,b^\mu{}_\nu$ is the rotation
generator of Eq.~\eqref{eq.4D_Fgen} and
$\Xi^{\mu\nu}=\Delta^{\mu\nu}-b^\mu b^\nu/b^2$ projects orthogonally to both
$u^\mu$ and $b^\mu$, so that $b^\mu{}_\lambda b^\lambda{}_\nu=-\Xi^\mu{}_\nu$
as used in Appendix~\ref{app:ohm},}
using normalized magnetic tensors for $b^2>0$ and the Hall-free limit at $b^2=0$,
\begin{align}
 (I-aF)^{-1}
 &=I-\frac{a^2b^2}{1+a^2b^2}\Xi
 +\frac{a}{1+a^2b^2}F,\notag\\
 \det_{u^\perp}(I-aF)&=1+a^2b^2>0,\notag\\
 d_{\rm SM}&\geq1-x+\frac{x}{W^2}
 =\frac{d_0+\lambda/W^2}{d_0+\lambda}>0.
 \label{eq.imex_hall_sm_bound}
\end{align}
{Here $d_{\rm SM}$ is the scalar Sherman--Morrison denominator of
Eq.~\eqref{eq.ohm_sm}, the factor by which the rank-one Amp\`ere feedback is
removed from the Hall operator, and Eq.~\eqref{eq.imex_hall_sm_bound} restates
the bound already established in Eq.~\eqref{eq.ohm_hall_invertibility}.  The
four sign conditions $\tau_J,w_e>0$, $\sigma_J\geq0$ and $\dtI>0$ suffice to
make both denominators positive without evaluating anything nonlinear: they
give $d_0>0$ and $\lambda\geq0$, hence $x=\lambda/d\in[0,1)$ and
$d_{\rm SM}\geq1-x+x/W^2>0$, while $\det_{u^\perp}(I-aF)=1+a^2b^2\geq1$
whatever the sign or magnitude of the Hall coefficient.  The current block can
therefore be declared invertible from the parameters and the trial state alone.}

To include every implicit moment, bound the closed stage map by
\begin{align}
 |\Pi|&\leq\Pi_+,&
 \sqrt{q^\mu q_\mu}&\leq q_+,\notag\\
 \sqrt{\pi^{\mu\nu}\pi_{\mu\nu}}&\leq\pi_+,&
 \sqrt{j^\mu j_\mu}&\leq J_+.
 \label{eq.imex_invariant_envelopes}
\end{align}
Here $j^\mu$ is the conduction current in
$\mathcal J^\mu=\rho_q u^\mu+j^\mu$, not the total four-current.
{The bounds may come from outward-rounded evaluations of the
local transport closure, from the causal limits of
Sec.~\ref{sec:causality}, or from both.}
With $d_A=\dtI J_+$, the Maxwell and dissipative displacements obey
\begin{align}
 \delta_E^{\rm EM}&=|\widehat E|d_A+\frac12d_A^2,&
 \delta_S^{\rm EM}&=|B|d_A,\notag\\
 \delta_E^{\rm diss}&=Z_+^2\Pi_++2W_+Z_+q_+\notag\\[-2pt]
 &\hspace{3em}+Z_+^2\pi_+,\notag\\
 \delta_S^{\rm diss}&=W_+Z_+\Pi_+
 +(W_+^2+Z_+^2)q_+\notag\\[-2pt]
 &\hspace{3em}+W_+Z_+\pi_+.
 \label{eq.imex_stage_budgets}
\end{align}
Define
$\widehat E_0=\tau+\Db-(\widehat E^2+B^2)/2$,
$\widehat m_i=S_i-(\widehat E\times B)_i$, and add the electromagnetic and
dissipative budgets.  Then a necessary condition for physicality is
\begin{align}
 {\widehat E_0+\delta_E^I\geq
 \sqrt{\Db^2+
 \bigl[\max(\widehat m-\delta_S^I,0)\bigr]^2}.}
 \label{eq.imex_widened_cone}
\end{align}

{The electron conserved mass and energy densities add two necessary interval}
conditions.  Denoting the mass-to-charge ratio by $c_e=|m_e/q_e|$ and assuming the bounds
$\rho_{e,-}\leq\rho_e\leq\rho_{e,+}$, we have
\begin{align}
 D_e&>0,\notag\\[-2pt]
 W_-\rho_{e,-}-c_eZ_+J_+
 &\leq D_e\leq W_+\rho_{e,+}+c_eZ_+J_+.
 \label{eq.imex_carrier_interval}
\end{align}
{The electron energy row admits a similar test.
The
stage equation bounds $|We_e-\widehat U_e|$ by}
\begin{align}
 {K_e:=c_e w_{e,+}Z_+J_++P_{e,+}\Theta_++\dtI J_+e_+,}
 \label{eq.imex_electron_budget}
\end{align}
{whose three terms bound in turn the advection of electron energy by the
current, the compression work, and Joule heating, with
$\Theta_+=\max|\dtI\theta|$ bounding the compression tracker and
$e_+=W_+(|\widehat E|+d_A)+Z_+|B|$ the comoving electric field.  Write
$\mathcal U_e^{\rm reach}:=[\widehat U_e-K_e,\widehat U_e+K_e]$ for the
enclosure of reachable values of $We_e$, and
$\mathcal U_e^{\rm adm}:=[W_-e_{e,-},W_+e_{e,+}]$ for those representable by an
admissible electron primitive at all. For our positive-charge convention, $U_e=We_e-c_e\alpha k^0$. The current term in $K_e$ accounts for this difference between stored and fluid energy. A root can exist only where the two
overlap,}
\begin{align}
 \mathcal U_e^{\rm reach}\cap\mathcal U_e^{\rm adm}\neq\varnothing.
 \label{eq.imex_electron_interval}
\end{align}
{Equations~\eqref{eq.imex_carrier_interval} and
\eqref{eq.imex_electron_interval} are one-sided in the same way as the cone
test above.  Each was obtained by letting one quantity range over its own
interval while the others sit at their extremes, whereas a physical electron
state has to satisfy all of them at once with a single $(\rho_e,e_e,j^\mu)$.}

\subsubsection{Sufficient conditions}

{In the orthonormal Eulerian frame, the momentum matrix of
Eq.~\eqref{eq.4D_Amatrix} is
\begin{align}
 A_i{}^j&=\Db h^*\delta_i{}^j+\Sigma_i{}^j/W,&
 h^*&=h_i+\Pi/\rhob,\notag\\
 \Sigma^{ij}&=\gamma^i{}_{\mu}\gamma^j{}_{\nu}\pi^{\mu\nu}.
 \label{eq.precall_momentum_eulerian}
\end{align}
The boost of the rest-frame shear gives
$\|\Sigma\|_2\leq W^2\pi_+$, with $\pi_+$ from
Eq.~\eqref{eq.imex_invariant_envelopes}.  If $h^*\geq h^*_->0$
throughout the allowed primitive region, a sufficient certificate is
\begin{align}
 a_-&:=\Db h^*_- -W_+\pi_+>0,\notag\\*
 \sigma_{\min}(A)&\geq a_-,&\|A^{-1}\|_2&\leq a_-^{-1}.
 \label{eq.precall_wald_momentum_bound}
\end{align}
Thus both the bulk contribution to enthalpy and the boosted shear enter.
For vanishing shear, only $\Db h^*_->0$ remains.
Under this condition the momentum rows may be written either as
fixed-point equations or as $({S^{\rm red}_i}-A_i{}^jz_j)/\Db=0$.
The coordinate form of Eq.~\eqref{eq.4D_Amatrix} still requires
$u_0\ne0$ and its own nonsingularity check.
}

We can further state a sufficient condition for a physical solution.  For {$X=X_5=(z_i,h_iW,w_eW)$, with $\rho_e$ reconstructed by the carrier equation, define}
$G_\lambda(X)=X-\Phi_{\lambda\dtI}(X)$.  On a parameter interval
$\Lambda_k$ and a primitive box $\mathcal X_k$, form
\begin{align}
 \boldsymbol K_k={}&\bar X_k
 -C_k\boldsymbol G_{\Lambda_k}(\bar X_k)\notag\\
 &+\left[I-C_k\partial_X\boldsymbol G_{\Lambda_k}(\mathcal X_k)\right]
 (\mathcal X_k-\bar X_k).
 \label{eq.precall_homotopy_krawczyk}
\end{align}
$\Phi_{\lambda\dtI}$ is the implicit stage map at the fraction
$\lambda\in[0,1]$ of the requested source time, so $G_\lambda=0$ is the
coupled matter--current stage equation.  Here $\mathcal X_k$ is a box of
{admissible velocities and ion/electron enthalpy variables,}
$\bar X_k$ its center, $\Lambda_k$ a subinterval of the continuation, and
$C_k$ an approximate inverse of $\partial_XG$ at the center.  Bold quantities
denote outward-rounded interval evaluations.
Assume positive box widths, a nonsingular point preconditioner and a stage map defined throughout the box, continuous in $\lambda$ for branch tracking. On smooth branches, the interval Jacobian must bound every derivative in $\mathcal X_k\times\Lambda_k$. At Lipschitz caps or EOS joins, it must bound all relevant secant slopes, rather than a single point derivative.
The inclusion
$\boldsymbol K_k\subset\operatorname{int}\mathcal X_k$ proves, for each
$\lambda\in\Lambda_k$, a unique root in the box, with nonsingular Jacobian on a continuously differentiable branch. At a nonsmooth join, a classical derivative need not exist.
Link neighboring tubes by bounding their root at the shared parameter value within a box contained in both tubes. Uniqueness then identifies the same root. A chain covering $[0,1]$ {and starting from} a verified physical $\dtI=0$ root establishes continuation along that branch.  They do not exclude a remote root outside the
tubes.  Excluding all other roots would require a cover of the full physical
box and its boundary.  The interval option also requires enclosure-capable EOS
and closure evaluations; its present implementation is much more expensive
than recovery and is disabled by default.

\subsection{Exact and lower-dimensional special cases}

\subsubsection{Vanishing dissipative sectors}

At $\dtI=0$, setting one physical dissipative field to zero imposes a
simple identity on its evolved moment:
\begin{align}
 \Pi=0:\quad&\Pi_*=Wy_\zeta,
 \label{eq.dt0_zero_bulk}\\
 q^\mu=0:\quad&Q_0=-y_\kappa Tz^2,\qquad
 Q_i=Wy_\kappa Tz_i,
 \label{eq.dt0_zero_heat}\\
 \pi^{\mu\nu}=0:\quad&\pi_*^{00}=-2y_\eta W,\qquad
 \pi_*^{0i}=-y_\eta z_i,\qquad \pi_*^{ij}=0.
 \label{eq.dt0_zero_shear}
\end{align}
These conditions may be imposed independently.  When all three hold, the
ion energy--momentum block is exactly nondissipative, although the finite-inertia current and electron energy remain.  When $\Pi=q^\mu=0$ but a
nonsingular shear moment remains, the shear-fixed construction below applies.
For a generic closure, the hatted predictors and
relaxation sources replace the source-free identities, so merely setting a
coefficient to zero does not create the same scalar reduction.

\subsubsection{Nondissipative bulk}
\label{app:dt0_bulk_bound}

For $\Pi=q^\mu=\pi^{\mu\nu}=0$, set
$R^2=m^2/\Db^2$.  With a $\Gamma$-law EOS,
$h=1+\Gamma\epsilon$, momentum conservation gives
\begin{align}
 W(h)&=\sqrt{1+\frac{R^2}{h^2}},&
 p(h)&=\frac{\Gamma-1}{\Gamma}\frac{\Db(h-1)}{W(h)}.
 \label{eq.dt0_bulk_hmap}
\end{align}
The represented matter energy is
\begin{align}
 E_{\rm f}(h)=\Db hW(h)-p(h)
 =\frac{\Db[h^2+(\Gamma-1)h+\Gamma R^2]}
 {\Gamma\sqrt{h^2+R^2}}.
 \label{eq.dt0_bulk_energy_h}
\end{align}
For $\Db>0$, $h\geq1$, and $1<\Gamma\leq2$,
\begin{align}
 \frac{dE_{\rm f}}{dh}=
 \frac{\Db\{h^3+[(2-\Gamma)h+\Gamma-1]R^2\}}
 {\Gamma(h^2+R^2)^{3/2}}>0.
 \label{eq.dt0_bulk_monotone}
\end{align}
Let $[h_-,h_+]$ be the enthalpy interval left by the density, thermal, and
velocity bounds.  The exact condition is
\begin{align}
 {\Db>0,\qquad h_-\leq h_+,\qquad
 E_{\rm f}(h_-)\leq E_0\leq E_{\rm f}(h_+).}
 \label{eq.dt0_bulk_box_condition}
\end{align}
Without upper bounds and with $h_-=1$, this becomes
$E_0\geq\sqrt{\Db^2+m^2}$.  A bracketed one-dimensional solve then constructs
the unique bulk primitive \cite{Kastaun:2020uxr,Wu:2017grhd}.

\subsubsection{Flow fixed by the shear moment}

A nonsingular source-free shear moment fixes the velocity algebraically.
Put
\begin{align}
 r=\frac12g_{\mu\nu}\pi_*^{\mu\nu}=y_\eta W,\qquad
 A_{ij}=\pi_*^{ij}-r\delta_{ij}.
\end{align}
For $y_\eta>0$ and $\det A\neq0$,
\begin{align}
 W=\frac{r}{y_\eta},\qquad
 z_i=W(A^{-1})_{ij}\pi_*^{0j}.
 \label{eq.dt0_shear_velocity}
\end{align}
The result must still obey $z^2=W^2-1$, the remaining shear component,
$W\in\mathcal W_D$, and the heat and current contractions in
Eq.~\eqref{eq.dt0_moment_compatibility}.  If $y_\eta=0$, exact reconstruction
instead requires $r=0$ and the subluminal solution of the same linear system.

{For the $\Gamma$-law EOS $p=(\Gamma-1)e_{\rm int}$, the ion internal
energy at fixed flow is explicit.}  With
\begin{align}
 {C_{\rm i}}:=E_0-\frac{z^im_i}{W}-\rhob,\qquad
 \mathcal D_{\rm i}:=1-
 \frac{y_\kappa(\Gamma-1)z^2}{\rhob W^2},
\end{align}
one obtains
\begin{align}
 {e_{\rm int}}=\frac{{C_{\rm i}}+Q_0/W^2}{\mathcal D_{\rm i}}.
 \label{eq.dt0_explicit_ion_energy}
\end{align}
Thus, $\mathcal D_{\rm i}\neq0$, the ion EOS bounds, and exact
energy--momentum reconstruction complete this branch.  The notation
$\mathcal D_{\rm i}$ is kept distinct from the derived charge below.

At fixed flow, the enthalpy-weighted current moment and electron EOS reduce
to one quadratic. Here $W\geq1$, $D_e>0$, and $1<\Gamma_e\leq2$. Define
\begin{align}
 x&:=w_e-1,&
 a&:=\frac{\Gamma_e-1}{\Gamma_e},\notag\\
 L&:=1+\frac{(2-\Gamma_e)W^2}{\Gamma_e-1},&
 E_*&:=\frac{m_e}{q_e}u^\nu M_\nu,\notag\\
 N&:=-\frac{m_e}{q_e}n^\nu M_\nu.
\end{align}
\begin{align}
 f_2x^2+f_1x+f_0&=0,\notag\\
 f_2&=D_eaL,\notag\\
 f_1&=D_eW^2+aL(D_e-E_*)\notag\\[-2pt]
 &\hspace{3em}-Wa(N+2WE_*),\notag\\
 f_0&=W^2(D_e-E_*).
 \label{eq.dt0_electron_quadratic}
\end{align}
For a real root,
\begin{align}
 \rho_e(x)=\frac{D_e(1+x)-E_*}{Wax}.
\end{align}
A simple root further requires $f_1^2-4f_2f_0>0$.

Because $f_2=D_eaL>0$, the sign of
$f_0=W^2(D_e-E_*)$ determines the root structure.  If $E_*>D_e$, then
$f_0<0$ and the two roots have opposite signs.
A hot electron state has $x=w_e-1>0$, leaving one positive algebraic candidate. If $N+2WE_*>0$, the quadratic is negative at $x=(E_*-D_e)/D_e$, so the positive root lies above the zero-density boundary. Otherwise, its density can be negative. The candidate must also satisfy the density, EOS, current, and causality bounds.

This condition depends on the normal conduction current.  Define
\begin{align}
 s:=\frac{n_\mu j^\mu}{q_m}=-\frac{\alpha j^0}{q_m}.
\end{align}
At $\dtI=0$, the carrier and energy rows satisfy
\begin{align}
 D_e&=W\rho_e+s,&
 E_*&=We_e+w_es,
\end{align}
and the gamma-law EOS gives
\begin{align}
 E_*-D_e
 &=x\left(\frac{W\rho_e}{\Gamma_e}+s\right),&
 f_0
 &=-W^2x\left(\frac{W\rho_e}{\Gamma_e}+s\right).
 \label{eq.dt0_electron_root_sign}
\end{align}
Consequently, the roots straddle the origin if and only if
$s>-W\rho_e/\Gamma_e$.  At equality the roots are $0$ and the physical
$x>0$. For $s<-W\rho_e/\Gamma_e$, both roots can be positive when the discriminant and
EOS bounds allow it. Subluminal drift may not distinguish the roots, but the full electron dominant energy condition does: $H_e\geq2\mathcal Q_e$ requires $\xi\leq1/2$, hence $s\geq-\rho_e\sqrt{W^2-1}/2$. For a hot physical state,
{
\begin{align}
 E_*-D_e&\geq x\rho_e\left(\frac{W}{\Gamma_e}
 -\frac{\sqrt{W^2-1}}{2}\right)>0,
 \label{eq.audit_dec_electron_root}
\end{align}}
for $1<\Gamma_e\leq2$. At fixed flow, at most one hot root satisfies the full electron DEC. This neither ensures that a root exists nor establishes uniqueness of the coupled flow inversion.

For vanishing normal current, $\alpha k^0=0$, one has
$D_e=W\rho_e$ and $E_*=W\rho_e(1+x/\Gamma_e)$, so the physical root is
\begin{align}
 x_{\rm ideal}=\Gamma_e\left(\frac{E_*}{D_e}-1\right).
 \label{eq.dt0_electron_ideal_anchor}
\end{align}
For $s\neq0$, however, this expression evaluates to
\begin{align}
 x_{\rm ideal}
 =x\frac{W\rho_e+\Gamma_es}{W\rho_e+s},
\end{align}
{and therefore cannot identify a branch.  We enumerate the real EOS roots
and test the carrier, current, DEC, causal and strong-hyperbolicity bounds.
At fixed flow at most one hot candidate survives the DEC test.}

{{Continuation to $\dtI>0$ starts from an admissible $\dtI=0$ root.
The condition $E_*>D_e$ is a necessary hot-DEC consistency check on the input-state candidate}
before any modification, and is reevaluated on the final
candidate.  If the final inversion outcome fails this criterion, we abort the process and
continue with our fallback inversion treatment as outlined in Appendix~\mbox{\ref{app:4D_ladder}}.}

{Furthermore, we do not impose $E_*>D_e$ through a general energy floor.
{Equation~\eqref{eq.dt0_electron_root_sign} allows algebraic hot states
with $E_*\leq D_e$, but they violate the electron DEC.} Increasing $E_*$ is
allowed only if the electrons are below the atmospheric temperature according to the electron EOS, and
$|E_*-D_e|\leq\epsilon_fD_e$; otherwise the conserved energy is left unchanged
and any branch ambiguity is handled explicitly.} 

Finally, direct evaluation of $\rho_e(x)$ is ill-conditioned for the cold
states, $x\ll1$, used by atmosphere and repair branches.  Let
\begin{align}
 \Delta&:=D_e-E_*,&
 \mathcal D&:=f_1^2-4f_2f_0,\notag\\
 S&:=-\frac12\left[f_1+\operatorname{sgn}(f_1)\sqrt{\mathcal D}\right],
 \notag\\
 x_0&:=\frac{S}{f_2},&
 x_1&:=\frac{f_0}{S}.
 \label{eq.dt0_electron_stable_roots}
\end{align}
Here $\operatorname{sgn}(0)$ is taken to be $+1$.
Division by $S$ requires $S\ne0$. At $x=0$, use the original moment rows, giving $E_*=D_e$ and $\rho_e=(N+2WE_*)/W^2$. These rows also apply at $S=0$, where $f_0=f_1=0$ and the cold root is repeated.
When $D_e$ and $E_*$ lie
within a factor of two, $\Delta$ is exact in binary arithmetic.
Substitution of
the stable root pair eliminates the cancelling numerator,
\begin{align}
 \rho_e(x_0)&=\frac{\Delta f_2+D_eS}{WaS},&
 \rho_e(x_1)&=\frac{S+D_eW^2}{W^3a}.
 \label{eq.dt0_electron_rest_stable}
\end{align}
The selected root is then checked in
Eq.~\eqref{eq.dt0_electron_quadratic}.  The carrier relation is already
satisfied by the reconstruction of $\rho_e$ and cannot by itself distinguish
the two roots.

We then check the full state against the moment constraints and the causality conditions for both species. We illustrate this as follows.

\begin{figure*}
 \centering
 \includegraphics[width=\textwidth]{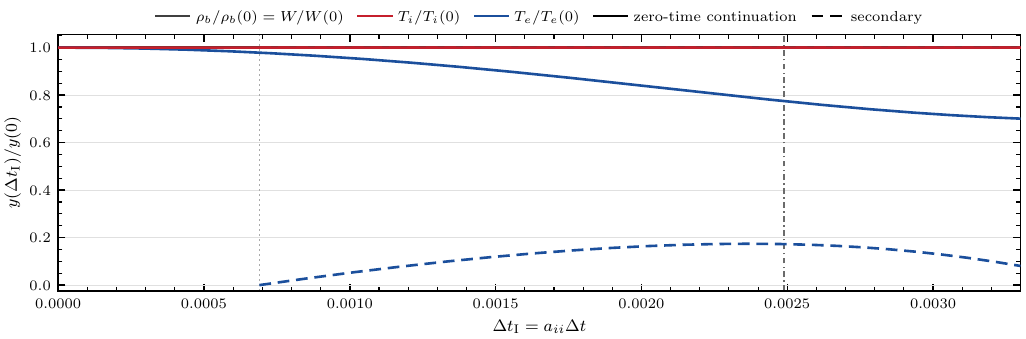}
 \caption{Example solution of the implicit equation illustrating the two branches.
 Only the top (solid) solution has a valid continuum limit for vanishing implicit timestep $\dtI \rightarrow 0$, whereas the lower one
 leads to unstable numerical evolutions.
 Shown are ion rest-mass density $\rho_b$, the ion temperature, $T_i$, and the electron
 temperature $T_e$.
 The dotted vertical line marks $w_e=1$ on the dashed branch. The dash-dotted line marks its characteristic-speed bound, which is weaker here than the full dominant energy condition.}
 \label{fig.wald_multiroot_branches}
\end{figure*}

Figure \ref{fig.wald_multiroot_branches} uses a stationary, flat-spacetime example without Hall rotation, with $\Gamma_i=\Gamma_e=4/3$, $D_b=\tau=1$, $D_e=10^{-4}$, $\mathcal E_{e*}=127/200000$, $\tauJ=1/4$, $q_e/m_e=10^5$, capped $\sigma_J=10^5$ and predictor $\widehat k_{*x}=40$. The energy-consistent temporal moment has magnitude $(q_e/m_e)\mathcal E_{e*}=127/2$. Magnetic field, conserved momentum, predictor electric field and ion dissipative moments vanish. With $t=\dtI$ and $w=w_e$, the current, electric field and electron energy equation are
{
\begin{align}
 d(w,t)&=(1/4+t)w+10^5t^2,\notag\\
 k_x&=\frac{10w}{d(w,t)},\qquad E_x=-\frac{10t}{d(w,t)},\notag\\
 F(w,t)&=\frac{15w-122}{200000}+\frac{100t^2}{d(w,t)^2}=0.
 \label{eq.audit_multiroot_fixture}
\end{align}}
At $t=0.0033$, the hot roots are $w\simeq1.5749744495$ and $w\simeq5.9997792020$, with drifts $\sqrt{j_\mu j^\mu}/[(q_e/m_e)\rho_e]\simeq0.672070$ and $0.383326$. The electron DEC requires this ratio to be at most $1/2$. Exact root counts and the absence of positive-time folds connect the upper branch to $122/15$. It satisfies the drift bound throughout the plotted interval; the lower branch fails. Both raw branches have $e^2>0$ and $b^2=0$ for every $t>0$, violating any finite bound $|e|\le M|b|$. The figure thus shows algebraic multiplicity before physicality projection, rather than two accepted states.

\subsection{Projection of a finite inadmissible state}
\label{app:coupled_energy_projection}

{Numerical error can produce an inadmissible conservative or predictor
state. Such states need to be addressed and limited. Based on our analysis of necessary and sufficient conditions for physicality, we provide a physicality projection for unphysical conserved states.}

\subsubsection{Number density repair}

The Eulerian charge $D_q:=-n_\mu\mathcal J^\mu$ is derived from the two evolved
species densities,
\begin{align}
 D_q&=\frac{q_e}{m_e}\left(\mu Z\Db-D_e\right),
 \qquad \mu:=\frac{m_e}{m_i},\notag\\
 Q^*&=\sqrt\gamma D_q
 =\frac{q_e}{m_e}\left(\mu ZD_b^*-N_e^*\right).
 \label{eq.dt0_density_charge_coupling}
\end{align}
 If the densities are negative, we instead derive the total charge density from the Gauss constraint, $Q^*=\sqrt\gamma\nabla_iE^i$.  
Let $q_m=q_e/m_e$, and let $D_{b,0}^*$ and $N_{e,0}^*$ denote the atmosphere and
magnetization lower bounds. %
The failed-cell transaction constructs the smallest pair on the fixed-charge
ray,
\begin{align}
 (D_b^*)^{\rm new}
 &=\max\!\left[D_{b,0}^*,
 \frac{N_{e,0}^*+Q^*/q_m}{\mu Z}\right],\notag\\
 (N_e^*)^{\rm new}
 &=\mu Z(D_b^*)^{\rm new}-\frac{Q^*}{q_m}.
 \label{eq:failed-c2p-gauss-pair}
\end{align}
A directed roundoff margin moves both densities inward along this neutral-pair ray without changing $Q^*$.

\subsubsection{\texorpdfstring{Coupled source-free ($\dtI=0$) energy projection}{Coupled source-free energy projection}}

After the structural densities are positive, mass, momentum,
electromagnetic fields, carrier, and, for an ordinary input, the full current
moment are held fixed.
{For the zero-ion-dissipation specialization with $\Gamma_c=\Gamma$, define}
\begin{align}
 R&:=\frac{|m|}{\Db},&
 C&:=\frac{\Gamma K\Db^{\Gamma-1}}{\Gamma-1},\notag\\
 h(W,\epsilon)&:=1+CW^{1-\Gamma}+\Gamma\epsilon.
 \label{eq.energy_projection_hW}
\end{align}

For $R>0$, momentum conservation is
\begin{align}
 R=h(W,\epsilon)\sqrt{W^2-1}.
 \label{eq.energy_projection_momentum}
\end{align}
Its right-hand side is strictly increasing for
$W>1$, $C\geq0$, $\epsilon\geq0$, and $1<\Gamma\leq2$:
\begin{align}
 \frac{d}{dW}\!\left[h\sqrt{W^2-1}\right]
 &=\frac{(1+\Gamma\epsilon)W+C(2-\Gamma)W^{2-\Gamma}}
 {\sqrt{W^2-1}}\notag\\
 &\quad+\frac{C(\Gamma-1)W^{-\Gamma}}
 {\sqrt{W^2-1}}>0.
 \label{eq.energy_projection_monotone}
\end{align}
The density and velocity bounds first give $W\in\mathcal W_D$.
Equation~\eqref{eq.energy_projection_momentum} at $\epsilon_+$ and
$\epsilon_-$ gives a single allowed interval $[W_-,W_+]$, which can be solved using a bracketed root finder.

For $R>0$, set
\begin{align}
 h(W)&=\frac{R}{\sqrt{W^2-1}},\notag\\
 p(W)&=\frac{\Gamma-1}{\Gamma}\frac{\Db}{W}[h(W)-1],\notag\\
 \tau_{\rm m}(W)&=\Db h(W)W-p(W)-\Db.
\end{align}
This energy is monotone on the fixed-momentum interval, so
\begin{align}
 \tau_-&=\tau_{\rm EM}+\tau_{\rm m}(W_+),&
 \tau_+&=\tau_{\rm EM}+\tau_{\rm m}(W_-),\notag\\
 \tau_{\rm EM}&:=\frac12(E^2+B^2).
 \label{eq.energy_projection_bounds}
\end{align}
An exterior $\tau$ is moved to its nearest endpoint.  An interior $\tau$
is left unchanged, and one bracketed solve recovers its $W$.  The algorithm
does not raise an already admissible energy.
{At $R=0$, use $W=1$ and $\tau_{\rm m}=\Db(h-1)/\Gamma$ directly;
the density window and thermal bounds still apply.}

The same flow gives the electron energy directly from the four-component
current moment,
\begin{align}
 U_{e,{\rm proj}}=\frac{m_e}{q_e}u^\nu M_\nu.
 \label{eq.energy_projection_electron}
\end{align}

{An admissible thermal root does not by itself make the independently
evolved dissipative electric current admissible.  A prolongation can preserve positive $D_e$ and
{$U_e$ while carrying $M_\mu$ outside the
combined current bound \eqref{eq:electron-current-combined-bound}}.
The physical interval of Eq.~\eqref{eq.precall_current_segment} is therefore
applied here as well, on the selected root and before the nonlinear map is
called.}

\subsubsection{\texorpdfstring{Projection with sources ($\dtI>0$)}{Projection with sources}}

At $\dtI>0$ the algorithm first constructs the source-free state with
Eqs.~\eqref{eq.energy_projection_clamp},
\eqref{eq.energy_projection_electron}, and
\eqref{eq.dt0_electron_quadratic}.  Production then solves the complete
five-coordinate target map, Eqs.~\eqref{eq.4D_hot_unknowns_five} and
\eqref{eq.4D_hot_residual_rows}, directly at the requested $\dtI$, starting
from the $\dtI=0$ solution.  If the direct solve fails to converge, the
bounded Davidenko continuation of Appendix~\ref{app:4D_ladder} advances the
same system in $\lambda\dtI$.

\subsubsection{Terminal projection sequence}

The terminal projection first enforces the kinematic and magnetization bounds.
With $z=\sqrt{z_iz^i}$, the velocity variable is rescaled according to
\begin{align}
 z_i&\longrightarrow z_i\min\!\left(1,\frac{z_{\max}}{z}\right),
 \qquad W=\sqrt{1+z_iz^i},
 \label{eq.appendix_terminal_kinematic}
\end{align}
where the rescaling factor is defined to be unity at $z=0$.  The baryon
density is then projected as

\begin{align}
 \rho_{\rm lo}&=\max(\rho_{\rm atmo},b^2/\sigma_{\rm limit}),\notag\\
 \rhob&\longrightarrow\min[\rho_{\max},\max(\rho_{\rm lo},\rhob)],
 \quad \rho_{\rm lo}\le\rho_{\max}.
 \label{eq.appendix_terminal_magnetization}
\end{align}

Here $b^2=b^\mu b_\mu$ is evaluated from the regulated velocity and the
Eulerian electromagnetic fields.

Because the two species densities determine the Eulerian charge, a change in
$D_b$ is accompanied by the charge-preserving update
\begin{align}
 D_e^{\rm new}=\mu ZD_b^{\rm new}-\frac{m_e}{q_e}D_q^{\rm in}.
 \label{eq:appendix-terminal-fixed-charge}
\end{align}
If this value lies below the electron atmosphere floor, $D_b^{\rm new}$ is
increased by the minimum additional amount required to make both species
densities admissible.  For $\dtI>0$, the electron thermodynamic state is then
obtained by solving
\begin{align}
 D_e&=W\rho_e-\frac{\kappa_e}{w_e},&
 \mathcal E_{e*}&=We_e-\kappa_e,&
 \kappa_e&=\alpha\frac{m_e}{q_e}k^0,\notag\\
 P_e&=P_e(e_e,\rho_e),&
 H_e&=e_e+P_e,&
 w_e&=\frac{H_e}{\rho_e}.
 \label{eq.appendix_terminal_electron_reclosure}
\end{align}
The current moment $k_{*\mu}$ is subsequently reconstructed with
Eq.~\eqref{eq:electron-code-kstar}.  If this system has no admissible solution,
the candidate is rejected and the recovery ladder advances further.

For pure inversion ($\dtI=0$), the fixed-charge projection instead retains the specific electron energy $\epsilon_e=e_e/\rho_e-1$ and sets
$e_e^{\rm new}=\rho_e^{\rm new}(1+\epsilon_e)$.

The carrier-dependent electron causal bound is imposed by moving the trial electron velocity toward the ion velocity while holding $D_e$ fixed:
\begin{align}
 v_e^i(s)=v_i^i+s\,[v_{e,{\rm tr}}^i-v_i^i],
 \qquad 0\leq s\leq1.
 \label{eq.precall_current_segment}
\end{align}
{For the reduced inviscid ion system, with zero heat conduction and
the bulk, shear and heat-flux variables omitted, a simplification is possible.
Vanishing background moments alone do not remove their relaxation waves.}
Set $x_i=c_{{\rm s}i}^2$ and $x=c_{\rm se}^2$.  Substitution of
$a=\pm\sqrt{x_i}$ in the electron cubic
\eqref{eq:main-electron-cubic} gives the exact acoustic-coincidence current,
\begin{align}
 J_{{\rm ei},W}^{(0)}
 &:=
 |q_e|\bar n_e\,
 \frac{\sqrt{x_i}\,|x-x_i|}
 {|x-(\Gamma_e-x)x_i|},&
 \sqrt{j_\mu j^\mu}&<0.99J_{{\rm ei},W}^{(0)}.
 \label{eq:wald-zero-transport-current-gap}
\end{align}
The factor $0.99$ places the numerical state inside the analytic boundary.
At $x=x_i$ the open gap closes, including at zero current.  At
$x_\star=\Gamma_e x_i/(1+x_i)$ the denominator vanishes and the acoustic pair
adds no current ceiling.  This bound is imposed with
Eq.~\eqref{eq:electron-current-combined-bound} using the physical
$\bar n_e$ and $j_\mu=k_\mu/w_e$.

To impose this, we first search the allowed electron-energy interval at fixed
current.  We first search for a sound contact on the hot side toward $x_\star$.  If no such state is available, the
whole segment \eqref{eq.precall_current_segment} is scanned because its
admissible set can be disconnected.  At every trial, $E^i$ is coupled through
Eq.~\eqref{eq.ohm_ampere}, the electron energy is reclosed with
Eq.~\eqref{eq.ohm_electron_energy}, and the largest certified $s$ is moved
slightly into its admissible component.  This map preserves $D_b$, $D_e$, and
hence $D_q$. $\rho_q=(D_q-\alpha j^0)/W$ and $M_\mu$ are reconstructed from
the same electron state.

The electric field is regulated along a connecting ray joining the trial and ideal
fields,

\begin{align}
 E^i(s)&=E_{\rm ideal}^i+s(E_{\rm tr}^i-E_{\rm ideal}^i),\notag\\
 s_E&=\sup\!\left\{s\in[0,1]:\mathcal G(t)\le0\right.\notag\\
 &\qquad\left.\text{for every }t\in[0,s]\right\},\notag\\
 {\mathcal G(t)}&{:=e_\mu(t)e^\mu(t)-\mathcal M^2b_\mu(t)b^\mu(t).}
 \label{eq.admissibility_terminal_electric}
\end{align}

This is done primarily since most of the states modeled are expected to be close to the ideal value in most parts of the domain.
When the trial field violates the configured ratio $\mathcal M$, bisection
determines $s_E$.  The current is reconstructed from the immutable stage
predictor, and the current and field projections are repeated if their
admissible intervals do not intersect.
Repeating the two projections does not guarantee a joint endpoint since
their intervals can be disjoint and alternating clipping can cycle. Require
a finite search budget and the final joint checks. If no joint state is
found, reject it or explicitly account for a repair that relaxes the
original stage equation. 
For $\dtI>0$, with undensitized saved field $\widehat E^i$ and $\mathcal R^i=(\widehat E^i-E^i)/\dtI$, this field segment gives
{
\begin{align}
 \mathcal R^i(s)&=\mathcal R^i_{\rm ideal}
       +s(\mathcal R^i_{\rm tr}-\mathcal R^i_{\rm ideal}),\notag\\
 z_i\mathcal R^i(s)&=z_i\mathcal R^i_{\rm ideal}
       +s\,z_i(\mathcal R^i_{\rm tr}-\mathcal R^i_{\rm ideal}).
 \label{eq.audit_repair_affine_ampere}
\end{align}}
The different ray $\mathcal R^i(t)=t\mathcal R^i_{\rm ideal}$ corresponds
to $E^i(t)=(1-t)\widehat E^i+tE^i_{\rm ideal}$. Precomputing its scalar contraction
avoids recomputing a dot product, but does not make floating arithmetic exact.
The dissipative and
electron causality bounds of Sec.~\ref{sec:causality} are then imposed.

The hot plasma-beta floor requires
\begin{align}
 e_{\rm th}&\geq
 \frac{\beta_{\rm floor}b^2}{2(\Gamma-1)}\,,
\end{align}
with $\beta_{\rm floor}=10^{-5}$.  Because this final thermal correction can
tighten the current bound, the current--field projection is repeated and
Eq.~\eqref{eq.ohm_electron_energy_projection} is reapplied at the final current
before constructing the conserved state and IMEX source.

If both ordinary and charge-supported positive-time matter projections fail,
the terminal atmosphere procedure then uses the initial input.  It
constructs a stationary carrier-supporting state with
$E^i=E^i_{\rm ideal}=0$. {If this also fails, returning the predictor
does not guarantee physicality. The state must pass the final admissibility tests,
or the stage requires a smaller timestep or further repair; see
Appendix~\ref{app:4D_floors}.}

\subsection{A summary for practical physicality enforcement}
\label{app:physicality_enforcement}

In order to ensure that a given conserved state is physical, we proceed as follows:
\begin{enumerate}
 \item Construct the Lorentz-factor interval
 $\mathcal W_D$ of Eq.~\eqref{eq.dt0_WD}.  If either conserved species
 density is nonpositive, apply the coupled Gauss-law repair of
 Eq.~\eqref{eq:failed-c2p-gauss-pair} before reconstructing the interval.
 \item {For zero ion dissipation, at fixed mass and momentum, we solve}
 Eq.~\eqref{eq.energy_projection_momentum} and form
 $[\tau_-,\tau_+]$ from Eq.~\eqref{eq.energy_projection_bounds}.  We apply only
 \begin{align}
  \tau\longrightarrow\min\!\left[\tau_+,\max(\tau,\tau_-)\right].
  \label{eq.energy_projection_clamp}
 \end{align}
 Then compute the compatible electron energy from
 Eq.~\eqref{eq.energy_projection_electron}, and then enumerate its roots with
 Eq.~\eqref{eq.dt0_electron_quadratic}.
 {With nonzero ion moments, use the full moment-constrained recovery
 instead of this scalar energy clamp.}
 \item Repeat the density window Eq.~\eqref{eq.dt0_WD}, the moment and
 composition tests Eqs.~\eqref{eq.dt0_moment_compatibility}--\eqref{eq.dt0_common_W},
 and the positive electron-root test following
 {Eq.~\eqref{eq.dt0_electron_quadratic}.}
 {For the shear-free closure, the positive-density and
 enthalpy floors also satisfy the momentum inverse through
 Eq.~\eqref{eq.4D_wald_momentum_certificate}.}
 \item {Verify on the projected state that
 \begin{align}
  E_*^{\rm proj}:=\frac{m_e}{q_e}u^\mu M_\mu>D_e,
  \label{eq.dt0_projected_unique_electron_root}
 \end{align}
 so that Eq.~\eqref{eq.dt0_electron_quadratic} has exactly one positive root.
 A state that fails this is rejected, and the cell is passed to the
 recovery ladder with its predictor restored.}
 \item Construct the unique $\dtI=0$ origin
 and solve the five coordinates $X_5=(z_i,h_iW,w_eW)$ directly at the target
 stage. The fixed-flow electron result does not prove uniqueness of the
 coupled inversion. If this solve is rejected, restore the projected predictor exactly
 before retrying the same equations.
 The bounded Davidenko continuation of Appendix~\ref{app:4D_ladder} is used
 only if the full implicit solution does not immediately converge.
 \item Require the original five residual rows and the nested carrier,
 current, Amp\`ere, electron-energy, thermodynamic, and causal checks to hold.  These are Eqs.~\eqref{eq.4D_residual}, \eqref{eq.4D_hot_residual_rows},
 \eqref{eq.4D_hot_carrier_map}, \eqref{eq:electron-B-summary} (after fluid projection, together with $u^\nu k_\nu=0$),
 \eqref{eq.ohm_ampere}, \eqref{eq.4D_hot_energy_map}, and
 \eqref{eq.4D_hot_cone}, together with Sec.~\ref{sec:causality}.
 Failure restores the exact input state before recovery.
 \item Apply the density, velocity, and
 magnetization projections of
 Eqs.~\eqref{eq.appendix_terminal_kinematic}--\eqref{eq.appendix_terminal_magnetization}.
 If either density changes, resolve the density at fixed input charge density using
 Eq.~\eqref{eq:appendix-terminal-fixed-charge}.  On the $\dtI=0$ slice, preserve the accepted electron
 specific energy and reapply the current regulator.
 \item 
 {Recompute the closure and enforce the electric-field bound
 \eqref{eq.admissibility_terminal_electric}.} Require
 $c_{\rm se}^2>0$ ($e_e>\rho_e$ for the gamma law), the electron bound
 \eqref{eq:electron-current-combined-bound}{, and either the primitive
 simple-root condition \eqref{eq.resgeneral.simple-reality-screen} with the
 current gap \eqref{eq:electron-ion-acoustic-current-limit}, or the exact
 causal and uniform-eigenbasis tests \eqref{eq.resgeneral.complete} and
 \eqref{eq.ev.full-resolvent-sh}.  The time floors
 \eqref{eq.caus.resistive-primitive-cone-floors} provide a separate
 sufficient cone bound, not a reality test.  For the reduced inviscid ion
 system, use
 \eqref{eq:wald-zero-transport-current-gap}.} Evaluate the
 field segment with Eq.~\eqref{eq.audit_repair_affine_ampere}.
 \item 
 Recover the electron state with
 Eq.~\eqref{eq.appendix_terminal_electron_reclosure}, reconstruct
 $j_\mu=k_\mu/w_e$, and apply the plasma-beta floor
\eqref{eq.betafloor}.  Search the allowed electron energy first; otherwise
 scan the full current segment \eqref{eq.precall_current_segment}.  Reclose
 Eqs.~\eqref{eq.ohm_ampere} and \eqref{eq.ohm_electron_energy} and repeat all
 current bounds after every thermal change.
 \item Finally, recompute the EOS, entropy, transport coefficients, conserved stage,
 and IMEX source from the final state.
 {Repeat the admissibility and strong-hyperbolicity tests for the final state.}
\end{enumerate}

\section{C2P inversion and implicit equation for resistive GR19M}
\label{app:funcs}
\label{app:4D}

We now describe the C2P inversion and the solution of the implicit equation in
detail.  Both recover the fluid, dissipative, and electron variables from the
same five physical unknowns.  For a C2P inversion we set $\dtI=0$, so that the
implicit source terms vanish.  For $\dtI:=\akk\alpha\Delta t>0$, the same
unknowns also satisfy the diagonally implicit stage of
Sec.~\ref{sec:implicit}.  The principal difficulty is to remain within the
physical domain while selecting the root connected continuously to the C2P
solution.

\subsection{Unknowns}
\label{app:4D_unknowns}

Once the fluid four-velocity and the ion and electron thermodynamic states are
specified, every remaining equation in Sec.~\ref{sec:implicit} is algebraic.
With $z_i:=u_i=Wv_i$ and $w_e:=h_e/m_e$, we choose the five unknowns
\begin{align}
  {X_{5}}&{=\left(z_i,h_iW,w_eW\right) \, .}
  \label{eq.4D_hot_unknowns_five}
\end{align}
For the Wald closure the ion viscous and heat sectors vanish, so the equations
for $(z_i,w_eW)$ are algebraically independent of the trial $h_iW$.
Nevertheless, we retain $h_iW$ as an independent unknown: the corresponding
enthalpy equation prevents otherwise regular first-stage states from being
rejected.  Its Jacobian column is sparse and is evaluated analytically.
Here $W=\sqrt{1+z_i\gamma^{ij}z_j}$ is the Lorentz factor and
$\rhob=\Db/W$ the ion rest-mass density.
Consistent with our discussion of physically admissible states in
Appendix~\ref{app:dt0_admissibility}, we enforce
\begin{align}
  {Wh_{i,\min}}& {\le h_iW\le Wh_{i,\max} \, ,}\notag\\
  {Ww_{e,\min}}& {\le w_eW\le Ww_{e,\max} \, ,}
  \label{eq.4D_hot_cone}
\end{align}
where both intervals are set by the EOS floors and ceilings.

\subsection{Algebraic recovery at fixed trial state}
\label{app:4D_inner}

For fixed $X_5$, the remaining primitive variables follow algebraically and
define the five residual equations.  {We first divide the stored densitized
variables by $\sqrt\gamma$.  In particular,
$Z^\kappa_\mu=\widetilde Z^\kappa_\mu/\sqrt\gamma=\alpha z^0{}_{\mu}$.}
We then invert
Eq.~\eqref{eq:electron-code-kstar} for $k_\mu$ and obtain
$j_\mu=k_\mu/w_e$.

\emph{(i) Kinematics and thermodynamics.}
The velocity $z_i$ determines $W$, $u^\mu$, $u_\mu$, and $\rhob$ as described
in Appendix~\ref{app:4D_unknowns}.  The ion enthalpy determines the internal
energy density {$e_{\rm int}$} through the $\Gamma$-law relation
\begin{align}
   &\notag\\
  {{e_{\rm int}}} &{= \frac{\rhob}{\Gamma} \left( \frac{h \, W}{W} - 1 \right)
   +\left(1-\frac{\Gamma_{\rm c}}{\Gamma}\right)e_{\rm c}} \, ,
  \label{eq.4D_epsfromhW}
\end{align}
where the cold contribution
$e_{\rm c}=\mc{K}\rhob^{\Gamma_{\rm c}}/(\Gamma_{\rm c}-1)$ is removed before
evaluating the thermal equation of state.
We use $p=(\Gamma_{\rm c}-1)e_{\rm c}+(\Gamma-1)e_{\rm th}$ and ${e_{\rm int}}=e_{\rm c}+e_{\rm th}$. The added term vanishes when $\Gamma_{\rm c}=\Gamma$.
We restrict
$e_{\rm th}\in[\epsilon_{\rm atmo}\rhob,\epsilon_{\max}\rhob]$, ensuring
physical pressure, temperature, and sound speed.

\emph{(ii) Electron sector.}
We recover $w_e=(w_eW)/W$ and define the electron fraction
$y:=\rho_e/(\mu\rhob)$.  The electron rest-mass density and comoving charge
then satisfy
\begin{align}
  \rho_e&=\mu\rhob y \, , &
  \rhoq&=\frac{q_e}{m_e}\left(\mu Z\rhob-\rho_e\right) \, .
  \label{eq.4D_hot_fixed_carrier}
\end{align}
{The inner electron--current solve determines $\rho_e$ and $\rhoq$ while
holding $w_e$ fixed.  The electron equation of state then supplies $e_e$,
$P_e$, and $T_e$.  Since the evolved current is $k_\mu=w_ej_\mu$, we define}
\begin{align}
  \kappa_e&:=\alpha\frac{m_e}{q_e}k^0 \, .
  \label{eq.4D_hot_scaled_current}
\end{align}
{Writing $D_e$ for the undensitized conservative carrier and
$\widehat{\mathcal E}_{e*}^{\,\ast}:=\gamma^{-1/2}\widetilde{\mathcal E}_{e*}^{\ast}$ for the undensitized electron-energy
predictor, carrier conservation and electron-energy conservation give}
\begin{align}
  D_e&=W\rho_e-\frac{\kappa_e}{w_e} \, ,
  \label{eq.4D_hot_carrier_map}\\
  e_e^{\rm rec}
  &=\frac{1}{W}\left[
  \widehat{\mathcal E}_{e*}^{\,\ast}+\kappa_e\right. \notag\\
  &\hspace{4em}\left.
  -\dtI P_e\theta+\dtI \, j^\mu e_\mu\right] \, ,
  \label{eq.4D_hot_energy_map}\\
  w_e^{\rm rec}&=w_e(e_e^{\rm rec},\rho_e) \, , \notag\\
  (w_eW)_{\rm rec}&=w_e^{\rm rec}W \, , \notag\\
  P_e^{\rm rec}&=P_e(e_e^{\rm rec},\rho_e) \, , \notag\\
  y_{\rm rec}&=\frac{D_e+\kappa_e/w_e}{W\mu\rhob} \, ,
  \label{eq.4D_hot_scalar_maps}
\end{align}
The electron equation of state, evaluated at
$(e_e^{\rm rec},\rho_e)$, supplies $w_e^{\rm rec}$ and closes the fifth
residual.  The Wald calculations do not replace this hot electron state by a
cold EOS value.

{The compression term in Eq.~\eqref{eq.4D_hot_energy_map} is obtained from
the density tracker through}
\begin{align}
 &\notag\\
 &(\dtI\theta)_{\rm used}=\begin{cases}
 W(y_{\rhob}^{\ast}/\rhob-1),&\dtI>0,\\
 0,&\dtI=0.
 \end{cases}
  \label{eq.4D_hot_theta_tracker}
\end{align}
which avoids evaluating a cell-centered spatial derivative of $u^\mu$ and
vanishes regularly in a C2P inversion.  The Joule term uses the electric field
obtained from the current equation; only the gyration geometry is evaluated at
the IMEX predictor field, as described in step~(v).

We first solve the current equation and Amp\`ere's law for $k_\mu$, hence
$j_\mu=k_\mu/w_e$, and then evaluate
Eq.~\eqref{eq.4D_hot_energy_map}.  Equation~\eqref{eq.4D_hot_carrier_map}
finally recovers the electron density and closes the five residual equations.

\emph{(iii) Transport coefficients.}
{The local closure supplies}
$\eta$, $\kappaq$, $\zeta$, $\sigma$, the four relaxation times, and the
{Hall transport coefficients of Eq.~\eqref{eq.transport_hall}}.
These closure values remain provisional until $\Pi$, $q^\mu$, and
$\pi^{\mu\nu}$ have been recovered.
{Require the electron bound \eqref{eq:electron-current-combined-bound}
and either the primitive simple-root condition
\eqref{eq.resgeneral.simple-reality-screen} with the current gap
\eqref{eq:electron-ion-acoustic-current-limit}, or the exact tests
\eqref{eq.resgeneral.complete} and \eqref{eq.ev.full-resolvent-sh}.
The time floors \eqref{eq.caus.resistive-primitive-cone-floors} give a
separate sufficient cone bound on the $M>0$ branch.  Reclose and repeat
these tests after any change to a time, moment or electron state.}
Raising a relaxation time
changes the corresponding gyration parameter, so $\dqB$ and $\dpiB$ are
rescaled together with it to hold the physical ion cyclotron frequencies
$\dqB/\tauq$ and $\dpiB/\tau_\pi$ fixed.  The current relaxation time $\tauJ$
is constrained by neither causality nor hyperbolicity and is unchanged.

\emph{(iv) Bulk viscous pressure.}
\myeqref{eq.Pibulk} is evaluated directly,

 \begin{align}
  \Pi = \frac{\widehat{\Pi}_{*} - W \, \widehat{y}_{\zeta}}
  {W + \dtI \, \Delta_{\Pi}} \, ,
  \qquad {\Delta_\Pi=\tau_\Pi^{-1}-(1-\dPiPi/\tau_\Pi)\theta} \, ,
  \label{eq.4D_Pi}
\end{align}
with $\widehat{y}_{\zeta}=\widehat{Y}_{\zeta}/\Db$ the expansion predictor of
\myeqref{eq.impltracker_generic}.
{The bulk pressure enters through $p+\Pi$ and the effective enthalpy
density $\rhob h^{*}:=\rhob+e_{\rm int}+p+\Pi$.}

\emph{(v) Comoving fields.}
The comoving electric and magnetic fields are built from the stage values of
$E^{i}$ and $B^{i}$,
\begin{align}
  e^{\mu} &= \Delta^{\mu}_{\phantom{\mu}\nu} \, W
  \left[ n^{\nu} \left( E^{k} v_{k} \right)
  + E^{\nu} + \left(v\times B\right)^{\nu} \right] \, ,
  \\
  b^{\mu} &= \Delta^{\mu}_{\phantom{\mu}\nu} \, W
  \left[ n^{\nu} \left( B^{k} v_{k} \right)
  + B^{\nu} - \left(v\times E\right)^{\nu} \right] \, ,
\end{align}
We use $10^{-100}\max(\rhob,\rho_{\rm atmo})$ as the lower numerical threshold
for $b^2$ and treat the cell as unmagnetized below it.  The corresponding
$\Xi^\mu{}_\nu$ and rotation generator are
\begin{align}
  \sqrt{b^{2}} \, b^{\mu}_{\phantom{\mu}\nu}
  = \levciv^{\mu\lambda\alpha\beta} \, b_{\alpha} \, u_{\beta} \,
  g_{\lambda\nu} \, .
  \label{eq.4D_Fgen}
\end{align}

Because $b^\mu$ depends on $E^i$, evaluating the Hall rotation with the unknown
electric field would make the current equation nonlinear.  We instead evaluate
$b^\mu$, $b^2$, $\Xi^\mu{}_\nu$, and the rotation generator
Eq.~\eqref{eq.4D_Fgen} at the IMEX predictor $E^{i*}$.  The Hall projector can
then be inverted analytically, while the electric response itself remains
fully implicit.

After obtaining the current, we recompute the comoving fields from the recovered
$E^i$.  These fields enter the electron Joule heating, heat flux, shear stress,
and subsequent physicality bounds.  The current residual retains the same
predictor Hall geometry used in the analytic inversion; only the rotation
geometry is held fixed.

As a consistency check, the Hall geometry can instead be iterated to a fixed
point with the damped Picard update
\begin{align}
  E^{i,(n+1)}_{\rm geom}=\left(1-\omega\right)E^{i,(n)}_{\rm geom}
  +\omega \, E^{i} \, ,
  \label{eq.4D_hot_inner_relax}
\end{align}
where $\omega$ is reduced if the residual ceases to decrease.
For $0<\omega\le1$, damping preserves fixed points and any existing contraction in a fixed norm. It need not make a noncontracting map converge.  A final
undamped evaluation verifies convergence.

\emph{(vi) Conduction current and electric field.}
The coupled implicit system for $\ju_{\mu}$ and $E^{i}$ is inverted in closed
form, as derived in Appendix~\ref{app:ohm}.  For ideal-MHD preconditioning we
impose $e^\mu=0$ by setting
$E^{i}=E^{i}_{\rm ideal}=\levciv^{ijk}B_jv_k$.  If the electron thermodynamics
is retained, the relaxation and Hall terms still determine $k_\mu$ and hence
$j_\mu=k_\mu/w_e$, but without the implicit Amp\`ere feedback.  This state
provides an initial guess for the hot recovery and is not itself returned.

\emph{(vii) Heat flux.}
After subtracting the transport-coefficient gradient contribution, the heat
flux follows from
\begin{align}
  \widehat{q}_{\nu} := \widehat{Q}_{\nu}
  - \widehat{Z}^{\kappa}_{\nu} \, T \, ,
\end{align}
after which $\widehat{q}$ is projected orthogonally to $u^{\mu}$, rotated by
the gyration projector and relaxed,
\begin{align}
  q_{\nu} = \frac{\mc{Q}^{\mu}_{\phantom{\mu}\nu} \,
  \widehat{q}_{\mu}}{W + \dtI \, \Delta_{\rm q}} \, ,
  \qquad
 {\Delta_{\rm q}=\tauq^{-1}-(1-\dqq/\tauq)\theta} \, ,
  \label{eq.4D_q}
\end{align}
with $\mc{Q}^{\mu}_{\phantom{\mu}\nu}$ of \myeqref{eq.heatfluxprojector}
evaluated at {$\varepsilon_{\rm q}=\dtI\dqB\sqrt{b^2}/[\tauq(W+\dtI\Delta_{\rm q})]$}.\\
\emph{(viii) Shear stress.}
The shear stress follows directly from
\begin{align}
  \pi_{(0)}^{\alpha\beta} =
  \frac{\Delta^{\alpha\beta}_{\phantom{\alpha\beta}\gamma\delta} \,
  \left(\widehat{\pi}_{*}\right)^{\gamma\delta}}
  {W + \dtI \, \Delta_{\pi}} \, ,
  \notag\\
  {\Delta_\pi=\tau_\pi^{-1}-(1-\dpipi/\tau_\pi)\theta} \, ,
  \label{eq.4D_pi}
\end{align}
where $\Delta^{\alpha\beta}_{\phantom{\alpha\beta}\gamma\delta}$ is the
symmetric, transverse, and traceless projector built from $u^\mu$.
We obtain the final shear from
$\pi^{\alpha\beta}=\mc P^{\alpha\beta}{}_{\gamma\delta}
\pi_{(0)}^{\gamma\delta}$, using
{$\varepsilon_\pi=\dtI\dpiB\sqrt{b^2}/[2\tau_\pi(W+\dtI\Delta_\pi)]$}, and remove the remaining
roundoff trace so that $\pi^\mu{}_\mu=0$.
At zero magnetic field, the Hall operator is the identity on the physical shear subspace and $\pi^{\alpha\beta}=\pi_{(0)}^{\alpha\beta}$.

\emph{(ix) Reduction to a purely hydrodynamic problem.}
{Once the electromagnetic and ion-dissipative variables are known, their
contributions can be removed from the conserved momentum and energy.  This
defines the purely hydrodynamic remainder}
\begin{align}
  S^{\rm hyd}_{j} &= S_{j}
  - \levciv_{jkl} \, E^{k} \, B^{l}
  - \alpha \, \pi^{0}_{\phantom{0}j}
  - \left( W \, q_{j} + \alpha \, q^{0} \, z_{j} \right) \, ,
  \label{eq.4D_Shyd} \\
  \tau^{\rm hyd} &= \tau
  - \frac{1}{2} \left( E^{2} + B^{2} \right)
  - 2 \, \alpha \, W \, q^{0}
  - \alpha^{2} \, \pi^{00} \, ,
  \label{eq.4D_tauhyd}
\end{align}
{where $-2\alpha W q^{0}=2W n_{\mu}q^{\mu}$ and
$\alpha^{2}\pi^{00}=n_{\mu}n_{\nu}\pi^{\mu\nu}$.  No electron term is
subtracted because the evolved bulk stress tensor does not contain
$T_e^{\mu\nu}$.}
Contracting the remaining
hydrodynamic stress tensor with $u^{\nu}$, using
$T^{\rm hyd}_{\mu\nu}u^\nu=-{(\rhob+e_{\rm int})}u_\mu$, gives the internal energy density
in closed form,
\begin{align}
  {e_{\rm int}} = \tau^{\rm hyd}
  {- \frac{z^i S^{\rm hyd}_i}{W}}
  + \rhob \, \frac{z^{2}}{1 + W} \, ,
  \label{eq.4D_energy}
\end{align}
{Here $z^{2}=W^2 v^2$.}
The equation of state then supplies the pressure and temperature, and the
recovered ion enthalpy is
\begin{align}
  (hW)_{\rm rec} &= \left( 1 + \frac{{e_{\rm int}}+p}{\rhob} \right) W \, ,
  \label{eq.4D_hW_reconstruct}
\end{align}

without the bulk-viscous contribution.

{The momentum residual uses $S^{\rm red}_j:=S^{\rm hyd}_j+\alpha\pi^0{}_{j}$,
with shear restored.  The energy equation~\eqref{eq.4D_energy} retains the
shear-subtracted momentum $S^{\rm hyd}_j$.}

\emph{(x) C2P inversion.}
For a C2P inversion, $\dtI=0$ and the dissipative variables reduce to the
explicit expressions
\begin{align}
  \Pi &= \frac{\widehat{\Pi}_{*}}{W} - y_{\zeta} \, , \\
  \pi^{\alpha\beta}
  &= \frac{1}{W} \left[ \left(\widehat{\pi}_{*}\right)^{\alpha\beta}
  + 2 \, y_{\eta} \, n^{\left(\alpha\right.} u^{\left.\beta\right)} \right] \, ,
  \\
  q_{\nu} &= \frac{1}{W} \left[ \widehat{Q}_{\nu}
  - \alpha \, y_{\kappa} \, T \, \Delta^{0}_{\phantom{0}\nu} \right] \, ,
\end{align}
which are exactly Eqs.~(\ref{eq.implPi})--(\ref{eq.implq}) at $\dtI=0$, while
$E^i$ and the stored current moment $k_{*\mu}$ keep their evolved values. We reconstruct the physical current $j_\mu=k_\mu/w_e$ from this moment and the coupled electron state. These expressions use the
evolved transport trackers rather than recomputing the closure.  The
hydrodynamic reduction and residual equations are otherwise identical to those
of the implicit equation.  The conserved-state physicality conditions are
derived in Appendix~\ref{app:dt0_admissibility}.

\subsection{Residual equations and momentum recovery}
\label{app:4D_residual}

The momentum equation of the reduced problem is linear in $z_{k}$ up to the
shear stress,
\begin{align}
  {S^{\rm red}_j}
  = \left( \rhob \, h^{*} \, W \, \delta^{k}_{\phantom{k}j}
  - \frac{\alpha}{u_{0}} \, \pi^{k}_{\phantom{k}j} \right) z_{k}
  =: A^{k}_{\phantom{k}j} \, z_{k} \, ,
  \label{eq.4D_Amatrix}
\end{align}
and is solved for $z_k$ by a {lower--upper (LU)} decomposition with partial pivoting of the
$3\times3$ matrix $A$.  Defining
$z_i^{\rm rec}=(A^{-1})^j{}_i{S^{\rm red}_j}$, the momentum and enthalpy
residuals are
\begin{align}
  F_{i} &= \frac{z_i^{\rm rec}-z_{i}}{s_{z_i}} \, ,
  &
  F_{4} &= \frac{\left(hW\right)_{\rm rec} - hW}
  {s_{hW}} \, ,
  \label{eq.4D_residual}
\end{align}
with scale factors
\begin{align}
  {s_{z_i}}& {=\max(1,|z_i^{\rm rec}|,|z_i|) \, ,}\notag\\
  {s_{hW}}& {=\max(1,|(hW)_{\rm rec}|,|hW|) \, .} \notag
\end{align}

The electron-energy equation adds a fifth residual to the four ion equations,
\begin{align}
  F_5&{=\frac{(w_eW)_{\rm rec}-w_eW}{s_e} \, ,}
  \label{eq.4D_hot_residual_rows}
\end{align}
where the positive normalization is fixed at the beginning of the solve,
\begin{align}
  s_e&{=\max\left(\left|(w_eW)_{\rm s}\right|,1\right) \, .}
  \label{eq.4D_hot_scales}
\end{align}
Keeping $s_e$ independent of the dissipative current keeps the fifth residual
well scaled near the root.  The electron density, current, and Amp\`ere
equations are solved together whenever $F_5$ is evaluated.

{For very small temperatures, we replace the three momentum
fixed-point residuals by the algebraically equivalent}
\begin{align}
 {R_i^{\rm dir}:=\frac{{S^{\rm red}_i}-A_i{}^jz_j}{\Db},
 \qquad {S^{\rm red}_i:=S_i^{\rm hyd}+\alpha\pi^0{}_{i}}.}
 \label{eq.4D_direct_momentum_residual}
\end{align}
Where $A$ is invertible, $R^{\rm dir}=0$ and $z^{\rm rec}-z=0$ have the same
roots.  The direct form avoids multiplying the momentum error by $A^{-1}$.
\begin{samepage}
In the limit of vanishing shear, $\pi_{\mu\nu}\rightarrow0$,
\begin{align}
 {A_i{}^j}&{=\Db h^*\delta_i{}^j,\qquad
 h^*=h_i+\Pi/\rhob,}\notag\\
 {\sigma_{\min}(A)}&{=\Db h^*
 \ge\Db h^*_->0.}
 \label{eq.4D_wald_momentum_certificate}
\end{align}
\end{samepage}
{The last inequality requires positive conserved baryon density and
an effective-enthalpy floor including the bulk pressure; an EOS thermal
floor alone is insufficient.}

\subsection{Nonlinear solution and acceptance}
\label{app:4D_ladder}

The overall recovery sequence and its two initial guesses are summarized in
Sec.~\ref{sec:c2p}.  The input checks, required conservative corrections, and
construction of the source-free C2P state follow the physicality-enforcement
procedure of Appendix~\ref{app:physicality_enforcement}.  Here we give only the
additional details of the nonlinear solve.

Let $F=(F_1,\ldots,F_5)$ denote the normalized residuals in
Eqs.~\eqref{eq.4D_residual} and \eqref{eq.4D_hot_residual_rows}, and define
$J_{ab}:=\partial F_a/\partial X_b$.
We write the numerical Jacobian at iteration $n$ as $\widehat J^{(n)}$, distinguishing it from the exact derivative $J(X^{(n)})$.
The initial five-dimensional Jacobian uses four finite-difference columns and the analytic ion-enthalpy column.  At iteration $n$, the quasi-Newton correction is obtained from
\begin{align}
 \sum_{b=1}^{5}\widehat J_{ab}^{(n)}\,\delta X_b^{(n)}
 =-F_a\!\left(X^{(n)}\right),\qquad a=1,\ldots,5.
 \label{eq.4D_quasi_newton_step}
\end{align}
To avoid finite differencing at every iteration, we apply the good-Broyden
rank-one update \cite{press2007numerical} between refreshes of the four
nonanalytic columns every four iterations.

The correction is first limited, in the scaled infinity norm, to three times
the Picard correction.  A backtracking line search then tests
$\alpha=1,1/2,\ldots,2^{-9}$.  With
$\varphi(X):=\tfrac12\sum_{a=1}^{5}F_a(X)^2$, a physical trial is accepted only
if
\begin{align}
 {\|F(X_{\rm t})\|_\infty}
 &{<\|F(X)\|_\infty,}\notag\\
 {\varphi(X_{\rm t})}
 &{\le\varphi(X)+10^{-4}F(X)^T\widehat J d.}
 \label{eq.4D_line_search}
\end{align}
Here $X_{\rm t}$ is the trial after physicality projection and $d:=X_{\rm t}-X$. Without projection, $d=\alpha\delta X$; clipping can change its direction. We evaluate the model and displacement in the same coordinates, with fixed station scales.
The directional derivative must be negative, and the decrease predicted by
the linear model must exceed
$64\epsilon_{\rm mach}\max[\varphi(X),10^{-300}]$.
The Broyden update uses the newly evaluated residual at the accepted iterate,
and the Wald production parameters require
$\|F\|_\infty\leq10^{-11}$ together with the nested checks listed below.

If no quasi-Newton proposal is accepted, we instead search along the Picard
direction $\delta X_{\rm P}=\Phi(X)-X$, where $\Phi$ denotes the algebraic
recovery defined above.  Picard convergence is guaranteed on a closed physical
box $\mathcal B$ only when
\begin{align}
 \Phi(\mathcal B)&\subseteq\mathcal B,
 &
 \sup_{X\in\mathcal B}
 \left\|\frac{\partial\Phi}{\partial X}\right\|_{\infty,s}&<1,
 \label{eq.4D_picard_contraction}
\end{align}
where the second norm uses the same coordinate scaling as the nonlinear solve.
The presence of a root inside $\mathcal B$ alone does not establish this
condition.  Because the production calculation does not evaluate an interval
bound on the derivative, the Picard step is treated as a safeguarded fallback
and must pass the same line search and final residual checks as a
quasi-Newton step.

The two initial guesses described in Sec.~\ref{sec:c2p} are tested
independently.  If both direct attempts at the implicit equation fail, the bounded Davidenko
continuation uses
\begin{align}
 F(X,\lambda)=0,\qquad
 F_X\frac{dX}{d\lambda}=-F_\lambda,\qquad 0\leq\lambda\leq1,
 \label{eq:hot-davidenko-path}
\end{align}
where $\lambda$ multiplies all implicit-stage terms.
The tangent uses the
five-dimensional Jacobian and a refined finite difference for $F_\lambda$.
After an Euler prediction, the original five equations are solved at the new
$\lambda$.  The corrected displacement is compared with the trapezoidal
integral of the endpoint tangents.  Failure of this comparison, an excessive
correction relative to the prediction, or departure from the physical domain
restores the last accepted state and reduces the step in $\lambda$.

At an electron EOS floor or ceiling, the fifth equation is restricted to the
allowed enthalpy interval,
\begin{align}
X_e&=\min\!\bigl[\max\!\bigl(\Phi_e(X),{W(\Phi_z)h_{e,\min}/m_e}\bigr),\notag\\&\qquad
 {W(\Phi_z)h_{e,\max}/m_e}\bigr].
 \label{eq:hot-davidenko-electron-face}
\end{align}
On an active face, we fix the specific electron enthalpy at its EOS bound and solve the other four equations. Since $X_e=W(z)h_{e,\mathrm{bound}}/m_e$ varies with the flow, the projected fifth residual must also vanish.
The bounds are $32$ attempted continuation steps, $16$ rejected intervals,
$64$ residual evaluations per step, $512$ evaluations in total.

The final solution is additionally checked to satisfy all relevant equations
\eqref{eq.4D_hot_carrier_map},
\eqref{eq:electron-B-summary} (after fluid projection, together with $u^\nu k_\nu=0$),\eqref{eq.ohm_ampere},
\eqref{eq.4D_hot_energy_map}, the EOS and causal
bounds (Sec. \ref{sec:causality}).
Otherwise, the candidate is rejected and the conserved predictor at $\dtI=0$
is restored.

\subsection{Atmosphere floors and causal regulation}
\label{app:4D_floors}

After a root has been found, we enforce the density and velocity ranges and the
magnetization ceiling.  A species pair that already satisfies these bounds is
unchanged.  If an atmosphere or magnetization floor raises $D_b$, the electron
density is adjusted so as to preserve the incoming net charge,
\begin{align}
 \Dq^{\rm in}&=\frac{q_e}{m_e}
 \left(\mu Z\Db^{\rm in}-D_e^{\rm in}\right),
 &
 D_e^{\rm new}&=\mu Z\Db^{\rm new}-\frac{m_e}{q_e}\Dq^{\rm in},
 \label{eq.4D_fixed_charge_floor}
\end{align}
and $D_b^{\rm new}$ is increased further only by the minimum amount needed to
support the positive electron floor.{ This increase is permitted only while $\rho_b\le\rho_{\max}$; otherwise reject the candidate and use the fallback.}  The adjusted pair is returned only if it
reproduces $\Dq^{\rm in}$ within roundoff.  This fixed-charge density floor is
distinct from the correction of an invalid input density, which is handled by
the grid-scale Gauss-law fallback.

For a C2P inversion, a changed density pair is recovered without invoking an
implicit Amp\`ere or electron-energy source.  We preserve electron specific internal energy
$\epsilon_e=e_e/\rho_e-1$ from the accepted primitive, reconstruct
$e_e^{\rm new}=\rho_e^{\rm new}(1+\epsilon_e)$, reapply the
{combined current cap \eqref{eq:electron-current-combined-bound}} in the conduction-current
representation (Sec. \ref{sec:causality}), and convert the complete state back to conserved form.  Values are returned only if the recovered $D_b$ and $D_e$ agree with the two
fixed-charge densities.  If this source-free density correction is rejected,
the candidate is discarded and the full C2P inversion is repeated from the
coupled, energy-projected initial guess.

The closure is then recomputed and $\pi^{\mu\nu}$ is projected to a transverse, symmetric, trace-free
tensor, while $q^\mu$ and $\ju^\mu$ are projected transverse to the flow.
The ion relaxation times are raised to their causal lower bounds, and the
dissipative state is moved toward equilibrium in the order $\pi^{\mu\nu}$, the
three transport ratios together, $q^\mu$, and $\Pi$.  The electric field and
physical current are reconstructed from the IMEX predictor.  If the
carrier-dependent drift bound is violated, the electron velocity is reduced
toward the ion velocity along the segment
Eq.~\eqref{eq.precall_current_segment}.  If that update is incompatible with
the electric-field bound, the field is moved toward its ideal-MHD value along
{the segment}
\begin{align}
 {E^i(\lambda_E)}
 &{=E^i_{\rm ideal}+\lambda_E(E^i_{\rm trial}-E^i_{\rm ideal}),}
 \notag\\
 {E^i_{\rm ideal}}&{=\levciv^{ijk}B_jv_k,\qquad 0\le\lambda_E\le1.}
 \label{eq.appendix_terminal_electric}
\end{align}

The plasma-beta and electron EOS floors are imposed only after the nonlinear
solve and followed by the same replay.  Thus, $E^i$, $\ju^\mu$, $D_e$, the
electron energy, the conserved state, and the implicit source all correspond
to one primitive state.
The source is
\begin{align}
 \mc{I}^{(k)}=
 \frac{\mc{V}^{(k)}-\mc{V}^{*,(k)}}{\akk\Delta t},
 \label{eq:imex-source-recovery}
\end{align}
and explicit pass-through rows receive zero stiff source.

If any final condition fails, the primitive values are rejected and the exact
conserved predictor and implicit source are restored.  If both the original and
charge-supported density corrections fail, the final atmosphere fallback uses
a stationary, carrier-supporting state with $E^i=E^i_{\rm ideal}=0$.  This state
must satisfy the same carrier, primitive-to-conserved, and source equations.
If it is also rejected, an evolution step that requires only conserved values
may retain the IMEX predictor with zero stiff source, provided $D_b>0$,
$D_e>0$, and the evolved electron energy is positive.

\section{Relaxation form of the evolved transport coefficients}
\label{app:relaxation}

The transport coefficients enter the Israel--Stewart equations through
divergences of the form $\nabla_{\mu}\left[\left(U+y_{0}\right)u^{\mu}\right]$,
where $U$ is a dissipative variable and $y_{0}$ the corresponding
{transport-ratio target (for example $y_0=\zeta/\tau_\Pi$ when $U=\Pi$).} Writing that combination in flux-conservative form would
require $y_{0}$, an algebraic function of the primitives, to be differentiated
across a shock. We avoid this by promoting $y_{0}$ to an independent, advected
scalar $y$ that relaxes to it. Starting from rest-mass conservation and the
equation to be solved,

 \begin{align}
  \nabla_{\mu} \left( \rhob \, u^{\mu} \right) &= 0 \, , \\
  \nabla_{\mu} \left[ \left( U + y_{0} \right) u^{\mu} \right]
  &{=u^{\mu}\nabla_{\mu}y_{0}-\tau^{-1}U
   +\left(1-\frac{\delta_{UU}}{\tau}\right)U\theta} \, ,
  \label{eq.relax_target}
\end{align}
we instead evolve
\begin{align}
  \nabla_{\mu} \left( \rhob \, y \, u^{\mu} \right) = \rhob \, \mc{I} \, ,
  \qquad
  \mc{I} := -\omega \, u^{0} \left( y - y_{0} \right) \, ,
  \label{eq.relax_tracker}
\end{align}
whose left-hand side is a genuine flux divergence. Rest-mass conservation turns
\myeqref{eq.relax_tracker} into $\mc{I}=u^{\mu}\nabla_{\mu}y$, so that schematically the
equation actually integrated,
\begin{align}
  \nabla_{\mu} \left[ \left( U + y \right) u^{\mu} \right]
  =\mc{I}-\tau^{-1}U
  +\left(1-\frac{\delta_{UU}}{\tau}\right)U\theta \, ,
\end{align}
reduces to \myeqref{eq.relax_target}
in the limit $\omega\rightarrow\infty$, in
which $y\rightarrow y_{0}$ and $\mc{I}\rightarrow u^{\mu}\nabla_{\mu}y_{0}$.
The gradient of the transport coefficient is thus never taken explicitly: it is
recovered as the relaxation source of an advected tracker. Since $\mc{I}$ is
stiff for large $\omega$, it belongs to the implicit sector, and the
$\omega\rightarrow\infty$ limit is taken analytically in
Eqs.~(\ref{eq.implyrhob})--(\ref{eq.implZkappanu}) rather than numerically. In fact, one can show that each helper source can be eliminated in favor of the difference between the
accepted primitive value and its stage predictor, which is why no explicit
$\omega$ appears in the implemented equations.
The $27$-mode characteristic and strong-hyperbolicity tests require this
algebraic elimination.  If the heat ratio and normal Cattaneo tracker
are instead evolved as independent finite-rate continuum fields, their
rest-frame principal rows are $\partial_t y_\kappa=0$ and
$\partial_t Z^\kappa_{\hat n}+\partial_{\hat n}y_\kappa=0$.
The resulting double material root has only one tracker eigenvector.
Here $\theta=\nabla_\mu u^\mu$ and $\delta_{UU}=\delta_{\Pi\Pi}$ for bulk viscosity. The expansion source recovers the chosen constitutive relation; omitting it assumes $\delta_{\Pi\Pi}=\tau_\Pi$.

\section{Implicit inversion of Ohm's law}
\label{app:ohm}

{The electron current and electric field form the stiffest local block of
the system.  Since we also evolve the electron energy, the conservative
current is the enthalpy-weighted variable $k_\mu=w_e\ju_\mu$, whereas Amp\`ere's law carries the physical current $\ju_\mu=k_\mu/w_e$.}

\subsection{{Electric current and electric field update}}
\label{app:ohm_hot}

{We use}
\begin{align}
  {w_e}& {:=\frac{h_e}{m_e}=\frac{H_e}{\rho_e} \, ,} &
  {\omega_e}& {:=\frac{q_e}{m_e}H_e \, ,} &
  {\chi_e}& {:=\frac{q_e}{m_e}P_e \, .}
\end{align}
{The undensitized current moment and its spatial flux are}
\begin{align}
  {\gamma^{-1/2}\widetilde{k}_{*\nu}}
  &{=W\left(k_\nu-\omega_eu_\nu\right)
  -\left(n_\mu k^\mu\right)u_\nu+\chi_en_\nu \, ,}
  \label{eq.ohm_hot_moment}\\
  {\mc F^i(k_{*\nu})}
  &{=\sqrt\gamma\bigl[W(\alpha v^i-\beta^i)
  \left(k_\nu-\omega_eu_\nu\right)} \notag\\
  &{\hspace{3em}
  +\alpha k^iu_\nu-\alpha\chi_e\delta^i{}_{\nu}\bigr] \, .}
  \label{eq.ohm_hot_flux}
\end{align}
{The corresponding connection contraction is}
\begin{align}
  {\Gamma\!\cdot\!\mc B
  =u\Gamma k+k\Gamma u-\chi_e\Gamma\Delta
  -(\omega_e-\chi_e)u\Gamma u \, .}
  \label{eq.ohm_hot_connection}
\end{align}
{Thus, the symmetric cross flux, electron inertia, and pressure term all
belong to the explicit finite-volume predictor.  The local implicit solve
contains only relaxation, electric driving, Hall rotation, and Amp\`ere
feedback.}

{Let $\widehat{k}_{*\nu}$ denote the undensitized predictor of
Eq.~\eqref{eq.ohm_hot_moment}, and define its transverse primitive part by}
\begin{align}
  {\widehat{k}_\nu
  :=\frac{1}{W}\Delta^\alpha{}_{\nu}
  \left(\widehat{k}_{*\alpha}-\chi_en_\alpha\right) \, .}
  \label{eq.ohm_projected_predictor}
\end{align}
{Projecting Eq.~\eqref{eq.implj} orthogonally to $u^\mu$ removes its
longitudinal source and gives}
\begin{align}
  {(\tauJ W+\dtI)k_\nu
  -\dtI\frac{\dJB}{w_e}\sqrt{b^2}\,
  b^\mu{}_{\nu}k_\mu}
  & {=\tauJ W\widehat{k}_\nu+\dtI\sigma_J e_\nu \, ,}
  \label{eq.ohm_start}
\end{align}
{where $\sigma_J=\tauJ\omega_{pe}^2$ and
$\dJB\sqrt{b^2}=\tauJ\omega_{ge}$.  Equivalently,}
\begin{align}
  {\frac{u^0\tauJ}{\akk\Delta t}
  (k_\nu-\widehat{k}_\nu)+k_\nu
  -\frac{\dJB}{w_e}\sqrt{b^2}\,b^\mu{}_{\nu}k_\mu
  =\sigma_J e_\nu \, .}
  \label{eq.ohm_physical}
\end{align}
However, note that because of Amp\`ere's law, the electric field does carry a $k^\mu$ dependence,
\begin{align}
  {e_\mu=\widehat e_\mu-W\dtI
  \left[\gamma_\mu{}^\nu+W^{-2}n_\mu v^\nu-v_\mu v^\nu\right]
  \frac{k_\nu}{w_e} \, ,}
  \label{eq.ohm_efield}
\end{align}
{where $\widehat e_\mu$ is built from the explicit field
$\widehat E^i$.  On the subspace orthogonal to $u^\mu$, the non-diagonal part
of this response is the rank-one operator}
\begin{align}
  {\mc R^\mu{}_{\nu}=x^\mu y_\nu
  =-\frac{1}{W^2}\bar n^\mu\bar n_\nu \, .}
  \label{eq.ohm_rankone}
\end{align}
Here $\bar n^\mu=\Delta^\mu{}_{\alpha}n^\alpha$, $x^\mu=\bar n^\mu$ and $y_\nu=-\bar n_\nu/W^2$. Projection makes each inverse act within $u^\perp$. Using $n^\mu n_\nu$ instead gives the same transverse equation but leaves a longitudinal residual.
{Substitution into Eq.~\eqref{eq.ohm_start} gives}
\begin{multline}
  {\biggl[\mc D_J\Delta^\mu{}_{\nu}
  -\dtI\frac{\dJB}{w_e}\sqrt{b^2}\,b^\mu{}_{\nu}}\\
  {-\frac{\dtI^2\sigma_J}{w_eW}
  \left[\bar n^\mu\bar n_\nu\right]\biggr]k_\mu
  =\tauJ W\widehat k_\nu+\dtI\sigma_J\widehat e_\nu \, ,}
  \label{eq.ohm_full}
\end{multline}
{with}
\begin{align}
  {\Delta_J}& {:=1+\frac{\sigma_J}{w_e}W\dtI \, ,} &
  {\mc D_J}& {:=\tauJ W+\dtI\Delta_J \, .}
\end{align}
{After division by $\mc D_J$,}
\begin{align}
  {\left[(\mc Q_J^{-1})^\mu{}_{\nu}
  +X_0\mc R^\mu{}_{\nu}\right]k_\mu}
  &{=\frac{\tauJ W\widehat k_\nu+
  \dtI\sigma_J\widehat e_\nu}{\mc D_J} \, ,}
  \label{eq.ohm_system}\\
  {(\mc Q_J^{-1})^\mu{}_{\nu}}
  &{:=\Delta^\mu{}_{\nu}-\varepsilon_Jb^\mu{}_{\nu} \, ,}
  \notag\\
  {\varepsilon_J}& {:=\frac{\dtI\dJB\sqrt{b^2}}{w_e\mc D_J} \, ,}
  \notag\\
  {X_0}&{:=\frac{\dtI^2\sigma_JW}{w_e\mc D_J} \, .}
\end{align}
{The displacement-current term in $\Delta_J$ keeps the solve regular as
$\sigma_J\Delta t\rightarrow\infty$.}

{Using
$b^\mu{}_{\lambda}b^\lambda{}_{\nu}=-\Xi^\mu{}_{\nu}$, the Hall operator is
inverted as}
\begin{align}
  {(\mc Q_J)^\mu{}_{\nu}
  =\Delta^\mu{}_{\nu}
  -\frac{\varepsilon_J^2}{1+\varepsilon_J^2}\Xi^\mu{}_{\nu}
  +\frac{\varepsilon_J}{1+\varepsilon_J^2}b^\mu{}_{\nu} \, .}
\end{align}
{The remaining rank-one term is removed with the Sherman--Morrison
identity,}
\begin{multline}
  {\left[(\mc Q_J^{-1}+X_0\mc R)^{-1}\right]^\mu{}_{\nu}
  =(\mc Q_J)^\mu{}_{\nu}}\\
  {-\frac{X_0(\mc Q_J)^\mu{}_{\alpha}x^\alpha
  y_\beta(\mc Q_J)^\beta{}_{\nu}}
  {1+X_0y_\beta(\mc Q_J)^\beta{}_{\alpha}x^\alpha} \, .}
  \label{eq.ohm_sm}
\end{multline}
This inverse is regular for any finite Hall coefficient.  With
$F^\mu{}_\nu=\sqrt{b^2}\,b^\mu{}_\nu$,
$a=\dtI\delta_{JB}/(w_e\mc D_J)$, and $x=X_0$,
\begin{align}
 \det_{u^\perp}(I-aF)
 &=1+a^2b^2>0,\\
 d_{\rm SM}
 &\ge1-x+\frac{x}{W^2}\\
 &=\frac{\tau_JW+\dtI+\dtI^2\sigma_J/(w_eW)}
 {\mc D_J}>0.
 \label{eq.ohm_hall_invertibility}
\end{align}
Hall rotation can therefore improve, but never destroy, the
Sherman--Morrison lower bound.  A nonpositive numerical denominator is a
failed map evaluation; it is not replaced by a clipped inverse.
{ We then recover
$\ju_\mu=k_\mu/w_e$ and update}
\begin{align}
  {E^i=\widehat E^i-\dtI
  \left(\jn^i-v^iv_k\jn^k\right) \, ,\qquad
  \jn^i=\gamma^{i\mu}\ju_\mu \, .}
  \label{eq.ohm_ampere}
\end{align}

\subsection{{Electron-energy recovery}}
\label{app:ohm_hot_recovery}

Alongside the electric-field update, we evolve the electron energy
implicitly because it enters the dynamical Ohm law:
\begin{align}
 &\notag\\
 &\widetilde{\mathcal E}_{e*}=\widetilde{\mathcal E}_{e*}^{\ast}+\dtI\sqrt\gamma(-P_e\theta+e_\mu\ju^\mu).
  \label{eq.ohm_electron_energy}\\
  {D_e}&{=W\rho_e-\frac{\kappa_e}{w_e} \, ,\qquad
  \kappa_e:=\alpha\frac{m_e}{q_e}k^0 \, .}
  \notag
\end{align}
{Writing $U_e:=\gamma^{-1/2}\widetilde{\mathcal E}_{e*}^{\ast}$ gives the scalar map}
\begin{align}
  e_e^{\rm map}
  &=\frac{U_e+\kappa_e-\dtI P_e\theta+ {\dtI e_\mu k^\mu}/w_e}{W} \, ,\\
  \rho_e
  &=\frac{D_e+\kappa_e/w_e}{W} \, .
  \label{eq:reduced-electron-energy-implicit}
\end{align}
{Every conversion between $k_\mu$ and $\ju_\mu$ uses the same trial
$w_e$; the equation-of-state result is returned to the outer residual and is
not inserted back into the same current solve.}

For the Wald closure, the nested carrier solve remains algebraic even at
nonzero Hall coefficient.  At fixed $(W,w_e)$ and frozen Hall geometry,
$\sigma_J=s\rho_e$ makes the current matrix and its right-hand side affine in
$\rho_e$.  Clearing the strictly positive determinant in
Eq.~\eqref{eq:reduced-electron-energy-implicit} gives a quartic on the
uncapped branch.  The conductivity-capped branch is affine, and both pieces
meet continuously at the cap.  All real roots are tested on their proper cap
side and checked against the current, Amp\`ere, energy, and causality rows.
The admissible set need not be connected, so this nested equation does not
define one universal Brent bracket.

Because the bulk stress tensor contains no electron stress, the fifth
residual is
\begin{align}
  {R_5}&{=\frac{w_e^{\rm map}W-w_eW}{s_e}=0 \, .}
  \label{eq:reduced-five-dimensional-root}
\end{align}
Before this positive-time map is evaluated, the ion energy is checked at
fixed mass and momentum. For the Wald closure with vanishing ion dissipation and the stated EOS, the density, velocity and thermal bounds give $[\tau_-,\tau_+]$ of
Eq.~\eqref{eq.energy_projection_bounds}.  We clamp an exterior value to the
nearest endpoint or use a bracketed scalar solve for an interior value.  On
the resulting flow,
 \begin{align}
 \mathcal E_{e*}\longrightarrow {\frac{m_e}{q_e}u^\nu M_\nu}
 \label{eq.ohm_electron_energy_projection}
\end{align}
{Here $M_\nu=\widetilde k_{*\nu}/\sqrt\gamma$ and $\mathcal E_{e*}$ are undensitized moments.}

For an ordinary positive-density input, this operation recloses the
electron energy without changing the carrier or current moment, and
Eq.~\eqref{eq.dt0_electron_quadratic} enumerates the physical electron roots.
Each root is tested against Eqs.~\eqref{eq:electron-current-combined-bound} and
\eqref{eq:wald-zero-transport-current-gap}.  In case of primitive inversion only this is a hard requirement.  In case of solving the implicit equation, a thermal-first and full current-segment replay
of Appendix~\ref{app:4D_floors} closes the current, Amp\`ere, electron-energy,
EOS, and ion-causality rows.
A structural density repair bypasses the quadratic and rebuilds a
{low-temperature, ion-comoving electron block, while strong hyperbolicity
still requires $e_e>\rho_e$.} Both constructions are Hall-independent at
$\dtI=0$.  The local root rule does not prove connection to an ideal branch.

\section{Braginskii Limit}
\label{app:braginskii}

For $b^2>0$, when magnetic rotation dominates the other shear terms,
$b^{\delta\mu} \, \pi^{\gamma}_{\phantom{\gamma}\mu} \,
\Delta^{\alpha\beta}_{\phantom{\alpha\beta}\gamma\delta} = 0$
\cite{Most:2021rhr}, which, using $u_{\mu} \, \pi^{\mu\nu} = 0$, leads to
\begin{align}
  b^{\mu\left(\alpha\right.} \pi^{\left.\beta\right)}_{\phantom{\beta\beta}\mu}
  = 0 \, ,
\end{align}
which can be reduced to
\begin{align}
  g^{\kappa\left(\alpha\right.} \pi^{\left.\beta\right)\mu} \, b_{\mu\kappa}
  = 0 \, ,
\end{align}
{and therefore the shear tensor is gyrotropic about $b^\mu$, orthogonal to $u^\mu$, and trace-free. Gyrotropy does not require $\pi^{\mu\nu}b_\nu=0$. We make the axisymmetric ansatz}
\begin{align}
  \pi^{\mu\nu} = A \, b^{\mu} \, b^{\nu} + B \, \Delta^{\mu\nu} \, .
\end{align}
Since $\pi^{\mu}_{\phantom{\mu}\mu} = 0$, we have
\begin{align}
  0 = A \, b^{2} + 3 \, B \implies B = -\frac{1}{3} \, A \, b^{2} \, ,
\end{align}
so,
\begin{align}
  \pi^{\mu\nu} = A \left( b^{\mu} \, b^{\nu}
  - \frac{1}{3} \, b^{2} \, \Delta^{\mu\nu} \right) \, .
\end{align}
Contracting both sides with $b_{\mu} \, b_{\nu}$ leads to
\begin{align}
    A = \frac{3}{2} \, b^{-4} \, b_{\mu} \, b_{\nu} \, \pi^{\mu\nu} \, ,
\end{align}
or, defining $\pi_{0} := -A$,
\begin{align}
  \pi^{\mu\nu} = \pi_{0} \left( -b^{\mu} \, b^{\nu}
  + \frac{1}{3} \, b^{2} \, \Delta^{\mu\nu} \right) \, .
\end{align}

\subsection{Braginskii limit: firehose instability}
\label{sec.firehose}

\begin{figure*}
  \centering
  \includegraphics[width=0.98\linewidth]{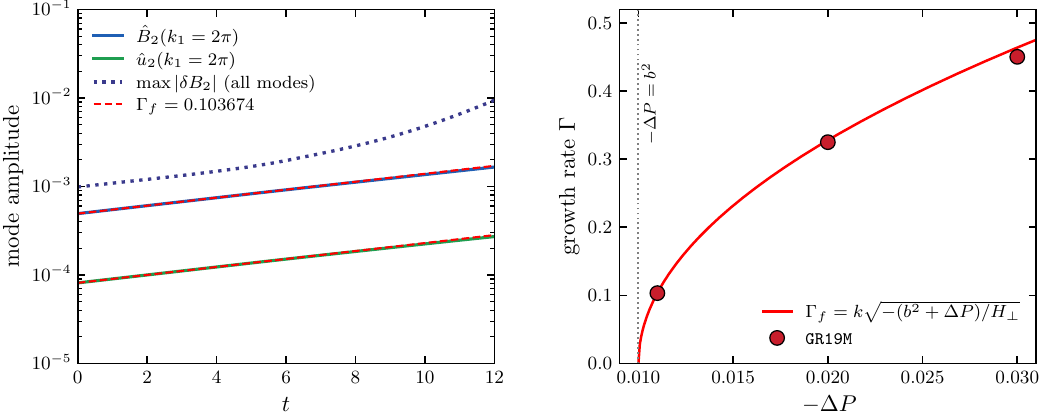}
  \caption{{Relativistic firehose instability.
  \emph{Left:} amplitude of the $k_{1}=2\pi$ Fourier mode of the transverse field
  $B^{2}$ (blue) and velocity $u^{2}$ (green) versus time; both grow at the analytic
  rate $\Gamma_{\!f}\simeq0.103674$ (dashed).
  The small-scale envelope $\max|\delta B_{2}|$ (dotted) is seeded by truncation error and grows faster, driving the eventual nonlinear takeover. \emph{Right:} measured growth rate (points) versus the seeded pressure
  anisotropy $-\Delta P$, on the analytic Braginskii firehose dispersion
  $\Gamma_{\!f}=k\sqrt{-(b^{2}+\Delta P)/H_\perp}$
  (solid); $H_\perp=\rhob+\Gamma u+b^{2}+\Delta P/3$. The instability switches on at the threshold $-\Delta P=b^{2}$ (dotted).}}
  \label{fig.firehose}
\end{figure*}

{
As a validation of the Braginskii viscous sector, we reproduce the relativistic
firehose instability of \citet{Chandra:2017auj}. When the
field-parallel pressure exceeds the perpendicular pressure by more than twice the magnetic pressure, $\Delta P < -b^{2}$, Alfv\'en waves propagating along $\bs{B}$ become
unstable and grow at a rate proportional to their wavenumber.

In Minkowski spacetime on the periodic domain $[0,1]^{2}$ with
$\Gamma=4/3$, we set a uniform background $\rhob=1$, $u=2$, guide field $B^{1}=0.1$
(so $b^{2}=0.01$), and a uniform, firehose-unstable pressure anisotropy
$\Delta P=-0.011<-b^{2}$. On this background, we seed a single linearly unstable shear-Alfv\'en
eigenmode,
\begin{align}
  u^{2}=A\sin(2\pi x^{1}),\qquad B^{2}=B\cos(2\pi x^{1}),
  \label{eq.firehose_mode}
\end{align}
with $A=0.1628\,\alpha$, $B=0.9867\,\alpha$, amplitude $\alpha=10^{-3}$, and the
eigenmode ratio $B/A=B^{1}k/\Gamma_{\!f}$ fixed by the induction equation. The
anisotropy is imposed through the gyrotropic Braginskii limit $\dpiB\to\infty$: the
magnetic-rotation term of the shear equation~\eqref{eq.comovpi} projects
$\pi^{\mu\nu}$ onto $-\Delta P\,(\hat{b}^{\mu}\hat{b}^{\nu}-\Delta^{\mu\nu}/3)$ with the
\emph{instantaneous} $\hat{b}^{\mu}$ at every implicit solve, while a long relaxation
time $\tau_{\pi}=\tauq=10^{6}$ approximately preserves its magnitude over this test (the background has no shear, so
its Braginskii target is $\Delta P_{0}\simeq0$). Induction is ideal, i.e., we set $e^\mu =0$. 
Linear theory for
this reduced (frozen-$\Delta P$) system gives the purely growing rate derived by \citet{Chandra:2017auj}
(including the anisotropic inertia, as in Eq.~(78) of
\citet{Chandra:2015iza})
\begin{align}
  &\Gamma_{\!f} = k\,\sqrt{-\frac{b^{2}+\Delta P}{\rhob+\Gamma u+b^{2}+\Delta P/3}}
  \notag\\
  &\phantom{\Gamma_{\!f}}=\frac{2\pi}{\sqrt{3673}}\simeq0.103674\label{eq.firehose_rate}
\end{align}
for $k=2\pi$, which also fixes $\Gamma=4/3$ (the choice $\Gamma=5/3$ would give
$0.095379$). We use $N=256^{2}$, $\cfl=0.4$, IMEX21 integration, and evolve
to $t=12$.
The transverse momentum includes $\delta\pi^{02}=(\Delta P/3)\delta u^{2}$, giving inertia $H_\perp:=\rhob+\Gamma u+b^{2}+\Delta P/3>0$. We freeze the anisotropy here; finite relaxation slowly changes the background.

\figref{fig.firehose} (left) shows the amplitude of the
$k_{1}=2\pi$ Fourier mode of $B^{2}$ and $u^{2}$: both grow exponentially at the
predicted rate, a linear fit giving $\Gamma=0.1036$ for each, in close agreement (within $0.12\%$ over $0.5\le t\le3.5$) with
\eqref{eq.firehose_rate}. Because the gyrotropic limit applies the field-aligned stress at all scales, truncation error seeds smaller-scale modes that grow faster (their growth rate $\propto k$); their envelope (dotted) rises above the imposed mode at late times and would drive the system nonlinear, exactly as reported by Ref. \mbox{\cite{Chandra:2017auj}}.
The right panel shows this behavior as a function of the seeded anisotropy and confirms that the measured growth rate follows the analytic dispersion relation, vanishing at the firehose threshold $-\Delta P=b^{2}$.
}

\def\eenergyoption{}
\section{Israel-Stewart Bondi solution}
\label{app:bondi-reference}
{In order to test our full {GR19M} system in curved spacetime against a reference solution, we here formulate stationary Bondi accretion for the ion
bulk, heat, and shear sector, as well as for the electron-energy system, generalizing the ideal solution \cite{Bondi:1952ni,Michel:1972}. This system is then solved numerically as a transcritical
{differential-algebraic equation (DAE) system posed as a boundary-value problem}.}

\subsection{Basic setup}

We solve the problem in Schwarzschild
coordinates, on the exterior chart $r>2M$, so $\mathcal A>0$,
\begin{align}
 {\rm d}s^2={}&-\mathcal A(r){\rm d}t^2\notag\\
 &+\mathcal A(r)^{-1}{\rm d}r^2+r^2{\rm d}\Omega^2,
 \qquad \mathcal A(r):=1-\frac{2M}{r}.
 \label{eq:bondi-schwarzschild-metric}
\end{align}

Let
\begin{align}
 v(r)&:=u^r<0, &
 R(r)&:=\sqrt{\mathcal A+v^2}=-u_t .
\end{align}
Normalization and future directedness then give
\begin{align}
 u^\mu&=\left(\frac{R}{\mathcal A},v,0,0\right), &
 s^\mu&=\left(\frac{v}{\mathcal A},R,0,0\right),
 \label{eq:bondi-us-basis}
\end{align}
where $s^\mu$ is the outward-pointing radial unit vector in the fluid rest
frame, with $s_\mu=\left(-v,R/\mathcal A,0,0\right)$,
$u^\mu u_\mu=-1$, $s^\mu s_\mu=1$ and $u^\mu s_\mu=0$.
{Stationarity and spherical symmetry force every
$u^\mu$-orthogonal vector \emph{field} compatible with the symmetry to be
proportional to $s^\mu$, hence}
\begin{align}
 q^\mu=q s^\mu,\qquad
 j^\mu=j s^\mu,\qquad
 e^\mu=E s^\mu,\qquad
 b^\mu=B s^\mu .
 \label{eq:bondi-radial-vectors}
\end{align}
The signed scalar $q$ is positive for an outward comoving heat flux.  A
monopolar magnetic field has constant radial flux,
\begin{align}
 B(r)=\frac{\Phi_B}{r^2}.
 \label{eq:bondi-monopole}
\end{align}
The field and all vector moments are parallel, so the explicit Hall and
gyro-rotation operators annihilate this background.
{For this reference the field therefore enters only through the
electromagnetic stress: the closure coefficients of
Sec.~\ref{app:bondi-closure} are functions of $\left(\rhob,T\right)$ alone.}

\subsection{Mass-flux elimination and kinematics}

Mass conservation enforces
\begin{align}
 \nabla_\mu(\rhob u^\mu)
 =\frac{1}{r^2}\frac{{\rm d}}{{\rm d}r}(r^2\rhob v)=0.
\end{align}
Defining a positive inward accretion rate gives
\begin{align}
 \dot M&=-4\pi r^2\rhob v>0, &
 \rhob&=-\frac{\dot M}{4\pi r^2v}, &
 \frac{\rhob'}{\rhob}&=-\frac{2}{r}-\frac{v'}{v}.
 \label{eq:bondi-mass-elimination}
\end{align}
Thus, $\rhob$ is eliminated algebraically and $\dot M$ becomes an
eigenparameter of the boundary-value problem.

The radial derivative of the redshift factor, the expansion, and the radial
eigenvalue of the kinematic shear are
\begin{align}
 R'&=\frac{M/r^2+vv'}{R},
 \label{eq:bondi-R-prime}\\
 \theta:=\nabla_\mu u^\mu&=v'+\frac{2v}{r},
 \label{eq:bondi-expansion}\\
 \sigma_\parallel:=s^\mu s^\nu\sigma_{\mu\nu}
 &=\frac{2}{3}\left(v'-\frac{v}{r}\right).
 \label{eq:bondi-shear-drive}
\end{align}
The tangential shear eigenvalues are $-\sigma_\parallel/2$.  Direct
evaluation of $a^\mu:=u^\nu\nabla_\nu u^\mu$ gives $a^\mu=R's^\mu$, so the
relativistic field-aligned heat drive reduces to a total derivative,
\begin{align}
 s^\mu\nabla_\mu T+T a_\mu s^\mu
 =RT'+TR'=(RT)'.
 \label{eq:bondi-heat-drive}
\end{align}

For a general equation of state $P=P(\rhob,T)$ and $e=e(\rhob,T)$, where $e$
includes rest-mass energy,
\begin{align}
 P'&=P_\rho\rhob'+P_TT', &
 e'&=e_\rho\rhob'+e_TT',
 \label{eq:bondi-eos-derivatives}
\end{align}
with $P_\rho:=\left(\partial P/\partial\rhob\right)_T$ and the analogous
definitions of $P_T$, $e_\rho$, $e_T$.  We here use a gamma-law equation of state with normalization $T=P/\rhob$,
\begin{align}
 P&=\rhob T,&
 e&=\rhob+\frac{P}{\Gamma-1},&
 c_s^2&=\frac{\Gamma P}{e+P},
 \label{eq:bondi-gamma-eos}
\end{align}
for which $P_\rho=T$, $P_T=\rhob$, $e_\rho=1+T/(\Gamma-1)$ and
$e_T=\rhob/(\Gamma-1)$.  Although the ideal seed has
$P=K_{\rm p}\rhob^\Gamma$, dissipative heating changes the entropy.

\subsection{Shear convention and directional pressures}

Introduce $\Delta^{\mu\nu}=g^{\mu\nu}+u^\mu u^\nu$ and the tangential
projector
\begin{align}
 \mathcal P_\perp^{\mu\nu}:=\Delta^{\mu\nu}-s^\mu s^\nu.
\end{align}
The unique symmetric, trace-free, flow-orthogonal stress compatible with
spherical symmetry is
\begin{align}
 \pi^{\mu\nu}
 =\varpi\left(s^\mu s^\nu-\frac{1}{2}\mathcal P_\perp^{\mu\nu}\right),
 \label{eq:bondi-shear-tensor}
\end{align}
with comoving eigenvalues $(\varpi,-\varpi/2,-\varpi/2)$.  Including the
bulk pressure $\Pi$, the material pressures parallel and perpendicular to the
radial magnetic field are
\begin{align}
 p_\parallel&=P+\Pi+\varpi,&
 p_\perp&=P+\Pi-\frac{\varpi}{2},
 \label{eq:bondi-directional-pressures}
\end{align}
so that, for the pressure-anisotropy convention used in the main text,
\begin{align}
 \Delta P:=p_\perp-p_\parallel=-\frac{3}{2}\varpi,
 \qquad
 \varpi=-\frac{2}{3}\Delta P.
 \label{eq:bondi-deltaP-map}
\end{align}

We can also express
\begin{align}
 \frac{p_\perp}{p_\parallel}
 &=\frac{P+\Pi-\varpi/2}{P+\Pi+\varpi},&
 \frac{q_\parallel}{p_\parallel}
 &=\frac{q}{P+\Pi+\varpi},
 \label{eq:bondi-directional-ratios}
\end{align}
where $q_\parallel=q^\mu s_\mu=q$.  Both ratios have a genuine pole wherever
$p_\parallel=0$; such a pole must not be interpreted as a large but finite
transport coefficient.

\subsection{Stress-energy conservation}

The material stress-energy tensor is
\begin{align}
 T_{\rm mat}^{\mu\nu}
 =e u^\mu u^\nu+(P+\Pi)\Delta^{\mu\nu}
 +q(u^\mu s^\nu+s^\mu u^\nu)+\pi^{\mu\nu}.
 \label{eq:bondi-material-stress}
\end{align}
For parallel comoving electric and magnetic fields the Poynting flux vanishes,
and with
\begin{align}
 \epsilon_{\rm EM}:=\frac{E^2+B^2}{2}
\end{align}
{the traceless
electromagnetic stress has energy density $+\epsilon_{\rm EM}$, radial pressure
$-\epsilon_{\rm EM}$ and tangential pressures $+\epsilon_{\rm EM}$.  Written out in components, we have}
\begin{align}
 T^r{}_t
 &=-vR(e+p_\parallel)-q(\mathcal A+2v^2),
 \label{eq:bondi-Trt}\\
 T^r{}_r
 &=\frac{ev^2+2qvR+p_\parallel R^2}{\mathcal A}
 -\epsilon_{\rm EM},
 \label{eq:bondi-Trr}\\
 T^t{}_t
 &=-\frac{eR^2+2qvR+p_\parallel v^2}{\mathcal A}
 -\epsilon_{\rm EM},
 \label{eq:bondi-Ttt}\\
T^\theta{}_\theta=T^\phi{}_\phi
&=p_\perp+\epsilon_{\rm EM}.
\label{eq:bondi-Tthth}
\end{align}
{The component $T^t{}_r=-T^r{}_t/\mathcal A^2$ is also nonzero; it drops out
of both rows below because $\Gamma^\mu{}_{\mu t}=0$ and $\partial_t=0$.}

The timelike Killing vector supplies an exact first integral,
\begin{align}
 \dot{\mathcal E}:=4\pi r^2T^r{}_t=\mathrm{constant},
 \qquad
 \frac{{\rm d}T^r{}_t}{{\rm d}r}+\frac{2}{r}T^r{}_t=0,
 \label{eq:bondi-energy-conservation}
\end{align}
and the radial projection of $\nabla_\mu T^\mu{}_\nu=0$ is
\begin{align}
 0={}&\left(T^r{}_r\right)'
 +\frac{2}{r}\left(T^r{}_r-T^\theta{}_\theta\right)\notag\\
 &+\frac{\mathcal A'}{2\mathcal A}
 \left(T^r{}_r-T^t{}_t\right),
 \qquad \mathcal A'=\frac{2M}{r^2}.
 \label{eq:bondi-radial-momentum}
\end{align}
{The derivatives $\left(T^r{}_t\right)'$ and $\left(T^r{}_r\right)'$ follow
by the product rule from Eqs.~\eqref{eq:bondi-mass-elimination},
\eqref{eq:bondi-R-prime} and \eqref{eq:bondi-eos-derivatives} together with
$\epsilon_{\rm EM}'=EE'-2B^2/r$, the last identity encoding $B'=-2B/r$.}

Note that, consistent with our main approximation of a small electron-to-ion mass ratio, we do not include the electron stress-energy here.

\subsection{Israel-Stewart closure}
\label{app:bondi-closure}

For a local closure, the constrained spherical reductions of the
flux-divergent Israel--Stewart equations take the scalar form
\begin{align}
 0={\cal R}_\Pi
 &:={\tau_\Pi}v\Pi'+\Pi+\zeta\theta
 +\delta_{\Pi\Pi}\Pi\theta,
 \label{eq:bondi-bulk-row}\\
 0={\cal R}_q
 &:={\tau_q}vq'+q+\kappa(RT)'
 +\delta_{qq}q\theta,
 \label{eq:bondi-heat-row}\\
 0={\cal R}_\pi
 &:={\tau_\pi}v\varpi'+\varpi+2\eta\sigma_\parallel
 +\delta_{\pi\pi}\varpi\theta,
 \label{eq:bondi-shear-row}
\end{align}
with instantaneous targets $\Pi_{\rm NS}=-\zeta\theta$,
$q_{\rm NS}=-\kappa(RT)'$ and $\varpi_{\rm NS}=-2\eta\sigma_\parallel$.
The compression coefficients are independent closure data.  For this
benchmark we choose
\begin{align}
 \delta_{\Pi\Pi}=\tau_\Pi,\qquad
 \delta_{qq}=\tau_q,\qquad
 \delta_{\pi\pi}=\tau_\pi.
 \label{eq:bondi-compression-coefficients}
\end{align}

For this test we further use
\begin{align}
 \tau_\Pi=\tau_q=\tau_\pi=\tau_i,\qquad \tau_J=\tau_e.
\end{align}
The collision time and numerical microphysics are Coulomb collision-inspired,
\begin{align}
 \tau_i&=\frac{12\pi^{3/2}}{\ln\Lambda}
 \left(\frac{q_e}{m_e}\right)^{-4}\mu^{-4}m_i^{-1}
 \frac{T^{3/2}}{Z^4\rhob},\notag\\
 \tau_e&=Z^2\sqrt{\frac{\mu}{2}}\,\tau_i,\notag\\
 m_i&=\mu=\frac{q_e}{m_e}=Z=1,\qquad \ln\Lambda=10\,,
 \label{eq:bondi-microphysics}
\end{align}
but other choices are equally admissible.

The reference family is parameterized by a dimensionless homotopy multiplier
$f\in[0,1]$:
\begin{align}
 {
 \kappa_f=fP\tau_i,\qquad
 \eta_f=fP\tau_i,\qquad
 \zeta_f=fP\tau_i .}
 \label{eq:bondi-transport-homotopy}
\end{align}

\subsubsection{{Electron sector}}
\label{app:bondi-charged}

Decompose the total current as
\begin{align}
 \mathcal J^\mu=\rhoq u^\mu+j s^\mu .
\end{align}
Strict stationarity of a radial electric field requires $\mathcal J^r=0$,
since otherwise the enclosed charge changes.  Hence,
\begin{align}
 \mathcal J^r=\rhoq v+jR=0,
 \qquad
 \mathcal J^t=-\frac{j}{v},
 \label{eq:bondi-stationary-current}
\end{align}
and, with the Maxwell sign convention of the evolution system, radial Gauss'
law is
\begin{align}
 E'+\frac{2E}{r}+\frac{j}{v}=0 .
 \label{eq:bondi-gauss-law}
\end{align}

\ifdefined\eenergyoption

Stationarity and spherical symmetry require every electron vector to be radial in the
same way as the other moments of \myeqref{eq:bondi-radial-vectors},
\begin{align}
 k^\mu&=ks^\mu, &
 \ju^\mu&=\frac{k}{w_e}s^\mu, \notag\\
 Q_e^\mu&=-Ks^\mu, &
 K&:=\frac{m_e}{q_e}k,
 \label{eq:bondi-ee-radial}
\end{align}
with $H_e=e_e+P_e$ and $w_e=H_e/(m_e\bar n_e)$.  Thus, $k$, and not the
physical amplitude $j=k/w_e$, is the current variable whose conservative
moment and flux are evolved.  Because $T_e^{\mu\nu}$ is omitted from the total
stress, Eqs.~\eqref{eq:bondi-Trt}--\eqref{eq:bondi-Tthth}, the Killing first
integral and the radial momentum equation are unchanged, and the electron
sector couples back only through the electromagnetic stress carried by $E$.

Electron energy conservation
\eqref{eq:electron-energy-summary}, with
$\nabla_\mu\left(Ks^\mu\right)=\left(KR\right)'+2KR/r$, reduces to
\begin{align}
 0=\mc{R}_{e_e}
 :=ve_e'+H_e\theta-\left(KR\right)'-\frac{2KR}{r}-\frac{Ek}{w_e},
 \label{eq:bondi-ee-energy-row}
\end{align}
where $Ek/w_e=e_\mu\ju^\mu$ is the Ohmic heating term.  Projecting
$\nabla_\mu\mc{B}^\mu{}_\nu$ on $s^\nu$ gives the momentum equation/Ohm's law,
\begin{align}
 0=\mc{R}_k
 &:=vk'+\theta k+kv'
 -\frac{q_e}{m_e}\left(H_eR'+RP_e'\right)
\notag\\&\quad +\frac{k}{\tauJ}-\omega_{pe}^2E .
 \label{eq:bondi-ee-ohm-row}
\end{align}

Using electron-number conservation and the gamma-law EOS in the energy row
gives the temperature equation \eqref{eq:bondi-main-temperature-row} quoted
in the main text.  Note that in writing the current row, we have used that the
Hall term vanishes because $b^\mu\parallel k^\mu$.

Gauss' law and the stationarity condition keep their physical-current form, so
that
\begin{align}
 E'+\frac{2E}{r}+\frac{k}{w_ev}&=0, &
 \rhoq&=-\frac{kR}{w_ev}, \notag\\
 \bar n_e=\frac{Z\rhob}{m_i}-\frac{\rhoq}{q_e}
 &=\frac{Z\rhob}{m_i}+\frac{kR}{q_e v\, w_e(\bar n_e)}, \notag\\
 P_e&=\left(\Gamma_e-1\right)\left(e_e-m_e\bar n_e\right).
 \label{eq:bondi-ee-closure}
\end{align}
{The carrier equation is implicit because $w_e$ depends on $\bar n_e$.}
For the gamma law it is solved in closed form: with
\begin{align}
 {n_Z}&:=\frac{Z\rhob}{m_i}, &
 \xi_e&:=\frac{m_ekR}{q_ev}, \notag\\
 \mathcal B_e&:=\Gamma_ee_e+\left(\Gamma_e-1\right)m_e {n_Z}-\xi_e, \notag\\
 \mathcal D_{\rm alg}&:=\mathcal B_e^2
  -4\Gamma_e\left(\Gamma_e-1\right)m_ee_e {n_Z},
 \label{eq:bondi-ee-carrier-defs}
\end{align}
the branch continuous with charge neutrality at $k=0$ is
\begin{align}
 \bar n_e=\frac{\mathcal B_e-\sqrt{\mathcal D_{\rm alg}}}
 {2\left(\Gamma_e-1\right)m_e},
 \label{eq:bondi-ee-carrier-root}
\end{align}
which tends to {$n_Z$} as $k\to0$ provided
$\Gamma_ee_e>\left(\Gamma_e-1\right)m_e {n_Z}$.  The other root approaches
$H_e=0$ and is not thermodynamically admissible.
For example, $\Gamma_e=5/3$, $e_e=5$, $m_e=q_e= {n_Z}=R=1$,
$v=-1/2$ and $k=-19/9$ give $\xi_e=38/9$ and two roots
$\bar n_e=3,25/6$. Their pressures are $4/3,5/9$, and their drift magnitudes
$|j|/(q_e\bar n_e)=1/3,19/50$ both satisfy the full electron DEC bound.
This is a local carrier example, not two global Bondi solutions.
For the selected branch require $\Gamma_e>1$, $m_e>0$, $\bar n_e>0$,
$e_e\ge m_e\bar n_e$ and the current bounds independently.
Differentiable carrier elimination further requires
$\mathcal D_{\rm alg}>0$; $H_e>0$ alone does not ensure positive pressure.  Once $\bar n_e$ is known,
$P_e$, $H_e$ and $w_e$ follow from $\left(e_e,\bar n_e\right)$ via the equation of state.

\fi

\subsection{Numerical approach}
\label{app:bondi-dae}

\subsubsection{Structure and index}

{After mass-flux elimination, the full Bondi system has 8 degrees of freedom,}
\begin{align}
 {\bm Y}{=\left(v,T,q,\varpi,\Pi,E,k,e_e\right)^{\rm T}}.
 \label{eq:bondi-unknowns}
\end{align}

In order to enforce positivity constraints on some of those variables and facilitate numerical solutions that can span multiple orders of magnitude, we choose to solve
\begin{align}
 {\widetilde{\bm Y}=\left(\ln(-v),\,\ln T,\,
 \frac{\varpi}{P},\,\frac{\Pi}{P},\,E,\,k,\,\ln e_e\right)^{\rm T}}\,.
 \label{eq:bondi-ee-nodal}
\end{align}

Because $T^r{}_t$ is linear in $q$, the Killing integral can be solved for the heat
flux at every radius,
\begin{align}
 q=-\frac{v R\left(e+P+{\Pi}+\varpi\right)
 +\dot{\mathcal E}/\left(4\pi r^{2}\right)}
 {\mathcal A+2v^{2}},
 \label{eq:bondi-heat-reconstruction}
\end{align}
{the radial total pressure $P_r=P+\Pi+\varpi$ of
Eq.~\eqref{eq:bondi-directional-pressures} appearing in the numerator.}  Thus
$q$ becomes a dependent variable, and we have eliminated it from the solution vector, ${\widetilde{\bm Y}}$.

{The Killing-energy and radial-momentum equations
\eqref{eq:bondi-energy-conservation} and
\eqref{eq:bondi-radial-momentum}, the ion dissipative equations
\eqref{eq:bondi-bulk-row}--\eqref{eq:bondi-shear-row}, Gauss' law
\eqref{eq:bondi-gauss-law}, and the electron equations
\eqref{eq:bondi-ee-energy-row} and \eqref{eq:bondi-ee-ohm-row} form the residual
$\bm{\mathcal R}(r,\bm Y,\bm Y')=0$.  Every closure coefficient is a local
algebraic function of $(r,\rhob,T,B)$, so this residual is affine in
$\bm Y'$,}
\begin{align}
 \mathsf M(r,\bm Y;\bm\lambda)\,\bm Y'
 =\bm S(r,\bm Y;\bm\lambda),
 \label{eq:bondi-affine-dae}
\end{align}
where $\bm\lambda$ collects $\dot M$, $\dot{\mathcal E}$, $\Phi_B$, $f$ and the
microscopic parameters.  
{The matrix $\mathsf M$ and source $\bm S$ can therefore be assembled
without numerical differentiation by evaluating the residual on derivative
basis vectors,}
\begin{align}
 \mathsf M_{ak}&=\bm{\mathcal R}_a(r,\bm Y,\bm e_k)
 -\bm{\mathcal R}_a(r,\bm Y,\bm 0), \notag \\
 S_a&=-\bm{\mathcal R}_a(r,\bm Y,\bm 0)\,.
 \label{eq:bondi-matrix-assembly}
\end{align}

{All eight rows carry a derivative.  Wherever $\mathsf M$ is invertible,
Eq.~\eqref{eq:bondi-affine-dae} is therefore the regular implicit {ordinary differential equation (ODE)}
$\bm Y'=\mathsf M^{-1}\bm S$.  Critical points occur where this derivative
matrix loses rank and becomes non-invertible, e.g., in the case of a perfect single fluid
at the ideal sonic point.}

Eliminating $q$ changes the eight-variable radial matrix. Set $x=\ln r$ and $\bm Y=\bm\Phi(x,\widetilde{\bm Y})$ using Eq.~\eqref{eq:bondi-heat-reconstruction}, and let $\mathsf L$ remove the Killing-energy row. Below, primes on $\widetilde{\bm Y}$ denote $\mathrm d/\mathrm dx$; primes on unreduced physical variables still denote $\mathrm d/\mathrm dr$. Then
\begin{align}
 {\bm Y'_r}
 &{=\frac{\bm\Phi_x+\bm\Phi_{\widetilde Y}
       \widetilde{\bm Y}'}{r}},\notag\\
 {\widetilde{\mathsf M}}
 &{=\frac{1}{r}\mathsf L\mathsf M\bm\Phi_{\widetilde Y}},\notag\\
 {\widetilde{\bm S}}
 &{=\mathsf L\left(\bm S-\frac{1}{r}\mathsf M\bm\Phi_x\right)},\notag\\
 {\widetilde{\bm{\mathcal R}}}
 &{=\widetilde{\mathsf M}\widetilde{\bm Y}'
       -\widetilde{\bm S}=0}.
 \label{eq:bondi-reduced-coordinate-map}
\end{align}
The term $\bm\Phi_x$ includes the radial dependence of the mass and Killing integrals. This seven-variable map is regular for $T,P,e_e>0$, $v<0$ and $\mathcal A+2v^2>0$.

\subsubsection{Critical points}

A stationary characteristic surface occurs where
\begin{align}
 \det\mathsf M=0 .
\end{align}

For a differentiable carrier branch, $e_T\ne0$, positive relaxation times and the exterior inflow assumptions above, Appendix~\ref{app:char_causality} gives
\begin{align}
 \det\mathsf M\ \propto\
 \mc Q_\parallel^c\!\left(\frac{v}{R}\right)
 \mc C_3\!\left(\frac{v}{R}\right).
 \label{eq:bondi-ee-critical-factor}
\end{align}

The first factor is the longitudinal ion quartic
$\mc Q_\parallel^c$ of Eq.~\eqref{eq.Qparc};
the second is the electron cubic \eqref{eq.Qe}, evaluated with
\begin{align}
 c_{\rm se}^2&:=P_{e,e}+\frac{\bar n_e}{H_e}P_{e,n}, &
 \sigma&:=\frac{\bar n_e}{H_e}P_{e,n}, &
 v_{\rm d}&:=\frac{k}{\omega_e},
 \label{eq:bondi-ee-critical-defs}
\end{align}
Here $\omega_e=(q_e/m_e)H_e$. The carrier-elimination factor is
\begin{align}
 {g_{\rm alg}}
 &{=1-\frac{R}{v}v_{\rm d}(1-\sigma)
   =\frac{\sqrt{\mathcal D_{\rm alg}}}{H_e}>0}
 \label{eq:bondi-carrier-derivative-domain}
\end{align}
on the minus branch. The determinant prefactor contains $1/g_{\rm alg}$, so Eq.~\eqref{eq:bondi-ee-critical-factor} does not give a regular eliminated matrix at $\mathcal D_{\rm alg}=0$.
The cubic roots $a_{e,j}$ place the electron critical surfaces at
\begin{align}
 v_{c,e,j}=a_{e,j}
 \sqrt{\frac{\mathcal A\left(r_{c,e,j}\right)}{1-a_{e,j}^2}} .
 \label{eq:bondi-ee-critical-velocity}
\end{align}
At zero drift $\mc C_3(a)=a\left(a^2-c_{\rm se}^2\right)$ and the nonmaterial
condition reduces to $v^2/R^2=c_{\rm se}^2$.  At finite drift all three roots
must be inspected, and only a subluminal root carrying the sign of the inflow
gives a stationary critical surface.

The ion critical speeds are the roots of Eq.~\eqref{eq.Qparc}; their reality
and interval conditions are Eqs.~\eqref{eq.Qpar.real} and
\eqref{eq.causpar}.

At such a point $\mathsf M$ drops rank and $\bm Y'$ is no longer determined by
Eq.~\eqref{eq:bondi-affine-dae}.  If $\bm\ell$ is a corresponding left null
vector, $\bm\ell^{\rm T}\mathsf M=0$.  Contracting
Eq.~\eqref{eq:bondi-affine-dae} with $\bm\ell$ annihilates the left-hand side,
so a solution exists only if
\begin{align}
  {\ \bm\ell^{\rm T}\bm S=0\ }
 \label{eq:bondi-critical-solvability}
\end{align}
there.  Equation~\eqref{eq:bondi-critical-solvability} is a solvability
condition; it fixes \emph{where} the surface may
be crossed, and it is what turns the critical radii into eigenparameters.
The slope itself is recovered by differentiating
Eq.~\mbox{\eqref{eq:bondi-affine-dae}} along the solution and projecting again with
$\bm\ell$.  Writing
$\mathcal D:=\mathrm{d}/\mathrm{d}x$ for the derivative along the trajectory,
\begin{align}
 \bm\ell_x^{\rm T}(\mathcal D\widetilde{\mathsf M})
   \widetilde{\bm Y}'
 +\bm\ell_x^{\rm T}\widetilde{\mathsf M}\mathcal D\widetilde{\bm Y}'
 =\bm\ell_x^{\rm T}\mathcal D\widetilde{\bm S},\end{align}
and the second term vanishes at the critical point for a twice-differentiable crossing.  The singular highest derivative is
thereby eliminated \emph{without} introducing a second derivative, and one
finite scalar condition on $\widetilde{\bm Y}'$ remains,
\begin{align}
 \bm\ell_x^{\rm T}
 \left.\frac{\mathrm{d}}{\mathrm{d}x}
 \widetilde{\bm{\mathcal R}}(x,\widetilde{\bm Y}(x),\widetilde{\bm Y}')
 \right|_{\widetilde{\bm Y}'\ \mathrm{fixed}}=0.\label{eq:bondi-finite-slope}
\end{align}
This is the differential-algebraic analogue of l'H\^opital's rule that needs to be used at the critical point.
At a simple critical point the reduced matrix has corank one. The scalar slope condition can be quadratic and can have several finite roots or none. For example, $yy'=x$ admits slopes $\pm1$ at the origin, while $x^2y'=x$ satisfies the undifferentiated solvability condition but has no finite crossing slope.

The seven-field solver uses constant coordinate scales $E/E_0$ and $k/k_0$, with $E_0=10^{-6}$ and $k_0=10^{-5}$. It divides the first four residuals by $(P/r,P,P,P)$ and applies additional positive local scales to the three electron/Maxwell rows. Finite nonzero row and column scalings preserve rank; the source uses the same row scaling. The solver finds a left singular vector by {singular-value decomposition (SVD)}, fixes its sign by its largest component, and transforms it back through the row scaling before projecting the raw residual. At corank one, a nonzero adjugate row also gives an exact left null vector. At higher corank the adjugate vanishes.
In the full solver,
Eq.~\eqref{eq:bondi-finite-slope} is evaluated as a centered difference in $x$
of step $10^{-5}$, displacing the state along $\widetilde{\bm Y}'$ and the
radius by $e^{\pm10^{-5}}$ at fixed $\widetilde{\bm Y}'$.

At a critical node the {$n_{\rm r}$} raw collocation equations have rank
{$n_{\rm r}-1$}, leaving one Hermite derivative unconstrained.  The redundant row is identified as the
component in which $\bm\ell_x$ for the same residual scaling is largest, and that single row --- and only that
row --- is replaced by Eq.~\eqref{eq:bondi-finite-slope}.

\subsubsection{Numerical implementation}

{The full solution is obtained on a mesh with $N=353$ points, using
cubic Hermite polynomials in $x=\ln r$ for the seven independent nodal values
and their slopes.  The three critical radii are unknowns and split
$[r_{\rm in},r_{\rm out}]$ into four radial segments; the critical points are
mesh points and move as the solution converges.}

With $h$ as the local grid spacing in $x$, the interpolant gives
\begin{align}
 \widetilde{\bm Y}_{i+1/2}
 &=\frac{\widetilde{\bm Y}_i+\widetilde{\bm Y}_{i+1}}{2}
 +\frac{h}{8}\left(\widetilde{\bm Y}'_i-\widetilde{\bm Y}'_{i+1}\right), \\
 \widetilde{\bm Y}'_{i+1/2}
 &=\frac{3}{2}\,
 \frac{\widetilde{\bm Y}_{i+1}-\widetilde{\bm Y}_i}{h}
 -\frac{\widetilde{\bm Y}'_i+\widetilde{\bm Y}'_{i+1}}{4},
 \label{eq:bondi-hermite-midpoint}
\end{align}
evaluated at $r_{i+1/2}=\sqrt{r_ir_{i+1}}$.  Imposing the DAE at both places
is what makes the node slopes genuine unknowns rather than finite-difference
reconstructions.

At the outer boundary, we impose the ideal solution {of
Sec.~\ref{sec.bondiDissipative}}, namely
$\ln\left[\rhob(r_{\rm out})/\rhob^{\rm ideal}\right]=0$ and
$\ln\left[T(r_{\rm out})/T_{\rm out}\right]=0$, together with
$\Pi(r_{\rm out})=\Pi_{\rm NS}=-\zeta\theta$ and the electron conditions
specified below.

{We solve the square Hermite collocation system with a sparse Newton
iteration and finite-difference Jacobians.  The accepted solution has a
maximum nodal residual $2.94\times10^{-11}$; the largest residual at four
additional points per Hermite interval is $3.69\times10^{-8}$.}

\subsection{Example solution used in the main text}

{For the reference solution presented in Sec.~\ref{sec.bondiDissipative}, we
use $N=353$ Hermite nodes in four moving segments, with 65, 65, 97, and 129 nodes before endpoint identification.  The
first three segments are uniform in $\ln r$.  In the last segment the uniform
coordinate $t\in[0,1]$ is mapped to $1-(1-t)^4$, in order to resolve the thin outer
electron skin-depth layer produced by the condition $k(r_{\rm out})=0$.}

{At $r_{\rm out}=160r_g$ we impose
{$\rhob=7.85513475790494\times10^{-3}$} and
$T_i=1.09395837263986\times10^{-1}$, and fix $\Pi$ to its local
Navier--Stokes value.  The corresponding monopole flux is
{$\Phi_B=8.82451710776$}.
We use {$\Gamma_e=5/3$}, set
$T_e(r_{\rm out})=T_i(r_{\rm out})$, and hence}
\begin{align}
 {P_{e,{\rm out}}}&{=\bar n_{e,{\rm out}}T_{i,{\rm out}}} ,\notag\\
 {e_{e,{\rm out}}}&{=m_e\bar n_{e,{\rm out}}
 +\frac{P_{e,{\rm out}}}{\Gamma_e-1}} ,\notag\\
 {k_{\rm out}}&{=0}.
 \label{eq:bondi-hot-outer-data}
\end{align}

{The heat flux is again fixed to its Navier--Stokes target and determines
$\dot{\mathcal E}$ through Eq.~\eqref{eq:bondi-heat-reconstruction}; 
$E$ is not prescribed at either boundary.  The boundary-value solve adjusts
its profile so that Gauss' law, the electron momentum row, and the critical
solvability condition hold simultaneously while $k(r_{\rm out})=0$.  It is
therefore the ambipolar field that balances the electron pressure and inertia
against the ion-supported flow.  The initial electron state is set from the neutral adiabat
$P_e/P_{e,{\rm out}}=(\bar n_e/\bar n_{e,{\rm out}})^{\Gamma_e}$.  
Substitution into Eq.~\eqref{eq:bondi-ee-energy-row} makes
$ve_e'+H_e\theta$ vanish identically when $E=k=0$.  
}

{The resulting eigenvalues and fitted critical radii are}
\begin{align}
 {\dot M}&{=2.788588317606},&
 {\dot{\mathcal E}}&{=3.975179490890},\notag\\
 {r_{c,e}/r_g}&{=3.340518319},&
 {r_{c,i}/r_g}&{=4.436458563},\notag \\
 {r_{c,q}/r_g}&{=18.055474904}.
 \label{eq:bondi-hot-reference-values}
\end{align}
This configuration has exactly three critical points:
$r_{c,e}$ is the electron-acoustic radius, $r_{c,i}$ the ion-acoustic radius,
and $r_{c,q}$ the ion second-sound radius associated with heat conduction.

\section{{GR19M} on axisymmetric grids}
\label{app:refmetric}

To mitigate the high-resolution requirements of some test problems, we adopt axisymmetry using the reference-metric formulation of relativistic hydrodynamics \cite{Montero:2013pca}. We briefly summarize this approach and then list the axis boundary conditions adopted in the code for future reference.
We adopt mesh coordinates $\left(R, z\right)$ with out-of-plane direction $\varphi$, where
$R$ denotes the cylindrical radius. 

\subsection{Reference metric formulation}

We introduce the flat reference three-metric in cylindrical coordinates
on the positive-radius chart $R>0$,
\begin{align}
  \hat{\gamma}_{ij} = {\rm diag} \left( 1, 1, R^{2} \right) \, ,
  \qquad \sqrt{\hat{\gamma}} = R \, ,
  \label{eq.refmetric}
\end{align}
and further introduce the weight-zero reference density
\begin{align}
  \hat{U} := \sqrt{\frac{\gamma}{\hat{\gamma}}} \, U
  = \frac{\widetilde{U}}{R} \,,
  \label{eq.refstate}
\end{align}
of a state vector $U$.
Because $\hat{\gamma}$ carries the entire coordinate degeneracy of the polar
map, $\hat{U}$ remains finite and generically nonzero on the axis. Care is required for tensors stored on the axis. While every quantity evolved in regular GRMHD is either a scalar or a vector, the shear is a true tensor. Therefore, its azimuthal component, stored as
\begin{align}
  \bar{\pi}^{\varphi\varphi} := R^{2} \, \pi^{\varphi\varphi} \, ,
  \label{eq.pizzbar}
\end{align}
requires an $R^2$ regularization factor.
This is the regularized azimuthal component. For a smooth axisymmetric
tensor, $\bar\pi^{\varphi\varphi}-\pi^{RR}=\mathcal O(R^2)$ near the axis;
parity alone does not enforce this condition. Divisions by $R$ or $R^2$ are
evaluated off axis, with axis values defined through regular limits.

We write the update as a Cartesian flux difference plus a pointwise connection:
\begin{align}
  \p_{t} \, \hat{U} + \p_{i} \, \hat{F}^{i} = \hat{S} + \mc{C} \, ,
  \label{eq.refsplit}
\end{align}
In the component formulas below, $F^i$ denotes the weight-zero flux
$\widehat F^i=\widetilde F^i/R$. For scalar balance rows,
$\widehat S=\widetilde S/R$. For rows with free indices, $\widehat S$
contains the remaining source after the displayed reference-index
connections have been extracted.
The connection $\mc{C}$ collects the terms generated by
\myeqref{eq.refstate}, which are differentiated explicitly with respect to the reference metric.  For a scalar reference density, it is simply
$\mc{C} = - F^{R}/R$.  Objects carrying a free index acquire additional
contributions.  Denoting by $F^{R}$ and $F^{\varphi}$ the radial and azimuthal
reference-weighted physical fluxes evaluated at the cell center, a lower covector
$\left( V_{R}, V_{z}, V_{\varphi} \right)$ --- which is the index structure of
$S_{j}$, $Q_{\nu}$, $Z^{\kappa}_{\nu}$, and {$k_{*\nu}$} --- picks up
\begin{subequations}
\begin{align}
  \mc{C}_{R} &= \frac{1}{R} \left( F^{\varphi}_{\varphi}
    - F^{R}_{R} \right) \, , \\
  \mc{C}_{z} &= - \frac{1}{R} \, F^{R}_{z} \, , \\
  \mc{C}_{\varphi} &= - R \, F^{\varphi}_{R} \, ,
\end{align}
\label{eq.covconn}
\end{subequations}
whereas an upper vector $\left( V^{R}, V^{z}, V^{\varphi} \right)$ --- the
structure of $B^{i}$, $E^{i}$, and $\pi^{t i}$ --- obtains
\begin{subequations}
\begin{align}
  \mc{C}^{R} &= - \frac{1}{R} \, F^{R\,R} + R \, F^{\varphi\,\varphi} \, , \\
  \mc{C}^{z} &= - \frac{1}{R} \, F^{R\,z} \, , \\
  \mc{C}^{\varphi} &= - \frac{1}{R} \left( 2 \, F^{R\,\varphi}
    + F^{\varphi\,R} \right) \, .
\end{align}
\label{eq.vecconn}
\end{subequations}
The symmetric spatial tensor $\pi^{ij}$ is treated in the same way, with
$\bar{\pi}^{\varphi\varphi}$ of \myeqref{eq.pizzbar} replacing
$\pi^{\varphi\varphi}$.  

The baryon density $\widetilde D_{\rm b}$, electron carrier
$\widetilde D_e$, and electron-energy density
$\widetilde{\mathcal E}_{e*}$ are time components of scalar conservation
currents. For all three, the
pointwise $-F^R/R$ term is replaced by the exact
annular finite-volume divergence for radial transport on a uniform grid with cell measure $R_{\rm c}\Delta R$,
\begin{align}
  \left.\p_t\hat U\right|_{\rm radial}
  = \frac{R_{\rm lo} \, F_{\rm lo}
    - R_{\rm hi} \, F_{\rm hi}}{R_{\rm c} \, \Delta R} \, ,
  \label{eq.annular}
\end{align}
which, given that the Cartesian face difference is applied separately, amounts
to the connection
\begin{align}
  \mc{C} = - \frac{F_{\rm lo} + F_{\rm hi}}{2 \, R_{\rm c}}
  \label{eq.annularconn}
\end{align}
built from the same numerical face fluxes rather than from a
resolution-dependent cell-centered approximation to $-F^R/R$.  Thus, the
carrier and electron-energy variables use one radial measure and
preserve constant
electron energy per carrier.

The same annular measure is used at AMR interfaces.  Spatial prolongation and
restriction act on $R\widehat D_{\rm b}$, $R\widehat D_e$, and
$R\widehat{\mathcal E}_{e*}$, and divide by the target cell radius only after
the conservative interpolation or average.  Their radial flux-register entries
are likewise $RF^R$, so the reflux correction is converted back to the public
weight-zero representation before it is committed.
Restriction and interpolation preserve proportional fields when both use the same linear or positively homogeneous limited operator. Across the reflected boundary, the oriented measure $R\hat U$ is odd for an even public scalar. This signed $R$ differs from the $|R|$ conversion used for native-field curls below.

\subsection{Boundary conditions on the axis}

A subtle aspect of axisymmetric evolution is the behavior on the axis. The axis both acts as a face term in the flux update and provides the interface to domain ghost cells beyond it; both roles require special treatment.

\emph{Ghost-cell boundary condition.} Every component of the state is filled by reflection about
$R=0$.  A component is reflected \emph{odd} if it carries an odd number of
radial indices and \emph{even} otherwise.  For the full {19-moment} system the
odd set is
\begin{align}
  \left\{ W v^{R}, \, B^{R}, \, E^{R}, \, q_{R}, \, \pi^{tR}, \,
  \pi^{Rz}, \, \pi^{R\varphi}, \, Z^{\kappa}_{R}, \, \ju_{R} \right\} \, ,
  \label{eq.axisodd}
\end{align}
and all remaining components, including $\Db$, $\tau$, the carrier
$\widetilde{D}_{e}$, both cleaning scalars and
$\bar{\pi}^{\varphi\varphi}$, are reflected even.  The same rule is applied to
the auxiliary reconstruction variables, of which only the radial component of
the electron four-current is odd. 

\emph{Flux boundary condition.} Parity of the reconstructed left state on the axis alone is not sufficient, because the reference
fluxes do not all vanish on the axis; the even radial momentum flux, for
instance, includes a non-vanishing pressure term. At the axis face we therefore form the
regular one-sided limit of the state by setting its odd components to zero,
and evaluate the physical radial flux $F_{\rm c}$ from that state at the first
cell center, $R_{\rm c} = \Delta R/2$.  
With $F_{1}$ the ordinary numerical flux at the
first interior face, the axis flux is closed as
\begin{align}
  F_{0} = 2 \, F_{\rm c} - F_{1} \, ,
  \label{eq.axisregular}
\end{align}
for which the paired difference and connection reduce identically to
$\Delta R^{-1} \left( F_{0} - F_{1} \right) - F_{\rm c}/R_{\rm c}
= - 2 \, F_{1} / \Delta R$, that is, to the annular finite-volume divergence
with zero native flux through $R=0$.

Care is required in the Maxwell update. The Levi-Civita terms in the
radial Maxwell flux leave the native $B^{\varphi}$ and $E^{\varphi}$ fluxes
finite on the axis, so dividing by $\sqrt{\hat{\gamma}} = R$ produces a
removable $1/R$ singularity in the reference flux.  Writing $F^{R} = A/R + C \, R + \mc{O}\left(R^{3}\right)$, the paired
operator $- \p_{R} F^{R} - F^{R}/R$ has a finite part $-2C$, which is recovered
by the quadratic even extrapolation
\begin{align}
  F_{0} = \frac{5}{3} \left( 2 \, F_{\rm c} - F_{1} \right) \, .
  \label{eq.axissingular}
\end{align}
{Equation~\eqref{eq.axissingular} preserves a uniform axial field exactly and reproduces the exact first-cell curl for $E_z=A+CR^2$.}
For unblended fluxes, the corresponding radial rates
are
{
\begin{align}
 \left.\p_t\hat U\right|_{\rm paired}
 &=\begin{cases}
 -2F_1/\Delta R,&\text{regular flux},\\
 4(F_{\rm c}-2F_1)/(3\Delta R),&\text{singular flux}.
 \end{cases}
 \label{eq.audit_axis_paired_rates}
\end{align}}
These rates are algebraically equivalent to
Eqs.~\mbox{\eqref{eq.axisregular}} and \mbox{\eqref{eq.axissingular}}; free-index
connections and physical sources remain separate. The singular formula has
$\mathcal O(\Delta R^2)$ error for a smooth quartic axial field.
Floating input-flux errors are not removed by the algebraic cancellation.

The single exception to this closure is $\bar{\pi}^{\varphi\varphi}$, whose inverse
reference-to-native mapping contributes $+F^{R}/R$ from its density
weight, in addition to the tensor-index connections, and which therefore retains
the parity closure, in which the axis flux is taken from the reflected
state for odd components and set to zero for even ones.

{
The reference density is the nonnegative radius
$\sqrt{\hat{\gamma}}=|R|$.  We therefore undo the reference densitization with
$|R|$, rather than signed $R$, when evaluating Amp\`ere's curl across the
reflected boundary.  The tensor parity is already carried by the ghost state.
Using signed $R$ would apply it twice and produce an
$\mathcal{O}(\Delta R^{-1})$ current in the first radial cell.

We also apply one strictly local regularization during implicit recovery.  If
an isolated failed cell $(0,j)$ next to the axis has no admissible electric field, while its two immediate $z$-neighbors have finite, physical values and lie outside the excision region (e.g., outside the black-hole horizon), we recover it by averaging:
\begin{align}
  E^{i}_{0,j} = \frac{1}{2}
  \left(E^{i}_{0,j-1}+E^{i}_{0,j+1}\right)\label{eq.axisendpointreconstruction}
\end{align}
rather than setting it to zero or to the local ideal value.
We recover the physical field using the failed cell's metric weight. Directly averaging $E^i$ agrees when all three weights agree, but generally differs otherwise. For a smooth stored field on equally spaced axial cells, this gives a second-order midpoint estimate. Neighboring fields may violate the failed cell's carrier, current, or electric-field bounds, so the candidate must pass local coupled recovery and stage-closure checks before use.
}

\section{The FIDO tetrad in Kerr--Schild coordinates}
\label{app:fidotetrad}

Measurements of local quantities, e.g., the reconnection rate or inflow velocities into a current sheet, are meaningful only in a locally orthonormal frame.
We therefore use a tetrad observer. For completeness, we list the tetrad used in our simulations for cylindrical Kerr--Schild coordinates
$\left(R, z, \phi\right)$. 
Expressed in 3+1 variables, the Kerr--Schild metric is defined as
\begin{subequations}
\begin{align}
  \gamma_{ij} &= \hat\gamma_{ij} + 2 H \, \ell_{i} \, \ell_{j} \, , \\
  \gamma^{ij} &= \hat\gamma^{ij}
    - \frac{2H}{1+2H} \, \ell^{i} \, \ell^{j} \, , \\
  \alpha &= \left(1+2H\right)^{-1/2} \, , \qquad
  \beta^{i} = \frac{2H}{1+2H} \, \ell^{i} \, , \\
  \sqrt{\gamma} &= R \, \sqrt{1+2H} = \frac{R}{\alpha} \, .
\end{align}
\label{eq.ksthreeplusone}
\end{subequations}
Here, we have used the Kerr--Schild
radius
\begin{align}
  r^{2} = \frac{1}{2} \left[ \left(R^{2}+z^{2}-a^{2}\right)
  + \sqrt{\left(R^{2}+z^{2}-a^{2}\right)^{2} + 4 \, a^{2} z^{2}} \, \right] ,
  \label{eq.ksradius}
\end{align}
as well as

\begin{align*}
{H = \frac{r^{3}}{r^{4}+a^{2}z^{2}} \, , \quad
  \ell_{R} = \frac{r R}{r^{2}+a^{2}} \, ,} \\
{\ell_{z} = \frac{z}{r} \, , \quad
{\ell_{\phi} = -\frac{a R^{2}}{r^{2}+a^{2}} \, ,}
}\end{align*}

in units $G=c=M=1$.  Raising with the flat reference metric
$\hat\gamma_{ij}={\rm diag}\left(1,1,R^{2}\right)$ of
\myeqref{eq.refmetric} gives $\ell^{R}=\ell_{R}$, $\ell^{z}=\ell_{z}$, and
{{$\ell^{\phi}=-a/\left(r^{2}+a^{2}\right)$}}, normalized so that
$\hat\gamma_{ij}\ell^{i}\ell^{j}=1$.

The FIDO is the observer whose four-velocity is the future unit normal to the
slice,
\begin{align}
  n_{\mu} = \left(-\alpha, 0, 0, 0\right) \, , \qquad
  n^{\mu} = \frac{1}{\alpha} \left(1, -\beta^{i}\right) \, ,
  \label{eq.fidonormal}
\end{align}
which sets the tetrad time direction to $e_{\left(0\right)}^{\mu}=n^{\mu}$.
The spatial tetrad vectors are
\begin{subequations}
\begin{align}
  e_{\left(1\right)}^{i} &= \frac{\delta^{i}_{\phantom{i}R}}
    {\sqrt{\gamma_{RR}}} \, ,
  \label{eq.tetradtangent} \\
  e_{\left(2\right)i} &= \frac{\delta_{i}^{\phantom{i}z}}
    {\sqrt{\gamma^{zz}}} \, ,
  \qquad \textrm{equivalently} \qquad
  e_{\left(2\right)}^{i} = \frac{\gamma^{iz}}{\sqrt{\gamma^{zz}}} \, ,
  \label{eq.tetradnormal} \\
  e_{\left(3\right)}^{i} &= \frac{1}{\sqrt{S}}
    \left( -\frac{\gamma_{R\phi}}{\gamma_{RR}}, \, 0, \, 1 \right) \, ,
  \qquad
  S = \gamma_{\phi\phi} - \frac{\gamma_{R\phi}^{2}}{\gamma_{RR}} \, .
  \label{eq.tetradbinormal}
\end{align}
\label{eq.fidotetrad}
\end{subequations}

\bibliography{inspires,non_inspire}

\begin{thebibliography}{175}%
\makeatletter
\providecommand \@ifxundefined [1]{%
 \@ifx{#1\undefined}
}%
\providecommand \@ifnum [1]{%
 \ifnum #1\expandafter \@firstoftwo
 \else \expandafter \@secondoftwo
 \fi
}%
\providecommand \@ifx [1]{%
 \ifx #1\expandafter \@firstoftwo
 \else \expandafter \@secondoftwo
 \fi
}%
\providecommand \natexlab [1]{#1}%
\providecommand \enquote  [1]{``#1''}%
\providecommand \bibnamefont  [1]{#1}%
\providecommand \bibfnamefont [1]{#1}%
\providecommand \citenamefont [1]{#1}%
\providecommand \href@noop [0]{\@secondoftwo}%
\providecommand \href [0]{\begingroup \@sanitize@url \@href}%
\providecommand \@href[1]{\@@startlink{#1}\@@href}%
\providecommand \@@href[1]{\endgroup#1\@@endlink}%
\providecommand \@sanitize@url [0]{\catcode `\\12\catcode `\$12\catcode
  `\&12\catcode `\#12\catcode `\^12\catcode `\_12\catcode `\%12\relax}%
\providecommand \@@startlink[1]{}%
\providecommand \@@endlink[0]{}%
\providecommand \url  [0]{\begingroup\@sanitize@url \@url }%
\providecommand \@url [1]{\endgroup\@href {#1}{\urlprefix }}%
\providecommand \urlprefix  [0]{URL }%
\providecommand \Eprint [0]{\href }%
\providecommand \doibase [0]{https://doi.org/}%
\providecommand \selectlanguage [0]{\@gobble}%
\providecommand \bibinfo  [0]{\@secondoftwo}%
\providecommand \bibfield  [0]{\@secondoftwo}%
\providecommand \translation [1]{[#1]}%
\providecommand \BibitemOpen [0]{}%
\providecommand \bibitemStop [0]{}%
\providecommand \bibitemNoStop [0]{.\EOS\space}%
\providecommand \EOS [0]{\spacefactor3000\relax}%
\providecommand \BibitemShut  [1]{\csname bibitem#1\endcsname}%
\let\auto@bib@innerbib\@empty
\bibitem [{\citenamefont {Davis}\ and\ \citenamefont
  {Tchekhovskoy}(2020)}]{Davis:2020wea}%
  \BibitemOpen
  \bibfield  {author} {\bibinfo {author} {\bibfnamefont {S.~W.}\ \bibnamefont
  {Davis}}\ and\ \bibinfo {author} {\bibfnamefont {A.}~\bibnamefont
  {Tchekhovskoy}},\ }\bibfield  {title} {\bibinfo {title}
  {{Magnetohydrodynamics Simulations of Active Galactic Nucleus Disks and
  Jets}},\ }\href {https://doi.org/10.1146/annurev-astro-081817-051905}
  {\bibfield  {journal} {\bibinfo  {journal} {Ann. Rev. Astron. Astrophys.}\
  }\textbf {\bibinfo {volume} {58}},\ \bibinfo {pages} {407} (\bibinfo {year}
  {2020})},\ \Eprint {https://arxiv.org/abs/2101.08839} {arXiv:2101.08839
  [astro-ph.HE]} \BibitemShut {NoStop}%
\bibitem [{\citenamefont {{Philippov}}\ and\ \citenamefont
  {{Kramer}}(2022)}]{2022ARA&A..60..495P}%
  \BibitemOpen
  \bibfield  {author} {\bibinfo {author} {\bibfnamefont {A.}~\bibnamefont
  {{Philippov}}}\ and\ \bibinfo {author} {\bibfnamefont {M.}~\bibnamefont
  {{Kramer}}},\ }\bibfield  {title} {\bibinfo {title} {{Pulsar Magnetospheres
  and Their Radiation}},\ }\href
  {https://doi.org/10.1146/annurev-astro-052920-112338} {\bibfield  {journal}
  {\bibinfo  {journal} {\araa}\ }\textbf {\bibinfo {volume} {60}},\ \bibinfo
  {pages} {495} (\bibinfo {year} {2022})}\BibitemShut {NoStop}%
\bibitem [{\citenamefont {Arons}(2012)}]{arons2012pulsar}%
  \BibitemOpen
  \bibfield  {author} {\bibinfo {author} {\bibfnamefont {J.}~\bibnamefont
  {Arons}},\ }\bibfield  {title} {\bibinfo {title} {Pulsar wind nebulae as
  cosmic pevatrons: a current sheet’s tale},\ }\href@noop {} {\bibfield
  {journal} {\bibinfo  {journal} {Space Science Reviews}\ }\textbf {\bibinfo
  {volume} {173}},\ \bibinfo {pages} {341} (\bibinfo {year}
  {2012})}\BibitemShut {NoStop}%
\bibitem [{\citenamefont {Andersson}\ and\ \citenamefont
  {Comer}(2021)}]{Andersson:2020phh}%
  \BibitemOpen
  \bibfield  {author} {\bibinfo {author} {\bibfnamefont {N.}~\bibnamefont
  {Andersson}}\ and\ \bibinfo {author} {\bibfnamefont {G.~L.}\ \bibnamefont
  {Comer}},\ }\bibfield  {title} {\bibinfo {title} {{Relativistic fluid
  dynamics: physics for many different scales}},\ }\href
  {https://doi.org/10.1007/s41114-021-00031-6} {\bibfield  {journal} {\bibinfo
  {journal} {Living Rev. Rel.}\ }\textbf {\bibinfo {volume} {24}},\ \bibinfo
  {pages} {3} (\bibinfo {year} {2021})},\ \Eprint
  {https://arxiv.org/abs/2008.12069} {arXiv:2008.12069 [gr-qc]} \BibitemShut
  {NoStop}%
\bibitem [{\citenamefont {Shen}\ \emph {et~al.}(2025)\citenamefont {Shen},
  \citenamefont {Chen}, \citenamefont {Huang}, \citenamefont {Ma},
  \citenamefont {Tang},\ and\ \citenamefont {Wang}}]{Shen:2025unr}%
  \BibitemOpen
  \bibfield  {author} {\bibinfo {author} {\bibfnamefont {D.}~\bibnamefont
  {Shen}}, \bibinfo {author} {\bibfnamefont {J.}~\bibnamefont {Chen}}, \bibinfo
  {author} {\bibfnamefont {X.-G.}\ \bibnamefont {Huang}}, \bibinfo {author}
  {\bibfnamefont {Y.-G.}\ \bibnamefont {Ma}}, \bibinfo {author} {\bibfnamefont
  {A.}~\bibnamefont {Tang}},\ and\ \bibinfo {author} {\bibfnamefont
  {G.}~\bibnamefont {Wang}},\ }\bibfield  {title} {\bibinfo {title} {{A Review
  of Intense Electromagnetic Fields in Heavy-Ion Collisions: Theoretical
  Predictions and Experimental Results}},\ }\href
  {https://doi.org/10.34133/research.0726} {\bibfield  {journal} {\bibinfo
  {journal} {Research}\ }\textbf {\bibinfo {volume} {8}},\ \bibinfo {pages}
  {0726} (\bibinfo {year} {2025})},\ \Eprint {https://arxiv.org/abs/2512.00739}
  {arXiv:2512.00739 [nucl-ex]} \BibitemShut {NoStop}%
\bibitem [{\citenamefont {Romatschke}\ and\ \citenamefont
  {Romatschke}(2019)}]{Romatschke:2017ejr}%
  \BibitemOpen
  \bibfield  {author} {\bibinfo {author} {\bibfnamefont {P.}~\bibnamefont
  {Romatschke}}\ and\ \bibinfo {author} {\bibfnamefont {U.}~\bibnamefont
  {Romatschke}},\ }\href {https://doi.org/10.1017/9781108651998} {\emph
  {\bibinfo {title} {{Relativistic Fluid Dynamics In and Out of
  Equilibrium}}}},\ Cambridge Monographs on Mathematical Physics\ (\bibinfo
  {publisher} {Cambridge University Press},\ \bibinfo {year} {2019})\ \Eprint
  {https://arxiv.org/abs/1712.05815} {arXiv:1712.05815 [nucl-th]} \BibitemShut
  {NoStop}%
\bibitem [{\citenamefont {Rezzolla}\ and\ \citenamefont
  {Zanotti}(2013)}]{rezzolla2013relativistic}%
  \BibitemOpen
  \bibfield  {author} {\bibinfo {author} {\bibfnamefont {L.}~\bibnamefont
  {Rezzolla}}\ and\ \bibinfo {author} {\bibfnamefont {O.}~\bibnamefont
  {Zanotti}},\ }\href@noop {} {\emph {\bibinfo {title} {Relativistic
  hydrodynamics}}}\ (\bibinfo  {publisher} {Oxford University Press},\ \bibinfo
  {year} {2013})\BibitemShut {NoStop}%
\bibitem [{\citenamefont {Anile}(1989)}]{Anile1989}%
  \BibitemOpen
  \bibfield  {author} {\bibinfo {author} {\bibfnamefont {A.~M.}\ \bibnamefont
  {Anile}},\ }\href@noop {} {\emph {\bibinfo {title} {Relativistic Fluids and
  Magneto-Fluids: With Applications in Astrophysics and Plasma Physics}}}\
  (\bibinfo  {publisher} {Cambridge University Press},\ \bibinfo {year}
  {1989})\BibitemShut {NoStop}%
\bibitem [{\citenamefont {Sironi}\ and\ \citenamefont
  {Spitkovsky}(2014)}]{Sironi:2014jfa}%
  \BibitemOpen
  \bibfield  {author} {\bibinfo {author} {\bibfnamefont {L.}~\bibnamefont
  {Sironi}}\ and\ \bibinfo {author} {\bibfnamefont {A.}~\bibnamefont
  {Spitkovsky}},\ }\bibfield  {title} {\bibinfo {title} {{Relativistic
  Reconnection: an Efficient Source of Non-Thermal Particles}},\ }\href
  {https://doi.org/10.1088/2041-8205/783/1/L21} {\bibfield  {journal} {\bibinfo
   {journal} {Astrophys. J. Lett.}\ }\textbf {\bibinfo {volume} {783}},\
  \bibinfo {pages} {L21} (\bibinfo {year} {2014})},\ \Eprint
  {https://arxiv.org/abs/1401.5471} {arXiv:1401.5471 [astro-ph.HE]}
  \BibitemShut {NoStop}%
\bibitem [{\citenamefont {Heinz}\ and\ \citenamefont
  {Snellings}(2013)}]{Heinz:2013th}%
  \BibitemOpen
  \bibfield  {author} {\bibinfo {author} {\bibfnamefont {U.}~\bibnamefont
  {Heinz}}\ and\ \bibinfo {author} {\bibfnamefont {R.}~\bibnamefont
  {Snellings}},\ }\bibfield  {title} {\bibinfo {title} {{Collective flow and
  viscosity in relativistic heavy-ion collisions}},\ }\href
  {https://doi.org/10.1146/annurev-nucl-102212-170540} {\bibfield  {journal}
  {\bibinfo  {journal} {Ann. Rev. Nucl. Part. Sci.}\ }\textbf {\bibinfo
  {volume} {63}},\ \bibinfo {pages} {123} (\bibinfo {year} {2013})},\ \Eprint
  {https://arxiv.org/abs/1301.2826} {arXiv:1301.2826 [nucl-th]} \BibitemShut
  {NoStop}%
\bibitem [{\citenamefont {Gershman}\ \emph {et~al.}(2024)\citenamefont
  {Gershman}, \citenamefont {Fuselier}, \citenamefont {Cohen}, \citenamefont
  {Turner}, \citenamefont {Liu}, \citenamefont {Chen}, \citenamefont {Phan},
  \citenamefont {Stawarz}, \citenamefont {DiBraccio}, \citenamefont {Masters}
  \emph {et~al.}}]{gershman2024magnetic}%
  \BibitemOpen
  \bibfield  {author} {\bibinfo {author} {\bibfnamefont {D.~J.}\ \bibnamefont
  {Gershman}}, \bibinfo {author} {\bibfnamefont {S.~A.}\ \bibnamefont
  {Fuselier}}, \bibinfo {author} {\bibfnamefont {I.~J.}\ \bibnamefont {Cohen}},
  \bibinfo {author} {\bibfnamefont {D.~L.}\ \bibnamefont {Turner}}, \bibinfo
  {author} {\bibfnamefont {Y.-H.}\ \bibnamefont {Liu}}, \bibinfo {author}
  {\bibfnamefont {L.-J.}\ \bibnamefont {Chen}}, \bibinfo {author}
  {\bibfnamefont {T.~D.}\ \bibnamefont {Phan}}, \bibinfo {author}
  {\bibfnamefont {J.~E.}\ \bibnamefont {Stawarz}}, \bibinfo {author}
  {\bibfnamefont {G.~A.}\ \bibnamefont {DiBraccio}}, \bibinfo {author}
  {\bibfnamefont {A.}~\bibnamefont {Masters}}, \emph {et~al.},\ }\bibfield
  {title} {\bibinfo {title} {Magnetic reconnection at planetary bodies and
  astrospheres},\ }\href@noop {} {\bibfield  {journal} {\bibinfo  {journal}
  {Space Science Reviews}\ }\textbf {\bibinfo {volume} {220}},\ \bibinfo
  {pages} {7} (\bibinfo {year} {2024})}\BibitemShut {NoStop}%
\bibitem [{\citenamefont {Akiyama}\ \emph {et~al.}(2019)\citenamefont {Akiyama}
  \emph {et~al.}}]{EventHorizonTelescope:2019dse}%
  \BibitemOpen
  \bibfield  {author} {\bibinfo {author} {\bibfnamefont {K.}~\bibnamefont
  {Akiyama}} \emph {et~al.} (\bibinfo {collaboration} {Event Horizon
  Telescope}),\ }\bibfield  {title} {\bibinfo {title} {{First M87 Event Horizon
  Telescope Results. I. The Shadow of the Supermassive Black Hole}},\ }\href
  {https://doi.org/10.3847/2041-8213/ab0ec7} {\bibfield  {journal} {\bibinfo
  {journal} {Astrophys. J. Lett.}\ }\textbf {\bibinfo {volume} {875}},\
  \bibinfo {pages} {L1} (\bibinfo {year} {2019})},\ \Eprint
  {https://arxiv.org/abs/1906.11238} {arXiv:1906.11238 [astro-ph.GA]}
  \BibitemShut {NoStop}%
\bibitem [{\citenamefont {Akiyama}\ \emph {et~al.}(2022)\citenamefont {Akiyama}
  \emph {et~al.}}]{EventHorizonTelescope:2022wkp}%
  \BibitemOpen
  \bibfield  {author} {\bibinfo {author} {\bibfnamefont {K.}~\bibnamefont
  {Akiyama}} \emph {et~al.} (\bibinfo {collaboration} {Event Horizon
  Telescope}),\ }\bibfield  {title} {\bibinfo {title} {{First Sagittarius A*
  Event Horizon Telescope Results. I. The Shadow of the Supermassive Black Hole
  in the Center of the Milky Way}},\ }\href
  {https://doi.org/10.3847/2041-8213/ac6674} {\bibfield  {journal} {\bibinfo
  {journal} {Astrophys. J. Lett.}\ }\textbf {\bibinfo {volume} {930}},\
  \bibinfo {pages} {L12} (\bibinfo {year} {2022})},\ \Eprint
  {https://arxiv.org/abs/2311.08680} {arXiv:2311.08680 [astro-ph.HE]}
  \BibitemShut {NoStop}%
\bibitem [{\citenamefont {Kunz}\ \emph {et~al.}(2014)\citenamefont {Kunz},
  \citenamefont {Schekochihin},\ and\ \citenamefont {Stone}}]{Kunz:2014qha}%
  \BibitemOpen
  \bibfield  {author} {\bibinfo {author} {\bibfnamefont {M.~W.}\ \bibnamefont
  {Kunz}}, \bibinfo {author} {\bibfnamefont {A.~A.}\ \bibnamefont
  {Schekochihin}},\ and\ \bibinfo {author} {\bibfnamefont {J.~M.}\ \bibnamefont
  {Stone}},\ }\bibfield  {title} {\bibinfo {title} {{Firehose and Mirror
  Instabilities in a Collisionless Shearing Plasma}},\ }\href
  {https://doi.org/10.1103/PhysRevLett.112.205003} {\bibfield  {journal}
  {\bibinfo  {journal} {Phys. Rev. Lett.}\ }\textbf {\bibinfo {volume} {112}},\
  \bibinfo {pages} {205003} (\bibinfo {year} {2014})},\ \Eprint
  {https://arxiv.org/abs/1402.0010} {arXiv:1402.0010 [astro-ph.HE]}
  \BibitemShut {NoStop}%
\bibitem [{\citenamefont {Kempski}\ \emph {et~al.}(2019)\citenamefont
  {Kempski}, \citenamefont {Quataert}, \citenamefont {Squire},\ and\
  \citenamefont {Kunz}}]{Kempski:2019tvo}%
  \BibitemOpen
  \bibfield  {author} {\bibinfo {author} {\bibfnamefont {P.}~\bibnamefont
  {Kempski}}, \bibinfo {author} {\bibfnamefont {E.}~\bibnamefont {Quataert}},
  \bibinfo {author} {\bibfnamefont {J.}~\bibnamefont {Squire}},\ and\ \bibinfo
  {author} {\bibfnamefont {M.~W.}\ \bibnamefont {Kunz}},\ }\bibfield  {title}
  {\bibinfo {title} {{Shearing-box simulations of MRI-driven turbulence in
  weakly collisional accretion discs}},\ }\href
  {https://doi.org/10.1093/mnras/stz1111} {\bibfield  {journal} {\bibinfo
  {journal} {Mon. Not. Roy. Astron. Soc.}\ }\textbf {\bibinfo {volume} {486}},\
  \bibinfo {pages} {4013} (\bibinfo {year} {2019})},\ \Eprint
  {https://arxiv.org/abs/1901.04504} {arXiv:1901.04504 [astro-ph.HE]}
  \BibitemShut {NoStop}%
\bibitem [{\citenamefont {Foucart}\ \emph {et~al.}(2017)\citenamefont
  {Foucart}, \citenamefont {Chandra}, \citenamefont {Gammie}, \citenamefont
  {Quataert},\ and\ \citenamefont {Tchekhovskoy}}]{Foucart:2017axc}%
  \BibitemOpen
  \bibfield  {author} {\bibinfo {author} {\bibfnamefont {F.}~\bibnamefont
  {Foucart}}, \bibinfo {author} {\bibfnamefont {M.}~\bibnamefont {Chandra}},
  \bibinfo {author} {\bibfnamefont {C.~F.}\ \bibnamefont {Gammie}}, \bibinfo
  {author} {\bibfnamefont {E.}~\bibnamefont {Quataert}},\ and\ \bibinfo
  {author} {\bibfnamefont {A.}~\bibnamefont {Tchekhovskoy}},\ }\bibfield
  {title} {\bibinfo {title} {{How important is non-ideal physics in simulations
  of sub-Eddington accretion on to spinning black holes?}},\ }\href
  {https://doi.org/10.1093/mnras/stx1368} {\bibfield  {journal} {\bibinfo
  {journal} {Mon. Not. Roy. Astron. Soc.}\ }\textbf {\bibinfo {volume} {470}},\
  \bibinfo {pages} {2240} (\bibinfo {year} {2017})},\ \Eprint
  {https://arxiv.org/abs/1706.01533} {arXiv:1706.01533 [astro-ph.HE]}
  \BibitemShut {NoStop}%
\bibitem [{\citenamefont {Dhruv}\ \emph {et~al.}(2025)\citenamefont {Dhruv},
  \citenamefont {Prather}, \citenamefont {Chandra}, \citenamefont {Joshi},\
  and\ \citenamefont {Gammie}}]{Dhruv:2025isy}%
  \BibitemOpen
  \bibfield  {author} {\bibinfo {author} {\bibfnamefont {V.}~\bibnamefont
  {Dhruv}}, \bibinfo {author} {\bibfnamefont {B.}~\bibnamefont {Prather}},
  \bibinfo {author} {\bibfnamefont {M.}~\bibnamefont {Chandra}}, \bibinfo
  {author} {\bibfnamefont {A.~V.}\ \bibnamefont {Joshi}},\ and\ \bibinfo
  {author} {\bibfnamefont {C.~F.}\ \bibnamefont {Gammie}},\ }\bibfield  {title}
  {\bibinfo {title} {{Electromagnetic Observables of Weakly Collisional Black
  Hole Accretion}},\ }\href {https://doi.org/10.3847/2041-8213/ae1236}
  {\bibfield  {journal} {\bibinfo  {journal} {Astrophys. J. Lett.}\ }\textbf
  {\bibinfo {volume} {993}},\ \bibinfo {pages} {L33} (\bibinfo {year}
  {2025})},\ \Eprint {https://arxiv.org/abs/2510.11365} {arXiv:2510.11365
  [astro-ph.HE]} \BibitemShut {NoStop}%
\bibitem [{\citenamefont {Gammie}\ \emph {et~al.}(2003)\citenamefont {Gammie},
  \citenamefont {McKinney},\ and\ \citenamefont {Toth}}]{Gammie:2003rj}%
  \BibitemOpen
  \bibfield  {author} {\bibinfo {author} {\bibfnamefont {C.~F.}\ \bibnamefont
  {Gammie}}, \bibinfo {author} {\bibfnamefont {J.~C.}\ \bibnamefont
  {McKinney}},\ and\ \bibinfo {author} {\bibfnamefont {G.}~\bibnamefont
  {Toth}},\ }\bibfield  {title} {\bibinfo {title} {{HARM: A Numerical scheme
  for general relativistic magnetohydrodynamics}},\ }\href
  {https://doi.org/10.1086/374594} {\bibfield  {journal} {\bibinfo  {journal}
  {Astrophys. J.}\ }\textbf {\bibinfo {volume} {589}},\ \bibinfo {pages} {444}
  (\bibinfo {year} {2003})},\ \Eprint {https://arxiv.org/abs/astro-ph/0301509}
  {arXiv:astro-ph/0301509} \BibitemShut {NoStop}%
\bibitem [{\citenamefont {De~Villiers}\ and\ \citenamefont
  {Hawley}(2003)}]{DeVilliers:2002ab}%
  \BibitemOpen
  \bibfield  {author} {\bibinfo {author} {\bibfnamefont {J.-P.}\ \bibnamefont
  {De~Villiers}}\ and\ \bibinfo {author} {\bibfnamefont {J.~F.}\ \bibnamefont
  {Hawley}},\ }\bibfield  {title} {\bibinfo {title} {{A Numerical method for
  general relativistic magnetohydrodynamics}},\ }\href
  {https://doi.org/10.1086/373949} {\bibfield  {journal} {\bibinfo  {journal}
  {Astrophys. J.}\ }\textbf {\bibinfo {volume} {589}},\ \bibinfo {pages} {458}
  (\bibinfo {year} {2003})},\ \Eprint {https://arxiv.org/abs/astro-ph/0210518}
  {arXiv:astro-ph/0210518} \BibitemShut {NoStop}%
\bibitem [{\citenamefont {Porth}\ \emph {et~al.}(2019)\citenamefont {Porth}
  \emph {et~al.}}]{EventHorizonTelescope:2019pcy}%
  \BibitemOpen
  \bibfield  {author} {\bibinfo {author} {\bibfnamefont {O.}~\bibnamefont
  {Porth}} \emph {et~al.} (\bibinfo {collaboration} {Event Horizon
  Telescope}),\ }\bibfield  {title} {\bibinfo {title} {{The Event Horizon
  General Relativistic Magnetohydrodynamic Code Comparison Project}},\ }\href
  {https://doi.org/10.3847/1538-4365/ab29fd} {\bibfield  {journal} {\bibinfo
  {journal} {Astrophys. J. Suppl.}\ }\textbf {\bibinfo {volume} {243}},\
  \bibinfo {pages} {26} (\bibinfo {year} {2019})},\ \Eprint
  {https://arxiv.org/abs/1904.04923} {arXiv:1904.04923 [astro-ph.HE]}
  \BibitemShut {NoStop}%
\bibitem [{\citenamefont {Zenitani}\ and\ \citenamefont
  {Hoshino}(2007)}]{Zenitani:2007}%
  \BibitemOpen
  \bibfield  {author} {\bibinfo {author} {\bibfnamefont {S.}~\bibnamefont
  {Zenitani}}\ and\ \bibinfo {author} {\bibfnamefont {M.}~\bibnamefont
  {Hoshino}},\ }\bibfield  {title} {\bibinfo {title} {{Particle Acceleration
  and Magnetic Dissipation in Relativistic Current Sheet of Pair Plasmas}},\
  }\href {https://doi.org/10.1086/522226} {\bibfield  {journal} {\bibinfo
  {journal} {Astrophys. J.}\ }\textbf {\bibinfo {volume} {670}},\ \bibinfo
  {pages} {702} (\bibinfo {year} {2007})}\BibitemShut {NoStop}%
\bibitem [{\citenamefont {{Barkov}}\ \emph {et~al.}(2014)\citenamefont
  {{Barkov}}, \citenamefont {{Komissarov}}, \citenamefont {{Korolev}},\ and\
  \citenamefont {{Zankovich}}}]{barkov2014twofluid}%
  \BibitemOpen
  \bibfield  {author} {\bibinfo {author} {\bibfnamefont {M.}~\bibnamefont
  {{Barkov}}}, \bibinfo {author} {\bibfnamefont {S.~S.}\ \bibnamefont
  {{Komissarov}}}, \bibinfo {author} {\bibfnamefont {V.}~\bibnamefont
  {{Korolev}}},\ and\ \bibinfo {author} {\bibfnamefont {A.}~\bibnamefont
  {{Zankovich}}},\ }\bibfield  {title} {\bibinfo {title} {{A multidimensional
  numerical scheme for two-fluid relativistic magneto-hydrodynamics}},\ }\href
  {https://doi.org/10.1093/mnras/stt2247} {\bibfield  {journal} {\bibinfo
  {journal} {\mnras}\ }\textbf {\bibinfo {volume} {438}},\ \bibinfo {pages}
  {704} (\bibinfo {year} {2014})},\ \Eprint {https://arxiv.org/abs/1309.5221}
  {arXiv:1309.5221 [astro-ph.HE]} \BibitemShut {NoStop}%
\bibitem [{\citenamefont {Zenitani}\ \emph
  {et~al.}(2009{\natexlab{a}})\citenamefont {Zenitani}, \citenamefont {Hesse},\
  and\ \citenamefont {Klimas}}]{Zenitani:2009bj}%
  \BibitemOpen
  \bibfield  {author} {\bibinfo {author} {\bibfnamefont {S.}~\bibnamefont
  {Zenitani}}, \bibinfo {author} {\bibfnamefont {M.}~\bibnamefont {Hesse}},\
  and\ \bibinfo {author} {\bibfnamefont {A.}~\bibnamefont {Klimas}},\
  }\bibfield  {title} {\bibinfo {title} {{Two-Fluid MHD Simulations of
  Relativistic Magnetic Reconnection}},\ }\href
  {https://doi.org/10.1088/0004-637X/722/1/968} {\bibfield  {journal} {\bibinfo
   {journal} {Astrophys. J.}\ }\textbf {\bibinfo {volume} {696}},\ \bibinfo
  {pages} {1385} (\bibinfo {year} {2009}{\natexlab{a}})},\ \bibinfo {note}
  {[Erratum: Astrophys.J. 722, 968--969 (2010)]},\ \Eprint
  {https://arxiv.org/abs/0902.2074} {arXiv:0902.2074 [astro-ph.HE]}
  \BibitemShut {NoStop}%
\bibitem [{\citenamefont {Koide}(2009)}]{Koide:2009yx}%
  \BibitemOpen
  \bibfield  {author} {\bibinfo {author} {\bibfnamefont {S.}~\bibnamefont
  {Koide}},\ }\bibfield  {title} {\bibinfo {title} {{Generalized Relativistic
  Magnetohydrodynamic Equations for Pair and Electron-Ion Plasmas}},\ }\href
  {https://doi.org/10.1088/0004-637X/696/2/2220} {\bibfield  {journal}
  {\bibinfo  {journal} {Astrophys. J.}\ }\textbf {\bibinfo {volume} {696}},\
  \bibinfo {pages} {2220} (\bibinfo {year} {2009})},\ \bibinfo {note}
  {[Erratum: Astrophys.J. 701, 2033 (2009)]},\ \Eprint
  {https://arxiv.org/abs/0902.4292} {arXiv:0902.4292 [astro-ph.HE]}
  \BibitemShut {NoStop}%
\bibitem [{\citenamefont {{Barkov}}\ and\ \citenamefont
  {{Komissarov}}(2016)}]{BarkovKomissarov:2016}%
  \BibitemOpen
  \bibfield  {author} {\bibinfo {author} {\bibfnamefont {M.~V.}\ \bibnamefont
  {{Barkov}}}\ and\ \bibinfo {author} {\bibfnamefont {S.~S.}\ \bibnamefont
  {{Komissarov}}},\ }\bibfield  {title} {\bibinfo {title} {{Relativistic
  tearing and drift-kink instabilities in two-fluid simulations}},\ }\href
  {https://doi.org/10.1093/mnras/stw362} {\bibfield  {journal} {\bibinfo
  {journal} {\mnras}\ }\textbf {\bibinfo {volume} {458}},\ \bibinfo {pages}
  {1939} (\bibinfo {year} {2016})},\ \Eprint {https://arxiv.org/abs/1602.02848}
  {arXiv:1602.02848 [astro-ph.HE]} \BibitemShut {NoStop}%
\bibitem [{\citenamefont {Chandra}\ \emph {et~al.}(2015)\citenamefont
  {Chandra}, \citenamefont {Gammie}, \citenamefont {Foucart},\ and\
  \citenamefont {Quataert}}]{Chandra:2015iza}%
  \BibitemOpen
  \bibfield  {author} {\bibinfo {author} {\bibfnamefont {M.}~\bibnamefont
  {Chandra}}, \bibinfo {author} {\bibfnamefont {C.~F.}\ \bibnamefont {Gammie}},
  \bibinfo {author} {\bibfnamefont {F.}~\bibnamefont {Foucart}},\ and\ \bibinfo
  {author} {\bibfnamefont {E.}~\bibnamefont {Quataert}},\ }\bibfield  {title}
  {\bibinfo {title} {{An Extended Magnetohydrodynamics Model for Relativistic
  Weakly Collisional Plasmas}},\ }\href
  {https://doi.org/10.1088/0004-637X/810/2/162} {\bibfield  {journal} {\bibinfo
   {journal} {Astrophys. J.}\ }\textbf {\bibinfo {volume} {810}},\ \bibinfo
  {pages} {162} (\bibinfo {year} {2015})},\ \Eprint
  {https://arxiv.org/abs/1508.00878} {arXiv:1508.00878 [astro-ph.HE]}
  \BibitemShut {NoStop}%
\bibitem [{\citenamefont {Foucart}\ \emph {et~al.}(2016)\citenamefont
  {Foucart}, \citenamefont {Chandra}, \citenamefont {Gammie},\ and\
  \citenamefont {Quataert}}]{Foucart:2015cws}%
  \BibitemOpen
  \bibfield  {author} {\bibinfo {author} {\bibfnamefont {F.}~\bibnamefont
  {Foucart}}, \bibinfo {author} {\bibfnamefont {M.}~\bibnamefont {Chandra}},
  \bibinfo {author} {\bibfnamefont {C.~F.}\ \bibnamefont {Gammie}},\ and\
  \bibinfo {author} {\bibfnamefont {E.}~\bibnamefont {Quataert}},\ }\bibfield
  {title} {\bibinfo {title} {{Evolution of Accretion Discs around a Kerr Black
  Hole using Extended Magnetohydrodynamics}},\ }\href
  {https://doi.org/10.1093/mnras/stv2687} {\bibfield  {journal} {\bibinfo
  {journal} {Mon. Not. Roy. Astron. Soc.}\ }\textbf {\bibinfo {volume} {456}},\
  \bibinfo {pages} {1332} (\bibinfo {year} {2016})},\ \Eprint
  {https://arxiv.org/abs/1511.04445} {arXiv:1511.04445 [astro-ph.HE]}
  \BibitemShut {NoStop}%
\bibitem [{\citenamefont {{Sironi}}\ \emph {et~al.}(2025)\citenamefont
  {{Sironi}}, \citenamefont {{Uzdensky}},\ and\ \citenamefont
  {{Giannios}}}]{2025ARA&A..63..127S}%
  \BibitemOpen
  \bibfield  {author} {\bibinfo {author} {\bibfnamefont {L.}~\bibnamefont
  {{Sironi}}}, \bibinfo {author} {\bibfnamefont {D.~A.}\ \bibnamefont
  {{Uzdensky}}},\ and\ \bibinfo {author} {\bibfnamefont {D.}~\bibnamefont
  {{Giannios}}},\ }\bibfield  {title} {\bibinfo {title} {{Relativistic Magnetic
  Reconnection in Astrophysical Plasmas: A Powerful Mechanism of Nonthermal
  Emission}},\ }\href {https://doi.org/10.1146/annurev-astro-020325-115713}
  {\bibfield  {journal} {\bibinfo  {journal} {\araa}\ }\textbf {\bibinfo
  {volume} {63}},\ \bibinfo {pages} {127} (\bibinfo {year} {2025})},\ \Eprint
  {https://arxiv.org/abs/2506.02101} {arXiv:2506.02101 [astro-ph.HE]}
  \BibitemShut {NoStop}%
\bibitem [{\citenamefont {Uzdensky}\ \emph {et~al.}(2010)\citenamefont
  {Uzdensky}, \citenamefont {Loureiro},\ and\ \citenamefont
  {Schekochihin}}]{Uzdensky:2010ts}%
  \BibitemOpen
  \bibfield  {author} {\bibinfo {author} {\bibfnamefont {D.~A.}\ \bibnamefont
  {Uzdensky}}, \bibinfo {author} {\bibfnamefont {N.~F.}\ \bibnamefont
  {Loureiro}},\ and\ \bibinfo {author} {\bibfnamefont {A.~A.}\ \bibnamefont
  {Schekochihin}},\ }\bibfield  {title} {\bibinfo {title} {{Fast magnetic
  reconnection in the plasmoid-dominated regime}},\ }\href
  {https://doi.org/10.1103/PhysRevLett.105.235002} {\bibfield  {journal}
  {\bibinfo  {journal} {Phys. Rev. Lett.}\ }\textbf {\bibinfo {volume} {105}},\
  \bibinfo {pages} {235002} (\bibinfo {year} {2010})},\ \Eprint
  {https://arxiv.org/abs/1008.3330} {arXiv:1008.3330 [astro-ph.SR]}
  \BibitemShut {NoStop}%
\bibitem [{\citenamefont {Guo}\ \emph {et~al.}(2015)\citenamefont {Guo},
  \citenamefont {Liu}, \citenamefont {Daughton},\ and\ \citenamefont
  {Li}}]{Guo:2015cua}%
  \BibitemOpen
  \bibfield  {author} {\bibinfo {author} {\bibfnamefont {F.}~\bibnamefont
  {Guo}}, \bibinfo {author} {\bibfnamefont {Y.-H.}\ \bibnamefont {Liu}},
  \bibinfo {author} {\bibfnamefont {W.}~\bibnamefont {Daughton}},\ and\
  \bibinfo {author} {\bibfnamefont {H.}~\bibnamefont {Li}},\ }\bibfield
  {title} {\bibinfo {title} {{Particle Acceleration and Plasma Dynamics during
  Magnetic Reconnection in the Magnetically-dominated Regime}},\ }\href
  {https://doi.org/10.1088/0004-637X/806/2/167} {\bibfield  {journal} {\bibinfo
   {journal} {Astrophys. J.}\ }\textbf {\bibinfo {volume} {806}},\ \bibinfo
  {pages} {167} (\bibinfo {year} {2015})},\ \Eprint
  {https://arxiv.org/abs/1504.02193} {arXiv:1504.02193 [astro-ph.HE]}
  \BibitemShut {NoStop}%
\bibitem [{\citenamefont {Palenzuela}\ \emph {et~al.}(2009)\citenamefont
  {Palenzuela}, \citenamefont {Lehner}, \citenamefont {Reula},\ and\
  \citenamefont {Rezzolla}}]{Palenzuela:2008sf}%
  \BibitemOpen
  \bibfield  {author} {\bibinfo {author} {\bibfnamefont {C.}~\bibnamefont
  {Palenzuela}}, \bibinfo {author} {\bibfnamefont {L.}~\bibnamefont {Lehner}},
  \bibinfo {author} {\bibfnamefont {O.}~\bibnamefont {Reula}},\ and\ \bibinfo
  {author} {\bibfnamefont {L.}~\bibnamefont {Rezzolla}},\ }\bibfield  {title}
  {\bibinfo {title} {{Beyond ideal MHD: towards a more realistic modeling of
  relativistic astrophysical plasmas}},\ }\href
  {https://doi.org/10.1111/j.1365-2966.2009.14454.x} {\bibfield  {journal}
  {\bibinfo  {journal} {Mon. Not. Roy. Astron. Soc.}\ }\textbf {\bibinfo
  {volume} {394}},\ \bibinfo {pages} {1727} (\bibinfo {year} {2009})},\ \Eprint
  {https://arxiv.org/abs/0810.1838} {arXiv:0810.1838 [astro-ph]} \BibitemShut
  {NoStop}%
\bibitem [{\citenamefont {Ripperda}\ \emph {et~al.}(2017)\citenamefont
  {Ripperda}, \citenamefont {Porth}, \citenamefont {Xia},\ and\ \citenamefont
  {Keppens}}]{Ripperda:2016sxe}%
  \BibitemOpen
  \bibfield  {author} {\bibinfo {author} {\bibfnamefont {B.}~\bibnamefont
  {Ripperda}}, \bibinfo {author} {\bibfnamefont {O.}~\bibnamefont {Porth}},
  \bibinfo {author} {\bibfnamefont {C.}~\bibnamefont {Xia}},\ and\ \bibinfo
  {author} {\bibfnamefont {R.}~\bibnamefont {Keppens}},\ }\bibfield  {title}
  {\bibinfo {title} {{Reconnection and particle acceleration in interacting
  flux ropes {\textendash} I. Magnetohydrodynamics and test particles in
  2.5D}},\ }\href {https://doi.org/10.1093/mnras/stx379} {\bibfield  {journal}
  {\bibinfo  {journal} {Mon. Not. Roy. Astron. Soc.}\ }\textbf {\bibinfo
  {volume} {467}},\ \bibinfo {pages} {3279} (\bibinfo {year} {2017})},\ \Eprint
  {https://arxiv.org/abs/1611.09966} {arXiv:1611.09966 [astro-ph.HE]}
  \BibitemShut {NoStop}%
\bibitem [{\citenamefont {Goodbred}\ and\ \citenamefont
  {Liu}(2022)}]{Goodbred:2022}%
  \BibitemOpen
  \bibfield  {author} {\bibinfo {author} {\bibfnamefont {M.}~\bibnamefont
  {Goodbred}}\ and\ \bibinfo {author} {\bibfnamefont {Y.-H.}\ \bibnamefont
  {Liu}},\ }\bibfield  {title} {\bibinfo {title} {{First-Principles Theory of
  the Relativistic Magnetic Reconnection Rate in Astrophysical Pair Plasmas}},\
  }\href {https://doi.org/10.1103/PhysRevLett.129.265101} {\bibfield  {journal}
  {\bibinfo  {journal} {Phys. Rev. Lett.}\ }\textbf {\bibinfo {volume} {129}},\
  \bibinfo {pages} {265101} (\bibinfo {year} {2022})},\ \Eprint
  {https://arxiv.org/abs/2301.12111} {arXiv:2301.12111 [astro-ph.HE]}
  \BibitemShut {NoStop}%
\bibitem [{\citenamefont {Moran}\ \emph {et~al.}(2025)\citenamefont {Moran},
  \citenamefont {Sironi}, \citenamefont {Levis}, \citenamefont {Ripperda},
  \citenamefont {Most},\ and\ \citenamefont {Selvi}}]{Moran:2025aqb}%
  \BibitemOpen
  \bibfield  {author} {\bibinfo {author} {\bibfnamefont {A.}~\bibnamefont
  {Moran}}, \bibinfo {author} {\bibfnamefont {L.}~\bibnamefont {Sironi}},
  \bibinfo {author} {\bibfnamefont {A.}~\bibnamefont {Levis}}, \bibinfo
  {author} {\bibfnamefont {B.}~\bibnamefont {Ripperda}}, \bibinfo {author}
  {\bibfnamefont {E.~R.}\ \bibnamefont {Most}},\ and\ \bibinfo {author}
  {\bibfnamefont {S.}~\bibnamefont {Selvi}},\ }\bibfield  {title} {\bibinfo
  {title} {{Effective Resistivity in Relativistic Reconnection: A Prescription
  Based on Fully Kinetic Simulations}},\ }\href
  {https://doi.org/10.3847/2041-8213/ada158} {\bibfield  {journal} {\bibinfo
  {journal} {Astrophys. J. Lett.}\ }\textbf {\bibinfo {volume} {978}},\
  \bibinfo {pages} {L45} (\bibinfo {year} {2025})},\ \Eprint
  {https://arxiv.org/abs/2501.04800} {arXiv:2501.04800 [astro-ph.HE]}
  \BibitemShut {NoStop}%
\bibitem [{\citenamefont {Ripperda}\ \emph {et~al.}(2026)\citenamefont
  {Ripperda}, \citenamefont {Grehan}, \citenamefont {Moran}, \citenamefont
  {Selvi}, \citenamefont {Sironi}, \citenamefont {Philippov}, \citenamefont
  {Bransgrove},\ and\ \citenamefont
  {Porth}}]{Ripperda:2026EffectiveResistivity}%
  \BibitemOpen
  \bibfield  {author} {\bibinfo {author} {\bibfnamefont {B.}~\bibnamefont
  {Ripperda}}, \bibinfo {author} {\bibfnamefont {M.~P.}\ \bibnamefont
  {Grehan}}, \bibinfo {author} {\bibfnamefont {A.}~\bibnamefont {Moran}},
  \bibinfo {author} {\bibfnamefont {S.}~\bibnamefont {Selvi}}, \bibinfo
  {author} {\bibfnamefont {L.}~\bibnamefont {Sironi}}, \bibinfo {author}
  {\bibfnamefont {A.}~\bibnamefont {Philippov}}, \bibinfo {author}
  {\bibfnamefont {A.}~\bibnamefont {Bransgrove}},\ and\ \bibinfo {author}
  {\bibfnamefont {O.}~\bibnamefont {Porth}},\ }\bibfield  {title} {\bibinfo
  {title} {Magnetic reconnection with a 0.1 rate: Effective resistivity in
  general relativistic magnetohydrodynamics},\ }\href@noop {} {\bibfield
  {journal} {\bibinfo  {journal} {arXiv e-prints}\ } (\bibinfo {year}
  {2026})},\ \Eprint {https://arxiv.org/abs/2601.02460} {arXiv:2601.02460
  [astro-ph.HE]} \BibitemShut {NoStop}%
\bibitem [{\citenamefont {Selvi}\ \emph {et~al.}(2023)\citenamefont {Selvi},
  \citenamefont {Porth}, \citenamefont {Ripperda}, \citenamefont {Bacchini},
  \citenamefont {Sironi},\ and\ \citenamefont {Keppens}}]{Selvi:2022uup}%
  \BibitemOpen
  \bibfield  {author} {\bibinfo {author} {\bibfnamefont {S.}~\bibnamefont
  {Selvi}}, \bibinfo {author} {\bibfnamefont {O.}~\bibnamefont {Porth}},
  \bibinfo {author} {\bibfnamefont {B.}~\bibnamefont {Ripperda}}, \bibinfo
  {author} {\bibfnamefont {F.}~\bibnamefont {Bacchini}}, \bibinfo {author}
  {\bibfnamefont {L.}~\bibnamefont {Sironi}},\ and\ \bibinfo {author}
  {\bibfnamefont {R.}~\bibnamefont {Keppens}},\ }\bibfield  {title} {\bibinfo
  {title} {{Effective Resistivity in Relativistic Collisionless
  Reconnection}},\ }\href {https://doi.org/10.3847/1538-4357/acd0b0} {\bibfield
   {journal} {\bibinfo  {journal} {Astrophys. J.}\ }\textbf {\bibinfo {volume}
  {950}},\ \bibinfo {pages} {169} (\bibinfo {year} {2023})},\ \Eprint
  {https://arxiv.org/abs/2211.04553} {arXiv:2211.04553 [astro-ph.HE]}
  \BibitemShut {NoStop}%
\bibitem [{\citenamefont {Bugli}\ \emph {et~al.}(2025)\citenamefont {Bugli},
  \citenamefont {Lopresti}, \citenamefont {Figueiredo}, \citenamefont
  {Mignone}, \citenamefont {Cerutti}, \citenamefont {Mattia}, \citenamefont
  {Del~Zanna}, \citenamefont {Bodo},\ and\ \citenamefont
  {Berta}}]{Bugli:2024fby}%
  \BibitemOpen
  \bibfield  {author} {\bibinfo {author} {\bibfnamefont {M.}~\bibnamefont
  {Bugli}}, \bibinfo {author} {\bibfnamefont {E.~F.}\ \bibnamefont {Lopresti}},
  \bibinfo {author} {\bibfnamefont {E.}~\bibnamefont {Figueiredo}}, \bibinfo
  {author} {\bibfnamefont {A.}~\bibnamefont {Mignone}}, \bibinfo {author}
  {\bibfnamefont {B.}~\bibnamefont {Cerutti}}, \bibinfo {author} {\bibfnamefont
  {G.}~\bibnamefont {Mattia}}, \bibinfo {author} {\bibfnamefont
  {L.}~\bibnamefont {Del~Zanna}}, \bibinfo {author} {\bibfnamefont
  {G.}~\bibnamefont {Bodo}},\ and\ \bibinfo {author} {\bibfnamefont
  {V.}~\bibnamefont {Berta}},\ }\bibfield  {title} {\bibinfo {title}
  {{Relativistic reconnection with effective resistivity - I. Dynamics and
  reconnection rate}},\ }\href {https://doi.org/10.1051/0004-6361/202452277}
  {\bibfield  {journal} {\bibinfo  {journal} {Astron. Astrophys.}\ }\textbf
  {\bibinfo {volume} {693}},\ \bibinfo {pages} {A233} (\bibinfo {year}
  {2025})},\ \Eprint {https://arxiv.org/abs/2410.20924} {arXiv:2410.20924
  [astro-ph.HE]} \BibitemShut {NoStop}%
\bibitem [{\citenamefont {{Bessho}}\ and\ \citenamefont
  {{Bhattacharjee}}(2007)}]{2007PhPl...14e6503B}%
  \BibitemOpen
  \bibfield  {author} {\bibinfo {author} {\bibfnamefont {N.}~\bibnamefont
  {{Bessho}}}\ and\ \bibinfo {author} {\bibfnamefont {A.}~\bibnamefont
  {{Bhattacharjee}}},\ }\bibfield  {title} {\bibinfo {title} {{Fast
  collisionless reconnection in electron-positron plasmasa)}},\ }\href
  {https://doi.org/10.1063/1.2714020} {\bibfield  {journal} {\bibinfo
  {journal} {Physics of Plasmas}\ }\textbf {\bibinfo {volume} {14}},\ \bibinfo
  {eid} {056503} (\bibinfo {year} {2007})}\BibitemShut {NoStop}%
\bibitem [{\citenamefont {Robbins}\ and\ \citenamefont
  {Spitkovsky}(2025)}]{Robbins:2025ofe}%
  \BibitemOpen
  \bibfield  {author} {\bibinfo {author} {\bibfnamefont {A.}~\bibnamefont
  {Robbins}}\ and\ \bibinfo {author} {\bibfnamefont {A.}~\bibnamefont
  {Spitkovsky}},\ }\bibfield  {title} {\bibinfo {title} {{Transition to
  Petschek Reconnection in Subrelativistic Pair Plasmas: Implications for
  Particle Acceleration}},\ }\href {https://doi.org/10.3847/1538-4357/ae1618}
  {\bibfield  {journal} {\bibinfo  {journal} {Astrophys. J.}\ }\textbf
  {\bibinfo {volume} {995}},\ \bibinfo {pages} {163} (\bibinfo {year}
  {2025})},\ \Eprint {https://arxiv.org/abs/2508.08533} {arXiv:2508.08533
  [physics.plasm-ph]} \BibitemShut {NoStop}%
\bibitem [{\citenamefont {Palenzuela}(2013)}]{Palenzuela:2012my}%
  \BibitemOpen
  \bibfield  {author} {\bibinfo {author} {\bibfnamefont {C.}~\bibnamefont
  {Palenzuela}},\ }\bibfield  {title} {\bibinfo {title} {{Modeling magnetized
  neutron stars using resistive MHD}},\ }\href
  {https://doi.org/10.1093/mnras/stt311} {\bibfield  {journal} {\bibinfo
  {journal} {Mon. Not. Roy. Astron. Soc.}\ }\textbf {\bibinfo {volume} {431}},\
  \bibinfo {pages} {1853} (\bibinfo {year} {2013})},\ \Eprint
  {https://arxiv.org/abs/1212.0130} {arXiv:1212.0130 [astro-ph.HE]}
  \BibitemShut {NoStop}%
\bibitem [{\citenamefont {Dionysopoulou}\ \emph {et~al.}(2013)\citenamefont
  {Dionysopoulou}, \citenamefont {Alic}, \citenamefont {Palenzuela},
  \citenamefont {Rezzolla},\ and\ \citenamefont
  {Giacomazzo}}]{Dionysopoulou:2012zv}%
  \BibitemOpen
  \bibfield  {author} {\bibinfo {author} {\bibfnamefont {K.}~\bibnamefont
  {Dionysopoulou}}, \bibinfo {author} {\bibfnamefont {D.}~\bibnamefont {Alic}},
  \bibinfo {author} {\bibfnamefont {C.}~\bibnamefont {Palenzuela}}, \bibinfo
  {author} {\bibfnamefont {L.}~\bibnamefont {Rezzolla}},\ and\ \bibinfo
  {author} {\bibfnamefont {B.}~\bibnamefont {Giacomazzo}},\ }\bibfield  {title}
  {\bibinfo {title} {{General-Relativistic Resistive Magnetohydrodynamics in
  three dimensions: formulation and tests}},\ }\href
  {https://doi.org/10.1103/PhysRevD.88.044020} {\bibfield  {journal} {\bibinfo
  {journal} {Phys. Rev. D}\ }\textbf {\bibinfo {volume} {88}},\ \bibinfo
  {pages} {044020} (\bibinfo {year} {2013})},\ \Eprint
  {https://arxiv.org/abs/1208.3487} {arXiv:1208.3487 [gr-qc]} \BibitemShut
  {NoStop}%
\bibitem [{\citenamefont {Dionysopoulou}\ \emph {et~al.}(2015)\citenamefont
  {Dionysopoulou}, \citenamefont {Alic},\ and\ \citenamefont
  {Rezzolla}}]{Dionysopoulou:2015tda}%
  \BibitemOpen
  \bibfield  {author} {\bibinfo {author} {\bibfnamefont {K.}~\bibnamefont
  {Dionysopoulou}}, \bibinfo {author} {\bibfnamefont {D.}~\bibnamefont
  {Alic}},\ and\ \bibinfo {author} {\bibfnamefont {L.}~\bibnamefont
  {Rezzolla}},\ }\bibfield  {title} {\bibinfo {title} {{General-relativistic
  resistive-magnetohydrodynamic simulations of binary neutron stars}},\ }\href
  {https://doi.org/10.1103/PhysRevD.92.084064} {\bibfield  {journal} {\bibinfo
  {journal} {Phys. Rev. D}\ }\textbf {\bibinfo {volume} {92}},\ \bibinfo
  {pages} {084064} (\bibinfo {year} {2015})},\ \Eprint
  {https://arxiv.org/abs/1502.02021} {arXiv:1502.02021 [gr-qc]} \BibitemShut
  {NoStop}%
\bibitem [{\citenamefont {Ripperda}\ \emph {et~al.}(2019)\citenamefont
  {Ripperda}, \citenamefont {Bacchini}, \citenamefont {Porth}, \citenamefont
  {Most}, \citenamefont {Olivares}, \citenamefont {Nathanail}, \citenamefont
  {Rezzolla}, \citenamefont {Teunissen},\ and\ \citenamefont
  {Keppens}}]{Ripperda:2019lsi}%
  \BibitemOpen
  \bibfield  {author} {\bibinfo {author} {\bibfnamefont {B.}~\bibnamefont
  {Ripperda}}, \bibinfo {author} {\bibfnamefont {F.}~\bibnamefont {Bacchini}},
  \bibinfo {author} {\bibfnamefont {O.}~\bibnamefont {Porth}}, \bibinfo
  {author} {\bibfnamefont {E.~R.}\ \bibnamefont {Most}}, \bibinfo {author}
  {\bibfnamefont {H.}~\bibnamefont {Olivares}}, \bibinfo {author}
  {\bibfnamefont {A.}~\bibnamefont {Nathanail}}, \bibinfo {author}
  {\bibfnamefont {L.}~\bibnamefont {Rezzolla}}, \bibinfo {author}
  {\bibfnamefont {J.}~\bibnamefont {Teunissen}},\ and\ \bibinfo {author}
  {\bibfnamefont {R.}~\bibnamefont {Keppens}},\ }\bibfield  {title} {\bibinfo
  {title} {{General relativistic resistive magnetohydrodynamics with robust
  primitive variable recovery for accretion disk simulations}},\ }\href
  {https://doi.org/10.3847/1538-4365/ab3922} {\bibfield  {journal} {\bibinfo
  {journal} {Astrophys. J. Suppl.}\ }\textbf {\bibinfo {volume} {244}},\
  \bibinfo {pages} {10} (\bibinfo {year} {2019})},\ \Eprint
  {https://arxiv.org/abs/1907.07197} {arXiv:1907.07197 [physics.comp-ph]}
  \BibitemShut {NoStop}%
\bibitem [{\citenamefont {Wright}\ and\ \citenamefont
  {Hawke}(2020)}]{Wright:2019blb}%
  \BibitemOpen
  \bibfield  {author} {\bibinfo {author} {\bibfnamefont {A.~J.}\ \bibnamefont
  {Wright}}\ and\ \bibinfo {author} {\bibfnamefont {I.}~\bibnamefont {Hawke}},\
  }\bibfield  {title} {\bibinfo {title} {{A resistive extension for ideal
  magnetohydrodynamics}},\ }\href {https://doi.org/10.1093/mnras/stz2779}
  {\bibfield  {journal} {\bibinfo  {journal} {Mon. Not. Roy. Astron. Soc.}\
  }\textbf {\bibinfo {volume} {491}},\ \bibinfo {pages} {5510} (\bibinfo {year}
  {2020})},\ \Eprint {https://arxiv.org/abs/1906.03150} {arXiv:1906.03150
  [physics.plasm-ph]} \BibitemShut {NoStop}%
\bibitem [{\citenamefont {Ripperda}\ \emph {et~al.}(2020)\citenamefont
  {Ripperda}, \citenamefont {Bacchini},\ and\ \citenamefont
  {Philippov}}]{Ripperda:2020bpz}%
  \BibitemOpen
  \bibfield  {author} {\bibinfo {author} {\bibfnamefont {B.}~\bibnamefont
  {Ripperda}}, \bibinfo {author} {\bibfnamefont {F.}~\bibnamefont {Bacchini}},\
  and\ \bibinfo {author} {\bibfnamefont {A.}~\bibnamefont {Philippov}},\
  }\bibfield  {title} {\bibinfo {title} {{Magnetic Reconnection and Hot Spot
  Formation in Black Hole Accretion Disks}},\ }\href
  {https://doi.org/10.3847/1538-4357/ababab} {\bibfield  {journal} {\bibinfo
  {journal} {Astrophys. J.}\ }\textbf {\bibinfo {volume} {900}},\ \bibinfo
  {pages} {100} (\bibinfo {year} {2020})},\ \Eprint
  {https://arxiv.org/abs/2003.04330} {arXiv:2003.04330 [astro-ph.HE]}
  \BibitemShut {NoStop}%
\bibitem [{\citenamefont {Shibata}\ \emph {et~al.}(2021)\citenamefont
  {Shibata}, \citenamefont {Fujibayashi},\ and\ \citenamefont
  {Sekiguchi}}]{Shibata:2021xmo}%
  \BibitemOpen
  \bibfield  {author} {\bibinfo {author} {\bibfnamefont {M.}~\bibnamefont
  {Shibata}}, \bibinfo {author} {\bibfnamefont {S.}~\bibnamefont
  {Fujibayashi}},\ and\ \bibinfo {author} {\bibfnamefont {Y.}~\bibnamefont
  {Sekiguchi}},\ }\bibfield  {title} {\bibinfo {title} {{Long-term evolution of
  neutron-star merger remnants in general relativistic resistive
  magnetohydrodynamics with a mean-field dynamo term}},\ }\href
  {https://doi.org/10.1103/PhysRevD.104.063026} {\bibfield  {journal} {\bibinfo
   {journal} {Phys. Rev. D}\ }\textbf {\bibinfo {volume} {104}},\ \bibinfo
  {pages} {063026} (\bibinfo {year} {2021})},\ \Eprint
  {https://arxiv.org/abs/2109.08732} {arXiv:2109.08732 [astro-ph.HE]}
  \BibitemShut {NoStop}%
\bibitem [{\citenamefont {Mattia}\ \emph {et~al.}(2023)\citenamefont {Mattia},
  \citenamefont {Del~Zanna}, \citenamefont {Bugli}, \citenamefont {Pavan},
  \citenamefont {Ciolfi}, \citenamefont {Bodo},\ and\ \citenamefont
  {Mignone}}]{Mattia:2023klx}%
  \BibitemOpen
  \bibfield  {author} {\bibinfo {author} {\bibfnamefont {G.}~\bibnamefont
  {Mattia}}, \bibinfo {author} {\bibfnamefont {L.}~\bibnamefont {Del~Zanna}},
  \bibinfo {author} {\bibfnamefont {M.}~\bibnamefont {Bugli}}, \bibinfo
  {author} {\bibfnamefont {A.}~\bibnamefont {Pavan}}, \bibinfo {author}
  {\bibfnamefont {R.}~\bibnamefont {Ciolfi}}, \bibinfo {author} {\bibfnamefont
  {G.}~\bibnamefont {Bodo}},\ and\ \bibinfo {author} {\bibfnamefont
  {A.}~\bibnamefont {Mignone}},\ }\bibfield  {title} {\bibinfo {title}
  {{Resistive relativistic MHD simulations of astrophysical jets}},\ }\href
  {https://doi.org/10.1051/0004-6361/202347126} {\bibfield  {journal} {\bibinfo
   {journal} {Astron. Astrophys.}\ }\textbf {\bibinfo {volume} {679}},\
  \bibinfo {pages} {A49} (\bibinfo {year} {2023})},\ \Eprint
  {https://arxiv.org/abs/2308.09477} {arXiv:2308.09477 [astro-ph.HE]}
  \BibitemShut {NoStop}%
\bibitem [{\citenamefont {Azizi}\ \emph {et~al.}(2026)\citenamefont {Azizi},
  \citenamefont {Shankar}, \citenamefont {M{\"o}sta}, \citenamefont {Haas},\
  and\ \citenamefont {Schnetter}}]{Azizi:2025nkk}%
  \BibitemOpen
  \bibfield  {author} {\bibinfo {author} {\bibfnamefont {S.}~\bibnamefont
  {Azizi}}, \bibinfo {author} {\bibfnamefont {S.}~\bibnamefont {Shankar}},
  \bibinfo {author} {\bibfnamefont {P.}~\bibnamefont {M{\"o}sta}}, \bibinfo
  {author} {\bibfnamefont {R.}~\bibnamefont {Haas}},\ and\ \bibinfo {author}
  {\bibfnamefont {E.}~\bibnamefont {Schnetter}},\ }\bibfield  {title} {\bibinfo
  {title} {{A resistive MHD module in the GPU-accelerated GRMHD code GRaM-X}},\
  }\href {https://doi.org/10.1088/1361-6382/ae99ce} {\bibfield  {journal}
  {\bibinfo  {journal} {Class. Quant. Grav.}\ }\textbf {\bibinfo {volume}
  {43}},\ \bibinfo {pages} {175009} (\bibinfo {year} {2026})},\ \Eprint
  {https://arxiv.org/abs/2510.18968} {arXiv:2510.18968 [astro-ph.HE]}
  \BibitemShut {NoStop}%
\bibitem [{\citenamefont {Beaudoin}\ \emph {et~al.}(2026)\citenamefont
  {Beaudoin}, \citenamefont {Ujevic}, \citenamefont {Jaeger}, \citenamefont
  {Neuweiler}, \citenamefont {Gieg}, \citenamefont {Kiuchi},\ and\
  \citenamefont {Dietrich}}]{Beaudoin:2026kka}%
  \BibitemOpen
  \bibfield  {author} {\bibinfo {author} {\bibfnamefont {M.}~\bibnamefont
  {Beaudoin}}, \bibinfo {author} {\bibfnamefont {M.}~\bibnamefont {Ujevic}},
  \bibinfo {author} {\bibfnamefont {R.}~\bibnamefont {Jaeger}}, \bibinfo
  {author} {\bibfnamefont {A.}~\bibnamefont {Neuweiler}}, \bibinfo {author}
  {\bibfnamefont {H.}~\bibnamefont {Gieg}}, \bibinfo {author} {\bibfnamefont
  {K.}~\bibnamefont {Kiuchi}},\ and\ \bibinfo {author} {\bibfnamefont
  {T.}~\bibnamefont {Dietrich}},\ }\href@noop {} {\bibinfo {title} {{Extending
  the infrastructure of the BAM code towards resistive general-relativistic
  magnetohydrodynamics: tests and first applications}}} (\bibinfo {year}
  {2026}),\ \Eprint {https://arxiv.org/abs/2607.11670} {arXiv:2607.11670
  [astro-ph.HE]} \BibitemShut {NoStop}%
\bibitem [{\citenamefont {Miravet-Ten{\'e}s}\ \emph {et~al.}(2026)\citenamefont
  {Miravet-Ten{\'e}s}, \citenamefont {Hawke}, \citenamefont {Matur},\ and\
  \citenamefont {Wright}}]{Miravet-Tenes:2026gno}%
  \BibitemOpen
  \bibfield  {author} {\bibinfo {author} {\bibfnamefont {M.}~\bibnamefont
  {Miravet-Ten{\'e}s}}, \bibinfo {author} {\bibfnamefont {I.}~\bibnamefont
  {Hawke}}, \bibinfo {author} {\bibfnamefont {R.}~\bibnamefont {Matur}},\ and\
  \bibinfo {author} {\bibfnamefont {A.~J.}\ \bibnamefont {Wright}},\
  }\href@noop {} {\bibinfo {title} {{An efficient approach to resistive GRMHD
  simulations of binary neutron star mergers}}} (\bibinfo {year} {2026}),\
  \Eprint {https://arxiv.org/abs/2609.12998} {arXiv:2609.12998 [astro-ph.HE]}
  \BibitemShut {NoStop}%
\bibitem [{\citenamefont {Braginskii}(1965)}]{braginskii1965transport}%
  \BibitemOpen
  \bibfield  {author} {\bibinfo {author} {\bibfnamefont {S.}~\bibnamefont
  {Braginskii}},\ }\bibfield  {title} {\bibinfo {title} {Transport processes in
  a plasma},\ }\href@noop {} {\bibfield  {journal} {\bibinfo  {journal}
  {Reviews of plasma physics}\ }\textbf {\bibinfo {volume} {1}} (\bibinfo
  {year} {1965})}\BibitemShut {NoStop}%
\bibitem [{\citenamefont {{Chew}}\ \emph {et~al.}(1956)\citenamefont {{Chew}},
  \citenamefont {{Goldberger}},\ and\ \citenamefont
  {{Low}}}]{1956RSPSA.236..112C}%
  \BibitemOpen
  \bibfield  {author} {\bibinfo {author} {\bibfnamefont {G.~F.}\ \bibnamefont
  {{Chew}}}, \bibinfo {author} {\bibfnamefont {M.~L.}\ \bibnamefont
  {{Goldberger}}},\ and\ \bibinfo {author} {\bibfnamefont {F.~E.}\ \bibnamefont
  {{Low}}},\ }\bibfield  {title} {\bibinfo {title} {{The Boltzmann Equation and
  the One-Fluid Hydromagnetic Equations in the Absence of Particle
  Collisions}},\ }\href {https://doi.org/10.1098/rspa.1956.0116} {\bibfield
  {journal} {\bibinfo  {journal} {Proceedings of the Royal Society of London
  Series A}\ }\textbf {\bibinfo {volume} {236}},\ \bibinfo {pages} {112}
  (\bibinfo {year} {1956})}\BibitemShut {NoStop}%
\bibitem [{\citenamefont {Ley}\ \emph {et~al.}(2026)\citenamefont {Ley},
  \citenamefont {Tran},\ and\ \citenamefont
  {Zweibel}}]{Ley:2026DoubleAdiabatic}%
  \BibitemOpen
  \bibfield  {author} {\bibinfo {author} {\bibfnamefont {F.}~\bibnamefont
  {Ley}}, \bibinfo {author} {\bibfnamefont {A.}~\bibnamefont {Tran}},\ and\
  \bibinfo {author} {\bibfnamefont {E.~G.}\ \bibnamefont {Zweibel}},\
  }\bibfield  {title} {\bibinfo {title} {On the double-adiabatic equations in
  the relativistic regime},\ }\bibfield  {journal} {\bibinfo  {journal} {arXiv
  e-prints}\ }\href {https://doi.org/10.48550/arXiv.2603.25594}
  {10.48550/arXiv.2603.25594} (\bibinfo {year} {2026}),\ \Eprint
  {https://arxiv.org/abs/2603.25594} {arXiv:2603.25594 [physics.plasm-ph]}
  \BibitemShut {NoStop}%
\bibitem [{\citenamefont {Wierzchucka}\ \emph {et~al.}(2026)\citenamefont
  {Wierzchucka}, \citenamefont {Bilbao}, \citenamefont {Thomas}, \citenamefont
  {Uzdensky},\ and\ \citenamefont
  {Schekochihin}}]{Wierzchucka:2026DoubleAdiabatic}%
  \BibitemOpen
  \bibfield  {author} {\bibinfo {author} {\bibfnamefont {A.}~\bibnamefont
  {Wierzchucka}}, \bibinfo {author} {\bibfnamefont {P.~J.}\ \bibnamefont
  {Bilbao}}, \bibinfo {author} {\bibfnamefont {A.~G.~R.}\ \bibnamefont
  {Thomas}}, \bibinfo {author} {\bibfnamefont {D.~A.}\ \bibnamefont
  {Uzdensky}},\ and\ \bibinfo {author} {\bibfnamefont {A.~A.}\ \bibnamefont
  {Schekochihin}},\ }\bibfield  {title} {\bibinfo {title} {Double-adiabatic
  equations of state for relativistic plasmas},\ }\bibfield  {journal}
  {\bibinfo  {journal} {arXiv e-prints}\ }\href
  {https://doi.org/10.48550/arXiv.2603.25669} {10.48550/arXiv.2603.25669}
  (\bibinfo {year} {2026}),\ \Eprint {https://arxiv.org/abs/2603.25669}
  {arXiv:2603.25669 [astro-ph.HE]} \BibitemShut {NoStop}%
\bibitem [{\citenamefont {Chandra}\ \emph {et~al.}(2017)\citenamefont
  {Chandra}, \citenamefont {Foucart},\ and\ \citenamefont
  {Gammie}}]{Chandra:2017auj}%
  \BibitemOpen
  \bibfield  {author} {\bibinfo {author} {\bibfnamefont {M.}~\bibnamefont
  {Chandra}}, \bibinfo {author} {\bibfnamefont {F.}~\bibnamefont {Foucart}},\
  and\ \bibinfo {author} {\bibfnamefont {C.~F.}\ \bibnamefont {Gammie}},\
  }\bibfield  {title} {\bibinfo {title} {{grim: A Flexible, Conservative Scheme
  for Relativistic Fluid Theories}},\ }\href
  {https://doi.org/10.3847/1538-4357/aa5f55} {\bibfield  {journal} {\bibinfo
  {journal} {Astrophys. J.}\ }\textbf {\bibinfo {volume} {837}},\ \bibinfo
  {pages} {92} (\bibinfo {year} {2017})},\ \Eprint
  {https://arxiv.org/abs/1702.01106} {arXiv:1702.01106 [astro-ph.IM]}
  \BibitemShut {NoStop}%
\bibitem [{\citenamefont {Chatterjee}\ \emph {et~al.}(2015)\citenamefont
  {Chatterjee}, \citenamefont {Elghozi}, \citenamefont {Novak},\ and\
  \citenamefont {Oertel}}]{Chatterjee:2014qsa}%
  \BibitemOpen
  \bibfield  {author} {\bibinfo {author} {\bibfnamefont {D.}~\bibnamefont
  {Chatterjee}}, \bibinfo {author} {\bibfnamefont {T.}~\bibnamefont {Elghozi}},
  \bibinfo {author} {\bibfnamefont {J.}~\bibnamefont {Novak}},\ and\ \bibinfo
  {author} {\bibfnamefont {M.}~\bibnamefont {Oertel}},\ }\bibfield  {title}
  {\bibinfo {title} {{Consistent neutron star models with magnetic field
  dependent equations of state}},\ }\href
  {https://doi.org/10.1093/mnras/stu2706} {\bibfield  {journal} {\bibinfo
  {journal} {Mon. Not. Roy. Astron. Soc.}\ }\textbf {\bibinfo {volume} {447}},\
  \bibinfo {pages} {3785} (\bibinfo {year} {2015})},\ \Eprint
  {https://arxiv.org/abs/1410.6332} {arXiv:1410.6332 [astro-ph.HE]}
  \BibitemShut {NoStop}%
\bibitem [{\citenamefont {Pimentel}\ \emph {et~al.}(2018)\citenamefont
  {Pimentel}, \citenamefont {Lora-Clavijo},\ and\ \citenamefont
  {Gonz{\'a}lez}}]{Pimentel:2018uwh}%
  \BibitemOpen
  \bibfield  {author} {\bibinfo {author} {\bibfnamefont {O.~M.}\ \bibnamefont
  {Pimentel}}, \bibinfo {author} {\bibfnamefont {F.~D.}\ \bibnamefont
  {Lora-Clavijo}},\ and\ \bibinfo {author} {\bibfnamefont {G.~A.}\ \bibnamefont
  {Gonz{\'a}lez}},\ }\bibfield  {title} {\bibinfo {title} {{Numerical general
  relativistic MHD with magnetically polarized matter}},\ }\href
  {https://doi.org/10.3847/1538-4357/aac6d0} {\bibfield  {journal} {\bibinfo
  {journal} {Astrophys. J.}\ }\textbf {\bibinfo {volume} {861}},\ \bibinfo
  {pages} {115} (\bibinfo {year} {2018})},\ \Eprint
  {https://arxiv.org/abs/1806.02266} {arXiv:1806.02266 [astro-ph.HE]}
  \BibitemShut {NoStop}%
\bibitem [{\citenamefont {Most}\ \emph {et~al.}(2025)\citenamefont {Most},
  \citenamefont {Peterson}, \citenamefont {Scurto}, \citenamefont {Pais},\ and\
  \citenamefont {Dexheimer}}]{Most:2025kqf}%
  \BibitemOpen
  \bibfield  {author} {\bibinfo {author} {\bibfnamefont {E.~R.}\ \bibnamefont
  {Most}}, \bibinfo {author} {\bibfnamefont {J.}~\bibnamefont {Peterson}},
  \bibinfo {author} {\bibfnamefont {L.}~\bibnamefont {Scurto}}, \bibinfo
  {author} {\bibfnamefont {H.}~\bibnamefont {Pais}},\ and\ \bibinfo {author}
  {\bibfnamefont {V.}~\bibnamefont {Dexheimer}},\ }\bibfield  {title} {\bibinfo
  {title} {{Impact of Magnetic-field-driven Anisotropies on the Equation of
  State Probed in Neutron Star Mergers}},\ }\href
  {https://doi.org/10.3847/2041-8213/adf62d} {\bibfield  {journal} {\bibinfo
  {journal} {Astrophys. J. Lett.}\ }\textbf {\bibinfo {volume} {989}},\
  \bibinfo {pages} {L29} (\bibinfo {year} {2025})},\ \Eprint
  {https://arxiv.org/abs/2506.21696} {arXiv:2506.21696 [astro-ph.HE]}
  \BibitemShut {NoStop}%
\bibitem [{\citenamefont {Galishnikova}\ \emph {et~al.}(2023)\citenamefont
  {Galishnikova}, \citenamefont {Philippov}, \citenamefont {Quataert},
  \citenamefont {Bacchini}, \citenamefont {Parfrey},\ and\ \citenamefont
  {Ripperda}}]{Galishnikova:2023}%
  \BibitemOpen
  \bibfield  {author} {\bibinfo {author} {\bibfnamefont {A.}~\bibnamefont
  {Galishnikova}}, \bibinfo {author} {\bibfnamefont {A.}~\bibnamefont
  {Philippov}}, \bibinfo {author} {\bibfnamefont {E.}~\bibnamefont {Quataert}},
  \bibinfo {author} {\bibfnamefont {F.}~\bibnamefont {Bacchini}}, \bibinfo
  {author} {\bibfnamefont {K.}~\bibnamefont {Parfrey}},\ and\ \bibinfo {author}
  {\bibfnamefont {B.}~\bibnamefont {Ripperda}},\ }\bibfield  {title} {\bibinfo
  {title} {Collisionless accretion onto black holes: Dynamics and flares},\
  }\href {https://doi.org/10.1103/PhysRevLett.130.115201} {\bibfield  {journal}
  {\bibinfo  {journal} {Physical Review Letters}\ }\textbf {\bibinfo {volume}
  {130}},\ \bibinfo {pages} {115201} (\bibinfo {year} {2023})},\ \Eprint
  {https://arxiv.org/abs/2212.02583} {arXiv:2212.02583 [astro-ph.HE]}
  \BibitemShut {NoStop}%
\bibitem [{\citenamefont {Zenitani}\ \emph
  {et~al.}(2009{\natexlab{b}})\citenamefont {Zenitani}, \citenamefont {Hesse},\
  and\ \citenamefont {Klimas}}]{Zenitani:2009di}%
  \BibitemOpen
  \bibfield  {author} {\bibinfo {author} {\bibfnamefont {S.}~\bibnamefont
  {Zenitani}}, \bibinfo {author} {\bibfnamefont {M.}~\bibnamefont {Hesse}},\
  and\ \bibinfo {author} {\bibfnamefont {A.}~\bibnamefont {Klimas}},\
  }\bibfield  {title} {\bibinfo {title} {{Relativistic Two-fluid Simulations of
  Guide Field Magnetic Reconnection}},\ }\href
  {https://doi.org/10.1088/0004-637X/705/1/907} {\bibfield  {journal} {\bibinfo
   {journal} {Astrophys. J.}\ }\textbf {\bibinfo {volume} {705}},\ \bibinfo
  {pages} {907} (\bibinfo {year} {2009}{\natexlab{b}})},\ \Eprint
  {https://arxiv.org/abs/0909.1955} {arXiv:0909.1955 [astro-ph.HE]}
  \BibitemShut {NoStop}%
\bibitem [{\citenamefont {Gorard}\ \emph {et~al.}(2025)\citenamefont {Gorard},
  \citenamefont {Juno},\ and\ \citenamefont {Hakim}}]{Gorard:2025grmhd}%
  \BibitemOpen
  \bibfield  {author} {\bibinfo {author} {\bibfnamefont {J.}~\bibnamefont
  {Gorard}}, \bibinfo {author} {\bibfnamefont {J.}~\bibnamefont {Juno}},\ and\
  \bibinfo {author} {\bibfnamefont {A.}~\bibnamefont {Hakim}},\ }\bibfield
  {title} {\bibinfo {title} {{Beyond GRMHD: A Robust Numerical Scheme for
  Extended, Non-Ideal General Relativistic Multifluid Simulations}},\
  }\href@noop {} {\bibfield  {journal} {\bibinfo  {journal} {arXiv e-prints}\ }
  (\bibinfo {year} {2025})},\ \Eprint {https://arxiv.org/abs/2510.26019}
  {arXiv:2510.26019 [gr-qc]} \BibitemShut {NoStop}%
\bibitem [{\citenamefont {Koide}(2020)}]{Koide:2020xts}%
  \BibitemOpen
  \bibfield  {author} {\bibinfo {author} {\bibfnamefont {S.}~\bibnamefont
  {Koide}},\ }\bibfield  {title} {\bibinfo {title} {{Generalized
  general-relativistic magnetohydrodynamic equations for plasmas of active
  galactic nuclei in the era of the Event Horizon Telescope}},\ }\href
  {https://doi.org/10.3847/1538-4357/aba743} {\bibfield  {journal} {\bibinfo
  {journal} {Astrophys. J.}\ }\textbf {\bibinfo {volume} {899}},\ \bibinfo
  {pages} {95} (\bibinfo {year} {2020})},\ \Eprint
  {https://arxiv.org/abs/2007.13029} {arXiv:2007.13029 [astro-ph.HE]}
  \BibitemShut {NoStop}%
\bibitem [{\citenamefont {Ripperda}\ \emph {et~al.}(2022)\citenamefont
  {Ripperda}, \citenamefont {Liska}, \citenamefont {Chatterjee}, \citenamefont
  {Musoke}, \citenamefont {Philippov}, \citenamefont {Markoff}, \citenamefont
  {Tchekhovskoy},\ and\ \citenamefont {Younsi}}]{Ripperda:2021zpn}%
  \BibitemOpen
  \bibfield  {author} {\bibinfo {author} {\bibfnamefont {B.}~\bibnamefont
  {Ripperda}}, \bibinfo {author} {\bibfnamefont {M.}~\bibnamefont {Liska}},
  \bibinfo {author} {\bibfnamefont {K.}~\bibnamefont {Chatterjee}}, \bibinfo
  {author} {\bibfnamefont {G.}~\bibnamefont {Musoke}}, \bibinfo {author}
  {\bibfnamefont {A.~A.}\ \bibnamefont {Philippov}}, \bibinfo {author}
  {\bibfnamefont {S.~B.}\ \bibnamefont {Markoff}}, \bibinfo {author}
  {\bibfnamefont {A.}~\bibnamefont {Tchekhovskoy}},\ and\ \bibinfo {author}
  {\bibfnamefont {Z.}~\bibnamefont {Younsi}},\ }\bibfield  {title} {\bibinfo
  {title} {{Black Hole Flares: Ejection of Accreted Magnetic Flux through 3D
  Plasmoid-mediated Reconnection}},\ }\href
  {https://doi.org/10.3847/2041-8213/ac46a1} {\bibfield  {journal} {\bibinfo
  {journal} {Astrophys. J. Lett.}\ }\textbf {\bibinfo {volume} {924}},\
  \bibinfo {pages} {L32} (\bibinfo {year} {2022})},\ \Eprint
  {https://arxiv.org/abs/2109.15115} {arXiv:2109.15115 [astro-ph.HE]}
  \BibitemShut {NoStop}%
\bibitem [{\citenamefont {Carter}(1989)}]{Carter:1987qr}%
  \BibitemOpen
  \bibfield  {author} {\bibinfo {author} {\bibfnamefont {B.}~\bibnamefont
  {Carter}},\ }\bibfield  {title} {\bibinfo {title} {{Covariant Theory of
  Conductivity in Ideal Fluid or Solid Media}},\ }\href@noop {} {\bibfield
  {journal} {\bibinfo  {journal} {Lect. Notes Math.}\ }\textbf {\bibinfo
  {volume} {1385}},\ \bibinfo {pages} {1} (\bibinfo {year} {1989})}\BibitemShut
  {NoStop}%
\bibitem [{\citenamefont {Rocha}\ \emph {et~al.}(2024)\citenamefont {Rocha},
  \citenamefont {Wagner}, \citenamefont {Denicol}, \citenamefont {Noronha},\
  and\ \citenamefont {Rischke}}]{Rocha:2023ilf}%
  \BibitemOpen
  \bibfield  {author} {\bibinfo {author} {\bibfnamefont {G.~S.}\ \bibnamefont
  {Rocha}}, \bibinfo {author} {\bibfnamefont {D.}~\bibnamefont {Wagner}},
  \bibinfo {author} {\bibfnamefont {G.~S.}\ \bibnamefont {Denicol}}, \bibinfo
  {author} {\bibfnamefont {J.}~\bibnamefont {Noronha}},\ and\ \bibinfo {author}
  {\bibfnamefont {D.~H.}\ \bibnamefont {Rischke}},\ }\bibfield  {title}
  {\bibinfo {title} {{Theories of Relativistic Dissipative Fluid Dynamics}},\
  }\href {https://doi.org/10.3390/e26030189} {\bibfield  {journal} {\bibinfo
  {journal} {Entropy}\ }\textbf {\bibinfo {volume} {26}},\ \bibinfo {pages}
  {189} (\bibinfo {year} {2024})},\ \Eprint {https://arxiv.org/abs/2311.15063}
  {arXiv:2311.15063 [nucl-th]} \BibitemShut {NoStop}%
\bibitem [{\citenamefont {Carter}\ and\ \citenamefont
  {Khalatnikov}(1994)}]{Carter:1993aq}%
  \BibitemOpen
  \bibfield  {author} {\bibinfo {author} {\bibfnamefont {B.}~\bibnamefont
  {Carter}}\ and\ \bibinfo {author} {\bibfnamefont {I.~M.}\ \bibnamefont
  {Khalatnikov}},\ }\bibfield  {title} {\bibinfo {title} {{Canonically
  covariant formulation of Landau's Newtonian superfluid dynamics}},\ }\href
  {https://doi.org/10.1142/S0129055X94000134} {\bibfield  {journal} {\bibinfo
  {journal} {Rev. Math. Phys.}\ }\textbf {\bibinfo {volume} {6}},\ \bibinfo
  {pages} {277} (\bibinfo {year} {1994})}\BibitemShut {NoStop}%
\bibitem [{\citenamefont {Carter}\ and\ \citenamefont
  {Langlois}(1998)}]{Carter:1998rn}%
  \BibitemOpen
  \bibfield  {author} {\bibinfo {author} {\bibfnamefont {B.}~\bibnamefont
  {Carter}}\ and\ \bibinfo {author} {\bibfnamefont {D.}~\bibnamefont
  {Langlois}},\ }\bibfield  {title} {\bibinfo {title} {{Relativistic models for
  superconducting superfluid mixtures}},\ }\href
  {https://doi.org/10.1016/S0550-3213(98)00430-1} {\bibfield  {journal}
  {\bibinfo  {journal} {Nucl. Phys. B}\ }\textbf {\bibinfo {volume} {531}},\
  \bibinfo {pages} {478} (\bibinfo {year} {1998})},\ \Eprint
  {https://arxiv.org/abs/gr-qc/9806024} {arXiv:gr-qc/9806024} \BibitemShut
  {NoStop}%
\bibitem [{\citenamefont {Langlois}\ \emph {et~al.}(1998)\citenamefont
  {Langlois}, \citenamefont {Sedrakian},\ and\ \citenamefont
  {Carter}}]{Langlois:1997bz}%
  \BibitemOpen
  \bibfield  {author} {\bibinfo {author} {\bibfnamefont {D.}~\bibnamefont
  {Langlois}}, \bibinfo {author} {\bibfnamefont {D.~M.}\ \bibnamefont
  {Sedrakian}},\ and\ \bibinfo {author} {\bibfnamefont {B.}~\bibnamefont
  {Carter}},\ }\bibfield  {title} {\bibinfo {title} {{Differential rotation of
  relativistic superfluid in neutron stars}},\ }\href
  {https://doi.org/10.1046/j.1365-8711.1998.01575.x} {\bibfield  {journal}
  {\bibinfo  {journal} {Mon. Not. Roy. Astron. Soc.}\ }\textbf {\bibinfo
  {volume} {297}},\ \bibinfo {pages} {1189} (\bibinfo {year} {1998})},\ \Eprint
  {https://arxiv.org/abs/astro-ph/9711042} {arXiv:astro-ph/9711042}
  \BibitemShut {NoStop}%
\bibitem [{\citenamefont {Gavassino}\ and\ \citenamefont
  {Antonelli}(2020)}]{Gavassino:2019wzd}%
  \BibitemOpen
  \bibfield  {author} {\bibinfo {author} {\bibfnamefont {L.}~\bibnamefont
  {Gavassino}}\ and\ \bibinfo {author} {\bibfnamefont {M.}~\bibnamefont
  {Antonelli}},\ }\bibfield  {title} {\bibinfo {title} {{Thermodynamics of
  uncharged relativistic multifluids}},\ }\href
  {https://doi.org/10.1088/1361-6382/ab5f23} {\bibfield  {journal} {\bibinfo
  {journal} {Class. Quant. Grav.}\ }\textbf {\bibinfo {volume} {37}},\ \bibinfo
  {pages} {025014} (\bibinfo {year} {2020})},\ \Eprint
  {https://arxiv.org/abs/1906.03140} {arXiv:1906.03140 [gr-qc]} \BibitemShut
  {NoStop}%
\bibitem [{\citenamefont {Andersson}\ \emph
  {et~al.}(2017{\natexlab{a}})\citenamefont {Andersson}, \citenamefont
  {Dionysopoulou}, \citenamefont {Hawke},\ and\ \citenamefont
  {Comer}}]{Andersson:2016wnc}%
  \BibitemOpen
  \bibfield  {author} {\bibinfo {author} {\bibfnamefont {N.}~\bibnamefont
  {Andersson}}, \bibinfo {author} {\bibfnamefont {K.}~\bibnamefont
  {Dionysopoulou}}, \bibinfo {author} {\bibfnamefont {I.}~\bibnamefont
  {Hawke}},\ and\ \bibinfo {author} {\bibfnamefont {G.~L.}\ \bibnamefont
  {Comer}},\ }\bibfield  {title} {\bibinfo {title} {{Beyond ideal
  magnetohydrodynamics: Resistive, reactive and relativistic plasmas}},\ }\href
  {https://doi.org/10.1088/1361-6382/aa6b3a} {\bibfield  {journal} {\bibinfo
  {journal} {Class. Quant. Grav.}\ }\textbf {\bibinfo {volume} {34}},\ \bibinfo
  {pages} {125002} (\bibinfo {year} {2017}{\natexlab{a}})},\ \Eprint
  {https://arxiv.org/abs/1610.00449} {arXiv:1610.00449 [gr-qc]} \BibitemShut
  {NoStop}%
\bibitem [{\citenamefont {Andersson}\ \emph
  {et~al.}(2017{\natexlab{b}})\citenamefont {Andersson}, \citenamefont {Hawke},
  \citenamefont {Dionysopoulou},\ and\ \citenamefont
  {Comer}}]{Andersson:2016fva}%
  \BibitemOpen
  \bibfield  {author} {\bibinfo {author} {\bibfnamefont {N.}~\bibnamefont
  {Andersson}}, \bibinfo {author} {\bibfnamefont {I.}~\bibnamefont {Hawke}},
  \bibinfo {author} {\bibfnamefont {K.}~\bibnamefont {Dionysopoulou}},\ and\
  \bibinfo {author} {\bibfnamefont {G.~L.}\ \bibnamefont {Comer}},\ }\bibfield
  {title} {\bibinfo {title} {{Beyond ideal magnetohydrodynamics: from fibration
  to 3 + 1 foliation}},\ }\href {https://doi.org/10.1088/1361-6382/aa6b39}
  {\bibfield  {journal} {\bibinfo  {journal} {Class. Quant. Grav.}\ }\textbf
  {\bibinfo {volume} {34}},\ \bibinfo {pages} {125003} (\bibinfo {year}
  {2017}{\natexlab{b}})},\ \Eprint {https://arxiv.org/abs/1610.00448}
  {arXiv:1610.00448 [gr-qc]} \BibitemShut {NoStop}%
\bibitem [{\citenamefont {Andersson}\ \emph {et~al.}(2021)\citenamefont
  {Andersson}, \citenamefont {Hawke}, \citenamefont {Celora},\ and\
  \citenamefont {Comer}}]{Andersson:2021kfk}%
  \BibitemOpen
  \bibfield  {author} {\bibinfo {author} {\bibfnamefont {N.}~\bibnamefont
  {Andersson}}, \bibinfo {author} {\bibfnamefont {I.}~\bibnamefont {Hawke}},
  \bibinfo {author} {\bibfnamefont {T.}~\bibnamefont {Celora}},\ and\ \bibinfo
  {author} {\bibfnamefont {G.~L.}\ \bibnamefont {Comer}},\ }\bibfield  {title}
  {\bibinfo {title} {{The physics of non-ideal general relativistic
  magnetohydrodynamics}},\ }\href {https://doi.org/10.1093/mnras/stab3257}
  {\bibfield  {journal} {\bibinfo  {journal} {Mon. Not. Roy. Astron. Soc.}\
  }\textbf {\bibinfo {volume} {509}},\ \bibinfo {pages} {3737} (\bibinfo {year}
  {2021})},\ \Eprint {https://arxiv.org/abs/2108.08732} {arXiv:2108.08732
  [astro-ph.HE]} \BibitemShut {NoStop}%
\bibitem [{\citenamefont {Hegade K.~R.}\ and\ \citenamefont
  {Stone}(2026)}]{HegadeKR:2026azf}%
  \BibitemOpen
  \bibfield  {author} {\bibinfo {author} {\bibfnamefont {A.}~\bibnamefont
  {Hegade K.~R.}}\ and\ \bibinfo {author} {\bibfnamefont {J.~M.}\ \bibnamefont
  {Stone}},\ }\href@noop {} {\bibinfo {title} {{Kinetic magnetohydrodynamics
  and Landau fluid closure in relativity}}} (\bibinfo {year} {2026}),\ \Eprint
  {https://arxiv.org/abs/2604.05058} {arXiv:2604.05058 [astro-ph.HE]}
  \BibitemShut {NoStop}%
\bibitem [{\citenamefont {Cercignani}\ and\ \citenamefont
  {Kremer}(2002)}]{CercignaniKremer2002}%
  \BibitemOpen
  \bibfield  {author} {\bibinfo {author} {\bibfnamefont {C.}~\bibnamefont
  {Cercignani}}\ and\ \bibinfo {author} {\bibfnamefont {G.~M.}\ \bibnamefont
  {Kremer}},\ }\href {https://doi.org/10.1007/978-3-0348-8165-4} {\emph
  {\bibinfo {title} {The Relativistic Boltzmann Equation: Theory and
  Applications}}},\ \bibinfo {series} {Progress in Mathematical Physics},
  Vol.~\bibinfo {volume} {22}\ (\bibinfo  {publisher} {Birkh{\"a}user Basel},\
  \bibinfo {year} {2002})\BibitemShut {NoStop}%
\bibitem [{\citenamefont {Denicol}\ \emph {et~al.}(2012)\citenamefont
  {Denicol}, \citenamefont {Niemi}, \citenamefont {Molnar},\ and\ \citenamefont
  {Rischke}}]{Denicol:2012cn}%
  \BibitemOpen
  \bibfield  {author} {\bibinfo {author} {\bibfnamefont {G.~S.}\ \bibnamefont
  {Denicol}}, \bibinfo {author} {\bibfnamefont {H.}~\bibnamefont {Niemi}},
  \bibinfo {author} {\bibfnamefont {E.}~\bibnamefont {Molnar}},\ and\ \bibinfo
  {author} {\bibfnamefont {D.~H.}\ \bibnamefont {Rischke}},\ }\bibfield
  {title} {\bibinfo {title} {{Derivation of transient relativistic fluid
  dynamics from the Boltzmann equation}},\ }\href
  {https://doi.org/10.1103/PhysRevD.85.114047} {\bibfield  {journal} {\bibinfo
  {journal} {Phys. Rev. D}\ }\textbf {\bibinfo {volume} {85}},\ \bibinfo
  {pages} {114047} (\bibinfo {year} {2012})},\ \bibinfo {note} {[Erratum:
  Phys.Rev.D 91, 039902 (2015)]},\ \Eprint {https://arxiv.org/abs/1202.4551}
  {arXiv:1202.4551 [nucl-th]} \BibitemShut {NoStop}%
\bibitem [{\citenamefont {Denicol}\ \emph {et~al.}(2018)\citenamefont
  {Denicol}, \citenamefont {Huang}, \citenamefont {Moln\'ar}, \citenamefont
  {Monteiro}, \citenamefont {Niemi}, \citenamefont {Noronha}, \citenamefont
  {Rischke},\ and\ \citenamefont {Wang}}]{Denicol:2018rbw}%
  \BibitemOpen
  \bibfield  {author} {\bibinfo {author} {\bibfnamefont {G.~S.}\ \bibnamefont
  {Denicol}}, \bibinfo {author} {\bibfnamefont {X.-G.}\ \bibnamefont {Huang}},
  \bibinfo {author} {\bibfnamefont {E.}~\bibnamefont {Moln\'ar}}, \bibinfo
  {author} {\bibfnamefont {G.~M.}\ \bibnamefont {Monteiro}}, \bibinfo {author}
  {\bibfnamefont {H.}~\bibnamefont {Niemi}}, \bibinfo {author} {\bibfnamefont
  {J.}~\bibnamefont {Noronha}}, \bibinfo {author} {\bibfnamefont {D.~H.}\
  \bibnamefont {Rischke}},\ and\ \bibinfo {author} {\bibfnamefont
  {Q.}~\bibnamefont {Wang}},\ }\bibfield  {title} {\bibinfo {title}
  {{Nonresistive dissipative magnetohydrodynamics from the Boltzmann equation
  in the 14-moment approximation}},\ }\href
  {https://doi.org/10.1103/PhysRevD.98.076009} {\bibfield  {journal} {\bibinfo
  {journal} {Phys. Rev. D}\ }\textbf {\bibinfo {volume} {98}},\ \bibinfo
  {pages} {076009} (\bibinfo {year} {2018})},\ \Eprint
  {https://arxiv.org/abs/1804.05210} {arXiv:1804.05210 [nucl-th]} \BibitemShut
  {NoStop}%
\bibitem [{\citenamefont {Denicol}\ \emph {et~al.}(2019)\citenamefont
  {Denicol}, \citenamefont {Moln\'ar}, \citenamefont {Niemi},\ and\
  \citenamefont {Rischke}}]{Denicol:2019iyh}%
  \BibitemOpen
  \bibfield  {author} {\bibinfo {author} {\bibfnamefont {G.~S.}\ \bibnamefont
  {Denicol}}, \bibinfo {author} {\bibfnamefont {E.}~\bibnamefont {Moln\'ar}},
  \bibinfo {author} {\bibfnamefont {H.}~\bibnamefont {Niemi}},\ and\ \bibinfo
  {author} {\bibfnamefont {D.~H.}\ \bibnamefont {Rischke}},\ }\bibfield
  {title} {\bibinfo {title} {{Resistive dissipative magnetohydrodynamics from
  the Boltzmann-Vlasov equation}},\ }\href
  {https://doi.org/10.1103/PhysRevD.99.056017} {\bibfield  {journal} {\bibinfo
  {journal} {Phys. Rev. D}\ }\textbf {\bibinfo {volume} {99}},\ \bibinfo
  {pages} {056017} (\bibinfo {year} {2019})},\ \Eprint
  {https://arxiv.org/abs/1902.01699} {arXiv:1902.01699 [nucl-th]} \BibitemShut
  {NoStop}%
\bibitem [{\citenamefont {Kushwah}\ and\ \citenamefont
  {Denicol}(2024)}]{Kushwah:2024zgd}%
  \BibitemOpen
  \bibfield  {author} {\bibinfo {author} {\bibfnamefont {K.}~\bibnamefont
  {Kushwah}}\ and\ \bibinfo {author} {\bibfnamefont {G.~S.}\ \bibnamefont
  {Denicol}},\ }\bibfield  {title} {\bibinfo {title} {{Relativistic dissipative
  magnetohydrodynamics from the Boltzmann equation for a two-component gas}},\
  }\href {https://doi.org/10.1103/PhysRevD.109.096021} {\bibfield  {journal}
  {\bibinfo  {journal} {Phys. Rev. D}\ }\textbf {\bibinfo {volume} {109}},\
  \bibinfo {pages} {096021} (\bibinfo {year} {2024})},\ \Eprint
  {https://arxiv.org/abs/2402.01597} {arXiv:2402.01597 [nucl-th]} \BibitemShut
  {NoStop}%
\bibitem [{\citenamefont {Kushwah}\ \emph {et~al.}(2026)\citenamefont
  {Kushwah}, \citenamefont {Denicol},\ and\ \citenamefont {V.~P.~de
  Brito}}]{Kushwah:2025jsb}%
  \BibitemOpen
  \bibfield  {author} {\bibinfo {author} {\bibfnamefont {K.}~\bibnamefont
  {Kushwah}}, \bibinfo {author} {\bibfnamefont {G.~S.}\ \bibnamefont
  {Denicol}},\ and\ \bibinfo {author} {\bibfnamefont {C.}~\bibnamefont
  {V.~P.~de Brito}},\ }\bibfield  {title} {\bibinfo {title} {{Relativistic
  resistive magnetohydrodynamics for a two-component plasma}},\ }\href
  {https://doi.org/10.1103/tl97-r2kh} {\bibfield  {journal} {\bibinfo
  {journal} {Phys. Rev. D}\ }\textbf {\bibinfo {volume} {113}},\ \bibinfo
  {pages} {036021} (\bibinfo {year} {2026})},\ \Eprint
  {https://arxiv.org/abs/2511.14787} {arXiv:2511.14787 [physics.plasm-ph]}
  \BibitemShut {NoStop}%
\bibitem [{\citenamefont {{Hammett}}\ \emph {et~al.}(1992)\citenamefont
  {{Hammett}}, \citenamefont {{Dorland}},\ and\ \citenamefont
  {{Perkins}}}]{1992PhFlB...4.2052H}%
  \BibitemOpen
  \bibfield  {author} {\bibinfo {author} {\bibfnamefont {G.~W.}\ \bibnamefont
  {{Hammett}}}, \bibinfo {author} {\bibfnamefont {W.}~\bibnamefont
  {{Dorland}}},\ and\ \bibinfo {author} {\bibfnamefont {F.~W.}\ \bibnamefont
  {{Perkins}}},\ }\bibfield  {title} {\bibinfo {title} {{Fluid models of phase
  mixing, Landau damping, and nonlinear gyrokinetic dynamics}},\ }\href
  {https://doi.org/10.1063/1.860014} {\bibfield  {journal} {\bibinfo  {journal}
  {Physics of Fluids B}\ }\textbf {\bibinfo {volume} {4}},\ \bibinfo {pages}
  {2052} (\bibinfo {year} {1992})}\BibitemShut {NoStop}%
\bibitem [{\citenamefont {{Wang}}\ \emph {et~al.}(2015)\citenamefont {{Wang}},
  \citenamefont {{Hakim}}, \citenamefont {{Bhattacharjee}},\ and\ \citenamefont
  {{Germaschewski}}}]{2015PhPl...22a2108W}%
  \BibitemOpen
  \bibfield  {author} {\bibinfo {author} {\bibfnamefont {L.}~\bibnamefont
  {{Wang}}}, \bibinfo {author} {\bibfnamefont {A.~H.}\ \bibnamefont {{Hakim}}},
  \bibinfo {author} {\bibfnamefont {A.}~\bibnamefont {{Bhattacharjee}}},\ and\
  \bibinfo {author} {\bibfnamefont {K.}~\bibnamefont {{Germaschewski}}},\
  }\bibfield  {title} {\bibinfo {title} {{Comparison of multi-fluid moment
  models with particle-in-cell simulations of collisionless magnetic
  reconnection}},\ }\href {https://doi.org/10.1063/1.4906063} {\bibfield
  {journal} {\bibinfo  {journal} {Physics of Plasmas}\ }\textbf {\bibinfo
  {volume} {22}},\ \bibinfo {eid} {012108} (\bibinfo {year} {2015})},\ \Eprint
  {https://arxiv.org/abs/1409.0262} {arXiv:1409.0262 [physics.plasm-ph]}
  \BibitemShut {NoStop}%
\bibitem [{\citenamefont {Romatschke}(2010)}]{Romatschke:2009im}%
  \BibitemOpen
  \bibfield  {author} {\bibinfo {author} {\bibfnamefont {P.}~\bibnamefont
  {Romatschke}},\ }\bibfield  {title} {\bibinfo {title} {{New Developments in
  Relativistic Viscous Hydrodynamics}},\ }\href
  {https://doi.org/10.1142/S0218301310014613} {\bibfield  {journal} {\bibinfo
  {journal} {Int. J. Mod. Phys. E}\ }\textbf {\bibinfo {volume} {19}},\
  \bibinfo {pages} {1} (\bibinfo {year} {2010})},\ \Eprint
  {https://arxiv.org/abs/0902.3663} {arXiv:0902.3663 [hep-ph]} \BibitemShut
  {NoStop}%
\bibitem [{\citenamefont {Most}\ \emph {et~al.}(2022)\citenamefont {Most},
  \citenamefont {Noronha},\ and\ \citenamefont {Philippov}}]{Most:2021uck}%
  \BibitemOpen
  \bibfield  {author} {\bibinfo {author} {\bibfnamefont {E.~R.}\ \bibnamefont
  {Most}}, \bibinfo {author} {\bibfnamefont {J.}~\bibnamefont {Noronha}},\ and\
  \bibinfo {author} {\bibfnamefont {A.~A.}\ \bibnamefont {Philippov}},\
  }\bibfield  {title} {\bibinfo {title} {{Modelling general-relativistic
  plasmas with collisionless moments and dissipative two-fluid
  magnetohydrodynamics}},\ }\href {https://doi.org/10.1093/mnras/stac1435}
  {\bibfield  {journal} {\bibinfo  {journal} {Mon. Not. Roy. Astron. Soc.}\
  }\textbf {\bibinfo {volume} {514}},\ \bibinfo {pages} {4989} (\bibinfo {year}
  {2022})},\ \Eprint {https://arxiv.org/abs/2111.05752} {arXiv:2111.05752
  [astro-ph.HE]} \BibitemShut {NoStop}%
\bibitem [{\citenamefont {Gavassino}(2026)}]{Gavassino:2026fil}%
  \BibitemOpen
  \bibfield  {author} {\bibinfo {author} {\bibfnamefont {L.}~\bibnamefont
  {Gavassino}},\ }\href@noop {} {\bibinfo {title} {{How acausal equations
  emerge from causal dynamics}}} (\bibinfo {year} {2026}),\ \Eprint
  {https://arxiv.org/abs/2604.07031} {arXiv:2604.07031 [nucl-th]} \BibitemShut
  {NoStop}%
\bibitem [{\citenamefont {Hiscock}\ and\ \citenamefont
  {Lindblom}(1983)}]{Hiscock:1983zz}%
  \BibitemOpen
  \bibfield  {author} {\bibinfo {author} {\bibfnamefont {W.~A.}\ \bibnamefont
  {Hiscock}}\ and\ \bibinfo {author} {\bibfnamefont {L.}~\bibnamefont
  {Lindblom}},\ }\bibfield  {title} {\bibinfo {title} {{Stability and causality
  in dissipative relativistic fluids}},\ }\href
  {https://doi.org/10.1016/0003-4916(83)90288-9} {\bibfield  {journal}
  {\bibinfo  {journal} {Annals Phys.}\ }\textbf {\bibinfo {volume} {151}},\
  \bibinfo {pages} {466} (\bibinfo {year} {1983})}\BibitemShut {NoStop}%
\bibitem [{\citenamefont {Hiscock}\ and\ \citenamefont
  {Lindblom}(1985)}]{Hiscock:1985zz}%
  \BibitemOpen
  \bibfield  {author} {\bibinfo {author} {\bibfnamefont {W.~A.}\ \bibnamefont
  {Hiscock}}\ and\ \bibinfo {author} {\bibfnamefont {L.}~\bibnamefont
  {Lindblom}},\ }\bibfield  {title} {\bibinfo {title} {{Generic instabilities
  in first-order dissipative relativistic fluid theories}},\ }\href
  {https://doi.org/10.1103/PhysRevD.31.725} {\bibfield  {journal} {\bibinfo
  {journal} {Phys. Rev. D}\ }\textbf {\bibinfo {volume} {31}},\ \bibinfo
  {pages} {725} (\bibinfo {year} {1985})}\BibitemShut {NoStop}%
\bibitem [{\citenamefont {M{\"u}ller}(1967)}]{muller1967paradoxon}%
  \BibitemOpen
  \bibfield  {author} {\bibinfo {author} {\bibfnamefont {I.}~\bibnamefont
  {M{\"u}ller}},\ }\bibfield  {title} {\bibinfo {title} {Zum paradoxon der
  w{\"a}rmeleitungstheorie},\ }\href@noop {} {\bibfield  {journal} {\bibinfo
  {journal} {Zeitschrift f{\"u}r Physik}\ }\textbf {\bibinfo {volume} {198}},\
  \bibinfo {pages} {329} (\bibinfo {year} {1967})}\BibitemShut {NoStop}%
\bibitem [{\citenamefont {Israel}\ and\ \citenamefont
  {Stewart}(1979)}]{Israel:1979wp}%
  \BibitemOpen
  \bibfield  {author} {\bibinfo {author} {\bibfnamefont {W.}~\bibnamefont
  {Israel}}\ and\ \bibinfo {author} {\bibfnamefont {J.~M.}\ \bibnamefont
  {Stewart}},\ }\bibfield  {title} {\bibinfo {title} {{Transient relativistic
  thermodynamics and kinetic theory}},\ }\href
  {https://doi.org/10.1016/0003-4916(79)90130-1} {\bibfield  {journal}
  {\bibinfo  {journal} {Annals Phys.}\ }\textbf {\bibinfo {volume} {118}},\
  \bibinfo {pages} {341} (\bibinfo {year} {1979})}\BibitemShut {NoStop}%
\bibitem [{\citenamefont {Bemfica}\ \emph {et~al.}(2018)\citenamefont
  {Bemfica}, \citenamefont {Disconzi},\ and\ \citenamefont
  {Noronha}}]{Bemfica:2017wps}%
  \BibitemOpen
  \bibfield  {author} {\bibinfo {author} {\bibfnamefont {F.~S.}\ \bibnamefont
  {Bemfica}}, \bibinfo {author} {\bibfnamefont {M.~M.}\ \bibnamefont
  {Disconzi}},\ and\ \bibinfo {author} {\bibfnamefont {J.}~\bibnamefont
  {Noronha}},\ }\bibfield  {title} {\bibinfo {title} {{Causality and existence
  of solutions of relativistic viscous fluid dynamics with gravity}},\ }\href
  {https://doi.org/10.1103/PhysRevD.98.104064} {\bibfield  {journal} {\bibinfo
  {journal} {Phys. Rev. D}\ }\textbf {\bibinfo {volume} {98}},\ \bibinfo
  {pages} {104064} (\bibinfo {year} {2018})},\ \Eprint
  {https://arxiv.org/abs/1708.06255} {arXiv:1708.06255 [gr-qc]} \BibitemShut
  {NoStop}%
\bibitem [{\citenamefont {Kovtun}(2019)}]{Kovtun:2019hdm}%
  \BibitemOpen
  \bibfield  {author} {\bibinfo {author} {\bibfnamefont {P.}~\bibnamefont
  {Kovtun}},\ }\bibfield  {title} {\bibinfo {title} {{First-order relativistic
  hydrodynamics is stable}},\ }\href {https://doi.org/10.1007/JHEP10(2019)034}
  {\bibfield  {journal} {\bibinfo  {journal} {JHEP}\ }\textbf {\bibinfo
  {volume} {2019}}\bibfield  {number} {\bibinfo  {number} { (10)},\ \bibinfo
  {pages} {034}},\ }\Eprint {https://arxiv.org/abs/1907.08191}
  {arXiv:1907.08191 [hep-th]} \BibitemShut {NoStop}%
\bibitem [{\citenamefont {Bemfica}\ \emph {et~al.}(2022)\citenamefont
  {Bemfica}, \citenamefont {Disconzi},\ and\ \citenamefont
  {Noronha}}]{Bemfica:2020zjp}%
  \BibitemOpen
  \bibfield  {author} {\bibinfo {author} {\bibfnamefont {F.~S.}\ \bibnamefont
  {Bemfica}}, \bibinfo {author} {\bibfnamefont {M.~M.}\ \bibnamefont
  {Disconzi}},\ and\ \bibinfo {author} {\bibfnamefont {J.}~\bibnamefont
  {Noronha}},\ }\bibfield  {title} {\bibinfo {title} {{First-Order
  General-Relativistic Viscous Fluid Dynamics}},\ }\href
  {https://doi.org/10.1103/PhysRevX.12.021044} {\bibfield  {journal} {\bibinfo
  {journal} {Phys. Rev. X}\ }\textbf {\bibinfo {volume} {12}},\ \bibinfo
  {pages} {021044} (\bibinfo {year} {2022})},\ \Eprint
  {https://arxiv.org/abs/2009.11388} {arXiv:2009.11388 [gr-qc]} \BibitemShut
  {NoStop}%
\bibitem [{\citenamefont {Armas}\ and\ \citenamefont
  {Camilloni}(2022)}]{Armas:2022wvb}%
  \BibitemOpen
  \bibfield  {author} {\bibinfo {author} {\bibfnamefont {J.}~\bibnamefont
  {Armas}}\ and\ \bibinfo {author} {\bibfnamefont {F.}~\bibnamefont
  {Camilloni}},\ }\bibfield  {title} {\bibinfo {title} {{A stable and causal
  model of magnetohydrodynamics}},\ }\href
  {https://doi.org/10.1088/1475-7516/2022/10/039} {\bibfield  {journal}
  {\bibinfo  {journal} {JCAP}\ }\textbf {\bibinfo {volume} {2022}}\bibfield
  {number} {\bibinfo  {number} { (10)},\ \bibinfo {pages} {039}},\ }\Eprint
  {https://arxiv.org/abs/2201.06847} {arXiv:2201.06847 [hep-th]} \BibitemShut
  {NoStop}%
\bibitem [{\citenamefont {Lier}\ \emph {et~al.}(2025)\citenamefont {Lier},
  \citenamefont {Jain}, \citenamefont {Armas},\ and\ \citenamefont
  {Porth}}]{Lier:2025wfw}%
  \BibitemOpen
  \bibfield  {author} {\bibinfo {author} {\bibfnamefont {R.}~\bibnamefont
  {Lier}}, \bibinfo {author} {\bibfnamefont {A.}~\bibnamefont {Jain}}, \bibinfo
  {author} {\bibfnamefont {J.}~\bibnamefont {Armas}},\ and\ \bibinfo {author}
  {\bibfnamefont {O.}~\bibnamefont {Porth}},\ }\bibfield  {title} {\bibinfo
  {title} {{Resistive relativistic magnetohydrodynamics without
  Amp{\`e}re{\textquoteright}s law}},\ }\href
  {https://doi.org/10.1103/stvc-tzfh} {\bibfield  {journal} {\bibinfo
  {journal} {Phys. Rev. D}\ }\textbf {\bibinfo {volume} {112}},\ \bibinfo
  {pages} {083059} (\bibinfo {year} {2025})},\ \Eprint
  {https://arxiv.org/abs/2501.04638} {arXiv:2501.04638 [astro-ph.HE]}
  \BibitemShut {NoStop}%
\bibitem [{\citenamefont {Figueras}\ \emph {et~al.}(2024)\citenamefont
  {Figueras}, \citenamefont {Held},\ and\ \citenamefont
  {Kov{\'a}cs}}]{Figueras:2024bba}%
  \BibitemOpen
  \bibfield  {author} {\bibinfo {author} {\bibfnamefont {P.}~\bibnamefont
  {Figueras}}, \bibinfo {author} {\bibfnamefont {A.}~\bibnamefont {Held}},\
  and\ \bibinfo {author} {\bibfnamefont {{\'A}.~D.}\ \bibnamefont
  {Kov{\'a}cs}},\ }\href@noop {} {\bibinfo {title} {{Well-posed initial value
  formulation of general effective field theories of gravity}}} (\bibinfo
  {year} {2024}),\ \Eprint {https://arxiv.org/abs/2407.08775} {arXiv:2407.08775
  [gr-qc]} \BibitemShut {NoStop}%
\bibitem [{\citenamefont {Bemfica}\ \emph
  {et~al.}(2019{\natexlab{a}})\citenamefont {Bemfica}, \citenamefont
  {Disconzi},\ and\ \citenamefont {Noronha}}]{Bemfica:2019cop}%
  \BibitemOpen
  \bibfield  {author} {\bibinfo {author} {\bibfnamefont {F.~S.}\ \bibnamefont
  {Bemfica}}, \bibinfo {author} {\bibfnamefont {M.~M.}\ \bibnamefont
  {Disconzi}},\ and\ \bibinfo {author} {\bibfnamefont {J.}~\bibnamefont
  {Noronha}},\ }\bibfield  {title} {\bibinfo {title} {{Causality of the
  Einstein-Israel-Stewart Theory with Bulk Viscosity}},\ }\href
  {https://doi.org/10.1103/PhysRevLett.122.221602} {\bibfield  {journal}
  {\bibinfo  {journal} {Phys. Rev. Lett.}\ }\textbf {\bibinfo {volume} {122}},\
  \bibinfo {pages} {221602} (\bibinfo {year} {2019}{\natexlab{a}})},\ \Eprint
  {https://arxiv.org/abs/1901.06701} {arXiv:1901.06701 [gr-qc]} \BibitemShut
  {NoStop}%
\bibitem [{\citenamefont {Bemfica}\ \emph
  {et~al.}(2019{\natexlab{b}})\citenamefont {Bemfica}, \citenamefont
  {Disconzi},\ and\ \citenamefont {Noronha}}]{Bemfica:2019knx}%
  \BibitemOpen
  \bibfield  {author} {\bibinfo {author} {\bibfnamefont {F.~S.}\ \bibnamefont
  {Bemfica}}, \bibinfo {author} {\bibfnamefont {M.~M.}\ \bibnamefont
  {Disconzi}},\ and\ \bibinfo {author} {\bibfnamefont {J.}~\bibnamefont
  {Noronha}},\ }\bibfield  {title} {\bibinfo {title} {{Nonlinear Causality of
  General First-Order Relativistic Viscous Hydrodynamics}},\ }\href
  {https://doi.org/10.1103/PhysRevD.100.104020} {\bibfield  {journal} {\bibinfo
   {journal} {Phys. Rev. D}\ }\textbf {\bibinfo {volume} {100}},\ \bibinfo
  {pages} {104020} (\bibinfo {year} {2019}{\natexlab{b}})},\ \Eprint
  {https://arxiv.org/abs/1907.12695} {arXiv:1907.12695 [gr-qc]} \BibitemShut
  {NoStop}%
\bibitem [{\citenamefont {Cordeiro}\ \emph {et~al.}(2024)\citenamefont
  {Cordeiro}, \citenamefont {Speranza}, \citenamefont {Ingles}, \citenamefont
  {Bemfica},\ and\ \citenamefont {Noronha}}]{Cordeiro:2023ljz}%
  \BibitemOpen
  \bibfield  {author} {\bibinfo {author} {\bibfnamefont {I.}~\bibnamefont
  {Cordeiro}}, \bibinfo {author} {\bibfnamefont {E.}~\bibnamefont {Speranza}},
  \bibinfo {author} {\bibfnamefont {K.}~\bibnamefont {Ingles}}, \bibinfo
  {author} {\bibfnamefont {F.~S.}\ \bibnamefont {Bemfica}},\ and\ \bibinfo
  {author} {\bibfnamefont {J.}~\bibnamefont {Noronha}},\ }\bibfield  {title}
  {\bibinfo {title} {{Causality Bounds on Dissipative General-Relativistic
  Magnetohydrodynamics}},\ }\href
  {https://doi.org/10.1103/PhysRevLett.133.091401} {\bibfield  {journal}
  {\bibinfo  {journal} {Phys. Rev. Lett.}\ }\textbf {\bibinfo {volume} {133}},\
  \bibinfo {pages} {091401} (\bibinfo {year} {2024})},\ \Eprint
  {https://arxiv.org/abs/2312.09970} {arXiv:2312.09970 [astro-ph.HE]}
  \BibitemShut {NoStop}%
\bibitem [{\citenamefont {Cordeiro}\ \emph
  {et~al.}(2026{\natexlab{a}})\citenamefont {Cordeiro}, \citenamefont
  {Speranza}, \citenamefont {Bemfica}, \citenamefont {Disconzi},\ and\
  \citenamefont {Noronha}}]{Cordeiro:2026}%
  \BibitemOpen
  \bibfield  {author} {\bibinfo {author} {\bibfnamefont {I.}~\bibnamefont
  {Cordeiro}}, \bibinfo {author} {\bibfnamefont {E.}~\bibnamefont {Speranza}},
  \bibinfo {author} {\bibfnamefont {F.~S.}\ \bibnamefont {Bemfica}}, \bibinfo
  {author} {\bibfnamefont {M.~M.}\ \bibnamefont {Disconzi}},\ and\ \bibinfo
  {author} {\bibfnamefont {J.}~\bibnamefont {Noronha}},\ }\href@noop {}
  {\bibinfo {title} {{Nonlinear Causality and Strong Hyperbolicity of
  Einstein-Israel-Stewart Theories of Transient Relativistic Fluid Dynamics}}}
  (\bibinfo {year} {2026}{\natexlab{a}}),\ \Eprint
  {https://arxiv.org/abs/2607.05639} {arXiv:2607.05639 [nucl-th]} \BibitemShut
  {NoStop}%
\bibitem [{\citenamefont {Plumberg}\ \emph {et~al.}(2022)\citenamefont
  {Plumberg}, \citenamefont {Almaalol}, \citenamefont {Dore}, \citenamefont
  {Noronha},\ and\ \citenamefont {Noronha-Hostler}}]{Plumberg:2021bme}%
  \BibitemOpen
  \bibfield  {author} {\bibinfo {author} {\bibfnamefont {C.}~\bibnamefont
  {Plumberg}}, \bibinfo {author} {\bibfnamefont {D.}~\bibnamefont {Almaalol}},
  \bibinfo {author} {\bibfnamefont {T.}~\bibnamefont {Dore}}, \bibinfo {author}
  {\bibfnamefont {J.}~\bibnamefont {Noronha}},\ and\ \bibinfo {author}
  {\bibfnamefont {J.}~\bibnamefont {Noronha-Hostler}},\ }\bibfield  {title}
  {\bibinfo {title} {{Causality violations in realistic simulations of
  heavy-ion collisions}},\ }\href
  {https://doi.org/10.1103/PhysRevC.105.L061901} {\bibfield  {journal}
  {\bibinfo  {journal} {Phys. Rev. C}\ }\textbf {\bibinfo {volume} {105}},\
  \bibinfo {pages} {L061901} (\bibinfo {year} {2022})},\ \Eprint
  {https://arxiv.org/abs/2103.15889} {arXiv:2103.15889 [nucl-th]} \BibitemShut
  {NoStop}%
\bibitem [{\citenamefont {Gavassino}\ \emph {et~al.}(2026)\citenamefont
  {Gavassino}, \citenamefont {Hirvonen}, \citenamefont {Paquet}, \citenamefont
  {Singh},\ and\ \citenamefont {Soares~Rocha}}]{Gavassino:2025mcq}%
  \BibitemOpen
  \bibfield  {author} {\bibinfo {author} {\bibfnamefont {L.}~\bibnamefont
  {Gavassino}}, \bibinfo {author} {\bibfnamefont {H.}~\bibnamefont {Hirvonen}},
  \bibinfo {author} {\bibfnamefont {J.-F.}\ \bibnamefont {Paquet}}, \bibinfo
  {author} {\bibfnamefont {M.}~\bibnamefont {Singh}},\ and\ \bibinfo {author}
  {\bibfnamefont {G.}~\bibnamefont {Soares~Rocha}},\ }\bibfield  {title}
  {\bibinfo {title} {{What happens in hydrodynamic simulations of heavy-ion
  collisions when causality is violated?}},\ }\href
  {https://doi.org/10.1051/epjconf/202636404003} {\bibfield  {journal}
  {\bibinfo  {journal} {EPJ Web Conf.}\ }\textbf {\bibinfo {volume} {364}},\
  \bibinfo {pages} {04003} (\bibinfo {year} {2026})},\ \Eprint
  {https://arxiv.org/abs/2511.07946} {arXiv:2511.07946 [nucl-th]} \BibitemShut
  {NoStop}%
\bibitem [{\citenamefont {Schoepe}\ \emph {et~al.}(2018)\citenamefont
  {Schoepe}, \citenamefont {Hilditch},\ and\ \citenamefont
  {Bugner}}]{Schoepe:2018}%
  \BibitemOpen
  \bibfield  {author} {\bibinfo {author} {\bibfnamefont {A.}~\bibnamefont
  {Schoepe}}, \bibinfo {author} {\bibfnamefont {D.}~\bibnamefont {Hilditch}},\
  and\ \bibinfo {author} {\bibfnamefont {M.}~\bibnamefont {Bugner}},\
  }\bibfield  {title} {\bibinfo {title} {{Revisiting Hyperbolicity of
  Relativistic Fluids}},\ }\href {https://doi.org/10.1103/PhysRevD.97.123009}
  {\bibfield  {journal} {\bibinfo  {journal} {Phys. Rev. D}\ }\textbf {\bibinfo
  {volume} {97}},\ \bibinfo {pages} {123009} (\bibinfo {year} {2018})},\
  \Eprint {https://arxiv.org/abs/1712.09837} {arXiv:1712.09837 [gr-qc]}
  \BibitemShut {NoStop}%
\bibitem [{\citenamefont {Dumbser}\ and\ \citenamefont
  {Zanotti}(2009)}]{Dumbser:2009hw}%
  \BibitemOpen
  \bibfield  {author} {\bibinfo {author} {\bibfnamefont {M.}~\bibnamefont
  {Dumbser}}\ and\ \bibinfo {author} {\bibfnamefont {O.}~\bibnamefont
  {Zanotti}},\ }\bibfield  {title} {\bibinfo {title} {{Very High Order P(N)
  P(M) Schemes on Unstructured Meshes for the Resistive Relativistic MHD
  Equations}},\ }\href {https://doi.org/10.1016/j.jcp.2009.06.009} {\bibfield
  {journal} {\bibinfo  {journal} {J. Comput. Phys.}\ }\textbf {\bibinfo
  {volume} {228}},\ \bibinfo {pages} {6991} (\bibinfo {year} {2009})},\ \Eprint
  {https://arxiv.org/abs/0903.4832} {arXiv:0903.4832 [gr-qc]} \BibitemShut
  {NoStop}%
\bibitem [{\citenamefont {Ressler}\ \emph {et~al.}(2015)\citenamefont
  {Ressler}, \citenamefont {Tchekhovskoy}, \citenamefont {Quataert},
  \citenamefont {Chandra},\ and\ \citenamefont {Gammie}}]{Ressler:2015ipa}%
  \BibitemOpen
  \bibfield  {author} {\bibinfo {author} {\bibfnamefont {S.~M.}\ \bibnamefont
  {Ressler}}, \bibinfo {author} {\bibfnamefont {A.}~\bibnamefont
  {Tchekhovskoy}}, \bibinfo {author} {\bibfnamefont {E.}~\bibnamefont
  {Quataert}}, \bibinfo {author} {\bibfnamefont {M.}~\bibnamefont {Chandra}},\
  and\ \bibinfo {author} {\bibfnamefont {C.~F.}\ \bibnamefont {Gammie}},\
  }\bibfield  {title} {\bibinfo {title} {{Electron Thermodynamics in GRMHD
  Simulations of Low-Luminosity Black Hole Accretion}},\ }\href
  {https://doi.org/10.1093/mnras/stv2084} {\bibfield  {journal} {\bibinfo
  {journal} {Mon. Not. Roy. Astron. Soc.}\ }\textbf {\bibinfo {volume} {454}},\
  \bibinfo {pages} {1848} (\bibinfo {year} {2015})},\ \Eprint
  {https://arxiv.org/abs/1509.04717} {arXiv:1509.04717 [astro-ph.HE]}
  \BibitemShut {NoStop}%
\bibitem [{\citenamefont {S{\k{a}}dowski}\ \emph {et~al.}(2017)\citenamefont
  {S{\k{a}}dowski}, \citenamefont {Wielgus}, \citenamefont {Narayan},
  \citenamefont {Abarca}, \citenamefont {McKinney},\ and\ \citenamefont
  {Chael}}]{Sadowski:2016fdh}%
  \BibitemOpen
  \bibfield  {author} {\bibinfo {author} {\bibfnamefont {A.}~\bibnamefont
  {S{\k{a}}dowski}}, \bibinfo {author} {\bibfnamefont {M.}~\bibnamefont
  {Wielgus}}, \bibinfo {author} {\bibfnamefont {R.}~\bibnamefont {Narayan}},
  \bibinfo {author} {\bibfnamefont {D.}~\bibnamefont {Abarca}}, \bibinfo
  {author} {\bibfnamefont {J.~C.}\ \bibnamefont {McKinney}},\ and\ \bibinfo
  {author} {\bibfnamefont {A.}~\bibnamefont {Chael}},\ }\bibfield  {title}
  {\bibinfo {title} {{Radiative, two-temperature simulations of low luminosity
  black hole accretion flows in general relativity}},\ }\href
  {https://doi.org/10.1093/mnras/stw3116} {\bibfield  {journal} {\bibinfo
  {journal} {Mon. Not. Roy. Astron. Soc.}\ }\textbf {\bibinfo {volume} {466}},\
  \bibinfo {pages} {705} (\bibinfo {year} {2017})},\ \Eprint
  {https://arxiv.org/abs/1605.03184} {arXiv:1605.03184 [astro-ph.HE]}
  \BibitemShut {NoStop}%
\bibitem [{\citenamefont {Chael}\ \emph {et~al.}(2019)\citenamefont {Chael},
  \citenamefont {Narayan},\ and\ \citenamefont {Johnson}}]{Chael:2018gzl}%
  \BibitemOpen
  \bibfield  {author} {\bibinfo {author} {\bibfnamefont {A.}~\bibnamefont
  {Chael}}, \bibinfo {author} {\bibfnamefont {R.}~\bibnamefont {Narayan}},\
  and\ \bibinfo {author} {\bibfnamefont {M.~D.}\ \bibnamefont {Johnson}},\
  }\bibfield  {title} {\bibinfo {title} {{Two-temperature, Magnetically
  Arrested Disc simulations of the jet from the supermassive black hole in
  M87}},\ }\href {https://doi.org/10.1093/mnras/stz988} {\bibfield  {journal}
  {\bibinfo  {journal} {Mon. Not. Roy. Astron. Soc.}\ }\textbf {\bibinfo
  {volume} {486}},\ \bibinfo {pages} {2873} (\bibinfo {year} {2019})},\ \Eprint
  {https://arxiv.org/abs/1810.01983} {arXiv:1810.01983 [astro-ph.HE]}
  \BibitemShut {NoStop}%
\bibitem [{\citenamefont {Most}\ and\ \citenamefont
  {Noronha}(2021)}]{Most:2021rhr}%
  \BibitemOpen
  \bibfield  {author} {\bibinfo {author} {\bibfnamefont {E.~R.}\ \bibnamefont
  {Most}}\ and\ \bibinfo {author} {\bibfnamefont {J.}~\bibnamefont {Noronha}},\
  }\bibfield  {title} {\bibinfo {title} {{Dissipative magnetohydrodynamics for
  nonresistive relativistic plasmas: An implicit second-order flux-conservative
  formulation with stiff relaxation}},\ }\href
  {https://doi.org/10.1103/PhysRevD.104.103028} {\bibfield  {journal} {\bibinfo
   {journal} {Phys. Rev. D}\ }\textbf {\bibinfo {volume} {104}},\ \bibinfo
  {pages} {103028} (\bibinfo {year} {2021})},\ \Eprint
  {https://arxiv.org/abs/2109.02796} {arXiv:2109.02796 [astro-ph.HE]}
  \BibitemShut {NoStop}%
\bibitem [{\citenamefont {Arnowitt}\ \emph {et~al.}(2008)\citenamefont
  {Arnowitt}, \citenamefont {Deser},\ and\ \citenamefont
  {Misner}}]{Arnowitt:1962hi}%
  \BibitemOpen
  \bibfield  {author} {\bibinfo {author} {\bibfnamefont {R.~L.}\ \bibnamefont
  {Arnowitt}}, \bibinfo {author} {\bibfnamefont {S.}~\bibnamefont {Deser}},\
  and\ \bibinfo {author} {\bibfnamefont {C.~W.}\ \bibnamefont {Misner}},\
  }\bibfield  {title} {\bibinfo {title} {{The Dynamics of general
  relativity}},\ }\href {https://doi.org/10.1007/s10714-008-0661-1} {\bibfield
  {journal} {\bibinfo  {journal} {Gen. Rel. Grav.}\ }\textbf {\bibinfo {volume}
  {40}},\ \bibinfo {pages} {1997} (\bibinfo {year} {2008})},\ \Eprint
  {https://arxiv.org/abs/gr-qc/0405109} {arXiv:gr-qc/0405109} \BibitemShut
  {NoStop}%
\bibitem [{\citenamefont {Baumgarte}\ and\ \citenamefont
  {Shapiro}(2003)}]{Baumgarte:2002vv}%
  \BibitemOpen
  \bibfield  {author} {\bibinfo {author} {\bibfnamefont {T.~W.}\ \bibnamefont
  {Baumgarte}}\ and\ \bibinfo {author} {\bibfnamefont {S.~L.}\ \bibnamefont
  {Shapiro}},\ }\bibfield  {title} {\bibinfo {title} {{General - relativistic
  MHD for the numerical construction of dynamical space - times}},\ }\href
  {https://doi.org/10.1086/346103} {\bibfield  {journal} {\bibinfo  {journal}
  {Astrophys. J.}\ }\textbf {\bibinfo {volume} {585}},\ \bibinfo {pages} {921}
  (\bibinfo {year} {2003})},\ \Eprint {https://arxiv.org/abs/astro-ph/0211340}
  {arXiv:astro-ph/0211340} \BibitemShut {NoStop}%
\bibitem [{\citenamefont {Dedner}\ \emph {et~al.}(2002)\citenamefont {Dedner},
  \citenamefont {Kemm}, \citenamefont {Kr{\"o}ner}, \citenamefont {Munz},
  \citenamefont {Schnitzer},\ and\ \citenamefont
  {Wesenberg}}]{dedner2002hyperbolic}%
  \BibitemOpen
  \bibfield  {author} {\bibinfo {author} {\bibfnamefont {A.}~\bibnamefont
  {Dedner}}, \bibinfo {author} {\bibfnamefont {F.}~\bibnamefont {Kemm}},
  \bibinfo {author} {\bibfnamefont {D.}~\bibnamefont {Kr{\"o}ner}}, \bibinfo
  {author} {\bibfnamefont {C.-D.}\ \bibnamefont {Munz}}, \bibinfo {author}
  {\bibfnamefont {T.}~\bibnamefont {Schnitzer}},\ and\ \bibinfo {author}
  {\bibfnamefont {M.}~\bibnamefont {Wesenberg}},\ }\bibfield  {title} {\bibinfo
  {title} {Hyperbolic divergence cleaning for the mhd equations},\ }\href@noop
  {} {\bibfield  {journal} {\bibinfo  {journal} {Journal of Computational
  Physics}\ }\textbf {\bibinfo {volume} {175}},\ \bibinfo {pages} {645}
  (\bibinfo {year} {2002})}\BibitemShut {NoStop}%
\bibitem [{\citenamefont {Komissarov}(2007)}]{Komissarov:2007wk}%
  \BibitemOpen
  \bibfield  {author} {\bibinfo {author} {\bibfnamefont {S.~S.}\ \bibnamefont
  {Komissarov}},\ }\bibfield  {title} {\bibinfo {title} {{Multi-dimensional
  Numerical Scheme for Resistive Relativistic MHD}},\ }\href
  {https://doi.org/10.1111/j.1365-2966.2007.12448.x} {\bibfield  {journal}
  {\bibinfo  {journal} {Mon. Not. Roy. Astron. Soc.}\ }\textbf {\bibinfo
  {volume} {382}},\ \bibinfo {pages} {995} (\bibinfo {year} {2007})},\ \Eprint
  {https://arxiv.org/abs/0708.0323} {arXiv:0708.0323 [astro-ph]} \BibitemShut
  {NoStop}%
\bibitem [{\citenamefont {Bucciantini}\ and\ \citenamefont
  {Del~Zanna}(2013)}]{Bucciantini:2012sm}%
  \BibitemOpen
  \bibfield  {author} {\bibinfo {author} {\bibfnamefont {N.}~\bibnamefont
  {Bucciantini}}\ and\ \bibinfo {author} {\bibfnamefont {L.}~\bibnamefont
  {Del~Zanna}},\ }\bibfield  {title} {\bibinfo {title} {{A fully covariant
  mean-field dynamo closure for numerical 3+1 resistive GRMHD}},\ }\href
  {https://doi.org/10.1093/mnras/sts005} {\bibfield  {journal} {\bibinfo
  {journal} {Mon. Not. Roy. Astron. Soc.}\ }\textbf {\bibinfo {volume} {428}},\
  \bibinfo {pages} {71} (\bibinfo {year} {2013})},\ \Eprint
  {https://arxiv.org/abs/1205.2951} {arXiv:1205.2951 [astro-ph.HE]}
  \BibitemShut {NoStop}%
\bibitem [{\citenamefont {Lier}\ \emph {et~al.}(2026)\citenamefont {Lier},
  \citenamefont {Armas},\ and\ \citenamefont {Porth}}]{Lier:2026owd}%
  \BibitemOpen
  \bibfield  {author} {\bibinfo {author} {\bibfnamefont {R.}~\bibnamefont
  {Lier}}, \bibinfo {author} {\bibfnamefont {J.}~\bibnamefont {Armas}},\ and\
  \bibinfo {author} {\bibfnamefont {O.}~\bibnamefont {Porth}},\ }\href@noop {}
  {\bibinfo {title} {{Finite-volume scheme for first-order viscoresistive
  relativistic magnetohydrodynamics}}} (\bibinfo {year} {2026}),\ \Eprint
  {https://arxiv.org/abs/2606.22691} {arXiv:2606.22691 [astro-ph.HE]}
  \BibitemShut {NoStop}%
\bibitem [{\citenamefont {Noronha}\ \emph {et~al.}(2022)\citenamefont
  {Noronha}, \citenamefont {Spali{\'n}ski},\ and\ \citenamefont
  {Speranza}}]{Noronha:2021syv}%
  \BibitemOpen
  \bibfield  {author} {\bibinfo {author} {\bibfnamefont {J.}~\bibnamefont
  {Noronha}}, \bibinfo {author} {\bibfnamefont {M.}~\bibnamefont
  {Spali{\'n}ski}},\ and\ \bibinfo {author} {\bibfnamefont {E.}~\bibnamefont
  {Speranza}},\ }\bibfield  {title} {\bibinfo {title} {{Transient Relativistic
  Fluid Dynamics in a General Hydrodynamic Frame}},\ }\href
  {https://doi.org/10.1103/PhysRevLett.128.252302} {\bibfield  {journal}
  {\bibinfo  {journal} {Phys. Rev. Lett.}\ }\textbf {\bibinfo {volume} {128}},\
  \bibinfo {pages} {252302} (\bibinfo {year} {2022})},\ \Eprint
  {https://arxiv.org/abs/2105.01034} {arXiv:2105.01034 [nucl-th]} \BibitemShut
  {NoStop}%
\bibitem [{\citenamefont {Cordeiro}\ \emph
  {et~al.}(2026{\natexlab{b}})\citenamefont {Cordeiro}, \citenamefont
  {Bemfica}, \citenamefont {Speranza},\ and\ \citenamefont
  {Noronha}}]{Cordeiro:2025diffusion}%
  \BibitemOpen
  \bibfield  {author} {\bibinfo {author} {\bibfnamefont {I.}~\bibnamefont
  {Cordeiro}}, \bibinfo {author} {\bibfnamefont {F.~S.}\ \bibnamefont
  {Bemfica}}, \bibinfo {author} {\bibfnamefont {E.}~\bibnamefont {Speranza}},\
  and\ \bibinfo {author} {\bibfnamefont {J.}~\bibnamefont {Noronha}},\
  }\bibfield  {title} {\bibinfo {title} {{Nonlinear causality of Israel-Stewart
  theory with diffusion}},\ }\href {https://doi.org/10.1103/dkcq-6s5b}
  {\bibfield  {journal} {\bibinfo  {journal} {Phys. Rev. D}\ }\textbf {\bibinfo
  {volume} {113}},\ \bibinfo {pages} {045024} (\bibinfo {year}
  {2026}{\natexlab{b}})},\ \Eprint {https://arxiv.org/abs/2507.20064}
  {arXiv:2507.20064 [nucl-th]} \BibitemShut {NoStop}%
\bibitem [{\citenamefont {Marti}\ and\ \citenamefont
  {Mueller}(1999)}]{Marti:1999wi}%
  \BibitemOpen
  \bibfield  {author} {\bibinfo {author} {\bibfnamefont {J.~M.}\ \bibnamefont
  {Marti}}\ and\ \bibinfo {author} {\bibfnamefont {E.}~\bibnamefont
  {Mueller}},\ }\bibfield  {title} {\bibinfo {title} {{Numerical hydrodynamics
  in special relativity}},\ }\href {https://doi.org/10.12942/lrr-1999-3}
  {\bibfield  {journal} {\bibinfo  {journal} {Living Rev. Rel.}\ }\textbf
  {\bibinfo {volume} {2}},\ \bibinfo {pages} {3} (\bibinfo {year} {1999})},\
  \Eprint {https://arxiv.org/abs/astro-ph/9906333} {arXiv:astro-ph/9906333}
  \BibitemShut {NoStop}%
\bibitem [{\citenamefont {Font}(2008)}]{Font:2008fka}%
  \BibitemOpen
  \bibfield  {author} {\bibinfo {author} {\bibfnamefont {J.~A.}\ \bibnamefont
  {Font}},\ }\bibfield  {title} {\bibinfo {title} {{Numerical Hydrodynamics and
  Magnetohydrodynamics in General Relativity}},\ }\href
  {https://doi.org/10.12942/lrr-2008-7} {\bibfield  {journal} {\bibinfo
  {journal} {Living Rev. Rel.}\ }\textbf {\bibinfo {volume} {11}},\ \bibinfo
  {pages} {7} (\bibinfo {year} {2008})}\BibitemShut {NoStop}%
\bibitem [{\citenamefont {Alic}\ \emph {et~al.}(2012)\citenamefont {Alic},
  \citenamefont {Mosta}, \citenamefont {Rezzolla}, \citenamefont {Zanotti},\
  and\ \citenamefont {Jaramillo}}]{Alic:2012df}%
  \BibitemOpen
  \bibfield  {author} {\bibinfo {author} {\bibfnamefont {D.}~\bibnamefont
  {Alic}}, \bibinfo {author} {\bibfnamefont {P.}~\bibnamefont {Mosta}},
  \bibinfo {author} {\bibfnamefont {L.}~\bibnamefont {Rezzolla}}, \bibinfo
  {author} {\bibfnamefont {O.}~\bibnamefont {Zanotti}},\ and\ \bibinfo {author}
  {\bibfnamefont {J.~L.}\ \bibnamefont {Jaramillo}},\ }\bibfield  {title}
  {\bibinfo {title} {{Accurate Simulations of Binary Black-Hole Mergers in
  Force-Free Electrodynamics}},\ }\href
  {https://doi.org/10.1088/0004-637X/754/1/36} {\bibfield  {journal} {\bibinfo
  {journal} {Astrophys. J.}\ }\textbf {\bibinfo {volume} {754}},\ \bibinfo
  {pages} {36} (\bibinfo {year} {2012})},\ \Eprint
  {https://arxiv.org/abs/1204.2226} {arXiv:1204.2226 [gr-qc]} \BibitemShut
  {NoStop}%
\bibitem [{\citenamefont {Toro}(2013)}]{toro2013riemann}%
  \BibitemOpen
  \bibfield  {author} {\bibinfo {author} {\bibfnamefont {E.~F.}\ \bibnamefont
  {Toro}},\ }\href@noop {} {\emph {\bibinfo {title} {Riemann solvers and
  numerical methods for fluid dynamics: a practical introduction}}}\ (\bibinfo
  {publisher} {Springer Science \& Business Media},\ \bibinfo {year}
  {2013})\BibitemShut {NoStop}%
\bibitem [{\citenamefont {Del~Zanna}\ \emph {et~al.}(2007)\citenamefont
  {Del~Zanna}, \citenamefont {Zanotti}, \citenamefont {Bucciantini},\ and\
  \citenamefont {Londrillo}}]{DelZanna:2007pk}%
  \BibitemOpen
  \bibfield  {author} {\bibinfo {author} {\bibfnamefont {L.}~\bibnamefont
  {Del~Zanna}}, \bibinfo {author} {\bibfnamefont {O.}~\bibnamefont {Zanotti}},
  \bibinfo {author} {\bibfnamefont {N.}~\bibnamefont {Bucciantini}},\ and\
  \bibinfo {author} {\bibfnamefont {P.}~\bibnamefont {Londrillo}},\ }\bibfield
  {title} {\bibinfo {title} {{ECHO: an Eulerian Conservative High Order scheme
  for general relativistic magnetohydrodynamics and magnetodynamics}},\ }\href
  {https://doi.org/10.1051/0004-6361:20077093} {\bibfield  {journal} {\bibinfo
  {journal} {Astron. Astrophys.}\ }\textbf {\bibinfo {volume} {473}},\ \bibinfo
  {pages} {11} (\bibinfo {year} {2007})},\ \Eprint
  {https://arxiv.org/abs/0704.3206} {arXiv:0704.3206 [astro-ph]} \BibitemShut
  {NoStop}%
\bibitem [{\citenamefont {Most}\ \emph {et~al.}(2019)\citenamefont {Most},
  \citenamefont {Papenfort},\ and\ \citenamefont {Rezzolla}}]{Most:2019kfe}%
  \BibitemOpen
  \bibfield  {author} {\bibinfo {author} {\bibfnamefont {E.~R.}\ \bibnamefont
  {Most}}, \bibinfo {author} {\bibfnamefont {L.~J.}\ \bibnamefont
  {Papenfort}},\ and\ \bibinfo {author} {\bibfnamefont {L.}~\bibnamefont
  {Rezzolla}},\ }\bibfield  {title} {\bibinfo {title} {{Beyond second-order
  convergence in simulations of magnetized binary neutron stars with realistic
  microphysics}},\ }\href {https://doi.org/10.1093/mnras/stz2809} {\bibfield
  {journal} {\bibinfo  {journal} {Mon. Not. Roy. Astron. Soc.}\ }\textbf
  {\bibinfo {volume} {490}},\ \bibinfo {pages} {3588} (\bibinfo {year}
  {2019})},\ \Eprint {https://arxiv.org/abs/1907.10328} {arXiv:1907.10328
  [astro-ph.HE]} \BibitemShut {NoStop}%
\bibitem [{\citenamefont {Mattia}\ and\ \citenamefont
  {Mignone}(2021)}]{Mattia:2021bwh}%
  \BibitemOpen
  \bibfield  {author} {\bibinfo {author} {\bibfnamefont {G.}~\bibnamefont
  {Mattia}}\ and\ \bibinfo {author} {\bibfnamefont {A.}~\bibnamefont
  {Mignone}},\ }\bibfield  {title} {\bibinfo {title} {{A comparison of
  approximate non-linear Riemann solvers for Relativistic MHD}},\ }\href
  {https://doi.org/10.1093/mnras/stab3373} {\bibfield  {journal} {\bibinfo
  {journal} {Mon. Not. Roy. Astron. Soc.}\ }\textbf {\bibinfo {volume} {510}},\
  \bibinfo {pages} {481} (\bibinfo {year} {2021})},\ \Eprint
  {https://arxiv.org/abs/2111.09369} {arXiv:2111.09369 [astro-ph.HE]}
  \BibitemShut {NoStop}%
\bibitem [{\citenamefont {Harten}\ \emph {et~al.}(1983)\citenamefont {Harten},
  \citenamefont {Lax},\ and\ \citenamefont {Leer}}]{harten1983upstream}%
  \BibitemOpen
  \bibfield  {author} {\bibinfo {author} {\bibfnamefont {A.}~\bibnamefont
  {Harten}}, \bibinfo {author} {\bibfnamefont {P.~D.}\ \bibnamefont {Lax}},\
  and\ \bibinfo {author} {\bibfnamefont {B.~v.}\ \bibnamefont {Leer}},\
  }\bibfield  {title} {\bibinfo {title} {On upstream differencing and
  godunov-type schemes for hyperbolic conservation laws},\ }\href@noop {}
  {\bibfield  {journal} {\bibinfo  {journal} {SIAM review}\ }\textbf {\bibinfo
  {volume} {25}},\ \bibinfo {pages} {35} (\bibinfo {year} {1983})}\BibitemShut
  {NoStop}%
\bibitem [{\citenamefont {Einfeldt}(1988)}]{einfeldt1988godunov}%
  \BibitemOpen
  \bibfield  {author} {\bibinfo {author} {\bibfnamefont {B.}~\bibnamefont
  {Einfeldt}},\ }\bibfield  {title} {\bibinfo {title} {On godunov-type methods
  for gas dynamics},\ }\href@noop {} {\bibfield  {journal} {\bibinfo  {journal}
  {SIAM Journal on numerical analysis}\ }\textbf {\bibinfo {volume} {25}},\
  \bibinfo {pages} {294} (\bibinfo {year} {1988})}\BibitemShut {NoStop}%
\bibitem [{\citenamefont {{Hu}}\ \emph {et~al.}(2013)\citenamefont {{Hu}},
  \citenamefont {{Adams}},\ and\ \citenamefont {{Shu}}}]{2013JCoPh.242..169H}%
  \BibitemOpen
  \bibfield  {author} {\bibinfo {author} {\bibfnamefont {X.~Y.}\ \bibnamefont
  {{Hu}}}, \bibinfo {author} {\bibfnamefont {N.~A.}\ \bibnamefont {{Adams}}},\
  and\ \bibinfo {author} {\bibfnamefont {C.-W.}\ \bibnamefont {{Shu}}},\
  }\bibfield  {title} {\bibinfo {title} {{Positivity-preserving method for
  high-order conservative schemes solving compressible Euler equations}},\
  }\href {https://doi.org/10.1016/j.jcp.2013.01.024} {\bibfield  {journal}
  {\bibinfo  {journal} {Journal of Computational Physics}\ }\textbf {\bibinfo
  {volume} {242}},\ \bibinfo {pages} {169} (\bibinfo {year} {2013})},\ \Eprint
  {https://arxiv.org/abs/1203.1540} {arXiv:1203.1540 [physics.flu-dyn]}
  \BibitemShut {NoStop}%
\bibitem [{\citenamefont {Van~Leer}(1979)}]{van1979towards}%
  \BibitemOpen
  \bibfield  {author} {\bibinfo {author} {\bibfnamefont {B.}~\bibnamefont
  {Van~Leer}},\ }\bibfield  {title} {\bibinfo {title} {Towards the ultimate
  conservative difference scheme. v. a second-order sequel to godunov's
  method},\ }\href@noop {} {\bibfield  {journal} {\bibinfo  {journal} {Journal
  of computational Physics}\ }\textbf {\bibinfo {volume} {32}},\ \bibinfo
  {pages} {101} (\bibinfo {year} {1979})}\BibitemShut {NoStop}%
\bibitem [{\citenamefont {{Borges}}\ \emph {et~al.}(2008)\citenamefont
  {{Borges}}, \citenamefont {{Carmona}}, \citenamefont {{Costa}},\ and\
  \citenamefont {{Don}}}]{2008JCoPh.227.3191B}%
  \BibitemOpen
  \bibfield  {author} {\bibinfo {author} {\bibfnamefont {R.}~\bibnamefont
  {{Borges}}}, \bibinfo {author} {\bibfnamefont {M.}~\bibnamefont {{Carmona}}},
  \bibinfo {author} {\bibfnamefont {B.}~\bibnamefont {{Costa}}},\ and\ \bibinfo
  {author} {\bibfnamefont {W.~S.}\ \bibnamefont {{Don}}},\ }\bibfield  {title}
  {\bibinfo {title} {{An improved weighted essentially non-oscillatory scheme
  for hyperbolic conservation laws}},\ }\href
  {https://doi.org/10.1016/j.jcp.2007.11.038} {\bibfield  {journal} {\bibinfo
  {journal} {Journal of Computational Physics}\ }\textbf {\bibinfo {volume}
  {227}},\ \bibinfo {pages} {3191} (\bibinfo {year} {2008})}\BibitemShut
  {NoStop}%
\bibitem [{\citenamefont {Pons}\ \emph {et~al.}(1998)\citenamefont {Pons},
  \citenamefont {Font}, \citenamefont {Ibanez}, \citenamefont {Marti},\ and\
  \citenamefont {Miralles}}]{Pons:1998ae}%
  \BibitemOpen
  \bibfield  {author} {\bibinfo {author} {\bibfnamefont {J.~A.}\ \bibnamefont
  {Pons}}, \bibinfo {author} {\bibfnamefont {J.~A.}\ \bibnamefont {Font}},
  \bibinfo {author} {\bibfnamefont {J.~M.}\ \bibnamefont {Ibanez}}, \bibinfo
  {author} {\bibfnamefont {J.~M.}\ \bibnamefont {Marti}},\ and\ \bibinfo
  {author} {\bibfnamefont {J.~A.}\ \bibnamefont {Miralles}},\ }\bibfield
  {title} {\bibinfo {title} {{General relativistic hydrodynamics with special
  relativistic Riemann solvers}},\ }\href@noop {} {\bibfield  {journal}
  {\bibinfo  {journal} {Astron. Astrophys.}\ }\textbf {\bibinfo {volume}
  {339}},\ \bibinfo {pages} {638} (\bibinfo {year} {1998})},\ \Eprint
  {https://arxiv.org/abs/astro-ph/9807215} {arXiv:astro-ph/9807215}
  \BibitemShut {NoStop}%
\bibitem [{\citenamefont {{White}}\ \emph {et~al.}(2016)\citenamefont
  {{White}}, \citenamefont {{Stone}},\ and\ \citenamefont
  {{Gammie}}}]{White:2016Athena}%
  \BibitemOpen
  \bibfield  {author} {\bibinfo {author} {\bibfnamefont {C.~J.}\ \bibnamefont
  {{White}}}, \bibinfo {author} {\bibfnamefont {J.~M.}\ \bibnamefont
  {{Stone}}},\ and\ \bibinfo {author} {\bibfnamefont {C.~F.}\ \bibnamefont
  {{Gammie}}},\ }\bibfield  {title} {\bibinfo {title} {{An Extension of the
  Athena++ Code Framework for General Relativistic Magnetohydrodynamics Based
  on Advanced Riemann Solvers and Staggered-Mesh Constrained Transport}},\
  }\href {https://doi.org/10.3847/0067-0049/225/2/22} {\bibfield  {journal}
  {\bibinfo  {journal} {The Astrophysical Journal Supplement Series}\ }\textbf
  {\bibinfo {volume} {225}},\ \bibinfo {pages} {22} (\bibinfo {year} {2016})},\
  \Eprint {https://arxiv.org/abs/1511.00943} {arXiv:1511.00943 [astro-ph.HE]}
  \BibitemShut {NoStop}%
\bibitem [{\citenamefont {Pareschi}\ and\ \citenamefont
  {Russo}(2005)}]{pareschi2005implicit}%
  \BibitemOpen
  \bibfield  {author} {\bibinfo {author} {\bibfnamefont {L.}~\bibnamefont
  {Pareschi}}\ and\ \bibinfo {author} {\bibfnamefont {G.}~\bibnamefont
  {Russo}},\ }\bibfield  {title} {\bibinfo {title} {Implicit--explicit
  runge--kutta schemes and applications to hyperbolic systems with
  relaxation},\ }\href@noop {} {\bibfield  {journal} {\bibinfo  {journal}
  {Journal of Scientific computing}\ }\textbf {\bibinfo {volume} {25}},\
  \bibinfo {pages} {129} (\bibinfo {year} {2005})}\BibitemShut {NoStop}%
\bibitem [{\citenamefont {Mignone}\ and\ \citenamefont
  {McKinney}(2007)}]{Mignone:2007fw}%
  \BibitemOpen
  \bibfield  {author} {\bibinfo {author} {\bibfnamefont {A.}~\bibnamefont
  {Mignone}}\ and\ \bibinfo {author} {\bibfnamefont {J.~C.}\ \bibnamefont
  {McKinney}},\ }\bibfield  {title} {\bibinfo {title} {{Equation of State in
  Relativistic Magnetohydrodynamics: variable versus constant adiabatic
  index}},\ }\href {https://doi.org/10.1111/j.1365-2966.2007.11849.x}
  {\bibfield  {journal} {\bibinfo  {journal} {Mon. Not. Roy. Astron. Soc.}\
  }\textbf {\bibinfo {volume} {378}},\ \bibinfo {pages} {1118} (\bibinfo {year}
  {2007})},\ \Eprint {https://arxiv.org/abs/0704.1679} {arXiv:0704.1679
  [astro-ph]} \BibitemShut {NoStop}%
\bibitem [{\citenamefont {Palenzuela}\ \emph {et~al.}(2015)\citenamefont
  {Palenzuela}, \citenamefont {Liebling}, \citenamefont {Neilsen},
  \citenamefont {Lehner}, \citenamefont {Caballero}, \citenamefont {O'Connor},\
  and\ \citenamefont {Anderson}}]{Palenzuela:2015dqa}%
  \BibitemOpen
  \bibfield  {author} {\bibinfo {author} {\bibfnamefont {C.}~\bibnamefont
  {Palenzuela}}, \bibinfo {author} {\bibfnamefont {S.~L.}\ \bibnamefont
  {Liebling}}, \bibinfo {author} {\bibfnamefont {D.}~\bibnamefont {Neilsen}},
  \bibinfo {author} {\bibfnamefont {L.}~\bibnamefont {Lehner}}, \bibinfo
  {author} {\bibfnamefont {O.~L.}\ \bibnamefont {Caballero}}, \bibinfo {author}
  {\bibfnamefont {E.}~\bibnamefont {O'Connor}},\ and\ \bibinfo {author}
  {\bibfnamefont {M.}~\bibnamefont {Anderson}},\ }\bibfield  {title} {\bibinfo
  {title} {{Effects of the microphysical Equation of State in the mergers of
  magnetized Neutron Stars With Neutrino Cooling}},\ }\href
  {https://doi.org/10.1103/PhysRevD.92.044045} {\bibfield  {journal} {\bibinfo
  {journal} {Phys. Rev. D}\ }\textbf {\bibinfo {volume} {92}},\ \bibinfo
  {pages} {044045} (\bibinfo {year} {2015})},\ \Eprint
  {https://arxiv.org/abs/1505.01607} {arXiv:1505.01607 [gr-qc]} \BibitemShut
  {NoStop}%
\bibitem [{\citenamefont {Kastaun}\ \emph {et~al.}(2021)\citenamefont
  {Kastaun}, \citenamefont {Kalinani},\ and\ \citenamefont
  {Ciolfi}}]{Kastaun:2020uxr}%
  \BibitemOpen
  \bibfield  {author} {\bibinfo {author} {\bibfnamefont {W.}~\bibnamefont
  {Kastaun}}, \bibinfo {author} {\bibfnamefont {J.~V.}\ \bibnamefont
  {Kalinani}},\ and\ \bibinfo {author} {\bibfnamefont {R.}~\bibnamefont
  {Ciolfi}},\ }\bibfield  {title} {\bibinfo {title} {{Robust Recovery of
  Primitive Variables in Relativistic Ideal Magnetohydrodynamics}},\ }\href
  {https://doi.org/10.1103/PhysRevD.103.023018} {\bibfield  {journal} {\bibinfo
   {journal} {Phys. Rev. D}\ }\textbf {\bibinfo {volume} {103}},\ \bibinfo
  {pages} {023018} (\bibinfo {year} {2021})},\ \Eprint
  {https://arxiv.org/abs/2005.01821} {arXiv:2005.01821 [gr-qc]} \BibitemShut
  {NoStop}%
\bibitem [{\citenamefont {Noble}\ \emph {et~al.}(2006)\citenamefont {Noble},
  \citenamefont {Gammie}, \citenamefont {McKinney},\ and\ \citenamefont
  {Del~Zanna}}]{Noble:2005gf}%
  \BibitemOpen
  \bibfield  {author} {\bibinfo {author} {\bibfnamefont {S.~C.}\ \bibnamefont
  {Noble}}, \bibinfo {author} {\bibfnamefont {C.~F.}\ \bibnamefont {Gammie}},
  \bibinfo {author} {\bibfnamefont {J.~C.}\ \bibnamefont {McKinney}},\ and\
  \bibinfo {author} {\bibfnamefont {L.}~\bibnamefont {Del~Zanna}},\ }\bibfield
  {title} {\bibinfo {title} {{Primitive variable solvers for conservative
  general relativistic magnetohydrodynamics}},\ }\href
  {https://doi.org/10.1086/500349} {\bibfield  {journal} {\bibinfo  {journal}
  {Astrophys. J.}\ }\textbf {\bibinfo {volume} {641}},\ \bibinfo {pages} {626}
  (\bibinfo {year} {2006})},\ \Eprint {https://arxiv.org/abs/astro-ph/0512420}
  {arXiv:astro-ph/0512420} \BibitemShut {NoStop}%
\bibitem [{\citenamefont {Siegel}\ \emph {et~al.}(2018)\citenamefont {Siegel},
  \citenamefont {M\"osta}, \citenamefont {Desai},\ and\ \citenamefont
  {Wu}}]{Siegel:2017sav}%
  \BibitemOpen
  \bibfield  {author} {\bibinfo {author} {\bibfnamefont {D.~M.}\ \bibnamefont
  {Siegel}}, \bibinfo {author} {\bibfnamefont {P.}~\bibnamefont {M\"osta}},
  \bibinfo {author} {\bibfnamefont {D.}~\bibnamefont {Desai}},\ and\ \bibinfo
  {author} {\bibfnamefont {S.}~\bibnamefont {Wu}},\ }\bibfield  {title}
  {\bibinfo {title} {{Recovery schemes for primitive variables in
  general-relativistic magnetohydrodynamics}},\ }\href
  {https://doi.org/10.3847/1538-4357/aabcc5} {\bibfield  {journal} {\bibinfo
  {journal} {Astrophys. J.}\ }\textbf {\bibinfo {volume} {859}},\ \bibinfo
  {pages} {71} (\bibinfo {year} {2018})},\ \Eprint
  {https://arxiv.org/abs/1712.07538} {arXiv:1712.07538 [astro-ph.HE]}
  \BibitemShut {NoStop}%
\bibitem [{\citenamefont {Wu}(2017)}]{Wu:2017grhd}%
  \BibitemOpen
  \bibfield  {author} {\bibinfo {author} {\bibfnamefont {K.}~\bibnamefont
  {Wu}},\ }\bibfield  {title} {\bibinfo {title} {{Design of Provably
  Physical-Constraint-Preserving Methods for General Relativistic
  Hydrodynamics}},\ }\href {https://doi.org/10.1103/PhysRevD.95.103001}
  {\bibfield  {journal} {\bibinfo  {journal} {Phys. Rev. D}\ }\textbf {\bibinfo
  {volume} {95}},\ \bibinfo {pages} {103001} (\bibinfo {year} {2017})},\
  \Eprint {https://arxiv.org/abs/1610.06274} {arXiv:1610.06274 [math.NA]}
  \BibitemShut {NoStop}%
\bibitem [{\citenamefont {Wu}\ and\ \citenamefont
  {Tang}(2017)}]{WuTang:2017rmhd}%
  \BibitemOpen
  \bibfield  {author} {\bibinfo {author} {\bibfnamefont {K.}~\bibnamefont
  {Wu}}\ and\ \bibinfo {author} {\bibfnamefont {H.}~\bibnamefont {Tang}},\
  }\bibfield  {title} {\bibinfo {title} {{Admissible States and
  Physical-Constraints-Preserving Schemes for Relativistic Magnetohydrodynamic
  Equations}},\ }\href {https://doi.org/10.1142/S0218202517500348} {\bibfield
  {journal} {\bibinfo  {journal} {Math. Models Methods Appl. Sci.}\ }\textbf
  {\bibinfo {volume} {27}},\ \bibinfo {pages} {1871} (\bibinfo {year}
  {2017})},\ \Eprint {https://arxiv.org/abs/1603.06660} {arXiv:1603.06660
  [math.NA]} \BibitemShut {NoStop}%
\bibitem [{\citenamefont {Most}\ and\ \citenamefont
  {Philippov}(2020)}]{Most:2020ami}%
  \BibitemOpen
  \bibfield  {author} {\bibinfo {author} {\bibfnamefont {E.~R.}\ \bibnamefont
  {Most}}\ and\ \bibinfo {author} {\bibfnamefont {A.~A.}\ \bibnamefont
  {Philippov}},\ }\bibfield  {title} {\bibinfo {title} {{Electromagnetic
  precursors to gravitational wave events: Numerical simulations of flaring in
  pre-merger binary neutron star magnetospheres}},\ }\href
  {https://doi.org/10.3847/2041-8213/ab8196} {\bibfield  {journal} {\bibinfo
  {journal} {Astrophys. J. Lett.}\ }\textbf {\bibinfo {volume} {893}},\
  \bibinfo {pages} {L6} (\bibinfo {year} {2020})},\ \Eprint
  {https://arxiv.org/abs/2001.06037} {arXiv:2001.06037 [astro-ph.HE]}
  \BibitemShut {NoStop}%
\bibitem [{\citenamefont {Most}\ and\ \citenamefont
  {Philippov}(2022)}]{Most:2022ojl}%
  \BibitemOpen
  \bibfield  {author} {\bibinfo {author} {\bibfnamefont {E.~R.}\ \bibnamefont
  {Most}}\ and\ \bibinfo {author} {\bibfnamefont {A.~A.}\ \bibnamefont
  {Philippov}},\ }\bibfield  {title} {\bibinfo {title} {{Electromagnetic
  precursor flares from the late inspiral of neutron star binaries}},\ }\href
  {https://doi.org/10.1093/mnras/stac1909} {\bibfield  {journal} {\bibinfo
  {journal} {Mon. Not. Roy. Astron. Soc.}\ }\textbf {\bibinfo {volume} {515}},\
  \bibinfo {pages} {2710} (\bibinfo {year} {2022})},\ \Eprint
  {https://arxiv.org/abs/2205.09643} {arXiv:2205.09643 [astro-ph.HE]}
  \BibitemShut {NoStop}%
\bibitem [{\citenamefont {Most}\ \emph
  {et~al.}(2024{\natexlab{a}})\citenamefont {Most}, \citenamefont {Kim},
  \citenamefont {Chatziioannou},\ and\ \citenamefont {Legred}}]{Most:2024eig}%
  \BibitemOpen
  \bibfield  {author} {\bibinfo {author} {\bibfnamefont {E.~R.}\ \bibnamefont
  {Most}}, \bibinfo {author} {\bibfnamefont {Y.}~\bibnamefont {Kim}}, \bibinfo
  {author} {\bibfnamefont {K.}~\bibnamefont {Chatziioannou}},\ and\ \bibinfo
  {author} {\bibfnamefont {I.}~\bibnamefont {Legred}},\ }\bibfield  {title}
  {\bibinfo {title} {{Nonlinear Alfv{\'e}n-wave Dynamics and Premerger Emission
  from Crustal Oscillations in Neutron Star Mergers}},\ }\href
  {https://doi.org/10.3847/2041-8213/ad785c} {\bibfield  {journal} {\bibinfo
  {journal} {Astrophys. J. Lett.}\ }\textbf {\bibinfo {volume} {973}},\
  \bibinfo {pages} {L37} (\bibinfo {year} {2024}{\natexlab{a}})},\ \Eprint
  {https://arxiv.org/abs/2407.17026} {arXiv:2407.17026 [astro-ph.HE]}
  \BibitemShut {NoStop}%
\bibitem [{\citenamefont {Zhang}\ \emph {et~al.}(2019)\citenamefont {Zhang},
  \citenamefont {Almgren}, \citenamefont {Beckner}, \citenamefont {Bell},
  \citenamefont {Blaschke}, \citenamefont {Chan}, \citenamefont {Day},
  \citenamefont {Friesen}, \citenamefont {Gott}, \citenamefont {Graves},
  \citenamefont {Katz}, \citenamefont {Myers}, \citenamefont {Nguyen},
  \citenamefont {Nonaka}, \citenamefont {Rosso}, \citenamefont {Williams},\
  and\ \citenamefont {Zingale}}]{Zhang2019}%
  \BibitemOpen
  \bibfield  {author} {\bibinfo {author} {\bibfnamefont {W.}~\bibnamefont
  {Zhang}}, \bibinfo {author} {\bibfnamefont {A.}~\bibnamefont {Almgren}},
  \bibinfo {author} {\bibfnamefont {V.}~\bibnamefont {Beckner}}, \bibinfo
  {author} {\bibfnamefont {J.}~\bibnamefont {Bell}}, \bibinfo {author}
  {\bibfnamefont {J.}~\bibnamefont {Blaschke}}, \bibinfo {author}
  {\bibfnamefont {C.}~\bibnamefont {Chan}}, \bibinfo {author} {\bibfnamefont
  {M.}~\bibnamefont {Day}}, \bibinfo {author} {\bibfnamefont {B.}~\bibnamefont
  {Friesen}}, \bibinfo {author} {\bibfnamefont {K.}~\bibnamefont {Gott}},
  \bibinfo {author} {\bibfnamefont {D.}~\bibnamefont {Graves}}, \bibinfo
  {author} {\bibfnamefont {M.~P.}\ \bibnamefont {Katz}}, \bibinfo {author}
  {\bibfnamefont {A.}~\bibnamefont {Myers}}, \bibinfo {author} {\bibfnamefont
  {T.}~\bibnamefont {Nguyen}}, \bibinfo {author} {\bibfnamefont
  {A.}~\bibnamefont {Nonaka}}, \bibinfo {author} {\bibfnamefont
  {M.}~\bibnamefont {Rosso}}, \bibinfo {author} {\bibfnamefont
  {S.}~\bibnamefont {Williams}},\ and\ \bibinfo {author} {\bibfnamefont
  {M.}~\bibnamefont {Zingale}},\ }\href {https://doi.org/10.21105/joss.01370}
  {\bibinfo {title} {Amrex: a framework for block-structured adaptive mesh
  refinement}} (\bibinfo {year} {2019})\BibitemShut {NoStop}%
\bibitem [{\citenamefont {{Zhang}}\ \emph {et~al.}(2020)\citenamefont
  {{Zhang}}, \citenamefont {{Myers}}, \citenamefont {{Gott}}, \citenamefont
  {{Almgren}},\ and\ \citenamefont {{Bell}}}]{2020arXiv200912009Z}%
  \BibitemOpen
  \bibfield  {author} {\bibinfo {author} {\bibfnamefont {W.}~\bibnamefont
  {{Zhang}}}, \bibinfo {author} {\bibfnamefont {A.}~\bibnamefont {{Myers}}},
  \bibinfo {author} {\bibfnamefont {K.}~\bibnamefont {{Gott}}}, \bibinfo
  {author} {\bibfnamefont {A.}~\bibnamefont {{Almgren}}},\ and\ \bibinfo
  {author} {\bibfnamefont {J.}~\bibnamefont {{Bell}}},\ }\bibfield  {title}
  {\bibinfo {title} {{AMReX: Block-Structured Adaptive Mesh Refinement for
  Multiphysics Applications}},\ }\href
  {https://doi.org/10.48550/arXiv.2009.12009} {\bibfield  {journal} {\bibinfo
  {journal} {arXiv e-prints}\ ,\ \bibinfo {eid} {arXiv:2009.12009}} (\bibinfo
  {year} {2020})},\ \Eprint {https://arxiv.org/abs/2009.12009}
  {arXiv:2009.12009 [cs.MS]} \BibitemShut {NoStop}%
\bibitem [{\citenamefont {{Myers}}\ \emph {et~al.}(2024)\citenamefont
  {{Myers}}, \citenamefont {{Zhang}}, \citenamefont {{Almgren}}, \citenamefont
  {{Antoun}}, \citenamefont {{Bell}}, \citenamefont {{Huebl}},\ and\
  \citenamefont {{Sinn}}}]{2024arXiv240312179M}%
  \BibitemOpen
  \bibfield  {author} {\bibinfo {author} {\bibfnamefont {A.}~\bibnamefont
  {{Myers}}}, \bibinfo {author} {\bibfnamefont {W.}~\bibnamefont {{Zhang}}},
  \bibinfo {author} {\bibfnamefont {A.}~\bibnamefont {{Almgren}}}, \bibinfo
  {author} {\bibfnamefont {T.}~\bibnamefont {{Antoun}}}, \bibinfo {author}
  {\bibfnamefont {J.}~\bibnamefont {{Bell}}}, \bibinfo {author} {\bibfnamefont
  {A.}~\bibnamefont {{Huebl}}},\ and\ \bibinfo {author} {\bibfnamefont
  {A.}~\bibnamefont {{Sinn}}},\ }\bibfield  {title} {\bibinfo {title} {{AMReX
  and pyAMReX: Looking Beyond ECP}},\ }\href
  {https://doi.org/10.48550/arXiv.2403.12179} {\bibfield  {journal} {\bibinfo
  {journal} {arXiv e-prints}\ ,\ \bibinfo {eid} {arXiv:2403.12179}} (\bibinfo
  {year} {2024})},\ \Eprint {https://arxiv.org/abs/2403.12179}
  {arXiv:2403.12179 [cs.DC]} \BibitemShut {NoStop}%
\bibitem [{\citenamefont {Bondi}(1952)}]{Bondi:1952ni}%
  \BibitemOpen
  \bibfield  {author} {\bibinfo {author} {\bibfnamefont {H.}~\bibnamefont
  {Bondi}},\ }\bibfield  {title} {\bibinfo {title} {{On spherically symmetrical
  accretion}},\ }\href {https://doi.org/10.1093/mnras/112.2.195} {\bibfield
  {journal} {\bibinfo  {journal} {Mon. Not. Roy. Astron. Soc.}\ }\textbf
  {\bibinfo {volume} {112}},\ \bibinfo {pages} {195} (\bibinfo {year}
  {1952})}\BibitemShut {NoStop}%
\bibitem [{\citenamefont {Michel}(1972)}]{Michel:1972}%
  \BibitemOpen
  \bibfield  {author} {\bibinfo {author} {\bibfnamefont {F.~C.}\ \bibnamefont
  {Michel}},\ }\bibfield  {title} {\bibinfo {title} {Accretion of matter by
  condensed objects},\ }\href {https://doi.org/10.1007/BF00649949} {\bibfield
  {journal} {\bibinfo  {journal} {Astrophysics and Space Science}\ }\textbf
  {\bibinfo {volume} {15}},\ \bibinfo {pages} {153} (\bibinfo {year}
  {1972})}\BibitemShut {NoStop}%
\bibitem [{\citenamefont {Kulsrud}(2020)}]{kulsrud2020plasma}%
  \BibitemOpen
  \bibfield  {author} {\bibinfo {author} {\bibfnamefont {R.~M.}\ \bibnamefont
  {Kulsrud}},\ }\href@noop {} {\emph {\bibinfo {title} {Plasma physics for
  astrophysics}}}\ (\bibinfo  {publisher} {Princeton University Press},\
  \bibinfo {year} {2020})\BibitemShut {NoStop}%
\bibitem [{\citenamefont {{K{\"a}ppeli}}(2022)}]{2022LRCA....8....2K}%
  \BibitemOpen
  \bibfield  {author} {\bibinfo {author} {\bibfnamefont {R.}~\bibnamefont
  {{K{\"a}ppeli}}},\ }\bibfield  {title} {\bibinfo {title} {{Well-balanced
  methods for computational astrophysics}},\ }\href
  {https://doi.org/10.1007/s41115-022-00014-6} {\bibfield  {journal} {\bibinfo
  {journal} {Living Reviews in Computational Astrophysics}\ }\textbf {\bibinfo
  {volume} {8}},\ \bibinfo {eid} {2} (\bibinfo {year} {2022})}\BibitemShut
  {NoStop}%
\bibitem [{\citenamefont {Bransgrove}\ \emph {et~al.}(2021)\citenamefont
  {Bransgrove}, \citenamefont {Ripperda},\ and\ \citenamefont
  {Philippov}}]{Bransgrove:2021heo}%
  \BibitemOpen
  \bibfield  {author} {\bibinfo {author} {\bibfnamefont {A.}~\bibnamefont
  {Bransgrove}}, \bibinfo {author} {\bibfnamefont {B.}~\bibnamefont
  {Ripperda}},\ and\ \bibinfo {author} {\bibfnamefont {A.}~\bibnamefont
  {Philippov}},\ }\bibfield  {title} {\bibinfo {title} {{Magnetic Hair and
  Reconnection in Black Hole Magnetospheres}},\ }\href
  {https://doi.org/10.1103/PhysRevLett.127.055101} {\bibfield  {journal}
  {\bibinfo  {journal} {Phys. Rev. Lett.}\ }\textbf {\bibinfo {volume} {127}},\
  \bibinfo {pages} {055101} (\bibinfo {year} {2021})}\BibitemShut {NoStop}%
\bibitem [{\citenamefont {Wald}(1974)}]{Wald:1974}%
  \BibitemOpen
  \bibfield  {author} {\bibinfo {author} {\bibfnamefont {R.~M.}\ \bibnamefont
  {Wald}},\ }\bibfield  {title} {\bibinfo {title} {Black hole in a uniform
  magnetic field},\ }\href {https://doi.org/10.1103/PhysRevD.10.1680}
  {\bibfield  {journal} {\bibinfo  {journal} {Physical Review D}\ }\textbf
  {\bibinfo {volume} {10}},\ \bibinfo {pages} {1680} (\bibinfo {year}
  {1974})}\BibitemShut {NoStop}%
\bibitem [{\citenamefont {Komissarov}\ and\ \citenamefont
  {McKinney}(2007)}]{Komissarov:2007rc}%
  \BibitemOpen
  \bibfield  {author} {\bibinfo {author} {\bibfnamefont {S.~S.}\ \bibnamefont
  {Komissarov}}\ and\ \bibinfo {author} {\bibfnamefont {J.~C.}\ \bibnamefont
  {McKinney}},\ }\bibfield  {title} {\bibinfo {title} {{Meissner effect and
  Blandford-Znajek mechanism in conductive black hole magnetospheres}},\ }\href
  {https://doi.org/10.1111/j.1745-3933.2007.00301.x} {\bibfield  {journal}
  {\bibinfo  {journal} {Mon. Not. Roy. Astron. Soc.}\ }\textbf {\bibinfo
  {volume} {377}},\ \bibinfo {pages} {L49} (\bibinfo {year} {2007})},\ \Eprint
  {https://arxiv.org/abs/astro-ph/0702269} {arXiv:astro-ph/0702269}
  \BibitemShut {NoStop}%
\bibitem [{\citenamefont {Komissarov}(2004)}]{Komissarov:2004ms}%
  \BibitemOpen
  \bibfield  {author} {\bibinfo {author} {\bibfnamefont {S.~S.}\ \bibnamefont
  {Komissarov}},\ }\bibfield  {title} {\bibinfo {title} {{Electrodynamics of
  black hole magnetospheres}},\ }\href
  {https://doi.org/10.1111/j.1365-2966.2004.07446.x} {\bibfield  {journal}
  {\bibinfo  {journal} {Mon. Not. Roy. Astron. Soc.}\ }\textbf {\bibinfo
  {volume} {350}},\ \bibinfo {pages} {407} (\bibinfo {year} {2004})},\ \Eprint
  {https://arxiv.org/abs/astro-ph/0402403} {arXiv:astro-ph/0402403}
  \BibitemShut {NoStop}%
\bibitem [{\citenamefont {Palenzuela}\ \emph {et~al.}(2010)\citenamefont
  {Palenzuela}, \citenamefont {Garrett}, \citenamefont {Lehner},\ and\
  \citenamefont {Liebling}}]{Palenzuela:2010xn}%
  \BibitemOpen
  \bibfield  {author} {\bibinfo {author} {\bibfnamefont {C.}~\bibnamefont
  {Palenzuela}}, \bibinfo {author} {\bibfnamefont {T.}~\bibnamefont {Garrett}},
  \bibinfo {author} {\bibfnamefont {L.}~\bibnamefont {Lehner}},\ and\ \bibinfo
  {author} {\bibfnamefont {S.~L.}\ \bibnamefont {Liebling}},\ }\bibfield
  {title} {\bibinfo {title} {{Magnetospheres of Black Hole Systems in
  Force-Free Plasma}},\ }\href {https://doi.org/10.1103/PhysRevD.82.044045}
  {\bibfield  {journal} {\bibinfo  {journal} {Phys. Rev. D}\ }\textbf {\bibinfo
  {volume} {82}},\ \bibinfo {pages} {044045} (\bibinfo {year} {2010})},\
  \Eprint {https://arxiv.org/abs/1007.1198} {arXiv:1007.1198 [gr-qc]}
  \BibitemShut {NoStop}%
\bibitem [{\citenamefont {Nathanail}\ and\ \citenamefont
  {Contopoulos}(2014)}]{Nathanail:2014aua}%
  \BibitemOpen
  \bibfield  {author} {\bibinfo {author} {\bibfnamefont {A.}~\bibnamefont
  {Nathanail}}\ and\ \bibinfo {author} {\bibfnamefont {I.}~\bibnamefont
  {Contopoulos}},\ }\bibfield  {title} {\bibinfo {title} {{Black Hole
  Magnetospheres}},\ }\href {https://doi.org/10.1088/0004-637X/788/2/186}
  {\bibfield  {journal} {\bibinfo  {journal} {Astrophys. J.}\ }\textbf
  {\bibinfo {volume} {788}},\ \bibinfo {pages} {186} (\bibinfo {year}
  {2014})},\ \Eprint {https://arxiv.org/abs/1404.0549} {arXiv:1404.0549
  [astro-ph.HE]} \BibitemShut {NoStop}%
\bibitem [{\citenamefont {Mahlmann}\ \emph {et~al.}(2018)\citenamefont
  {Mahlmann}, \citenamefont {Cerd{\'a}-Dur{\'a}n},\ and\ \citenamefont
  {Aloy}}]{Mahlmann:2018ukr}%
  \BibitemOpen
  \bibfield  {author} {\bibinfo {author} {\bibfnamefont {J.~F.}\ \bibnamefont
  {Mahlmann}}, \bibinfo {author} {\bibfnamefont {P.}~\bibnamefont
  {Cerd{\'a}-Dur{\'a}n}},\ and\ \bibinfo {author} {\bibfnamefont {M.~A.}\
  \bibnamefont {Aloy}},\ }\bibfield  {title} {\bibinfo {title} {{Numerically
  solving the relativistic Grad-Shafranov equation in Kerr spacetimes:
  Numerical techniques}},\ }\href {https://doi.org/10.1093/mnras/sty858}
  {\bibfield  {journal} {\bibinfo  {journal} {Mon. Not. Roy. Astron. Soc.}\
  }\textbf {\bibinfo {volume} {477}},\ \bibinfo {pages} {3927} (\bibinfo {year}
  {2018})},\ \Eprint {https://arxiv.org/abs/1802.00815} {arXiv:1802.00815
  [astro-ph.HE]} \BibitemShut {NoStop}%
\bibitem [{\citenamefont {Kim}\ \emph {et~al.}(2024)\citenamefont {Kim},
  \citenamefont {Most}, \citenamefont {Throwe}, \citenamefont {Teukolsky},\
  and\ \citenamefont {Deppe}}]{Kim:2024mau}%
  \BibitemOpen
  \bibfield  {author} {\bibinfo {author} {\bibfnamefont {Y.}~\bibnamefont
  {Kim}}, \bibinfo {author} {\bibfnamefont {E.~R.}\ \bibnamefont {Most}},
  \bibinfo {author} {\bibfnamefont {W.}~\bibnamefont {Throwe}}, \bibinfo
  {author} {\bibfnamefont {S.~A.}\ \bibnamefont {Teukolsky}},\ and\ \bibinfo
  {author} {\bibfnamefont {N.}~\bibnamefont {Deppe}},\ }\bibfield  {title}
  {\bibinfo {title} {{General relativistic force-free electrodynamics with a
  discontinuous Galerkin-finite difference hybrid method}},\ }\href
  {https://doi.org/10.1103/PhysRevD.109.123019} {\bibfield  {journal} {\bibinfo
   {journal} {Phys. Rev. D}\ }\textbf {\bibinfo {volume} {109}},\ \bibinfo
  {pages} {123019} (\bibinfo {year} {2024})},\ \Eprint
  {https://arxiv.org/abs/2404.01531} {arXiv:2404.01531 [astro-ph.IM]}
  \BibitemShut {NoStop}%
\bibitem [{\citenamefont {Parfrey}\ \emph {et~al.}(2019)\citenamefont
  {Parfrey}, \citenamefont {Philippov},\ and\ \citenamefont
  {Cerutti}}]{Parfrey:2019}%
  \BibitemOpen
  \bibfield  {author} {\bibinfo {author} {\bibfnamefont {K.}~\bibnamefont
  {Parfrey}}, \bibinfo {author} {\bibfnamefont {A.}~\bibnamefont {Philippov}},\
  and\ \bibinfo {author} {\bibfnamefont {B.}~\bibnamefont {Cerutti}},\
  }\bibfield  {title} {\bibinfo {title} {First-principles plasma simulations of
  black-hole jet launching},\ }\href
  {https://doi.org/10.1103/PhysRevLett.122.035101} {\bibfield  {journal}
  {\bibinfo  {journal} {Physical Review Letters}\ }\textbf {\bibinfo {volume}
  {122}},\ \bibinfo {pages} {035101} (\bibinfo {year} {2019})},\ \Eprint
  {https://arxiv.org/abs/1810.03613} {arXiv:1810.03613 [astro-ph.HE]}
  \BibitemShut {NoStop}%
\bibitem [{\citenamefont {Tchekhovskoy}\ and\ \citenamefont
  {Spitkovsky}(2013)}]{Tchekhovskoy:2012hm}%
  \BibitemOpen
  \bibfield  {author} {\bibinfo {author} {\bibfnamefont {A.}~\bibnamefont
  {Tchekhovskoy}}\ and\ \bibinfo {author} {\bibfnamefont {A.}~\bibnamefont
  {Spitkovsky}},\ }\bibfield  {title} {\bibinfo {title} {{Time-Dependent 3D
  Magnetohydrodynamic Pulsar Magnetospheres: Oblique Rotators}},\ }\href
  {https://doi.org/10.1093/mnrasl/slt076} {\bibfield  {journal} {\bibinfo
  {journal} {Mon. Not. Roy. Astron. Soc.}\ }\textbf {\bibinfo {volume} {435}},\
  \bibinfo {pages} {1} (\bibinfo {year} {2013})},\ \Eprint
  {https://arxiv.org/abs/1211.2803} {arXiv:1211.2803 [astro-ph.HE]}
  \BibitemShut {NoStop}%
\bibitem [{\citenamefont {Most}\ \emph
  {et~al.}(2024{\natexlab{b}})\citenamefont {Most}, \citenamefont
  {Beloborodov},\ and\ \citenamefont {Ripperda}}]{Most:2024qgc}%
  \BibitemOpen
  \bibfield  {author} {\bibinfo {author} {\bibfnamefont {E.~R.}\ \bibnamefont
  {Most}}, \bibinfo {author} {\bibfnamefont {A.~M.}\ \bibnamefont
  {Beloborodov}},\ and\ \bibinfo {author} {\bibfnamefont {B.}~\bibnamefont
  {Ripperda}},\ }\bibfield  {title} {\bibinfo {title} {{Monster Shocks,
  Gamma-Ray Bursts, and Black Hole Quasi-normal Modes from Neutron-star
  Collapse}},\ }\href {https://doi.org/10.3847/2041-8213/ad7e1f} {\bibfield
  {journal} {\bibinfo  {journal} {Astrophys. J. Lett.}\ }\textbf {\bibinfo
  {volume} {974}},\ \bibinfo {pages} {L12} (\bibinfo {year}
  {2024}{\natexlab{b}})},\ \Eprint {https://arxiv.org/abs/2404.01456}
  {arXiv:2404.01456 [astro-ph.HE]} \BibitemShut {NoStop}%
\bibitem [{\citenamefont {Chatterjee}\ \emph {et~al.}(2026)\citenamefont
  {Chatterjee}, \citenamefont {Philippov}, \citenamefont {Beloborodov},
  \citenamefont {Parfrey}, \citenamefont {Ripperda},\ and\ \citenamefont
  {Most}}]{Chatterjee:2026tdp}%
  \BibitemOpen
  \bibfield  {author} {\bibinfo {author} {\bibfnamefont {K.}~\bibnamefont
  {Chatterjee}}, \bibinfo {author} {\bibfnamefont {A.}~\bibnamefont
  {Philippov}}, \bibinfo {author} {\bibfnamefont {A.~M.}\ \bibnamefont
  {Beloborodov}}, \bibinfo {author} {\bibfnamefont {K.}~\bibnamefont
  {Parfrey}}, \bibinfo {author} {\bibfnamefont {B.}~\bibnamefont {Ripperda}},\
  and\ \bibinfo {author} {\bibfnamefont {E.~R.}\ \bibnamefont {Most}},\
  }\bibfield  {title} {\bibinfo {title} {{Relativistic Magnetohydrodynamic
  Simulations of Giant Magnetar Bursts}},\ }\href
  {https://doi.org/10.3847/2041-8213/ae8997} {\bibfield  {journal} {\bibinfo
  {journal} {Astrophys. J. Lett.}\ }\textbf {\bibinfo {volume} {1006}},\
  \bibinfo {pages} {L45} (\bibinfo {year} {2026})},\ \Eprint
  {https://arxiv.org/abs/2602.17755} {arXiv:2602.17755 [astro-ph.HE]}
  \BibitemShut {NoStop}%
\bibitem [{\citenamefont {{Lynden-Bell}}(1996)}]{1996MNRAS.279..389L}%
  \BibitemOpen
  \bibfield  {author} {\bibinfo {author} {\bibfnamefont {D.}~\bibnamefont
  {{Lynden-Bell}}},\ }\bibfield  {title} {\bibinfo {title} {{Magnetic
  collimation by accretion discs of quasars and stars}},\ }\href
  {https://doi.org/10.1093/mnras/279.2.389} {\bibfield  {journal} {\bibinfo
  {journal} {\mnras}\ }\textbf {\bibinfo {volume} {279}},\ \bibinfo {pages}
  {389} (\bibinfo {year} {1996})}\BibitemShut {NoStop}%
\bibitem [{\citenamefont {{Loureiro}}\ \emph {et~al.}(2007)\citenamefont
  {{Loureiro}}, \citenamefont {{Schekochihin}},\ and\ \citenamefont
  {{Cowley}}}]{2007PhPl...14j0703L}%
  \BibitemOpen
  \bibfield  {author} {\bibinfo {author} {\bibfnamefont {N.~F.}\ \bibnamefont
  {{Loureiro}}}, \bibinfo {author} {\bibfnamefont {A.~A.}\ \bibnamefont
  {{Schekochihin}}},\ and\ \bibinfo {author} {\bibfnamefont {S.~C.}\
  \bibnamefont {{Cowley}}},\ }\bibfield  {title} {\bibinfo {title}
  {{Instability of current sheets and formation of plasmoid chains}},\ }\href
  {https://doi.org/10.1063/1.2783986} {\bibfield  {journal} {\bibinfo
  {journal} {Physics of Plasmas}\ }\textbf {\bibinfo {volume} {14}},\ \bibinfo
  {pages} {100703} (\bibinfo {year} {2007})},\ \Eprint
  {https://arxiv.org/abs/astro-ph/0703631} {arXiv:astro-ph/0703631 [astro-ph]}
  \BibitemShut {NoStop}%
\bibitem [{\citenamefont {Samtaney}\ \emph {et~al.}(2009)\citenamefont
  {Samtaney}, \citenamefont {Loureiro}, \citenamefont {Uzdensky}, \citenamefont
  {Schekochihin},\ and\ \citenamefont {Cowley}}]{Samtaney:2009bq}%
  \BibitemOpen
  \bibfield  {author} {\bibinfo {author} {\bibfnamefont {R.}~\bibnamefont
  {Samtaney}}, \bibinfo {author} {\bibfnamefont {N.~F.}\ \bibnamefont
  {Loureiro}}, \bibinfo {author} {\bibfnamefont {D.~A.}\ \bibnamefont
  {Uzdensky}}, \bibinfo {author} {\bibfnamefont {A.~A.}\ \bibnamefont
  {Schekochihin}},\ and\ \bibinfo {author} {\bibfnamefont {S.~C.}\ \bibnamefont
  {Cowley}},\ }\bibfield  {title} {\bibinfo {title} {{Formation of Plasmoid
  Chains in Magnetic Reconnection}},\ }\href
  {https://doi.org/10.1103/PhysRevLett.103.105004} {\bibfield  {journal}
  {\bibinfo  {journal} {Phys. Rev. Lett.}\ }\textbf {\bibinfo {volume} {103}},\
  \bibinfo {pages} {105004} (\bibinfo {year} {2009})},\ \Eprint
  {https://arxiv.org/abs/0903.0542} {arXiv:0903.0542 [astro-ph.SR]}
  \BibitemShut {NoStop}%
\bibitem [{\citenamefont {{Bhattacharjee}}\ \emph {et~al.}(2009)\citenamefont
  {{Bhattacharjee}}, \citenamefont {{Huang}}, \citenamefont {{Yang}},\ and\
  \citenamefont {{Rogers}}}]{2009PhPl...16k2102B}%
  \BibitemOpen
  \bibfield  {author} {\bibinfo {author} {\bibfnamefont {A.}~\bibnamefont
  {{Bhattacharjee}}}, \bibinfo {author} {\bibfnamefont {Y.-M.}\ \bibnamefont
  {{Huang}}}, \bibinfo {author} {\bibfnamefont {H.}~\bibnamefont {{Yang}}},\
  and\ \bibinfo {author} {\bibfnamefont {B.}~\bibnamefont {{Rogers}}},\
  }\bibfield  {title} {\bibinfo {title} {{Fast reconnection in
  high-Lundquist-number plasmas due to the plasmoid Instability}},\ }\href
  {https://doi.org/10.1063/1.3264103} {\bibfield  {journal} {\bibinfo
  {journal} {Physics of Plasmas}\ }\textbf {\bibinfo {volume} {16}},\ \bibinfo
  {eid} {112102} (\bibinfo {year} {2009})},\ \Eprint
  {https://arxiv.org/abs/0906.5599} {arXiv:0906.5599 [physics.plasm-ph]}
  \BibitemShut {NoStop}%
\bibitem [{\citenamefont {Goldreich}\ and\ \citenamefont
  {Julian}(1969)}]{Goldreich:1969sb}%
  \BibitemOpen
  \bibfield  {author} {\bibinfo {author} {\bibfnamefont {P.}~\bibnamefont
  {Goldreich}}\ and\ \bibinfo {author} {\bibfnamefont {W.~H.}\ \bibnamefont
  {Julian}},\ }\bibfield  {title} {\bibinfo {title} {{Pulsar
  electrodynamics}},\ }\href {https://doi.org/10.1086/150119} {\bibfield
  {journal} {\bibinfo  {journal} {Astrophys. J.}\ }\textbf {\bibinfo {volume}
  {157}},\ \bibinfo {pages} {869} (\bibinfo {year} {1969})}\BibitemShut
  {NoStop}%
\bibitem [{\citenamefont {Comisso}\ and\ \citenamefont
  {Asenjo}(2021)}]{Comisso:2020ykg}%
  \BibitemOpen
  \bibfield  {author} {\bibinfo {author} {\bibfnamefont {L.}~\bibnamefont
  {Comisso}}\ and\ \bibinfo {author} {\bibfnamefont {F.~A.}\ \bibnamefont
  {Asenjo}},\ }\bibfield  {title} {\bibinfo {title} {{Magnetic Reconnection as
  a Mechanism for Energy Extraction from Rotating Black Holes}},\ }\href
  {https://doi.org/10.1103/PhysRevD.103.023014} {\bibfield  {journal} {\bibinfo
   {journal} {Phys. Rev. D}\ }\textbf {\bibinfo {volume} {103}},\ \bibinfo
  {pages} {023014} (\bibinfo {year} {2021})},\ \Eprint
  {https://arxiv.org/abs/2012.00879} {arXiv:2012.00879 [astro-ph.HE]}
  \BibitemShut {NoStop}%
\bibitem [{\citenamefont {Newman}\ \emph {et~al.}(1965)\citenamefont {Newman},
  \citenamefont {Couch}, \citenamefont {Chinnapared}, \citenamefont {Exton},
  \citenamefont {Prakash},\ and\ \citenamefont {Torrence}}]{Newman:1965my}%
  \BibitemOpen
  \bibfield  {author} {\bibinfo {author} {\bibfnamefont {E.~T.}\ \bibnamefont
  {Newman}}, \bibinfo {author} {\bibfnamefont {E.}~\bibnamefont {Couch}},
  \bibinfo {author} {\bibfnamefont {K.}~\bibnamefont {Chinnapared}}, \bibinfo
  {author} {\bibfnamefont {A.}~\bibnamefont {Exton}}, \bibinfo {author}
  {\bibfnamefont {A.}~\bibnamefont {Prakash}},\ and\ \bibinfo {author}
  {\bibfnamefont {R.}~\bibnamefont {Torrence}},\ }\bibfield  {title} {\bibinfo
  {title} {{Metric of a Rotating, Charged Mass}},\ }\href
  {https://doi.org/10.1063/1.1704351} {\bibfield  {journal} {\bibinfo
  {journal} {J. Math. Phys.}\ }\textbf {\bibinfo {volume} {6}},\ \bibinfo
  {pages} {918} (\bibinfo {year} {1965})}\BibitemShut {NoStop}%
\bibitem [{\citenamefont {{Lyutikov}}\ and\ \citenamefont
  {{McKinney}}(2011)}]{2011PhRvD..84h4019L}%
  \BibitemOpen
  \bibfield  {author} {\bibinfo {author} {\bibfnamefont {M.}~\bibnamefont
  {{Lyutikov}}}\ and\ \bibinfo {author} {\bibfnamefont {J.~C.}\ \bibnamefont
  {{McKinney}}},\ }\bibfield  {title} {\bibinfo {title} {{Slowly balding black
  holes}},\ }\href {https://doi.org/10.1103/PhysRevD.84.084019} {\bibfield
  {journal} {\bibinfo  {journal} {\prd}\ }\textbf {\bibinfo {volume} {84}},\
  \bibinfo {eid} {084019} (\bibinfo {year} {2011})},\ \Eprint
  {https://arxiv.org/abs/1109.0584} {arXiv:1109.0584 [astro-ph.HE]}
  \BibitemShut {NoStop}%
\bibitem [{\citenamefont {MacDonald}\ and\ \citenamefont
  {Thorne}(1982)}]{MacDonald:1982zz}%
  \BibitemOpen
  \bibfield  {author} {\bibinfo {author} {\bibfnamefont {D.}~\bibnamefont
  {MacDonald}}\ and\ \bibinfo {author} {\bibfnamefont {K.~S.}\ \bibnamefont
  {Thorne}},\ }\bibfield  {title} {\bibinfo {title} {{Black-hole
  electrodynamics - an absolute-space/universal-time formulation}},\
  }\href@noop {} {\bibfield  {journal} {\bibinfo  {journal} {Mon. Not. Roy.
  Astron. Soc.}\ }\textbf {\bibinfo {volume} {198}},\ \bibinfo {pages} {345}
  (\bibinfo {year} {1982})}\BibitemShut {NoStop}%
\bibitem [{\citenamefont {Selvi}\ \emph {et~al.}(2025)\citenamefont {Selvi},
  \citenamefont {Porth}, \citenamefont {Ripperda},\ and\ \citenamefont
  {Sironi}}]{Selvi:2025dur}%
  \BibitemOpen
  \bibfield  {author} {\bibinfo {author} {\bibfnamefont {S.}~\bibnamefont
  {Selvi}}, \bibinfo {author} {\bibfnamefont {O.}~\bibnamefont {Porth}},
  \bibinfo {author} {\bibfnamefont {B.}~\bibnamefont {Ripperda}},\ and\
  \bibinfo {author} {\bibfnamefont {L.}~\bibnamefont {Sironi}},\ }\bibfield
  {title} {\bibinfo {title} {{On the Universality of the Split Monopole Black
  Hole Magnetosphere}},\ }\href {https://doi.org/10.3847/2041-8213/adeb6c}
  {\bibfield  {journal} {\bibinfo  {journal} {Astrophys. J. Lett.}\ }\textbf
  {\bibinfo {volume} {988}},\ \bibinfo {pages} {L33} (\bibinfo {year}
  {2025})},\ \Eprint {https://arxiv.org/abs/2504.10367} {arXiv:2504.10367
  [astro-ph.HE]} \BibitemShut {NoStop}%
\bibitem [{\citenamefont {Selvi}\ \emph {et~al.}(2024)\citenamefont {Selvi},
  \citenamefont {Porth}, \citenamefont {Ripperda},\ and\ \citenamefont
  {Sironi}}]{Selvi:2024lsh}%
  \BibitemOpen
  \bibfield  {author} {\bibinfo {author} {\bibfnamefont {S.}~\bibnamefont
  {Selvi}}, \bibinfo {author} {\bibfnamefont {O.}~\bibnamefont {Porth}},
  \bibinfo {author} {\bibfnamefont {B.}~\bibnamefont {Ripperda}},\ and\
  \bibinfo {author} {\bibfnamefont {L.}~\bibnamefont {Sironi}},\ }\bibfield
  {title} {\bibinfo {title} {{Current Sheet Alignment in Oblique Black Hole
  Magnetospheres: A Black Hole Pulsar?}},\ }\href
  {https://doi.org/10.3847/2041-8213/ad4a5b} {\bibfield  {journal} {\bibinfo
  {journal} {Astrophys. J. Lett.}\ }\textbf {\bibinfo {volume} {968}},\
  \bibinfo {pages} {L10} (\bibinfo {year} {2024})},\ \Eprint
  {https://arxiv.org/abs/2402.16055} {arXiv:2402.16055 [astro-ph.HE]}
  \BibitemShut {NoStop}%
\bibitem [{\citenamefont {Kim}\ \emph {et~al.}(2025)\citenamefont {Kim},
  \citenamefont {Most}, \citenamefont {Beloborodov},\ and\ \citenamefont
  {Ripperda}}]{Kim:2024fuy}%
  \BibitemOpen
  \bibfield  {author} {\bibinfo {author} {\bibfnamefont {Y.}~\bibnamefont
  {Kim}}, \bibinfo {author} {\bibfnamefont {E.~R.}\ \bibnamefont {Most}},
  \bibinfo {author} {\bibfnamefont {A.~M.}\ \bibnamefont {Beloborodov}},\ and\
  \bibinfo {author} {\bibfnamefont {B.}~\bibnamefont {Ripperda}},\ }\bibfield
  {title} {\bibinfo {title} {{Black Hole Pulsars and Monster Shocks as Outcomes
  of Black Hole{\textendash}Neutron Star Mergers}},\ }\href
  {https://doi.org/10.3847/2041-8213/adbff9} {\bibfield  {journal} {\bibinfo
  {journal} {Astrophys. J. Lett.}\ }\textbf {\bibinfo {volume} {982}},\
  \bibinfo {pages} {L54} (\bibinfo {year} {2025})},\ \Eprint
  {https://arxiv.org/abs/2412.05760} {arXiv:2412.05760 [astro-ph.HE]}
  \BibitemShut {NoStop}%
\bibitem [{\citenamefont {Meringolo}\ \emph {et~al.}(2025)\citenamefont
  {Meringolo}, \citenamefont {Camilloni},\ and\ \citenamefont
  {Rezzolla}}]{Meringolo:2025bdu}%
  \BibitemOpen
  \bibfield  {author} {\bibinfo {author} {\bibfnamefont {C.}~\bibnamefont
  {Meringolo}}, \bibinfo {author} {\bibfnamefont {F.}~\bibnamefont
  {Camilloni}},\ and\ \bibinfo {author} {\bibfnamefont {L.}~\bibnamefont
  {Rezzolla}},\ }\bibfield  {title} {\bibinfo {title} {{Electromagnetic Energy
  Extraction from Kerr Black Holes: Ab Initio Calculations}},\ }\href
  {https://doi.org/10.3847/2041-8213/ae06a6} {\bibfield  {journal} {\bibinfo
  {journal} {Astrophys. J. Lett.}\ }\textbf {\bibinfo {volume} {992}},\
  \bibinfo {pages} {L8} (\bibinfo {year} {2025})},\ \Eprint
  {https://arxiv.org/abs/2507.08942} {arXiv:2507.08942 [gr-qc]} \BibitemShut
  {NoStop}%
\bibitem [{\citenamefont {Sironi}\ \emph {et~al.}(2021)\citenamefont {Sironi},
  \citenamefont {Rowan},\ and\ \citenamefont {Narayan}}]{Sironi:2020mzd}%
  \BibitemOpen
  \bibfield  {author} {\bibinfo {author} {\bibfnamefont {L.}~\bibnamefont
  {Sironi}}, \bibinfo {author} {\bibfnamefont {M.~E.}\ \bibnamefont {Rowan}},\
  and\ \bibinfo {author} {\bibfnamefont {R.}~\bibnamefont {Narayan}},\
  }\bibfield  {title} {\bibinfo {title} {{Reconnection-driven particle
  acceleration in relativistic shear flows}},\ }\href
  {https://doi.org/10.3847/2041-8213/abd9bc} {\bibfield  {journal} {\bibinfo
  {journal} {Astrophys. J. Lett.}\ }\textbf {\bibinfo {volume} {907}},\
  \bibinfo {pages} {L44} (\bibinfo {year} {2021})},\ \Eprint
  {https://arxiv.org/abs/2009.11877} {arXiv:2009.11877 [astro-ph.HE]}
  \BibitemShut {NoStop}%
\bibitem [{\citenamefont {Teukolsky}(2026)}]{Teukolsky:2026}%
  \BibitemOpen
  \bibfield  {author} {\bibinfo {author} {\bibfnamefont {S.~A.}\ \bibnamefont
  {Teukolsky}},\ }\bibfield  {title} {\bibinfo {title} {{Characteristic
  Decomposition for Relativistic Numerical Simulations: II.
  Magnetohydrodynamics}},\ }\href {https://doi.org/10.1103/g9cn-8kmj}
  {\bibfield  {journal} {\bibinfo  {journal} {Phys. Rev. D}\ }\textbf {\bibinfo
  {volume} {113}},\ \bibinfo {pages} {124016} (\bibinfo {year} {2026})},\
  \Eprint {https://arxiv.org/abs/2511.13837} {arXiv:2511.13837 [gr-qc]}
  \BibitemShut {NoStop}%
\bibitem [{\citenamefont {Press}(2007)}]{press2007numerical}%
  \BibitemOpen
  \bibfield  {author} {\bibinfo {author} {\bibfnamefont {W.~H.}\ \bibnamefont
  {Press}},\ }\href@noop {} {\emph {\bibinfo {title} {Numerical recipes 3rd
  edition: The art of scientific computing}}}\ (\bibinfo  {publisher}
  {Cambridge university press},\ \bibinfo {year} {2007})\BibitemShut {NoStop}%
\bibitem [{\citenamefont {Montero}\ \emph {et~al.}(2014)\citenamefont
  {Montero}, \citenamefont {Baumgarte},\ and\ \citenamefont
  {M{\"u}ller}}]{Montero:2013pca}%
  \BibitemOpen
  \bibfield  {author} {\bibinfo {author} {\bibfnamefont {P.~J.}\ \bibnamefont
  {Montero}}, \bibinfo {author} {\bibfnamefont {T.~W.}\ \bibnamefont
  {Baumgarte}},\ and\ \bibinfo {author} {\bibfnamefont {E.}~\bibnamefont
  {M{\"u}ller}},\ }\bibfield  {title} {\bibinfo {title} {{General relativistic
  hydrodynamics in curvilinear coordinates}},\ }\href
  {https://doi.org/10.1103/PhysRevD.89.084043} {\bibfield  {journal} {\bibinfo
  {journal} {Phys. Rev. D}\ }\textbf {\bibinfo {volume} {89}},\ \bibinfo
  {pages} {084043} (\bibinfo {year} {2014})},\ \Eprint
  {https://arxiv.org/abs/1309.7808} {arXiv:1309.7808 [gr-qc]} \BibitemShut
  {NoStop}%
\end{thebibliography}%
\label{lastpage}
\end{document}